\documentclass[oneside,letterpaper,oldfontcommands]{memoir}

\usepackage{citthesis}

\usepackage[T1]{fontenc}
\usepackage[]{lmodern}
\usepackage[sc]{mathpazo}

\usepackage{stmaryrd}

\usepackage[colorlinks,linkcolor=blue,filecolor=blue,citecolor=blue,urlcolor=blue,backref=page]{hyperref}

\usepackage{cite}
\usepackage{amsmath,amssymb,amsfonts}
\usepackage{algorithmic}
\usepackage{graphicx}
\usepackage{textcomp}
\usepackage{xcolor}

\usepackage{multirow}
\usepackage{tabularx}
\usepackage{gensymb} 
\usepackage{diagbox}
\usepackage{multirow}
\usepackage{amssymb}
\usepackage{pifont}

\usepackage{mdwlist}
\usepackage[ruled,vlined,shortend, linesnumbered]{algorithm2e} 
\usepackage[]{algorithmic} 
\usepackage{booktabs}
\usepackage{color}

\usepackage{amsthm,amssymb}

\usepackage[]{caption}
\usepackage[labelformat=simple]{subcaption}

\usepackage{multirow}
\usepackage{mathrsfs}
\usepackage{amsbsy}
\usepackage[mathscr]{eucal} 
\usepackage[flushleft]{threeparttable}
\usepackage{comment}
\usepackage{paralist}
\usepackage{pifont}

\usepackage{bbding} 

\usepackage[colorinlistoftodos]{todonotes}

\usepackage{mdframed}  
\theoremstyle{definition}
\newtheorem{definition}{Definition}[section]
\newtheorem{problemdefinition}{Problem Definition}[section]

\newenvironment{problem}
  {   
    \begin{mdframed}[  
      linewidth=1pt,       
      linecolor=black,     
      roundcorner=5pt,     
      leftmargin=0cm,      
      rightmargin=0cm,     
      innerleftmargin=0.5cm,  
      innerrightmargin=0.5cm, 
      innertopmargin=0.5cm,   
      innerbottommargin=0.5cm,
      backgroundcolor=yellow!20  
    ]
    \begin{problemdefinition}  
  }
  {   
    \end{problemdefinition}   
    \end{mdframed}            
  }

\newtheorem{modeldefinition}{Model Definition}[]
\newenvironment{model}
  {   
    \begin{mdframed}[  
      linewidth=1pt,       
      linecolor=black,     
      roundcorner=5pt,     
      leftmargin=0cm,      
      rightmargin=0cm,     
      innerleftmargin=0.5cm,  
      innerrightmargin=0.5cm, 
      innertopmargin=0.5cm,   
      innerbottommargin=0.5cm,
      backgroundcolor=yellow!20  
    ]
    \begin{modeldefinition}  
  }
  {   
    \end{modeldefinition}   
    \end{mdframed}            
  }

\newtheorem{derivedef}{Derivation}[]
\newenvironment{derivation}
  {   
    \begin{mdframed}[  
      linewidth=1pt,       
      linecolor=black,     
      roundcorner=5pt,     
      leftmargin=0cm,      
      rightmargin=0cm,     
      innerleftmargin=0.5cm,  
      innerrightmargin=0.5cm, 
      innertopmargin=0.5cm,   
      innerbottommargin=0.5cm,
      backgroundcolor=gray!10  
    ]
    \begin{derivedef}  
  }
  {   
    \end{derivedef}   
    \end{mdframed}            
  }

\newenvironment{mytheorem}
  {   
    \begin{mdframed}[  
      linewidth=1pt,       
      linecolor=black,     
      roundcorner=5pt,     
      leftmargin=0cm,      
      rightmargin=0cm,     
      innerleftmargin=0.5cm,  
      innerrightmargin=0.5cm, 
      innertopmargin=0.5cm,   
      innerbottommargin=0.5cm,
      backgroundcolor=yellow!20  
    ]
    \begin{theorem}  
  }
  {   
    \end{theorem}   
    \end{mdframed}            
  }

\usepackage{bm}
\usepackage{amsthm}
\usepackage{amsmath}
\usepackage{amssymb}
\usepackage{thmtools}
\usepackage{thm-restate}
\declaretheorem[name=Theorem,numberwithin=section]{theorem}
\DeclareMathOperator{\Exp}{\mathbb{E}}

\newtheorem{assumption}{Assumption}[section]
\newtheorem{knowledge}{Domain knowledge}[]

\usepackage{mathtools}

\newsubfloat{figure}

\begin{document}

\frontmatter

\thetitlepage
\copyrightpage

\section*{Acknowledgements}
To my dear advisor and the person I always dream of being,  Larry Pileggi, for guiding me through an incredibly fruitful academic journey at CMU. Without his mentorship, it would have been impossible to achieve both the depth and breadth of my research. Over the years, his remarkable high-level thinking to unify research problems that might otherwise seem unrelated has been a constant driving source through every stage of my work. I've also witnessed from him the magic power of encouragement which makes miracles happen. Whether facing paper rejections, missing out on awards, or feeling anxious during the job search, he consistently lit me up with his optimism, reassuring me that my research was strong and that there was always time to learn and grow. As I reflect on the past several years, I realized how much I've grown towards a better researcher, presenter, and communicator. The transformation is huge, yet the journey itself was joyful, thanks to the relaxed, open environment he fostered—one where I could grow up without pressure, but only boundless encouragement. Beyond research, Larry’s passion for poker brought a lot of “poker nights” where I enjoyed the group. As a source of direction to enjoy life, this has inspired me to pick up painting as a hobby. 

To my committee members  Vyas Sekar,  Soummya Kar,  Bryan Hooi and  Guannan Qu for their guidance towards achieving the goals within this thesis.

To my dear career mentor and collaborator,  Vyas Sekar, whose guidance has been a guiding light in my journey toward a future academic career. His encouragement made a transformative impact in my life plan, providing the confidence and energy to pursue a faculty position, which I once felt completely out of reach. With hands-on instructions and practical advice, he kindly gave me a roadmap which eventually made me succeed in job search. As a collaborator, he is a true master of the art in storytelling. The bright and clear research logistics learned from him will have a lifelong impact to drive me towards always bringing forth high-quality works, like what he has been doing.

To my dear collaborator,  Bryan Hooi, for mentoring my first PhD project and being the first guiding light in my research. His expertise has given me interdisciplinary insights at the intersection of power grids and AI. As a remarkable collaborator, he made me achieve the first publishable work early in the PhD journey, building the foundational knowledge and confidence for my research afterwards.

To my dear mentor,  Amritanshu Pandey, for being my best collaborator and the one whom I worked most closely with over the years. He is always a source of encouragement and inspiration, and a source of help in all aspects, shaping the style of my paper writing, presenting and communicating. His efforts made my research journey smooth and rewarding at every stage.

To my dear mentor,  Guannan Qu, for giving comprehensive support in every dimension of the job search. The research insights, application advice, interview tips, and practice opportunities for the job talk were essential in leading to a job offer.

To my dear collaborator,  Soummya Kar, for helping me refine research ideas for the job search and identify potential future directions that promise exciting extensions beyond the current work.

To many other warm professors at CMU — Ramteen Sioshansi,  Julia Fanti,  Valerie Karplus, and Granger Morgan — for their valuable feedback, warm advice, and numerous opportunities to practice and refine my job talk, which were essential for academic growth.

To other kind mentors, Jan Drgona and  Nawaf Nazir, for providing a fantastic internship experience at PNNL, which enabled the exploration of new topics beyond the thesis and fostered connections with more experts in the field.

To my dear colleagues, Amrit, Aayushya, Elizabeth, Marko and Tim, for creating a welcoming and supportive environment that made me a better presenter and communicator year after year. Being in the same group with them made the journey enjoyable and fulfilling.

To my dear friend, Shuo Zhou, for being an amazing friend and incredible companion filling every weekend with happy moments. She has been my ultimate “ally” in countless adventures across the virtual world of games. The “Honor of Kings” has witnessed more than 3,000 battles where we have fought together to defend our “glory”. Our shared milestones, both in the game and in life, have recorded thousands of victories against the fierce opponents (and, of course, some defeats).

Finally, to my wonderful family, Mom and Dad, who are the foundation of every achievement in my life. Their love and support made everything possible.

\newpage
\section*{Abstract}
The growing threats of variations, uncertainties, anomalies, and sophisticated cyberattacks on power grids are driving a critical need to advance situational awareness which allows system operators to form a complete and accurate picture of the present and future state of the system. Simulation and estimation are foundational tools in this process. However, existing tools lack the robustness and efficiency required to achieve the level of situational awareness needed for the ever-evolving threat landscape. Industry-standard (steady-state) simulators are not robust to blackouts, often leading to non-converging or non-actionable results in cases where blackout induces an “infeasible system”. Similarly, estimation tools lack robustness to anomalous data, returning erroneous system states and failing to accurately locate root causes of real-time anomalies. Efficiency is the other major concern as nonlinearities and scalability issues make large systems slow to converge. Data-driven alternatives to these physics-based tools have been widely developed using machine learning (ML) for fast speed; but even state-of-the-art models find it hard to ensure the satisfaction of necessary network constraints on large systems. Together, these limitations make system operators struggle in responding effectively to threat incidents.

This thesis addresses these robustness and efficiency gaps through a dual-fold contribution. We first address the inherent limitations in the existing physics-based and data-driven worlds; and then transcend the boundaries of conventional algorithmic design in the direction of a new paradigm -- Physics-ML Synergy -- which draws on and integrates the strengths of the two worlds. Our simulation and convex estimation approaches are built on circuit formulation which provides a unified framework that applies to both transmission and distribution. Sparse optimization acts as the key enabler to make these tools intrinsically robust and immune to random threats. By enforcing sparse “threat indicators”, our methods produce actionable outcomes to pinpoint dominant sources of (random) blackouts and data errors, meanwhile suggesting easy corrections for threat mitigation. Further, we explore sparsity-exploiting optimizations to develop lightweight ML models whose prediction and detection capabilities are a complement to physics-based tools; and whose lightweight designs advance generalization and scalability. Finally, Physics-ML Synergy brings robustness and efficiency further against targeted cyberthreats, by interconnecting our physics-based tools with lightweight ML. We explore synergy designs where ML-generated predictions feed prior knowledge into simulators and estimators, in the form of warm starting points and prior distributions.  

We validate these significant advancements on large-scale power system simulation and estimation tasks, spanning a variety of threats. As a result, our robust simulators and estimators pinpoint dominant blackout sources on the 80k-bus Eastern Interconnection grid; and identify a mixture of random bad data and topology errors on >25k-bus systems, with significantly faster speed than standard solvers. Our lightweight ML generalizes well to dynamic graphs and achieves time-series anomaly detection at the millisecond scale for 60K-node systems (approximately 75\% of the Eastern Interconnection) per time tick per sensor.  Finally, our Physics-ML Synergy further advanced simulation efficiency under MadIoT cyberattacks with >3x faster speed on a 2000-bus power grid; and advanced estimation robustness against false data injection attacks while scaling almost linearly.

\newpage
\tableofcontents 

\newpage
\listoftables
\listoffigures

\mainmatter

\newcommand{\hide}[1]{}
\newcommand{\atn}[1]{\textcolor{red}{#1}}
\newcommand{\notice}[1]{{\textsf{\textcolor{green}{{\em [#1]}}}}}
\newcommand{\reminder}[1]{{\textsf{\textcolor{blue}{[#1]}}}}
\newcommand{\vectornorm}[1]{\left|\left|#1\right|\right|}

\newcommand{\mkclean}{
    \renewcommand{\reminder}{\hide}
}

\newcommand{\bit}{\begin{compactitem}}
\newcommand{\eit}{\end{compactitem}}
\newcommand{\ben}{\begin{compactenum}}
\newcommand{\een}{\end{compactenum}}
\newtheorem{observation}{Observation}
\newtheorem{iproblem}{Informal Problem}
\newcommand{\method}{\textsc{DynWatch}\xspace}
\newcommand{\methodls}{\textsc{DynWatch-Local}\xspace}
\newcommand{\codeurl}{https://github.com/bhooi/dynamic.git}
\newcommand{\mycomm}[1]{\textcolor{blue}{$\vartriangleright$#1}}
\newcommand{\G}{\mathcal{G}}
\renewcommand{\S}{\mathcal{S}}
\newcommand{\T}{\mathcal{T}}
\newcommand{\V}{\mathcal{V}}
\newcommand{\E}{\mathcal{E}}
\newcommand{\N}{\mathcal{N}}
\newcommand{\datasmal}{\textsc{case2383}\xspace}
\newcommand{\datasmall}{\textsc{case2869}\xspace}
\newcommand{\datalarge}{\textsc{case9241}\xspace}
\newcommand{\dataxlarge}{\textsc{ACTIVSg25k}\xspace}

\chapter{Introduction and Motivation}
Emerging sources of disruption and instability necessitate fundamental research to re-envision situational awareness \cite{SA_overview} algorithms for safety-critical cyber-physical systems. The ability to perceive both current and future system states and respond to incidents is essential to safeguard the security, reliability, and resilience of new distributed, digitally controlled infrastructures, as well as legacy systems increasingly retrofitted with smart capabilities. The electric power grid, which serves as the backbone of modern industries, particularly calls for new situational awareness paradigms to meet this need. Today, extreme weather events \cite{extreme-weather-resilience}\cite{extreme-weather-events2014}\cite{extreme-weather-outage} and targeted attacks \cite{lee2017crashoverride}\cite{ukrain2015}\cite{grid-cybersecurity-Vyas} have intensified the frequency and severity of blackouts, posing serious operational risks. Meanwhile, new cyberthreats specifically target data integrity \cite{fdia-acse}\cite{fdia-review}\cite{FDIA_DC}\cite{grid-cybersecurity-Vyas}, injecting anomalous data to complicate grid operators’ real-time data processing and decision-making. Additionally, transmission and distribution systems must now cope with rapid fluctuations in power flows from microgrids \cite{microgrid-integration-review}, distributed generation \cite{DER-challenges-review}, and energy trades \cite{energy-trading-risk}. Together, these evolving threats make it imperative to advance situational awareness, ensuring that the electric power grid can respond to threat incidents, prepare for amidst infrastructure and power electronics upgrades, as well as support market trading and renewable penetration to power an increasingly electrified economy in a sustainable, low carbon transition.

Simulation and estimation are fundamental tools for routine situational awareness. Estimation problem identifies system states and anomaly conditions in real time based on observed measurement data, while simulation models "what-if" scenarios to assess their impacts. These two core functionalities cover both steady-state analysis (typically in the frequency domain, such as power flow simulation \cite{traditionalPF} and steady-state estimation \cite{WLS-SE}) and transient analysis (typically in the time domain, such as transient stability analysis \cite{transient-stability-analysis-review}, electromechanical transient simulation \cite{gridEMT-review}, and dynamic state estimation \cite{dynamicSE-review}\cite{dynamicSE-role}). Among these analyses, steady-state simulation \cite{traditionalPF} and estimation \cite{WLS-SE} are primarily used for daily operations and planning, enabling operators to identify the steady-state operating point in real time or in response to specific contingency scenarios. This thesis focuses on steady-state situational awareness, which is the primary horizon to recognize most real-world cyber-physical anomalies. In certain cases, though, steady-state analysis is complemented by computationally intensive transient analysis to provide deeper insights into specific or local conditions.

However, today’s industry-standard simulation \cite{traditionalPF} and estimation \cite{WLS-SE} tools lack the robustness and efficiency required to achieve comprehensive situational awareness. As physics-based tools rely on fundamental physical laws, they are known for their accuracy but often struggle with efficiency, particularly in large systems where convergence is slow due to the nonlinear nature of power grid problems. Furthermore, a critical gap in robustness restricts these tools' ability to handle diverse threat conditions. For example, standard simulation tools \cite{traditionalPF} are not sufficiently robust to blackout scenarios, often resulting in divergence  or non-actionable outcomes \cite{powerflow-diverge} when the scenario mathematically represents an "infeasible system" whose solution does not exist. Similarly, standard estimation tools \cite{WLS-SE}\cite{traditional-BDI}\cite{TE-WLSE-BDI} are not robust to anomalous data, frequently producing erroneous system states and failing to accurately identify the root causes of anomalies.

Recent research has led to advanced methods aimed at reformulating the problems to advance robustness and efficiency of situational awareness tools, but significant limitations remain. To improve efficiency, convexified methods \cite{convexTESE-SDP-weng} made relaxation using techniques like semidefinite programming, which simplify the problem but still struggle with scalability on large systems. Linear estimators \cite{LAVSE-PMU-abur}\cite{LAVSE-PMU-abur2}\cite{abur-robustSE-PMU}, which naturally create convex problems, have also been developed, by assuming the grid is observable via modern data sources such as synchrophasors (PMUs). However, the assumption is currently unrealistic. In terms of robustness, several methods aim to address specific threat conditions. Advanced simulation techniques \cite{infeasibility-quantified-pf-trad}\cite{SUGAR-pf}\cite{sugar-powerflow-amrit} quantify infeasibility in blackout scenarios, allowing operators to converge to a solution that measures the severity of the blackout. However, this quantified infeasibility is often messily distributed across the network, providing limited insight into root causes. Robust estimators \cite{LAVSE-PMU-abur}\cite{LAVSE-PMU-abur2}\cite{abur-robustSE-PMU}\cite{LAVSE-dc-Abur-splitform} have also been developed to handle random bad data within continuous measurements. Robust “generalized” state estimators \cite{GSE-AC-nonlinear}\cite{GSE-DC-substation}\cite{TESE-GSE}\cite{TESE-GSE-PMUabur}\cite{convexTESE-SDP-weng} bring robustness further by integrating switches directly into the estimation model, enhancing robustness against topology errors that could otherwise distort system states. However, the Simplex method \cite{simplex} and semidefinite programming \cite{large-scale-SDP},  solvers in these robust estimators, are not scalable on large systems; and the robustness provided in these approaches are limited to random anomalous data, but not the false data injected in a targeted way \cite{fdia-acse}.  

As attempts to resolve gaps of efficiency, data-driven methods leveraging machine learning (ML) have been widely studied. The use of ML has been proven particularly effective in capturing abnormal behaviors  \cite{keogh2007finding}\cite{breunig2000lof}\cite{liu2008isolation}\cite{yi2017grouped}\cite{ramaswamy2000efficient}\cite{jones2014anomaly}\cite{akoglu2010oddball}\cite{chen2012community}\cite{mongiovi2013netspot}\cite{araujo2014com2}\cite{akoglu2010event}\cite{ranshous2016scalable}. Meanwhile, data-driven alternatives to physics-based tools are also actively explored \cite{ppf-dnn}\cite{pf-dnn-topo}\cite{gnn-pf}\cite{unrolled-se-2018}\cite{unrolled-se}\cite{unrolled-gnu-se}. These alternatives primarily aim at faster speed but comparable accuracy, unintentionally creating an adversarial relationship between the data-driven and physics-based worlds. Overall, many existing ML approaches are insufficient for the level of situational awareness we need for power grids, and they generally are not designed to deal with the robustness gaps we mentioned above. Moreover, key obstacles \cite{towards-practical-ML-Li}—including limited generalization to unseen conditions, scalability to large networks, and interpretability of results—are not fully resolved even for state-of-the-art ML models \cite{dc3}\cite{neuromancer}\cite{hamiltonianNN}\cite{lagrangianNN}\cite{hooi2018gridwatch}\cite{pmu-graph-spatial}\cite{pmu-graph-temporal}, raising significant concerns for applying ML in heavily-constrained safety-critical infrastructures like the real-world power grids. 

Clearly, the current paradigm for designing simulation and estimation algorithms has become an either-or choice: use either physics-based or data-driven methods, given the distinct advantages of each approach. However, based on their pros and cons as discussed, it is apparent that neither of these approaches alone provides the necessary robustness and efficiency to handle the complexities of real-world power grids. Physics-based methods are known for accuracy but can suffer from lower speed and difficult modeling of anomalies. Data-driven methods offer speed and better anomaly detectability, but are limited in accuracy and generalization. Alternatively, combining the complementary strengths of both approaches offers a promising path toward improved situational awareness. 

This thesis aims to fill the efficiency and robustness gaps by exploring the collaborative relationships between the two worlds. We not only address the inherent limitations within physics-based and data-driven categories, but also transcend the boundaries of conventional algorithmic design in the direction of a new paradigm -- Physics-ML-Synergy -- which draws on and integrates the strengths of each category. We aim to demonstrate that neither world should be disregarded, and that interconnecting physics-based and data-driven worlds can have a transformative impact, enabling robust and efficient situational awareness capabilities that are impossible with either approach alone.


The first part of this thesis addresses the inherent limitations within each world. The key enablers to this end are circuit-based formulation and the exploitation of sparse structures. Circuit-based modeling provides a unified framework that integrates both transmission and distribution networks for simulation \cite{SUGAR-pf}\cite{sugar-powerflow-amrit} and estimation \cite{sugar-SE-Hug}\cite{sugar-SE-Hug2}\cite{SUGAR-SE-Alex}, leveraging the natural circuit structure of power systems. In Chapter \ref{ch: ECF}, we extend circuit-based modeling techniques (which have been already explored for simulation \cite{SUGAR-pf}\cite{sugar-powerflow-amrit} and optimal power flow studies \cite{SUGAR-opf}) to the estimation purpose, developing measurement-based device models and a circuit-based estimation method that is naturally convex, scalable, and fast with closed-form solution \cite{sugar-se-L2}. Moreover,  independent current sources are inserted across the circuit system as “threat indicators”, working for both simulation and estimation, providing a baseline level of robustness by allowing us to quantify blackout conditions and measurement noises.

The second key enabler is the exploiting of sparsity to foster inherent robustness and efficiency in situational awareness tools. Sparsity \cite{sparsePFref10-lasso}\cite{sparsePFref9-fusedlasso}\cite{robust-estimation-Huber}\cite{robust-regression-Huber} in vectors or matrices refers to a structure with only a small fraction of non-zero values carrying essential information, while most of the entries are zero which can be ignored without losing significant accuracy. This often leads to sparse explanations and sparse recommendations, which are ubiquitous in engineering systems and the exploiting of which can foster efficient or effective decision-making. For example, root causes of faults, errors, and deficiencies are oftentimes sparsely distributed across the network; the mitigation of which should focus interventions on sparse recommendation of key areas. In terms of ML, a key issue is the generalization to different distributions of data. From a structure-exploiting perspective, relevant data are sparsely distributed among the set of all historical data, suggesting that a sparse recommendation of the most relevant features / data are needed to make the optimal single-point prediction with minimal generalization error. Moreover, the power system only has a few branches adjacent to each node, representing a sparsely-connected graphical model that can be leveraged to reduce model complexity and account for network topology.  In Chapter \ref{ch: physics}, we develop sparse optimization techniques to produce threat indicators that pinpoint dominant sources of blackout and data anomalies, building intrinsic robustness into the simulation \cite{SUGAR-sparsePF-Li} and estimation tools \cite{SUGAR_SE_WLAV}\cite{SUGAR_GSE_WLAV}\cite{SUGAR-SE-TnD}. While in Chapter \ref{ch: ML}, we leverage temporal and spatial sparsity-exploiting optimization for lightweight ML that gives scalable, interpretable and generalizable ML designs for prediction and detection purposes.

The second part of this thesis centers on the development of Physics-ML Synergy -- where physics-based tools “chat with” ML to merge their benefits. The goal is to show if interconnecting these tools together can bring efficiency and robustness further against targeted cyberthreats. Central to this synergy is the information transfer between the two tools to address specific situational awareness needs. Chapter \ref{ch: synergy} explores two critical applications: simulation and estimation, where the physics-based tools developed in Chapter \ref{ch: physics} and lightweight ML models from Chapter \ref{ch: ML} are interconnected to bridge remaining performance gaps. In the first application, ML predictions provide a warm starting point for circuit-based simulation, significantly accelerating the overall simulation process \cite{gridwarm}. In the second application, time-series ML contributes prior knowledge as regularization in circuit-based estimation, creating an ML-augmented estimation tool that is robust to cyberattack-induced false data. The enhanced efficiency and robustness—unachievable with physics-based tools alone—demonstrates the transformative potential of Physics-ML Synergy in advancing situational awareness for complex, safety-critical systems.

To validate the advancements presented in this thesis, we conduct extensive experiments that demonstrate robustness across various threats (blackouts, bad data, topology error, MadIoT attack and false data injection attacks) in the simulation and estimation tasks. We also evaluate speed and scalability of our methods and large-scale systems ranging from thousands of nodes to over 80k-node systems. Lastly, this thesis concludes with guidelines for future work, suggesting directions to extend Physics-ML Synergy to handling a broader range of threats, complementing missing physics in partially observable conditions, and extended applications to general cyber-physical systems.

\chapter{Background}\label{ch: bkg}

\section{Simulation and estimation key to situational awareness} \label{sec: intro SA}

Situational awareness (SA) \cite{SA_overview} means the operator’s understanding and perception of the dynamic situation. The concept is initially defined by Tolk and Keether in the aircraft domain: “the ability to envision the current and future disposition of both red and blue aircraft and surface threats''. Endsley extended the early definition beyond fighter aircraft while still highlighting the \textbf{"present and future"}. The understanding of SA is also advanced by the operational methods and performance measures defined to evaluate the operator’s SA. Examples include the situational awareness rating technique (SART) and the SA global assessment technique (SAGAT). 

This thesis discusses steady-state situational awareness in the context of the power grid as a characteristic of operators under the changing grid conditions. It refers to the grid operator's ability to perceive and understand the system’s \textbf{present} state and project its \textbf{future} behavior. Present situational awareness provides operators with a real-time snapshot of system conditions, while future situational awareness enables them to anticipate potential challenges and make proactive decisions.

Estimation (and detection) tools are fundamental for present situational awareness which focuses on providing a real-time snapshot of system conditions for real-time understanding of the current conditions. This involves recognizing the present power flows, voltages, equipment status, and any ongoing disturbances based on real-time data (e.g., through state estimation). And the goal is to provide operators with an accurate and up-to-date representation of the system's status, allowing them to maintain control and respond to current issues. Below we informally define the estimation problem for present situational awareness. For a comprehensive understanding of the present operating conditions, estimation problem needs to go beyond the estimating states of the system, but also ongoing anomalies:

\begin{problem}[Power grid estimation for present  situational awareness]\label{def-informal:estimation} \noindent\\
\textbf{Input:} network parameters and real-time measurement data;
measurement models depicting how data relates to the variables to estimate; 
(optional) historical, (optional) past events and switching history.\\
\noindent\textbf{Output (for time $t$):} estimated system state variables (typically bus voltages which can be further used to calculate line currents, and power flows), updated topology (an accurate view of the network connectivity), and anomaly alerts 
\end{problem}

Simulation is the fundamental tool for future situational awareness which is to anticipate and project potential future states of the system based on current and forecasted conditions. This could include answering "what-if" questions on how the system states will evolve under changing loads, equipment failures, or other contingencies. The goal here is to help operators evaluate whether the system can survive changes and serving current or forecasted loads, so that they can foresee potential risks and prepare for them in advance to ensure system reliability. Below we informally define the simulation problem for future situational awareness of power grid:

\begin{problem}[Power grid simulation for future steady-state situational awareness]\label{def-informal:simulation}  \noindent\\
\textbf{Input:} the state at time $t_1$ (e.g., obtained from estimation), possible "what-if" scenarios for a future time $t_2$ (e.g., changes in load, generation forecasts, equipment failures, other contingencies, etc.), and system models in the form of network balance equations.\\
\noindent\textbf{Output (for a future time $t_2$):} system behavior (typically bus voltages which can be used to further calculate currents, power flows, etc)  
\end{problem}

The goal of this thesis is to build high-quality situational awareness to handle a variety of threats, ranging from physical to cyber indents;  from natural faults to malicious attacks; and from random perturbations to targeted or interactive threat behaviors. Table \ref{tab: threats} summarizes threats that represent key obstacles in power system situational awareness. 

Cyberattacks are of particularly high interest in this case. Recent years have witnessed and documented a large variety of cyber attack methods capable of causing different consequences. Some de-energize the power grid in a brute-force way via intrusion into the substation, power plants or control rooms to open circuit breakers  (e.g., power outage in 2015\cite{ukrain2015} and 2016 Ukrainian attacks\cite{lee2017crashoverride}). Some target IoT-controlled load devices on the distribution grid for a manipulation of demand to destabilize the power grid (e.g., MadIoT attack\cite{madiot}\cite{MadIoTRebuttal}\cite{MadIoTOnNY}). Some aim at data loss and delay by attacking the communication channels (e.g., jitter attack\cite{jitter_attack}). And some inject targeted false data to mislead the operators (e.g. false data injection attack\cite{FDIA_AC}\cite{fdia-acse}). Unlike traditional bad data, false data generated by a cyber attack are usually assumed to be targeted and interactive, meaning they are inter-correlated or even carefully designed by some relationship to maximize the impact. 

Being aware of these threats, advanced situational awareness tools are needed to defend against them. Real-time estimation needs to identify data errors and maintain "present" situational awareness regardless of them. Advanced simulation tools are needed to proactively simulate contingencies and blackouts, so as to gain future situational awareness of potential failures and build resiliency against them.

\begin{table}[]
\caption{Threats that represent key obstacles in power system situational awareness}
\label{tab: threats}
\begin{tabular}{l|l|l}
\hline
\textbf{type}                                     & \textbf{threat}                                                                                              & \textbf{possible causes}                                                                                                           \\ \hline
\multicolumn{1}{c|}{\multirow{4}{*}{traditional}} & \begin{tabular}[c]{@{}l@{}}random data error \\ and low data quality\end{tabular}                            & \begin{tabular}[c]{@{}l@{}}large measurement noise, meter mis-calibration, \\ communication error, etc\end{tabular}                \\ \cline{2-3} 
\multicolumn{1}{c|}{}                             & \begin{tabular}[c]{@{}l@{}}topology error \\ (wrong switch status data)\end{tabular}                         & \begin{tabular}[c]{@{}l@{}}circuit breaker fault, communication error or latency, \\ switch data mis-synchronization\end{tabular}  \\ \cline{2-3} 
\multicolumn{1}{c|}{}                             & \begin{tabular}[c]{@{}l@{}}traditional contingency \\ (outage of \\ generator/line/transformer)\end{tabular} & \begin{tabular}[c]{@{}l@{}}overloading, physical damage,  mechanical issues,\\ circuit breaker faults, etc\end{tabular}            \\ \cline{2-3} 
\multicolumn{1}{c|}{}                             & blackout                                                                                                     & high demand conditions, (cascading) contingencies                                                                                  \\ \hline
\multirow{3}{*}{modern}                           & \begin{tabular}[c]{@{}l@{}}interactive data errors \\ induced by cyber attacks\end{tabular}                  & \begin{tabular}[c]{@{}l@{}}cyber intrusion into meters / communication / data center \\ to inject false data, \\jamming / spoofing attacks on modern synchrophasors, etc \end{tabular}          \\ \cline{2-3} 
                                                  & supply / load perturbations                                                                                   & MadIoT attack, renewable energy variations, etc.                                                                                   \\ \cline{2-3} 
                                                  & \begin{tabular}[c]{@{}l@{}}contingency and blackout\\ induced by cyberattacks\end{tabular}                   & \begin{tabular}[c]{@{}l@{}}cyber intrusion into substations / power plants / \\ control room to disconnect generators, etc\end{tabular} \\ \hline
\end{tabular}
\end{table}

\section{Robustness and efficiency gaps}
 
The word "robustness" has many connotations in different fields. This thesis discusses the lack of robustness in simulation and estimation tools, meaning that, under some threat conditions, they result in erroneous, meaningless and non-actionable outputs or even cannot reach a solution. In the rest of this Chapter, Section \ref{sec: simulation gap} discusses existing simulation methods and their lacking robustness to blackout failures; and Section \ref{sec: estimation gap} discusses the existing estimation approaches and their limited robustness to anomalous data. 

Whereas, the efficiency gap refers to the lack of time-efficiency in existing tools. The challenge stems from the nonlinearity introduced by analytical models, specifically nonlinear network models in simulation, and nonlinear measurement models in estimation, respectively. These nonlinear models result in nonlinear programming problems that are NP-hard and large systems are slow to converge especially when the solver does not have a good initial point to start with.

Next in Section \ref{sec: simulation gap} and  \ref{sec: estimation gap}, we will go through the robustness and efficiency gaps in each tool, with a review of traditional and state-of-the-art approaches in literature. 

\subsection{Gaps in steady-state simulation tools}\label{sec: simulation gap}
\subsubsection{Traditional power flow simulation}
Simulation evaluates whether a system can survive "what-if" scenarios while serving the present or forecasted loads. Traditional steady-state simulation method \cite{traditionalPF} for this purpose outputs the bus voltage solution by iteratively solving a set of nonlinear network balance equations. Let $ N_{bus}$  be the number of buses (nodes), $ V_i $ be the voltage magnitude at bus $ i $, $ \theta_i $ be the voltage phase angle at bus $ i $, 
\( P_{G_i} \) be the active power supply generated at bus \( i \), \( P_{\text{load}_i} \) be the active power consumed by the load at bus \( i \), \( Q_{G_i} \) be the reactive power generated at bus \( i \), \( Q_{\text{load}_i} \) be the reactive power consumed by the load at bus \( i \).
$ Y_{ij} = G_{ij} + jB_{ij} $ be the element of the admittance matrix $ Y $, where $ G_{ij} $ and $ B_{ij} $ are the conductance and susceptance between buses $ i $ and $ j $, respectively. Each node $i$ is constrained by two network equations in (\ref{eq:powerbalance}) that enforce power balance for both the active (real) and reactive (imaginary) components of the power. They intuitively mean that the total amount of power flowing into the bus must equal the total amount of power flowing out of the bus.
\begin{align} 
F_i(V, \theta) =
\begin{cases}
P_{G_i} - P_{\text{load}_i} - V_i \sum_{j=1}^{N} V_j \left( G_{ij} \cos(\theta_i - \theta_j) + B_{ij} \sin(\theta_i - \theta_j) \right) = 0 \\
Q_{G_i} - Q_{\text{load}_i} - V_i \sum_{j=1}^{N} V_j \left( G_{ij} \sin(\theta_i - \theta_j) - B_{ij} \cos(\theta_i - \theta_j) \right) = 0 
\end{cases}
\label{eq:powerbalance}
\end{align}

So the traditional power flow simulation problem can be mathematically defined as:

\begin{problem}[Traditional power flow simulation]\label{def: trad simulation} evaluates "what-if" scenarios by solving bus voltage solution $V, \theta$ from a set of nonlinear equations: 
\begin{equation}
    F_i(V, \theta)=0 \text{ as in  (\ref{eq:powerbalance}), for }  i=1,2,...,N_{bus} 
\end{equation}
\end{problem}
Such a traditional power flow simulation aims to converge to a so-called "feasible" solution of the system. The efficiency gap arises in the process of reaching convergence, as the nonlinear function $F$ makes it a nonlinear programming problem that is generally considered NP-hard. Newton-Raphson method is the common method to solve it where the nonlinear equations are linearized iteratively, and solutions are updated until convergence. But there is no general guarantee that all instances can be solved efficiently, particularly for large and complex networks.

And this traditional method fails completely if no feasible solution exists for the system, which represents the scenario of power system blackout. Often, due to severe contingencies, heavy loading, and other limitations that make power supply unable to meet the demand, simulation indicates network collapse, which corresponds to a grid that has likely blacked out \cite{powerblackout}. What-if scenarios related to blackouts are of high interest to both operation and planning, since they result in severe consequences ranging from economic loss, physical damage of infrastructure, and severe social disruptions. 

Traditional methods are not robust to such failures. As defined in Problem \ref{def: trad simulation}, this collapsed grid state mathematically represents an "infeasible system" with no solution to the equations $F(V, \theta)=0$. The methodology diverges \cite{powerflow-diverge}, resulting in no useful information from the simulation output. 

\subsubsection{Infeasibility-quantified simulation to converge under blackouts}
To avoid divergence without providing information, improved simulation method \cite{SUGAR-pf-marko},\cite{infeasibility-quantified-pf-trad},\cite{SUGAR-pf} has been developed to return meaningful power flow solutions for such collapsed grid states. The main idea is to quantify the "infeasibility" of these traditionally unsolvable cases and reformulate simulation into a constrained optimization problem:

\begin{problem}
[Infeasibility-quantified simulation]\label{def: sugar simulation} evaluates "what-if" scenarios while considering possible blackouts by solving $x=[V, \theta]$ and compensation terms $n$ from: 
\begin{align}
&\min_{x,n} \frac{1}{2}||n||_2^2\notag\\
\text{s.t. }
   & F_i(x)+n_i=0 \text{ for }  i=1,2,...,N_{bus} 
\end{align}
\end{problem}

\noindent Specifically, this new formulation introduces additional slack variables $n_i$ at each bus $i$ with intuitive meanings:
\begin{itemize}
    \item $n_i$ is a "infeasibility" indicator" that indicates power deficiency at each location that causes system blackout
    \item $n_i$ also represents compensation terms that capture how much additional flow needs to be injected at each bus (node) to make the network balance conditions hold again
\end{itemize}
\noindent So that at each bus $i$, the traditional system model $F_i(x)=0$ with $x=[V,\theta]$ can be extended to a new set of equations $F_i(x)+n_i=0$. This modified system with compensations mathematically represents a feasible system with infinite solutions. By minimizing the sum of squares (i.e., L2 norm) of the compensation terms, the method reaches the optimal solution that minimizes the "energy of reconstruction error", considering $n$ as the error that prevents the system from being feasible. 

The variables $n$ are "infeasibility indicators" which extend traditional simulation to account for both feasible and infeasible systems. The values of $n$ can clearly distinguish a feasible case from an infeasible one. Convergence with zero $n$ everywhere denotes system balance satisfied and feasible solutions reached. Convergence with nonzero $n_i$ at any node $i$ denotes an infeasible system with specific power flow deficiency at each bus $i$. The larger the $n$ values, the more severe the blackout. Whereas, if divergence happens, it is totally due to insufficient convergence techniques of the algorithm.

Such "infeasibility-quantified" simulation has been proposed and developed for both the traditional power flow simulation \cite{infeasibility-quantified-pf-trad} which models power balance as in (\ref{eq:powerbalance}), and newer circuit formulations of power system \cite{SUGAR-pf-marko}\cite{SUGAR-pf} which models current balance under rectangular coordinate, the later allows application to three-phase distribution grid without loss of generality \cite{SUGAR-pf}.

\subsection{Gaps in steady-state estimation (and detection) tools} \label{sec: estimation gap}

\subsubsection{Estimation input: conventional and modern data}

As informally defined in Problem \ref{def-informal:estimation}, the  present situational awareness aims to retrieve real-time system information from measurement data. These data come from a large number of meters, i.e., sensors, installed across the system. These meters collect measurements over time on different types of buses and branches (transmission lines and transformers). The type and quality of measurement data depend highly on the type of measurement device. Table \ref{tab: data collection} shows the realistic setting of data collection with a detailed comparison.

Conventional monitoring technologies, like the supervisory control and data acquisition (SCADA) system, collect steady-state measurements of circuit-breaker status (or switch status), voltage magnitude $|V|$, current magnitude $|I|$, real and reactive power flows $P,Q$. These data usually have the measurement accuracy of $\pm 2\%$ as required by PJM\cite{pjm-m01}. 
Remote terminal units (RTU), also called remote telemetry units, act as the interface in the SCADA system to gather all data from field devices and transmit them to the master station in a central control system. Any control signal issued by the control center will also be received by RTU and further delivered to the field devices (e.g., relays). In this work, we use RTU to represent the measurement device in SCADA. 

Modern networked phasor measurement units (PMU) collect voltage and current phasors, frequency, etc, with high speed and accuracy. The transient data measured by PMUs are synchronized by the Global Positioning System (GPS) and transferred to the control center via a hierarchy of local and regional phasor data concentrators (PDCs) as well as the underlying synchrophasor communication networks (ranging from wired to wireless networks). A review of PMU communication technologies can be found in \cite{review-pmu-tech}.
Unlike the magnitude signals collected by conventional SCADA, synchrophasors are characterized by both the real-time magnitude and phase angle of an electrical phasor quantity (voltage or current). Each phase angle is with respect to a consistent reference across the system, whereas SCADA data only enable computing a local phase angle between voltage and current. 
Usually, PMU is more than 100 times faster than the conventional SCADA system, with a sampling rate of over 600Hz, a reporting rate of more than 30 samples per second, and an accuracy of $\pm 1\%$ or better for all measurements.
The last decade has seen PMUs installed on both the transmission and distribution grid (distribution-level PMU, micro-PMU or $\mu$PMU on distribution grid), allowing for real-time transient monitoring, data analysis and operation. Today's industrial practices and guidelines\cite{pmu_guideline} require that PMUs be placed near significant generating plants generally above 100 MVA (including wind and solar), large load buses, and grid control devices to capture safety-critical incidents. Reports\cite{pmu_coverage2022} showed the number of PMU deployment in the US grid bulk power system (i.e., the generation and transmission system) was over 2,500 in 2019 and over 3,000 in 2022. 

\begin{table}[]
\caption{Realistic settings of data collection on power grid}
\label{tab: data collection}
\small
\begin{tabular}{llllll}
\hline
\multicolumn{1}{c|}{\textbf{location}}                                                                          & \multicolumn{1}{c|}{\textbf{device}}                                              & \multicolumn{1}{c|}{\textbf{data}}                                                                                                                               & \multicolumn{1}{c|}{\textbf{type}}                                                & \multicolumn{1}{c|}{\textbf{\begin{tabular}[c]{@{}c@{}}frequency of \\ acquisition\end{tabular}}}     & \multicolumn{1}{c}{\textbf{accuracy}}                                                    \\ \hline
\multicolumn{1}{l|}{\multirow{2}{*}{\begin{tabular}[c]{@{}l@{}}transmission\\ and\\ distribution\end{tabular}}} & \multicolumn{1}{l|}{RTU}                                                          & \multicolumn{1}{l|}{switch status}                                                                                                                               & \multicolumn{1}{l|}{\begin{tabular}[c]{@{}l@{}}discrete \\ (on/off)\end{tabular}} & \multicolumn{1}{l|}{\begin{tabular}[c]{@{}l@{}}upon change in PJM,\\ every 4 seconds in\\ ISO-NE\end{tabular}} & \multicolumn{1}{c}{-}                                                                    \\ \cline{2-6} 
\multicolumn{1}{l|}{}                                                                                           & \multicolumn{1}{l|}{RTU}                                                          & \multicolumn{1}{l|}{\begin{tabular}[c]{@{}l@{}}voltage/current magnitude,\\ real/reactive power,\\ frequency\end{tabular}}                                       & \multicolumn{1}{l|}{\begin{tabular}[c]{@{}l@{}}analog\\steady-state\end{tabular}}                                                 & \multicolumn{1}{l|}{2-10 seconds}                                                                     & \begin{tabular}[c]{@{}l@{}}$\pm2\%$\\ in PJM\end{tabular}                                \\ \hline
\multicolumn{1}{l|}{transmission}                                                                               & \multicolumn{1}{l|}{PMU}                                                          & \multicolumn{1}{l|}{\begin{tabular}[c]{@{}l@{}}voltage/current phasor,\\ frequency,\\ rate of change of frequency,\\ dynamic reactive device power\end{tabular}} & \multicolumn{1}{l|}{\begin{tabular}[c]{@{}l@{}}analog\\transient\end{tabular}}                                                    & \multicolumn{1}{l|}{\begin{tabular}[c]{@{}l@{}}reporting rate $\geq$ 30\\ samples/second\end{tabular}}  & \begin{tabular}[c]{@{}l@{}}$\pm 1\%$\\ (angle $\pm1\degree$ \\ or $\pm1\%$)\end{tabular} \\ \hline
\multicolumn{1}{l|}{\multirow{2}{*}{distribution}}                                                              & \multicolumn{1}{l|}{$\mu$PMU}                                                    & \multicolumn{1}{l|}{\begin{tabular}[c]{@{}l@{}}voltage/current phasor,\\ frequency, etc\end{tabular}}                                                            & \multicolumn{1}{l|}{\begin{tabular}[c]{@{}l@{}}analog\\transient\end{tabular}}                                                   & \multicolumn{1}{l|}{\begin{tabular}[c]{@{}l@{}}reporting rate 10-120 \\samples/second\end{tabular}}     & \begin{tabular}[c]{@{}l@{}}$\pm1\degree$ \\ or $\pm1\%$\end{tabular}                     \\ \cline{2-6} 
\multicolumn{1}{l|}{}                                                                                           & \multicolumn{1}{l|}{\begin{tabular}[c]{@{}l@{}}smart meter, \\ IoTs\end{tabular}} & \multicolumn{1}{l|}{loads, line flows, etc}                                                                                                                      & \multicolumn{1}{l|}{\begin{tabular}[c]{@{}l@{}}analog\\steady-state\end{tabular}}                                                 & \multicolumn{1}{c|}{-}                                                                                & \multicolumn{1}{c}{-}                                                                    \\ \hline
\multicolumn{6}{l}{\begin{tabular}[x]{@{}l@{}}
* This table mainly includes regular electronic updates. Some significant events like changes in transformer tap \\position, dynamic ratings are not included (more details in \cite{pjm-m01}. But the real-world control room will be notified \\immediately upon these events. \\
* RTUs and PMUs provide measurements for buses, transmission lines, transformers and reactive devices (like\\ synchronous condensers).
\end{tabular}}                                       \end{tabular}
\end{table}

\textbf{Observability issues and pseudo-measurements}: The transmission system requires collected measurements to guarantee system observability. The concept of observability was initially defined by F. Schweppe and Wildes in \cite{WLS-SE} as being able to uniquely determine all the state variables (bus voltage magnitudes and phase angles) from the given set of measurements using the control room's state estimation (SE) algorithm. This requires proper meter placement depending on the type, location and number of measurements. However, in real-time operation, a temporary unavailability of data might occur and cause an unobservable system or unobservable islands. In this case, we can add pseudo-measurements which will be handled the same way as real data in the estimation problem. These measurements can be nominal load/generation settings, estimated coefficients or patterns such as the daily load cycle, zero-injection buses, etc. 

Taking these measurements as input, the estimation task for present situational awareness (as informally defined in Problem \ref{def-informal:estimation}) is traditionally split into 3 processes: 1) network topology processor for topology estimation, 2) state estimator (SE) for estimating state variables (voltages), and 3) bad data detection which is typically implemented by hypothesis tests immediately after the state estimation. Later methods of robust state estimation merge 2) and 3) together into one unit. And generalized state estimation (GSE) merges 1) and 2) together to jointly estimate topology and voltages. And a robust version of GSE merges 1), 2) and 3) together into one. Next we will discuss existing methods in terms of their formulations and limitations. In general, the inclusion of traditional is the main source of nonlinearity and that results in efficiency challenges to estimation. And the random and targeted data errors, as illustrated in Table \ref{tab: threats}, results in the gap of robustness. 

\subsubsection{Network Topology Processor} \label{sec: NTP}

\begin{figure}[h]
  \begin{subfigure}[t]{0.53\linewidth}
      \centering	\includegraphics[width=\textwidth]{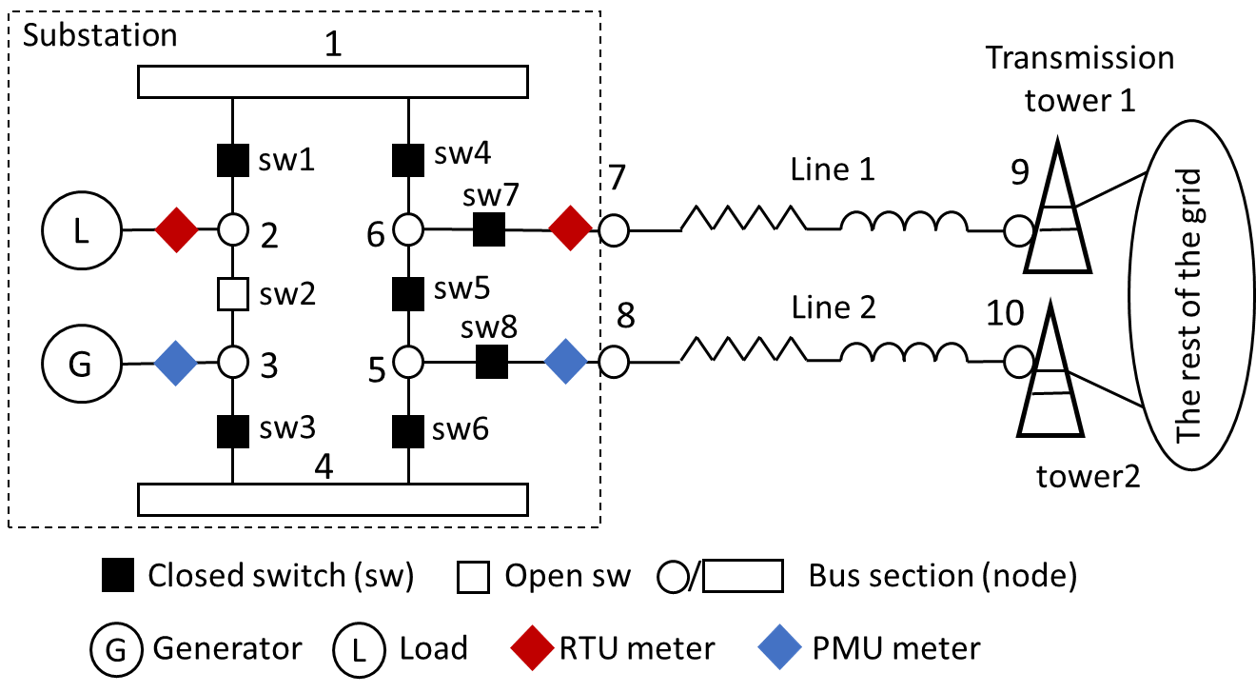}
	\caption{Node-breaker (NB) model of a grid sub-network.}
	\label{fig:NB model}
  \end{subfigure}
  \hfill
    \begin{subfigure}[t]{0.43\linewidth}
      	\centering	\includegraphics[width=\textwidth]{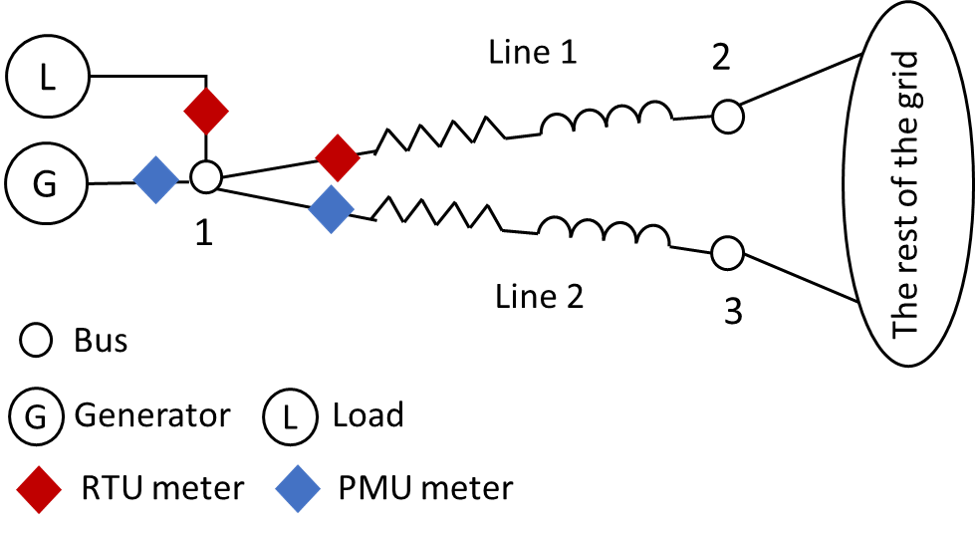}
	\caption{Bus-branch (BB) model: substation buses are merged.}
	\label{fig:BB model}
    \end{subfigure}
    \caption{Network topology processor converts node-breaker (NB) model to bus-branch (BB) model, and identifies network topology.}
    \label{fig: NB to BB}
  \end{figure}

 A network topology processor \cite{TE-NTP-book}\cite{TE-NTP-tracking} provides situational awareness of the network topology.   In today's control room, NTP converts the NB model into a simplified bus-branch (BB) format to be used by state estimation, contingency analysis and optimal control. As Figure \ref{fig: NB to BB} shows, the node-breaker (NB) model includes all inactive and active switch components that exist on the system, whereas the bus-branch model only contains active bus and branch components which significantly decrease the size of the network. 

 The standard algorithm for NTP \cite{TE-NTP-book} is as follows:
\begin{enumerate}
    \item \textit{Raw data processing} converts raw data into normalized units and performs simple checks to verify operating limits, rate of change of operating variables, and other data consistency checks (e.g., confirm zero flows on open switches and zero voltage across closed switches).
    \item \textit{Bus section processing} recursively merges any two nodes connected by a closed switch into one bus.
    \item \textit{Connectivity analysis} identifies active network topology from the switch status data and reassigns locations of metering devices on the bus-branch model.
\end{enumerate}
As only minor or no topology changes occur most of the time between subsequent NTP runs, a real-world NTP increases its efficiency by operating in tracking mode \cite{TE-NTP-tracking} where only the groups of bus sections with switch status changes are processed recursively. 

When processing on the NB model, NTP assumes all its input status data are correct. However, switch status data  can be corrupted due to telemetry error, operator entry error, physical damage (e.g., a line is tripped but the disconnection is not reflected on the circuit breaker status) or even a cyber-attack. In case of such incorrect statuses, NTP can falsely merge/split buses and output an erroneous grid topology. 

\subsubsection{State estimation and bad data detection}\label{sec: SE}

Taking the network topology and measurement data as input, 
grid operators run AC state estimation (ACSE) at minute scale (every 5 minutes in ERCOT and US Midwest ISO (MISO), every 3 minutes in the U.K. grid and ISO-NE, and every 1-2 minutes in PJM\cite{pjm-m12}) to gain situational awareness of the AC grid state variables and bad data alarms.
The AC state variables include voltage magnitude and phase angle at every bus. They are used to estimate the generation and load configurations together with network topology and parameters, to provide situational awareness of the up-to-date AC power flow model (i.e., system model description) to be used in contingency analysis, optimal power flow and other studies. 

Traditional method  \cite{WLS-SE} for this purpose is a weighted least squares (WLS) method:

\begin{problem}[Traditional AC state estimation]\label{def: trad SE} Given a vector of measurement $z$, state estimation solves  bus voltages $x=[|V|,\theta]$ by minimizing the weighted least-squares (WLS) of measurement error:
    \begin{equation}
    \min_x \sum_i w_i(z_i-f_i(x))^2
    \label{eq: trad SE}
    \end{equation}
\end{problem}

\noindent where $x$ is the vector of state variables containing  voltage magnitudes $|V|$ and phase angles $\theta$, $z_i$ is the $i$-th measurement, and $f_i(x)$ denotes the relationship between measurement $z_i$ and state $x$. 

The function $f_i(x)$ for $\forall i$ is determined by Kirchoff’s law and power flow equations. The efficiency gap arises as $f_i(x)$ is a nonlinear equation for some conventional SCADA measurements like the real and reactive power, e.g.,
\begin{equation}
    P_{i}  = V_i \sum_{j=1}^{N} V_j \left( G_{ij} \cos(\theta_i - \theta_j) + B_{ij} \sin(\theta_i - \theta_j) \right) 
\end{equation}

\noindent making the objective function nonlinear and non-convex. Thus, this minimization problem does not have a closed-form solution and must be solved iteratively. 

The robustness gap also arises when facing data errors as listed in Table \ref{tab: threats}. 
The most classical family of methods to figure out bad data are hypothesis tests based on post-processing of the weighted least square (WLS) problem residuals \cite{traditional-BDI},\cite{robustSE-reweight},\cite{robustSE-modify}. More specifically:
\begin{problem}[Traditional bad data detection (BDD)]\label{trad BDD}
    Assuming that the measurement noise satisfy Gaussian distribution, the WLS residual $J_i=(z_i-f_i(\hat{x}))^2, \forall i$ (obtained from traditional state estimation Problem \ref{def: trad SE}) satisfies Chi-square distribution whose critical value $\tau$ can be used as a threshold to detect bad data.
    \begin{equation}
        z_i \text{ is a bad data if } J_i>\tau
    \end{equation}
\end{problem}
 Based on hypothesis testing, bad data is usually handled by adjusting suspicious measurements and rerunning ACSE until residuals are satisfactory. Possible adjustments include removing bad data, modifying \cite{robustSE-modify} bad data or the iterative reweighted least square method \cite{robustSE-reweight}, where weights on suspicious measurements are set to lower values. Clearly, the iterative re-running of the ACSE problem incurs a computational burden. Other robust estimators include least median of squares \cite{robustSE-median}, yet this is not widely adopted due to numerical difficulty in handling the median.

The family of \textbf{robust state estimators} has been developed as the performance of residual-based hypothesis tests can be unsatisfactory, especially when multiple bad-data exist. Built upon the assumption that bad data are typically sparse, the problem formulation is typically based on minimizing the weighted least absolute value (WLAV) in the objective function \cite{LAVSE-dc-soliman}\cite{LAVSE-dc-Vidya}\cite{LAVSE-PMU-abur}\cite{LAVSE-PMU-abur2}. These methods 
enable automatic rejection of bad data, eliminating the need for iterative re-runs while providing clear identification of suspicious measurements from the sparse residual vector:
\begin{problem}[Robust state estimation]\label{def: WLAV SE} Given a vector of measurement $z$, robust SE solves  bus voltages $x=[|V|,\theta]$ by minimizing the weighted least absolute value (WLAV) of measurement error, resulting in a sparse residual to capture and reject (random) bad data:
    \begin{equation}
    \min_x \sum_i w_i|z_i-f_i(x)|
    \label{eq: WLAV SE}
    \end{equation}
\end{problem}

Despite this desirable property, the non-differentiable L1-norm terms in the WLAV approach introduces additional complexity. At present, two approaches to WLAV-ACSE exist: i) those that solve highly non-convex ACSE due to inclusion of power measurements from conventional RTUs \cite{LAVSE-dc-Singh-splitform}; and ii) those that are formulated as a linear programming (LP) problem \cite{LAVSE-PMU-abur}\cite{LAVSE-PMU-abur2} with the assumption that network only consists of PMUs. These convex WLAV formulations based solely on PMUs take advantage of the linear relationship between phasor data and grid states; however, assuming PMU-observability of the network is unrealistic for grids today. And Simplex method\cite{simplex}, which is typically used to solve linear programming problems in these linear SE tasks, is not scalable to large systems.

To be practical for the reality, ACSE must include RTUs (i.e., conventional SCADA measurements). Unfortunately, most methods that include conventional RTUs (hybrid or otherwise) correspond to ACSE methods with highly complex and non-convex solution space that presents convergence challenges and high residuals. To address these problems, there has been some work on convex relaxations of ACSE methods (applicable to WLAV as well). For instance, \cite{convexTESE-SDP-weng} convexified the problem by reformulating the voltage magnitude $|V|$, phase angle $\delta$, and product of $V$ and $I$, in semidefinite programming (SDP) form, through matrix transformation; however this method fails to effectively scale  to large-scale networks. More recently to overcome the convergence and high residual limitations, the researchers in \cite{SUGAR_SE_L2}\cite{SUGAR-SE-Alex} proposed a circuit theoretic ACSE approach that reduces the problem to an equality-constrained quadratic programming (QP) problem by creating linear measurement models following a "sensitivity" based mapping of conventional grid measurements. This approach provides a convex hybrid ACSE formulation scalable to large networks, but it is not implicitly resilient to bad data.

\subsubsection{Generalized State Estimation} \label{sec: GSE}

Despite a simple checking of data consistency, NTP does not include a rigorous analysis to correct wrong switch statuses. Thus, NTP's topology output can be inaccurate. The SE, which takes the erroneous topology as input, always assumes the given topology is correct and the bad-data detection algorithms cannot separate bad data from topology errors. Thus, any topology error will harm the operator's situational awareness by falsifying the SE solution while remaining undetectable. 

There have been attempts to build robustness against wrong switch statuses and resulting topology errors\cite{GSE-DC-substation}\cite{GSE-AC-nonlinear}\cite{TE-WLSE-BDI}\cite{TESE-GSE-PMUabur}\cite{convexTESE-SDP-weng}. One approach is a variant of ACSE bad-data detection (BDD) \cite{TE-WLSE-BDI} which runs ACSE for a set of possible topologies and then determines the optimal topology based on the one with the smallest residual. However, the problem becomes combinatorial and expensive to solve when multiple wrong switch statuses exist.

A more solvable family of methods is generalized state estimation (GSE), also known as generalized topology processing. The classical generalized state estimation (GSE)\cite{GSE-DC-substation} was demonstrated on a small substation network using a linear DC grid model with node-breaker representation. The algorithm is as follows:
\begin{enumerate}
    \item \textit{Switch modeling}: For any switch $(sw)$ that connects node $i,j$ in the NB model, \cite{GSE-DC-substation} creates a pseudo measurement of zero power flow ($P_{ij}=0$) if it is open, and zero angle difference ($\theta_{ij}=0$) if it is closed. Other elements are modeled similarly to the bus-branch model.
    \item \textit{Estimation}: These pseudo measurements for discrete states, along with the analog measurements, are then used to run state estimation on the network.
    \item \textit{Bad data detection}: A hypothesis test is performed on the residual of each switch and analog measurement to check if any data value is wrong. 
\end{enumerate}
 While a DC model \cite{GSE-DC-substation} provides the desirable properties of linearity and problem convexity, it does not have the expressiveness or fidelity to represent the AC system accurately. Therefore, \cite{GSE-AC-nonlinear} extended the DC-GSE \cite{GSE-DC-substation} to AC-network constrained GSE (AC-GSE). But this presents challenges as well. Due to nonlinear branch flow equations, AC-GSE results in a non-convex formulation with significant drawbacks in performance \cite{GSE-AC-nonlinear}. More recently, advanced GSE methods \cite{TESE-GSE-PMUabur}\cite{convexTESE-SDP-weng}\cite{TESE-GSE} have applied WLAV method on node breaker (NB) model to perform the joint estimation of AC states and topology for the entire AC power grid, allowing traditional bad data and topology errors to be effectively identified and separated. Below gives a general formulation of these approaches:
 
 \begin{problem}[Robust generalized state estimation]\label{def: WLAV GSE} Given a vector of continuous measurement $z_{cont}$ and a vector of switch statuses $s$, GSE defines slack variables $y_{i,j}$ representing (current or power) flow on each switching device $(i,j)$, and crease pseudo measurements $z_{sw}$ according to its measured status. Then solve bus voltages $x=[|V|,\theta]$ and flows $y$ from an extended measurement vector $z=[z_{cont},z_{sw}]$ by minimizing the weighted least absolute value (WLAV) of measurement error, resulting in a sparse residual to capture and reject (random) bad data:
    \begin{equation}
    \min_{x,y} \sum_i w_i|z_i-f_i(x)-M_i\cdot y|
    \label{eq: WLAV GSE}
    \end{equation}
\end{problem}

However, due to possible nonlinear $f_i$ functions, similar drawbacks as in robust state estimation also limit the time-efficiency of GSE approaches. \cite{TESE-GSE-PMUabur} assumed full observability of the network model using PMU data alone, so that $f_i$ functions are linear, but the assumption remains unrealistic in most countries and regions where traditional SCADA RTU meters are still dominant.  \cite{convexTESE-SDP-weng} used semidefinite programming (SDP) relaxation to obtain a convex GSE formulation, but SDP does not scale well to large-scale systems \cite{large-scale-SDP}.

\subsection{Data-driven learning-based situational awareness tools}

\subsubsection{Function approximators of simulation and estimation}

Finding data-driven alternatives to physics-based simulators or estimators has been widely studied in recent years. The common strategy is to leverage neural networks to approximate the physical functions in power system analysis. The application includes not only power flow (PF) \cite{ppf-dnn}\cite{pf-dnn-topo}\cite{gnn-pf} and state estimation (SE) \cite{unrolled-se-2018}\cite{unrolled-se}\cite{unrolled-gnu-se}, but also optimal power flow problems, ranging from DC optimal power flow (DCOPF)\cite{dcopf-dnn} to ACOPF \cite{dc3}\cite{gnn-acopf}\cite{gnn-acopf-warm}. 

These methods are generally based on (usually supervised) learning of the input-to-solution mapping using historical system operational data or synthetic data. 
A popular type of method is to learn a 'one-step' mapping function. Some use deep neural network (DNN) architectures \cite{ppf-dnn}\cite{pf-dnn-topo}\cite{dc3}\cite{dcopf-dnn} to learn high-dimensional input-output mappings, some use recurrent neural nets (RNNs) \cite{rnn-dse} to capture grid dynamics, and others apply graph neural networks (GNN) \cite{gnn-pf}\cite{gnn-acopf}\cite{gnn-se} to capture the exact topological structure of power grid. One the other hand, another type of method particularly designed for state estimation tasks is unrolled neural networks \cite{unrolled-se-2018}\cite{unrolled-se}\cite{unrolled-gnu-se}, whose layers mimic the iterative updates to solve state estimation problems using first-order optimization methods (i.e., gradient descent methods), based on quadratic approximations of the original problem.

To promote \textit{physical} feasibility of the solution, many works impose equality or inequality system constraints by i) encoding hard constraints inside NN layers (e.g. using sigmoid layer to encode technical limits of upper and lower bounds), ii) applying prior on the NN architecture (e.g., Hamiltonian \cite{hamiltonianNN} and Lagrangian
neural networks \cite{lagrangianNN}), iii) augmenting the objective function with penalty terms in a supervised \cite{ppf-dnn} or unsupervised \cite{dc3}\cite{pf-dnn-topo} way, and iv) projecting outputs \cite{dcopf-dnn} to the feasible domain. Later works use homotopy-based meta optimization heuristics \cite{homotopy-pnnl} to enhance the equality and inequality constraints. 
In all these methods, incorporating (nonlinear) system constraints remains a challenge, even with state-of-the-art toolboxes \cite{neuromancer}, and most popular strategies lack rigorous guarantees of nonlinear constraint satisfaction. 

While these methods have advanced the state-of-the-art in physics-informed ML for power grid applications, critical limitations in terms of generalization, interpretation, and scalability exist \cite{towards-practical-ML-Li}. We discuss these further:

\textbf{Limited generalization:} Many existing methods do not adapt well to changing grid conditions. Take changes in network topology as an example. Many current works are built on non-graphical architectures without any topology-related inputs. These, once trained, only work for one fixed topology and cannot generalize to dynamic grid conditions. More recently, some works have begun to encode topology information. Graph model-based methods (e.g., GNN \cite{gnn-pf}\cite{gnn-acopf}\cite{gnn-se}) naturally impose topology as a hard constraint and thus can account for topology changes. Alternatively, work in \cite{pf-dnn-topo} encodes the topology information into the penalty term (as a soft constraint) through the admittance and adjacency matrix, and \cite{unrolled-gnu-se} accounts for topology in NN implicitly by applying a topology-based prior through a penalty term. While these methods lead to better topology adaptiveness, they also have some risks: the use of penalty terms \cite{pf-dnn-topo}\cite{unrolled-gnu-se} to embed topology information as a soft constraint can lead to limited precision; and, for problems (like OPF) where information needs to be exchanged between far-away graph locations, the use of GNNs requires carefully designed global context vectors to output predictions with global-level considerations. 

\textbf{Limited interpretability:} Despite that many ML models (NN, decision trees, K-nearest-neighbors) are universal approximators, interpretations of their functionality from a physically meaningful perspective are still very limited. The general field of \textit{model interpretability} \cite{interpretability} focuses on explaining \textit{how a model works}, to mitigate fears of the unknown.
Broadly, investigations of interpretability have been categorized into transparency (also called ad-hoc interpretations) and post-hoc interpretations. The former aims to elucidate the mechanism by which the ML \textit{blackbox} works before any training begins by considering the notions of \textit{simulatability} (Can a human work through the model from input to output, in reasonable time steps through every calculation required to
produce a prediction?), \textit{decomposability} (Can we attach an intuitive explanation to each part of a model: each input, parameter, and calculation?), and \textit{algorithmic transparency} (Does the learning algorithm itself confer guarantees on convergence, error surface even for unseen data/problems?). And \textit{post-hoc interpretation} aims to inspect a learned model after training by considering its natural language explanations, visualizations
of learned representations/models, or explanations of empirical
examples. However, none of these concepts in the field of ML model interpretability formally evaluates \textit{how a ML model makes predictions in a physically meaningful way} when it is used on an industrial system like the power grid.  Some recent works have explored the physical meaningfulness of their models from the power system perspective. Still, interpretations are made in conceptually different ways without uniform metric: Unrolled neural networks (which has been used as data-driven state estimation for power grid \cite{unrolled-gnu-se}\cite{unrolled-se-2018}) are more decomposable and interpretable in a way that the layers mimic the iterations in the physical solvers, yet these models \cite{unrolled-se}\cite{unrolled-se-2018}\cite{unrolled-gnu-se} mainly unroll first-order solvers instead of the second-order (Newton-Raphson) realistic solvers.
GNN-based models \cite{gnn-pf}\cite{gnn-acopf}\cite{gnn-se} naturally enable better interpretability in terms of representing the graph structure. 
Work in \cite{ppf-dnn} provides some interpretation of its DNN model for PF by matching the gradients with power sensitivities and subsequently accelerating the training by pruning out unimportant gradients. \cite{pf-dnn-topo} learns a weight matrix that can approximate the bus admittance matrix; however, with only limited precision. To summarize, due to the limited interpretability, ML models still have some opacity and blackbox-ness, when compared with the purely physics-based models (e.g., power flow equations).

\textbf{Scalability issues}:
In the case of large-scale systems, models (like DNNs) that learn the mapping from high-dimensional input-output pairs will inevitably require larger and deeper designs of model architecture and thereafter, massive data to learn such mappings. This can affect the practical use in real-world power grid analytics.

\subsubsection{Anomaly detection and localization methods}

Beyond approximating simulators and state estimators, data-driven methods have been proved efficient in detecting and locating anomalies from data. Compared with bad data detection, anomaly detection presents a much stronger detectability to identify anomalous data, outages, and modern attacks, etc. A broad variety of machine learning techniques have been shown applicable to developing data-driven methods for this purpose, although with limitations in certain conditions. 

\textbf{Time Series Anomaly Detection:} 
Numerous univariate methods (hypothesis testings, auto-regressive and moving average models, etc.)  exist for anomaly detection in time series~\cite{keogh2007finding}. These methods apply to multivariate data when used together with dimension reduction techniques (principal component analysis, independent component analysis, etc.). Other methods have a multivariate nature and can learn spatial correlations in data.  Local outlier factor (LOF)~\cite{breunig2000lof} uses a local density approach. Isolation Forests~\cite{liu2008isolation}, as an ensemble method, partition the data using a set of trees for anomaly detection. Support vector machine (SVM) uses a classification method to make direct anomaly decisions.
More recently, many approaches use neural networks~\cite{yi2017grouped}, distance-based~\cite{ramaswamy2000efficient}, and exemplars~\cite{jones2014anomaly}. 

\textbf{Anomaly Detection in Temporal Graphs:} \cite{akoglu2010oddball} found anomalous changes in graphs using an egonet (i.e. neighborhood) based approach, while \cite{chen2012community} uses a community-based approach. \cite{mongiovi2013netspot} found connected regions with high anomalousness. \cite{araujo2014com2} detected large and/or transient communities using Minimum Description Length. \cite{akoglu2010event} found change points in dynamic graphs, while other partition-based~\cite{aggarwal2011outlier} and sketch-based~\cite{ranshous2016scalable} works also exist for anomaly detection. However, these methods are not applicable to power grid anomaly detection because they focus on detecting unusual communities or connections, and power grid anomalies are detected from sensor data points.

Most of these general-purpose techniques, however, suffer from a lack of domain knowledge, as they ignore the grid-specific temporal and spatial patterns. This can limit their practicality. The localization of anomaly sources can be rough and inaccurate without using accurate and complete system knowledge; e.g., without the knowledge of substation breaker configuration and its operation during a contingency, the fault on switching devices cannot be localized precisely.
Methods based on non-graphical models (e.g., auto-regressive models, neural networks) cannot consider network topology and can result in false positive outcomes when applied to a realistic dynamic graph. Also, neighborhood based methods significantly degrade in performance when faced with high-dimensionality which is common in power systems.

To address these limitations, many works have explored the inclusion of domain knowledge into machine learning. Attempts range from creating better features \cite{hooi2018gridwatch} to enforcing constraints \cite{pmu-graph-spatial}\cite{pmu-graph-temporal}\cite{hooi2018gridwatch} (e.g., topology information, spatial correlations, etc)..

\chapter{Equivalent circuit formulation bringing opportunity for efficiency}\label{ch: ECF}

\section{A circuit viewpoint of simulation and estimation}

Both the transmission and distribution sectors of the electrical power grid inherently exhibit circuit characteristics, making circuit-based formulations naturally applicable to simulation \cite{SUGAR-pf-marko}\cite{SUGAR-pf}, estimation \cite{SUGAR-SE-Alex}\cite{SUGAR-SE-TnD}, and optimization \cite{SUGAR-opf}. However, industry-standard formulations like 'PQV' for transmission and current injection methods for distribution operate on distinct principles, prohibiting seamless analysis across transmission and distribution boundaries, particularly as distributed energy integration blurs these distinctions. Table \ref{tab: PQV vs ECF} shows the difference between the two formulations:

\begin{table}[h]
\caption{Comparison between PQV and Circuit-based Formulation}
\label{tab: PQV vs ECF}
\begin{tabular}{c|c|c}
\hline
                     & \textbf{PQV Formulation}                                                                                                  & \textbf{Circuit-based Formulation}                                                                                          \\ \hline
Coordinate           & Polar coordinate                                                                                                          & Rectangular coordinate                                                                                                       \\ \hline
State variable       & \begin{tabular}[c]{@{}c@{}}Voltage magnitude and phase angle\\ $x=[|V_1|, \theta_1, ...., |V_N|, \theta_N]$\end{tabular}  & \begin{tabular}[c]{@{}c@{}}Real and imaginary voltage\\ $x=[V_1^R, V_1^I, ..., V_N^R, V_N^I]$\end{tabular}                 \\ \hline
Governing equations  & \begin{tabular}[c]{@{}c@{}}Nonlinear power balance equations\\ (active and reactive power balance) \\ See (\ref{eq:powerbalance})\end{tabular}    & \begin{tabular}[c]{@{}c@{}}Current balance equations  \\ by Kirchhoff's current law (KCL)  \\ See (\ref{eq: kcl example start})-(\ref{eq: kcl example end})\end{tabular} \\ \hline
Nonlinear components & \begin{tabular}[c]{@{}c@{}}Transmission lines and shunt elements \\ (leads to nonlinear power flow equations)\end{tabular}  & \begin{tabular}[c]{@{}c@{}}Generation (bus with PV control) \\ and Load (bus with PQ control)\end{tabular} \\ \hline
Linear components    & \begin{tabular}[c]{@{}c@{}}Slack bus, Load (bus with PQ control), \\ Generation (bus with PV control)\end{tabular}        & \begin{tabular}[c]{@{}c@{}}Slack bus, Transmission lines, and \\ Shunt elements (admittance)\end{tabular} \\ \hline
\end{tabular}
\end{table}

In this chapter, we would like to provide a road map of adopting circuit-based formulations in situational awareness tasks, spanning simulation and estimation. We start from illustrating how a circuit viewpoint can provide new insights into the efficiency and robustness of simulation and estimation problems; so as to inspire new opportunities for advancements.

Let us start from a toy example in Figure \ref{fig: ECF, case4}. In the left figure, a power system consists of generation units, loads, and branches (transmission lines or transformers) to deliver supply to load demands. Traditional RTU devices from the SCADA system and modern PMU devices are installed across the network to monitor the system in real time. 
\begin{figure}[h]
         \centering         \includegraphics[width=\linewidth]{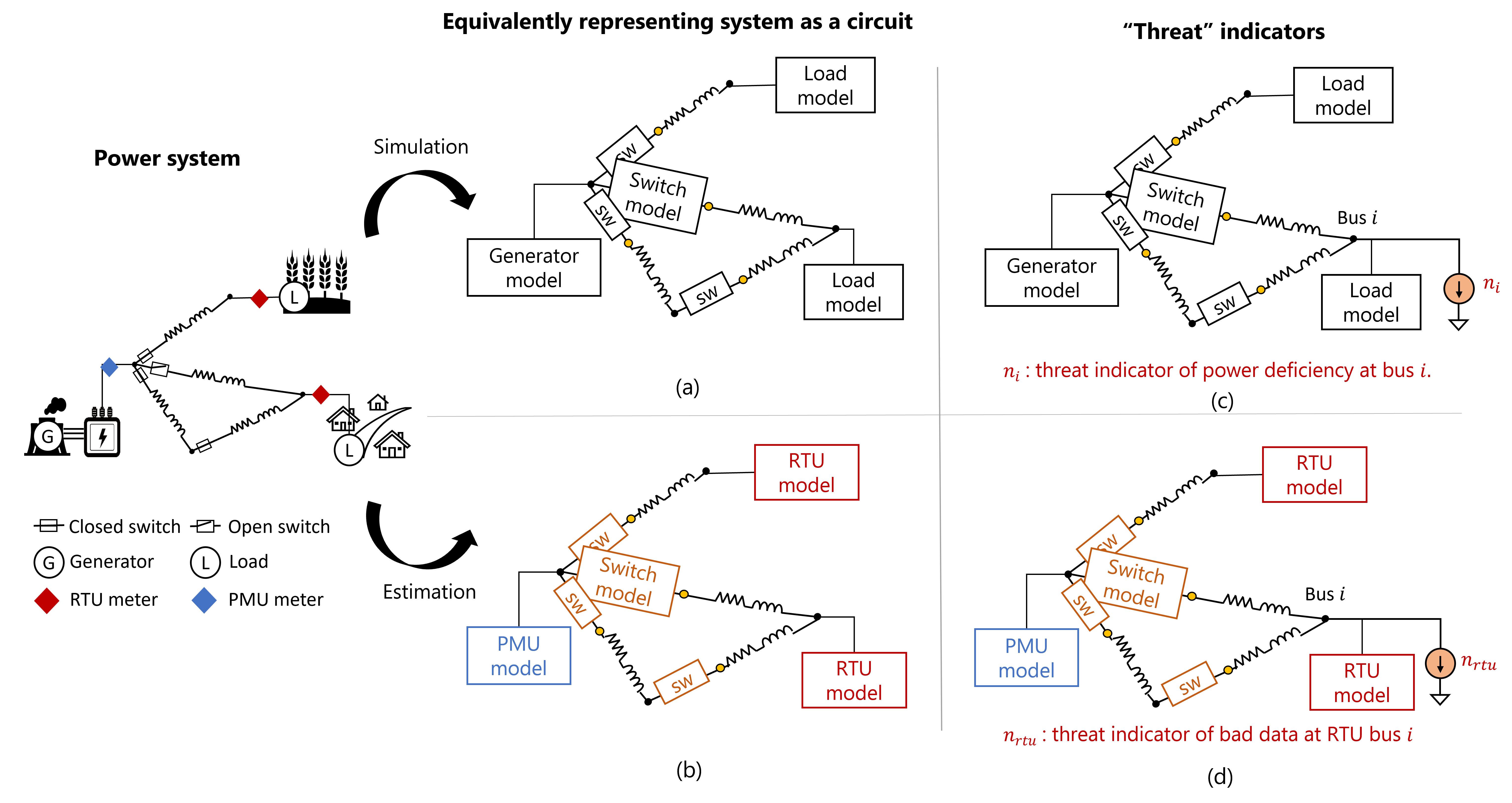}
         \caption{Circuit-based viewpoint for simulation and estimation on power system.}
         \label{fig: ECF, case4}
     \end{figure}

For numerical analysis of the system, each component / device on the system need to have models, representing either fundamental physical laws or observed behaviors:
\begin{itemize}
    \item Simulation: Each component can be modeled using physics-based principles. For steady-state analysis, loads are represented by static $PQ$ models, and generators by static $P$ and $|V|$ control. Previous work \cite{SUGAR-pf}\cite{SUGAR-opf} shows that power system components, including PQ and PV buses, can be directly mapped to equivalent circuits based on current-voltage relationships.
    \item Estimation: We propose using measurement-based models for each observed component, treating measurements as constraints on the system's operating point. For example, consider a measurement of a generator's current to be the value of 1 per unit (p.u.), it represents a constraint that the generator is equivalent to some circuit element whose current equals 1 p.u., if the measurement is accurate. Then based on substitution theorem\cite{larry_book}, this generator can be equivalently replaced by an independent current source of 1 p.u., which is a linear equivalent circuit model of the measured generator.\cite{larry_book}.
\end{itemize}

Following this idea, we can aggregate the component-wise models together, and represent the power system as a circuit system. As shown in the middle of Figure \ref{fig: ECF, case4}, both simulation and estimation tasks have similar objective of finding solutions to a feasible system:
\begin{itemize}
    \item Simulation: find state variables that can make this (circuit (a) in Figure \ref{fig: ECF, case4}) a feasible system that can deliver generation to load as pre-defined. In case of system collapse (i.e.,blackout), this (circuit system) becomes infeasible. 
    \item Estimation: find state variables that can make this (circuit (b) in Figure \ref{fig: ECF, case4}) a feasible system that is exactly consistent with what observed in data. In case of data errors, this (circuit system) becomes infeasible.
\end{itemize}

Let's first assume that a feasible system exist. Circuit-based modeling provides opportunities to advance efficiency of problem-solving. Decades of research in circuit simulation have developed mature heuristics\cite{sparsePFref15-spice} that can solve large-scale (nonlinear) circuit (with even billions of nodes) with fast speed. These heuristics includes but are not limited to variable limiting, homotopy methods, etc. For simulation problem, branches are linear with circuit-based formulation, so that nonlinearity only locally exists at bus locations of loads and generations. Prior works \cite{SUGAR-pf}\cite{SUGAR-pf} of circuit-based power flow analysis have successfully adopted these heuristics to simulate transmission and distribution systems. For estimation problem, apart from linear branch models and various heuristics, opportunities exist in improving linearity of the measurement-based models so that the entire circuit system can be highly linear. This thesis will develop measurement models to this end in Section \ref{sec: sugar meter models}, that can result in a completely linear circuit to represent the entire power system for highly-efficient steady-state estimation at a time $t$. 

Let's then discuss threat scenarios (e.g., blackout, measurement noises or errors) which makes the cases "infeasible". On a circuit system, these threats represent unwanted perturbations and spatial inconsistencies that make the circuit an infeasible system. How do we handle this? To restore feasibility, we can introduce independent current sources at nodes where nodal balance is violated. By optimally determining the location and magnitude of these current compensations, we can generate "threat indicators" for the system:

\begin{itemize}
    \item An independent current source $n_i$ acts as a threat indicator for power deficiency at node $i$. By optimizing the set ${n_1, n_2, ...}$, we can assess blackout severity and the corrective actions required at each node.
    \item The current source $n_{rtu}$ quantifies noise at RTU measurements. Optimizing ${n_{rtu}, n_{pmu}, ...}$ reveals data noise levels, enabling corrections to align with the true system state. 
\end{itemize}


Clearly, this provides opportunities to build intrinsic robustness of simulation and estimation tools by quantifying threats as indicators. To determine optimal threat levels, we can formulate simulation and estimation tools into constrained optimization problem
\begin{align}
    \min_{x,n} L(n)&\text{: a loss function to minimize threat indicators}\\
    \text{s.t.} H(x,n)=0&\text{: KCL equations of the 'circuit' system}
\end{align}

The level of robustness we can achieve depends highly on the choice of objective function. As previously defined in Problem \ref{def: sugar simulation}, minimizing the L2-norm (or weighted least squares) of threat indicators leads to  ensuring convergence and quantifying severity under blackout. This chapter extends this idea to convex error-quantified estimation, while later chapters explore sparsity-promoting objectives to enhance robustness.

In the rest of this chapter, we focus on developing a circuit-based estimation that is highly efficient with its convexity and closed-form solution. First, Section \ref{sec: sugar meter models} will develop linear circuit models for all types of measurement devices. Then, Section \ref{sec: linear circuit and KCL} maps the power system into an aggregated equivalent circuit that can capture the system's stead-state operation point at the time of measurements; and shows that affine network constraints can be written. Next, Section \ref{sec: sugarSE} formulates an convex error-quantified circuit-based estimation problem, which proves to guarantee convexity and closed-form solutions. Its advantage over traditional estimation is discussed from both the optimization and probabilistic viewpoint in Section \ref{sec: sugar SE probabilistic view}.  Finally, Section \ref{sec: sugarSE result} demonstrates its advantage in speed by compare with traditional estimation method via experiments.
The estimation problems presented in this thesis have been also applied to combined transmission and distribution systems \cite{SUGAR-SE-TnD}.

\section{Circuit-inspired linear measurement models}\label{sec: sugar meter models}
For power system estimation problems under steady state, we consider realistic grid settings with both continuous measurements from RTUs and PMUs, as well as discrete status measurements (from circuit-breakers or switching devices). We build a linear circuit model for every measured component that captures its physics at the current operating point. Such a measurement-based model is updated recursively based on the latest data so that the up-to-date system physics is captured accurately. 

Illustrating PMU and RTU measurements into these linear circuit models has been explored early in prior works of state estimation \cite{SUGAR-SE-Alex} and extended in our approaches in this thesis \cite{SUGAR_SE_L2}\cite{SUGAR_SE_WLAV}\cite{SUGAR_GSE_WLAV}.

\subsection{Remote terminal unit (RTU) model}

We start with how circuit modeling transforms nonlinear measurement relationships in a physically meaningful way for convex and linear constraint formulation.

\begin{figure}[h]
	\centering
	\includegraphics[width=0.3\linewidth]{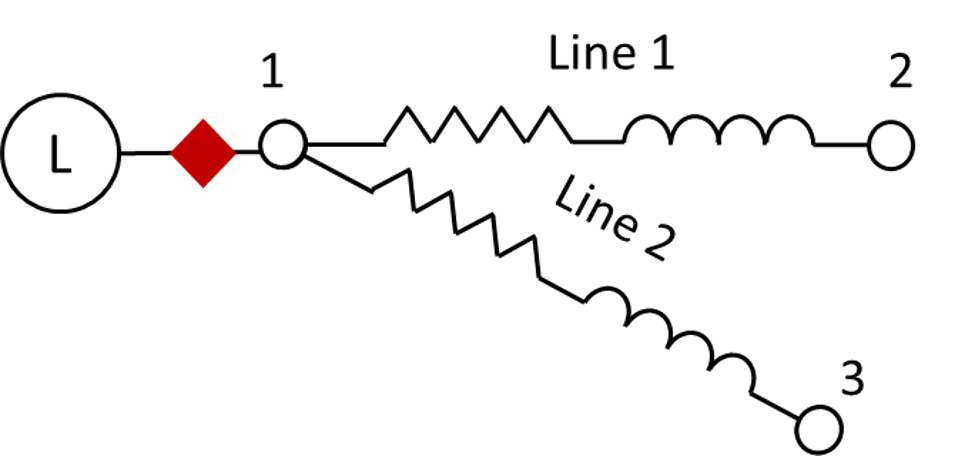}
	\caption{Suppose the power injection at a load bus is measured by an RTU device (in purple); the measurements are $P_{rtu}$, $Q_{rtu}$ and $|V|_{rtu}$.}
	\label{fig:explain rtu}
\end{figure}

RTU provides measurements of voltage magnitude $|V_{rtu}|$, real power injection $P_{rtu}$ and reactive power injection $Q_{rtu}$ at a measured bus, taking an example as in Figure \ref{fig:explain rtu}.

In traditional modeling, these power injection measurements result in nonlinear models under polar coordinate, with voltage magnitude $|V|$ and phase angle $|\theta|$ being the state variables:
\begin{derivation}[Nonlinear RTU measurement models for estimation on Figure \ref{fig:explain rtu}]  the relationship between measurements and state variables $|V|, \theta$ is written under polar coordinate:
    $$
P_{rtu} = \sum_{k\in\{2,3\}} |V_1||V_k|(G_{line(1,k)} cos\theta_{1k}+B_{line(1,k)} sin\theta_{1k}) + n_P
$$
 $$
 Q_{rtu} = \sum_{k\in\{2,3\}} |V_1||V_k|(G_{line(1,k)} sin\theta_{1k}-B_{line(1,k)} cos\theta_{1k}) + n_Q
$$
$$|V|_{rtu} = |V_1| + n_V$$
\end{derivation}

Now, we seek a linear model that can reduce the nonlinearity challenges.  
Despite no direct phasor measurements, these observations can be mapped into linear formulation \cite{SUGAR-SE-Alex} using the following relationship between bus voltages and injection currents:
\begin{align}
    I^R=\frac{P}{|V|^2}  V^R+\frac{Q}{|V|^2}  V^I\\
I^I=\frac{P}{|V|^2}  V^I-\frac{Q}{|V|^2}  V^R
\end{align}
and a linear RTU model can be developed accordingly, with by mapping original measurements to two new sensitivity parameters:
\begin{equation}
     G_{rtu}=\frac{P_{rtu}}{|V|_{rtu}^2};
     B_{rtu}=-\frac{Q_{rtu}}{|V|_{rtu}^2}
\end{equation}

\begin{figure}[h]
         \centering         \includegraphics[width=0.5\linewidth]{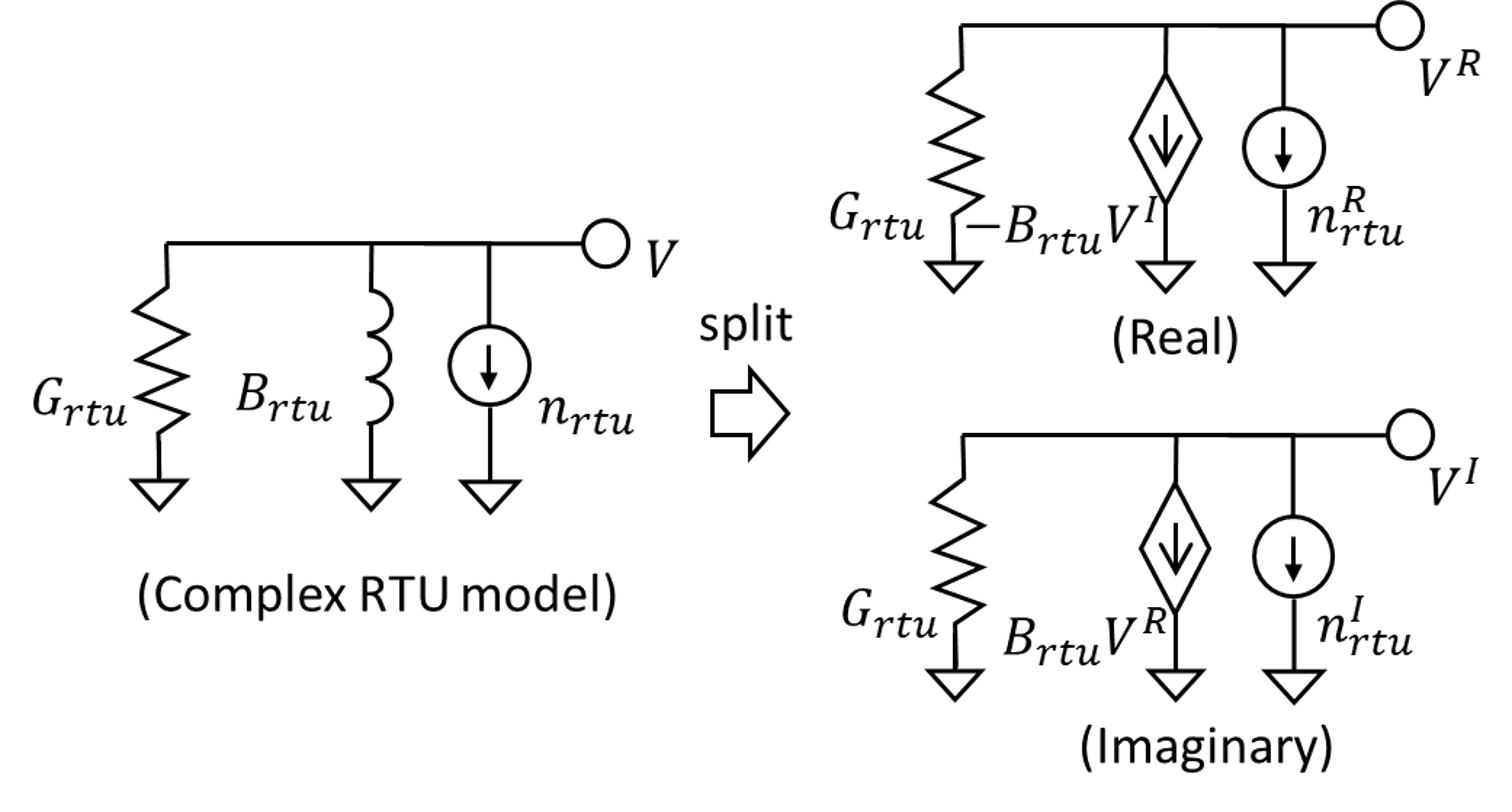}
         \caption{Linear RTU model: measurements are mapped to sensitivities.}
         \label{fig:RTU}
     \end{figure}

Considering that measurement can have noises and errors, we add an additional current source  $n_{rtu}^R,n_{rtu}^I$  to capture the possible noises, so that we have linear models:
\begin{model}[Linear RTU model] creates a linear model for convey the original RTU measurement information in a linear way under rectangular coordinate. 
    \begin{align}
I_{rtu}^R=G_{rtu}V^R+B_{rtu}V^I+n_{rtu}^R	\\
	I_{rtu}^I=G_{rtu}V^I-B_{rtu}V^R+n_{rtu}^I	
\end{align}
This model represents a linear circuit in Figure \ref{fig:RTU}. Original measurements imposes constraint that the measured component is an element whose voltage magnitude is $|V_{rtu}|$, real power is $P_{rtu}$ and reactive power is $Q_{rtu}$; whereas this new model conveys the same information in an equivalent way that the measured component is a circuit element whose (local) admittance is $G_{rtu}+jB_{rtu}$. Intuitively, the "pseudo" measurements of local equivalent admittance $G_{rtu}, B_{rtu}$ are given by a real-time measurement device.
\end{model}

Notably, one might question whether the linear $G_{rtu}$ and $B_{rtu}$ parameters used to model the RTU measurements at load buses are relaxations of the actual $P_{rtu}$ and $Q_{rtu}$ measurements. However, from an equivalent circuit perspective, we argue that our circuit modeling is \textbf{not a relaxation}. From a single measurement, one cannot gauge whether the measured load is constant power, constant impedance or ZIP load. So \textbf{all characterizations of loads are valid and equivalent representations of the current steady-state operating point}.

Upon building a linear RTU model, these $G_{rtu},B_{rtu}$ can form linear constraints for estimation using RTU data. For computational analyticity, the complex variables and functions are split into real and imaginary parts, and constraints are written separately. 

\begin{figure}[h]
         \centering         \includegraphics[width=0.8\linewidth]{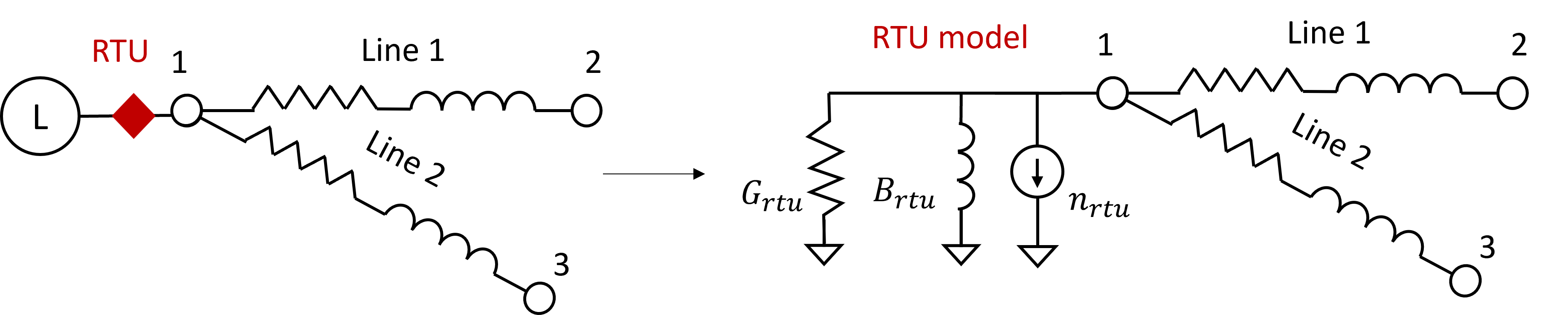}
         \caption{Connecting RTU model to the load bus: the load is now described by its "pseudo" measurements of local admittance.}
         \label{fig:explainRTU2}
\end{figure}

\begin{derivation}[Linear RTU model giving linear constraints for estimation on Figure \ref{fig:explain rtu}] With a linear RTU model conveying the measurement information, we install the RTU model to the measured load component and obtain a circuit model locally, as in Figure \ref{fig:explainRTU2}.
Then the constraints at this bus can be written as linear KCL equations under rectangular coordinate, with real and imaginary voltage $V^R, V^I$ as state variables:
\begin{align}
  G_{rtu1} V_2^R-B_{rtu1}V_2^I+n_{rtu1}^R +G_{line1}(V_1^R-V_{2}^R) - B_{line2}(V_1^I-V_{2}^I)+G_{line2}(V_1^R-V_{3}^R) - B_{line2}(V_1^I-V_{3}^I)= 0  \\
G_{rtu1}V_2^I+B_{rtu1}V_2^R+n_{rtu1}^I + G_{line1}(V_1^I-V_{2}^I) + B_{line1}(V_1^R-V_{2}^R) + G_{line2}(V_1^I-V_{3}^I) + B_{line2}(V_1^R-V_{3}^R)=0
\end{align}
\end{derivation}

\subsection{Phasor measurement unit (PMU) model}

\begin{figure}[h]
         \centering         \includegraphics[width=0.6\linewidth]{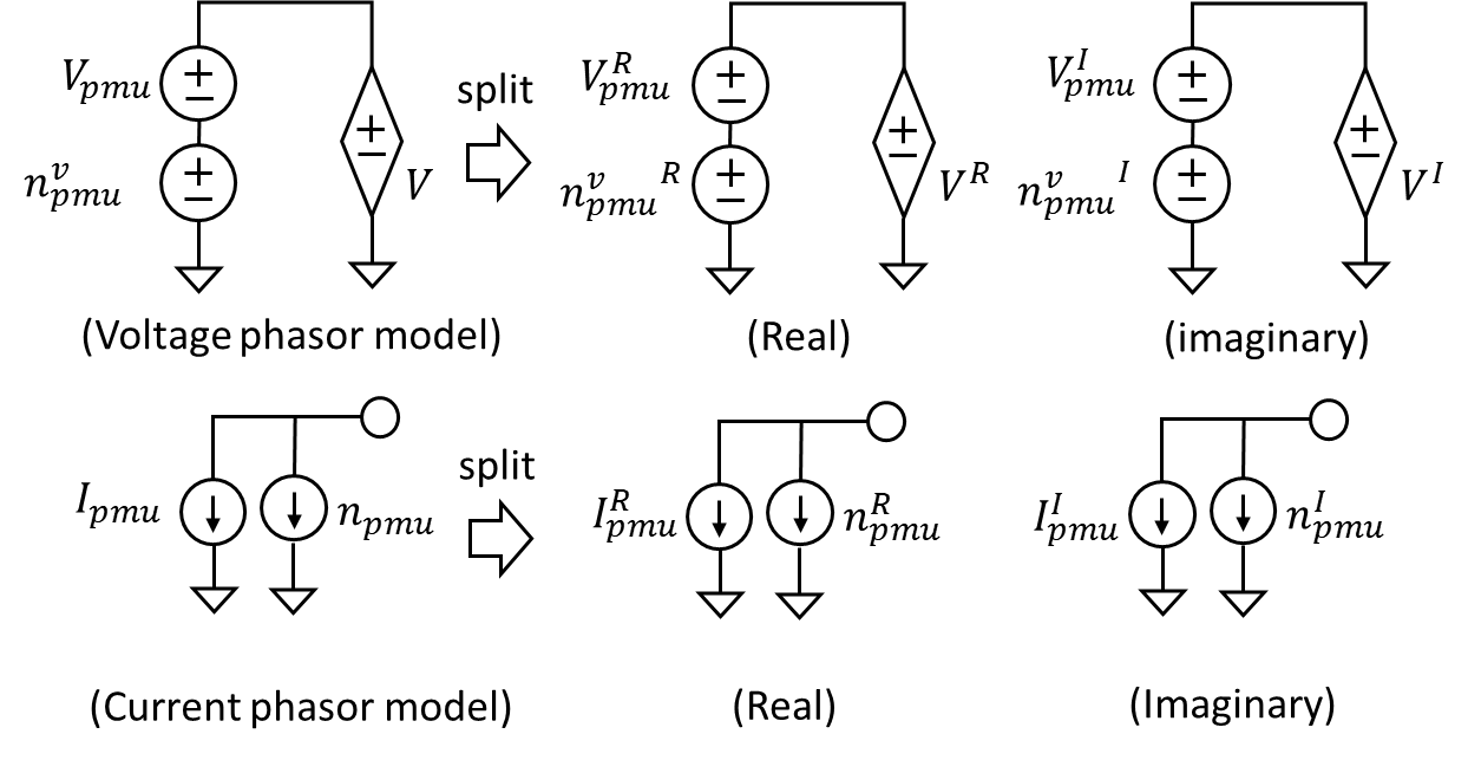}
         \caption{Linear PMU model: measurements have a linear nature under circuit-based formulation.}
         \label{fig:PMU}
     \end{figure}

As a synchronized measurement device, a PMU can provide real-time meter readings of voltage and injection current phasors in rectangular coordinates: $(V_{pmu}=V_{pmu}^{R}+jV_{pmu}^{I})$ 
 and $(I_{pmu}=I_{pmu}^{R}+jI_{pmu}^{I})$.
By substitution theorem, we can safely develop a new PMU model. The model is represented by independent current sources taking the value of the measured real and imaginary current. To further consider measurement errors such that KCL is always satisfied, we attach additional slack current sources $n_{pmu}^R,n_{pmu}^I$ to represent the measurement noise or error. Meanwhile, the observation of voltage phasor indicates the state variables $V^R,V^I$ at this bus should be close to $V_{pmu}^R,V_{pmu}^I$, with the additional independent voltage sources $n_{pmu}^R,n_{pmu}^I$ representing measurement errors. This can be illustrated in linear models:
\begin{model}[Linear PMU model]\label{cc model: PMU}
    \begin{align}
    I^R=I_{pmu}^R+n_{pmu}^R; &
    I^I=I_{pmu}^I+n_{pmu}^I\\
    V^R=V_{pmu}^R+{n_{pmu}^v}^R; & 
    V^I=V_{pmu}^I+{n_{pmu}^v}^I
\end{align}
These linear models represent linear circuits as in Figure \ref{fig:PMU}. 
\end{model}

\subsection{PMU and RTU line flow models}
As an extension, models for PMU line measurement models have been created similarly. To account for measurement errors,  models include slack variables $n_{line, pmu}^R, n_{line, pmu}^I, n_{line, rtu}^R, n_{line, rtu}^I$ to capture the measurement error. 

For example, a PMU device can also measure flow $I_{line, pmu}^R+jI_{line, pmu}^I$ on a line adjacent to the bus. We establish linear models in a way consistent with injection models. 
\begin{model}[Linear PMU line flow models]  
    \begin{align}
     I_{line}^R=I_{line, pmu}^R+n_{line, pmu}^R\\
    I_{line}^I=I_{line, pmu}^I+n_{line, pmu}^I
\end{align}
This measurement-based model is implemented in Figure \ref{fig: line PMU, RTU} as an additional linear control circuit coupled with the main circuit.
\end{model}
As in Figure \ref{fig: line PMU, RTU}, the control circuits are coupled with the main circuit by controlled sources, e.g., $I_{line}=Y_{line}(V_i-V_j)$ for line $(i,j)$ represents a voltage controlled current sources where the source $I_{line}$ on the control circuit is controlled by voltage $V_i$ and $V_j$ on the main circuit. RTU flow model is also developed similarly, where model parameters $G_{line, rtu}=\frac{P_{line,rtu}}{|V_{rtu}|^2},B_{line, rtu}=-\frac{Q_{line,rtu}}{|V_{rtu}|^2}$.

\begin{figure}[h]
	\centering
	\includegraphics[width=0.7\linewidth]{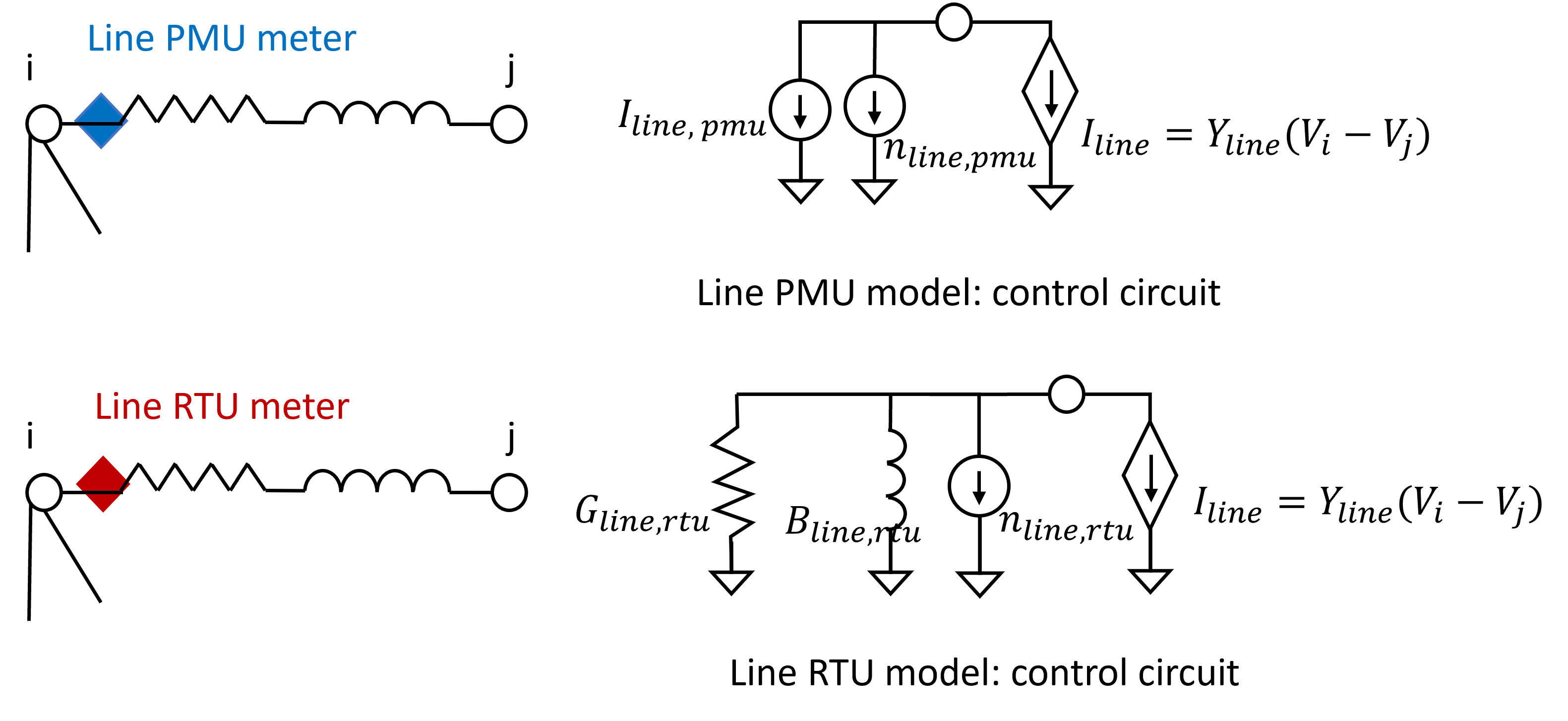}
	\caption{Linear model for line flow meters: implemented as control circuits coupled with the main circuit via voltage controlled current sources $I_{line}$.}
	\label{fig: line PMU, RTU}
\end{figure}

\subsection{Linear models for Switching Devices}
To perform GSE on an AC-network constrained NB model, this paper introduces two new models: i) open switch and ii) closed switch. The model considers the possibility of \textit{wrong} switch status (i.e., open switch reported as close and vice versa) and includes noise terms to estimate the correct switch status.

\begin{figure}[h]
	\centering
	\includegraphics[width=0.5\linewidth]{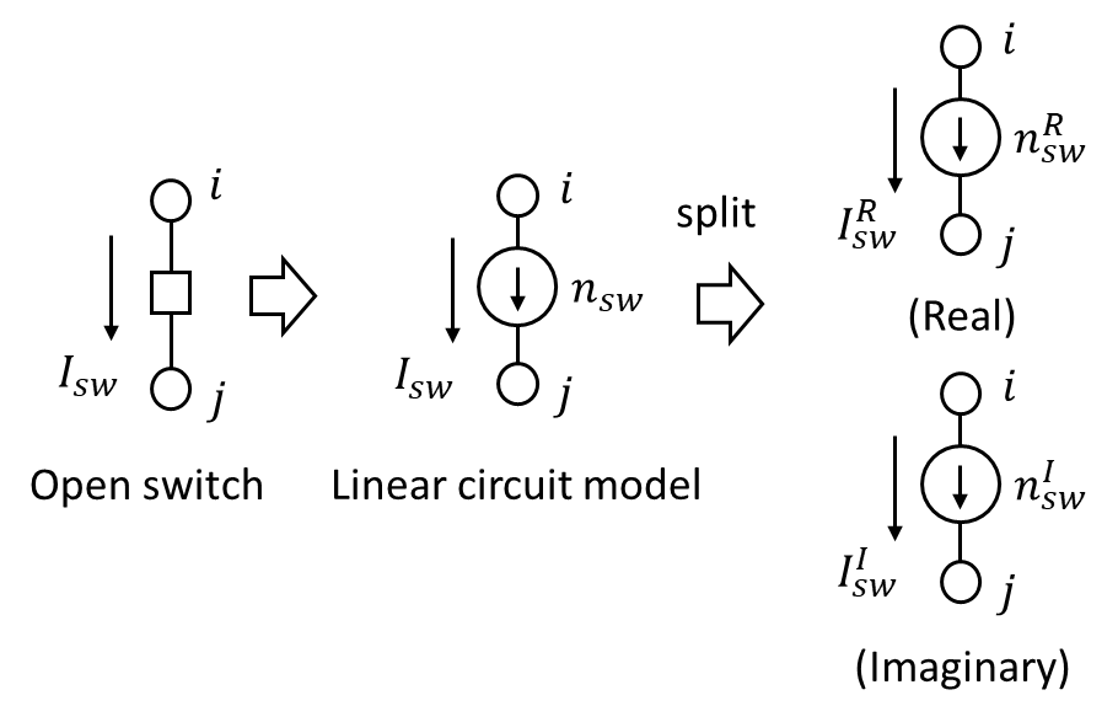}
	\caption{Open switch model: $n_{sw}$ close to zero if the status is correct; $n_{sw}$ compensates the current flow on the branch if the status is wrong.}
	\label{fig:open sw model}
\end{figure}

An open switch is simply modeled by an open circuit and as such no current can flow through it, i.e., the total current flow $I_{sw}=I_{sw}^R+jI_{sw}^I=0$. However, to account for a possibly wrong status, we add a slack current source in parallel, i.e. a noise term $n_{sw}=n_{sw}^R+jn_{sw}^I$, to compensate for the current that would otherwise flow through the switch in case it was actually closed. 
\begin{model}[Linear open switch model]Figure \ref{fig:open sw model} and (\ref{eq:open sw}) show the model.
    \begin{align}\label{eq:open sw}
    I_{sw}^R=n_{sw}^R\\ I_{sw}^I=n_{sw}^I
\end{align}
\end{model}

A closed switch is modeled as a low impedance branch (reactance $x_{sw} \approx 0.0001$ p.u.), since a closed switch is ideally a short circuit with zero voltage drop across it. Similarly, to account for possibly wrong status, we add a slack current source (i.e., a 'noise' term) in parallel. In case the closed switch is actually open, the noise term $n_{sw}$ will provide sufficient current to nullify the current flowing through the closed switch model, such that the total current flow between the from and to node of the switch is zero, effectively representing an open switch. 

\begin{figure}[h]
	\centering
	\includegraphics[width=0.65\linewidth]{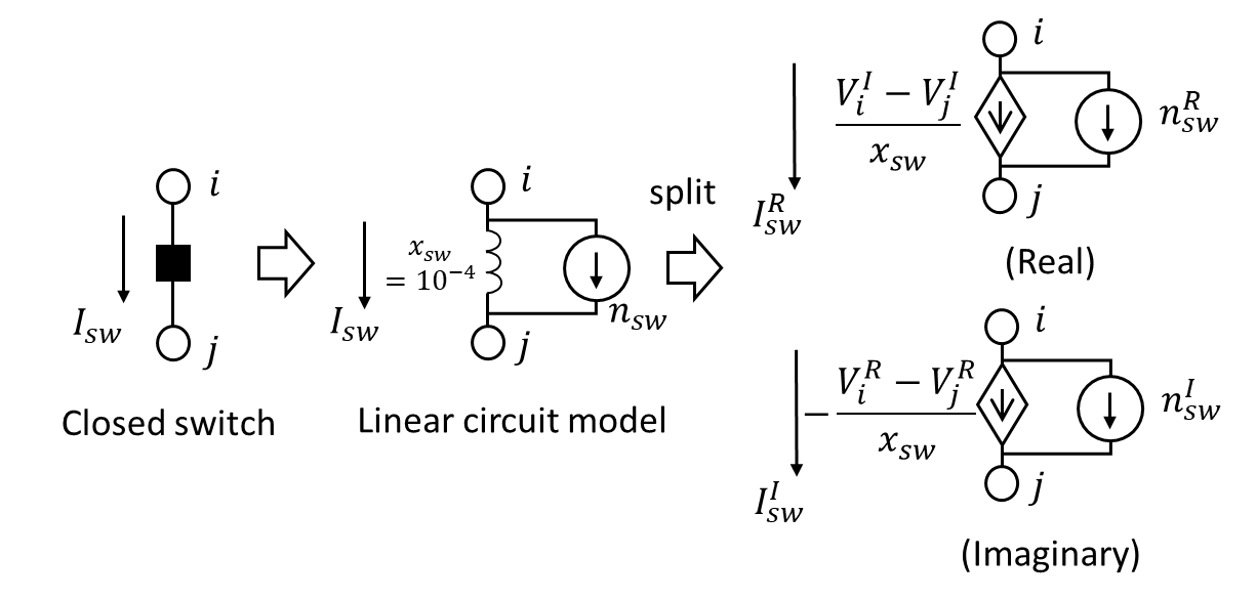}
	\caption{Closed switch model: $n_{sw}$ close to zero if the status is correct; $n_{sw}$ will offset the current flow on the branch if the status is wrong. $x_{sw}$ is in p.u.}
	\label{fig:closed sw model}
\end{figure}

\begin{model}[Linear closed switch model] Figure \ref{fig:closed sw model} shows the closed switch model, which is mathematically expressed in (\ref{eq:closed sw}).
    \begin{equation}\label{eq:closed sw}
    I_{sw}^R=\frac{1}{x_{sw}}(V_i^I-V_j^I) + n_{sw}^R\\  I_{sw}^I=-\frac{1}{x_{sw}}(V_i^R-V_j^R) + n_{sw}^I
\end{equation}
\end{model}

\subsection{Other grid devices}\label{sec: physical_models}

Physical devices such as transformers, lines and shunts are linear and their derivations and construction are covered in detail in \cite{sugar-powerflow-amrit}. Any nonlinear physical model (i.e., load or generation), for the purposes of estimation, is replaced by its equivalent linear circuit model from the measurement devices described in Section above following the substitution theorem in circuit-theory \cite{larry_book}.

\section{Forming equivalent circuit network for estimation}\label{sec: linear circuit and KCL}

\begin{figure}[h]
	\centering	\includegraphics[width=0.95\linewidth]{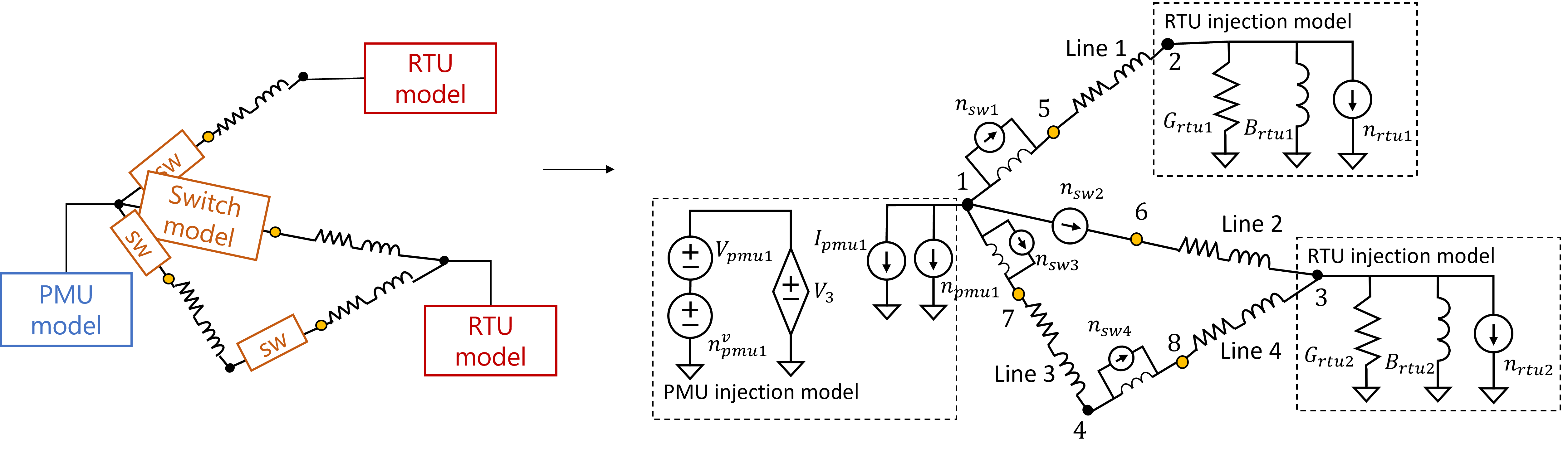}
	\caption{Equivalently representing a power system steady-state operating point with an aggregated circuit model.}
	\label{fig: case4 convert2EC}
\end{figure}

With the circuit models described above,  Figure \ref{fig: case4 convert2EC} shows the equivalent circuit representation of a power system, where we replace all switches and measured components with the established linear circuit models. The resulting aggregated circuit consisting of the main circuit and a set of control circuits captures the information from measurement data from an equivalent circuit-theoretic viewpoint. The main circuit captures the \textit{non-redundant set} of measurements including the AC network constraints at zero-injection nodes, whereas the set of control circuits captures the information from remaining redundant measurement data.

In this way, the entire system is mapped to a linear circuit whose network constraints represented by Kirchhoff's current law (KCL) on all nodes, are a set of affine constraints in the form of \begin{equation}
        [Y,B]\begin{bmatrix}
    x\\n
\end{bmatrix} = J
\end{equation}

Taking the simple 4-bus system in Figure \ref{fig: case4 convert2EC} as an example, the affine constraints are KCL equations derived from Modified Nodal Analysis:

\begin{derivation}[Affine network constraints a 4-bus toy example in Figure \ref{fig: ECF, case4}]

 \noindent Node 1: (connects PMU and switches)
\begin{equation}
\begin{split}
    I_{pmu1}^R + n_{pmu1}^R +I_{sw1}^R+ I_{sw2}^R+ I_{sw3}^R =0\label{eq: kcl example start}\\
     I_{pmu1}^I + n_{pmu1}^I +I_{sw1}^I+I_{sw2}^I+ I_{sw3}^I =0
\end{split}
\end{equation}
PMU voltage at node 1: (control circuit)
\begin{equation}
\begin{split}
    V_{pmu1}^R + n_{pmu1}^R - V_1^R = 0\\
    V_{pmu1}^I + n_{pmu1}^I - V_1^I = 0
\end{split}
\end{equation}
Node 2: (connects RTU and transmission line 1 whose admittance is $G_{line1}+jB_{line1}$)
\begin{equation}
\begin{split}
    G_{rtu1} V_2^R-B_{rtu1}V_2^I+n_{rtu1}^R \\+ G_{line1}(V_2^R-V_5^R) - B_{line1}(V_2^I-V_5^I) = 0\\
G_{rtu1}V_2^I+B_{rtu1}V_2^R+n_{rtu1}^I \\+  G_{line1}(V_2^I-V_5^I) + B_{line1)}(V_2^R-V_5^R)= 0
\end{split}
\end{equation}
Node 3: (connects RTU and transmission line 2 and line 4)
\begin{equation}
\begin{split}
    G_{rtu2} V_3^R-B_{rtu2}V_3^I+n_{rtu2}^R \\+ G_{line2}(V_3^R-V_6^R) - B_{line2}(V_3^I-V_6^I) \\
    + G_{line4}(V_3^R-V_8^R) - B_{line4}(V_3^I-V_8^I)
    = 0\\
G_{rtu2}V_2^I+B_{rtu1}V_2^R+n_{rtu1}^I \\+  G_{line2}(V_3^I-V_6^I) + B_{line2)}(V_3^R-V_6^R) \\+  G_{line4}(V_3^I-V_8^I) + B_{line4)}(V_3^R-V_8^R)= 0
\end{split}
\end{equation}
Node 4: (zero-injection node, connects line 3 and switch)
\begin{equation}
\begin{split}
    I_{sw4}^R+ G_{line3}(V_4^R-V_7^R) - B_{line3}(V_4^I-V_7^I) = 0\\
I_{sw4}^I +  G_{line3}(V_4^I-V_7^I) + B_{line3)}(V_4^R-V_7^R)= 0
\end{split}
\end{equation}
Pseudo node 6: (connects line 2 and open switch $sw2$)
\begin{equation}
\begin{split}
    -I_{sw2}^R+ G_{line2}(V_6^R-V_3^R) - B_{line2}(V_6^I-V_3^I) = 0\\
-I_{sw2}^I +  G_{line2}(V_6^I-V_3^I) + B_{line2)}(V_6^R-V_3^R)= 0
\end{split}
\end{equation}
Pseudo node $i=5,7,8$: (connect closed switch $swk$ and line k, line is between (i,j))
\begin{equation}
\begin{split}
    -I_{swk}^R+ G_{linek}(V_i^R-V_j^R) - B_{linek}(V_i^I-V_j^I) = 0\\
-I_{swk}^I +  G_{linek}(V_i^I-V_j^I) + B_{linek)}(V_i^R-V_j^R)= 0
\end{split}
\end{equation}
Open switches $sw_2$:  (control circuit):
\begin{equation}
    I_{sw2}^R = n_{sw2}^R, I_{sw2}^I = n_{sw2}^I
\end{equation}
Closed switches $sw_1,sw_3,sw_4:$ (control circuit for $swk=(i,j)$, $x_{sw}=10^{-4}$)
\begin{equation}
\begin{split}
    I_{swk}^R = \frac{1}{x_{sw}}(V_i^I-V_j^I) + n_{swk}^R,\\ 
    I_{swk}^I = -\frac{1}{x_{sw}}(V_i^R-V_j^R) + n_{swk}^I\label{eq: kcl example end}
\end{split}
\end{equation}
    
\end{derivation}

\section{Convex and error-quantified estimation}\label{sec: sugarSE}

Now that the steady-state condition of power system can be equivalently represented as a linear aggregated circuit, giving affine constraints in the form of $[Y,B]\begin{bmatrix}
    x\\n
\end{bmatrix} = J$, 
the voltage state variables can be obtained by solving the network equations of the constructed linear circuit. Mathematically, there exist infinitely many solutions as the introduction of slack variables $n^R, n^I$ results in an under-determined system of equations. Therefore, we next formulate an optimization problem to estimate a unique grid state, which provides a good estimate of grid states. Without any prior knowledge on any specific structures in the noises, we can reasonably assume that measurement noises or errors satisfy Gaussian distribution. And thus we can formulate the optimization with a weighted least-square (WLS) algorithm below. Section \ref{sec: sugar SE probabilistic view} further provides a probabilistic viewpoint of the method showing that the method is equivalently an M-estimator with Gaussian likelihood functions.

\begin{problem}[Error-quantified circuit-based (generalized) state estimation]\label{def: sugar-SE} Given a real-time set of continues measurements $z_{cont}$, status measurements $z_{sw}$ on switch set $\{sw_1,...sw_K\}$, the estimation problem solves state $x$ with measurement noise (error) quantified in $n=[n_{pmu}, n_{rtu}, n_{sw}]$. This is achieved by a weighted least squares (WLS) estimation problem subject to AC-network constraints at all nodes:
\begin{equation}
    \min_{x,n} \frac{1}{2}n^TW n\notag\\ 
\end{equation}
    \text{s.t. (linear) KCL equations at all nodes: } 
    \begin{equation}
        [Y,B]\begin{bmatrix}
    x\\n
\end{bmatrix} = J \label{eq: sugar SE}
    \end{equation}
This is a Quadratic programming problem with closed-form solution, with $\lambda$ being the vector of Lagrangian multipliers for the equality constraints in (\ref{eq: sugar SE}):
\begin{equation}
    \label{sol: sugarSE closed form solution}
    \begin{bmatrix}
    x^*\\n^*\\\lambda^*
    \end{bmatrix} = \begin{bmatrix}
    Y& B& 0 \\0 & 0 & Y^T \\0 & W & B^T\\
    \end{bmatrix}^{-1} \begin{bmatrix}
    J\\0\\0
    \end{bmatrix}
\end{equation}
Notably, this problem definition applies to both robust state estimation (switch set is empty) and generalized state estimation (switch set is non-empty). 
\end{problem}

Weight selection of the estimator is discussed in the later Chapter (see Section \ref{sec: sensitivity}) together with the robust estimator development. 

Due to the affine network constraints and quadratic objective function, the circuit-based estimation problem is proved to be a convex problem, or more specifically, an equality-constrained quadratic programming (QP) that always converges to a global optimum of the loss function. The closed-form solution to this problem can be derived as follows:

\begin{derivation}[Closed-form solution to the quadratic programming problem for error-quantified circuit-based estimation]
The quadratic programming problem defined can be solved with the Lagrangian method. 
We introduce Lagrangian multipliers $\lambda$ and the Lagrangian function can be written as:
\begin{equation}
    L(x,n,\lambda) = \frac{1}{2}n^TWn+\lambda^T(Yx+Bn-J)
\end{equation}
Then calculating the partial derivatives with respect to $[x,n,\lambda]$, we obtain the first-order optimality conditions:
\begin{align}
    \frac{\partial L}{\partial\lambda} = 0 &\Rightarrow Yx+Bn=J \\
    \frac{\partial L}{\partial x} = 0 &\Rightarrow Y^T\lambda = 0\\
    \frac{\partial L}{\partial n} = 0 &\Rightarrow Wn + B^T\lambda = 0  
\end{align}
Rewriting the equations in matrix form, we have that $x^*$ is the optimal solution if and only if there exists a $\lambda^*$ such that:
\begin{equation}
     \begin{bmatrix}
    Y& B& 0 \\0 & 0 & Y^T \\0 & W & B^T\\
    \end{bmatrix}\begin{bmatrix}
    x^*\\n^*\\\lambda^*
    \end{bmatrix}  = \begin{bmatrix}
    J\\0\\0
    \end{bmatrix}
    \label{eq: sugerSE optimality}
\end{equation}
Given matrix $Y$ and $W$ are non-singular, the optimal solution can be written in closed-form as 
\begin{equation}
    \begin{bmatrix}
    x^*\\n^*\\\lambda^*
    \end{bmatrix} = \begin{bmatrix}
    Y& B& 0 \\0 & 0 & Y^T \\0 & W & B^T\\
    \end{bmatrix}^{-1} \begin{bmatrix}
    J\\0\\0
    \end{bmatrix}
\end{equation}
\end{derivation}

\section{Discussion from an optimization and probabilistic viewpoint: circuit-based estimator better than MLE} \label{sec: sugar SE probabilistic view}

Next, we compare the traditional estimation method  (Problem \ref{def: trad SE}) and the circuit-based estimation (Problem \ref{def: sugar-SE}) from an optimization and probabilistic view. We discuss based on the assumption that the measurement errors satisfy independent Gaussian distribution $n_i\sim \bm{N}(0,\sigma_i^2)$. And for simplicity, we look at the scenario with an empty switch set (reducing problems to state estimation instead of generalized state estimation), so that we can better distinguish zero-injection buses with the other bus locations where a measurement model is connected.

If we converge to the global optima of the defined non-convex optimization in first place, solution of the traditional estimation method (Problem \ref{def: trad SE}) is a maximum likelihood estimation (MLE), with its Lagrangian function equal to a log-likelihood:
\begin{equation}
    L_{\text{traditional estimation}} = \sum_i w_i(z_i-f_i(x))^2
\end{equation}

\noindent Mathematically, MLE finds the model that is most likely to generate the data. This intrinsic property of MLE makes the traditional WLS method (a classical MLE) purely measurement-dependent: it will always make estimations purely based on measurements.

In contrast, the circuit-based estimation (Problem \ref{def: sugar-SE}) leads to a slightly different Lagrangian. According to the affine constraints $Yx+Bn=J$, the error terms in $n$ can be expressed linearly with $x$. Thus, we can safely replace $n$ with its linear function of $x$. Then the circuit-based estimation problem is translated to the following equivalent form:
\begin{align}
    &\min_{x} \sum_{i\in \text{measurement buses}} w_i||J_i-Y_ix||_2^2\notag\\
    \text{s.t. }
    & Y_ix=0, \forall i\in \text{zero-injection buses}
\end{align} 
\noindent where $Y_i$ denote the rows of matrix $Y$ that correspond to the bus $i$ on the circuit model, similarly for $J_i$. And more precisely, the set of zero-injection buses here are not necessarily equal to the power system's zero-injection buses which typically means buses connected with no generation or load, but only branches. Instead, here they represent aggregated circuit's zero-injection buses. They should not connect with any PMU, RTU or switch measurement model, otherwise independent current injection $n_i$ exit. But if we narrow down to the state estimation problem, where the switch set is empty, the two sets are typically equal. 

This formulation provides a clear layout for comparing the traditional and circuit-based methods:

\textbf{Optimization view:} the circuit-based method is mathematically equivalent to a traditional WLS method plus some additional constraints. These constraints are independent of measurement data and are accurate system topology information that reflects the network balance on zero-injection buses (and also any unmeasured buses with forecasted pseudo-measurements). Mathematically, the additional constraints imposed on the WLS problem shrinks the feasible space to a smaller physical region that contains an optimal solution satisfying network balance (at zero-injection buses). As the system grows larger with more zero-injection buses and other pseudo-measurements, the circuit-based method imposes an increased number of constraints to effectively localize the feasible space to a physical region. Consequently, the solution ends up being more physically meaningful. 

And we can further write out the Lagrangian function of the circuit-based method:
\begin{equation}
    L_\text{circuit-based estimation} = \sum_{i\in \text{measurement buses}} w_i||J_i-Y_ix||_2^2 + \sum_{j\in\text{zero-injection buses}}\lambda_jY_jx
\end{equation}

\textbf{Probabilistic view:} circuit-based estimation maximizes a log likelihood plus a topology-based term. This additional term serves to include some accurate prior knowledge into the estimate, which is conceptually similar to a Bayesian treatment. 
In general, in the non-Gaussian case, the topology-based terms may be viewed as a regularized M-estimator \cite{M-estimator} that promotes physically relevant solutions (estimates). Thus, compared with the traditional one which is an MLE, the circuit-based method turns out to consider more physics and avoid extreme conclusions like those in (non-regularized) MLE. Importantly, this viewpoint suggests a potential advantage for large-scale systems and distribution networks. As the system grows larger, the increased incorporation of accurate real-world system physics will make more contributions to resulting in physically meaningful solutions.

\section{Experiments to evaluate circuit-based estimation: highly efficient with closed-form solution}\label{sec: sugarSE result}

The proposed circuit-based estimation  applies to both transmission and distribution network \cite{SUGAR-SE-TnD}, and works for both state estimation \cite{SUGAR_SE_WLAV} and generalized state estimation \cite{SUGAR_GSE_WLAV}. 
In this Section, focus on evaluating efficiency instead of robustness. So, assumption is made that no bad data or topology errors exist. We  experimentally compare the proposed circuit-based estimators with traditional estimator. We aim to demonstrate high performance of the our approach based on speed and accuracy:
\begin{itemize}
    \item \textbf{Speed}: we evaluate the number of iterations each method takes to converge. Since the traditional method is developed in MATPOWER\cite{matpower} on Matlab, whereas our circuit-based approach is implemented on Python, it could be unfair to compare the work time directly. Instead, the number of iterations is a reasonable metric because the linear or linearized equations solved in the two methods have comparable size, so that computation complexity of one iteration is considered to be at the same order.  
    \item \textbf{Accuracy}: we evaluate the accuracy of voltage solutions by calculating the root mean squared error (RMSE) from the true solution:
    \begin{equation}
        RMSE=\sqrt{\frac{||\hat{x}-x_{true}||_2^2}{N_{bus}}}
    \end{equation}
\end{itemize}

And here are the assumptions and settings for creating the experiments:
\begin{itemize}
    \item \textbf{Traditional data}: we create traditional RTU measurements on each injection bus, measuring bus voltage magnitude $|V|$, real power injection $P$ and reactive power injection $Q$
    \item \textbf{Gaussian measurement error}: we create measurements by adding Gaussian noise to the ground truth solutions obtained by power flow simulation
    \item \textbf{Good data quality}: no bad data or topology errors exist. Since this Chapter focuses on efficiency instead of robustness, the proposed method in this Chapter is not robust to data errors. Later Chapters further extend the method to a robust estimator, and evaluate bad data and topology errors. Since topology is always correct, this Section experiments on state estimation only, instead of generalized state estimation.
\end{itemize}

Table \ref{tab: sugarSE results, speed} shows the results on power system cases of different sizes:

\begin{table}[h]
\caption{Result of state estimation: traditional VS circuit-based method}
\label{tab: sugarSE results, speed}
\begin{tabular}{c|ccc|ccc}
\hline
\multirow{2}{*}{\textbf{Case name}} & \multicolumn{3}{c|}{\textbf{Num of iterations to converge}}                                                                                                                                                                 & \multicolumn{3}{c}{\textbf{Estimation error}}                                                                                                                                                                              \\ \cline{2-7} 
                                    & \multicolumn{1}{c|}{\textbf{\begin{tabular}[c]{@{}c@{}}Traditional,\\ case start\end{tabular}}} & \multicolumn{1}{c|}{\textbf{\begin{tabular}[c]{@{}c@{}}Traditional,\\ flat start\end{tabular}}} & \textbf{Circuit-based} & \multicolumn{1}{c|}{\textbf{\begin{tabular}[c]{@{}c@{}}Traditional,\\ case start\end{tabular}}} & \multicolumn{1}{c|}{\textbf{\begin{tabular}[c]{@{}c@{}}Traditional,\\ flat start\end{tabular}}} & \textbf{Circuit-based} \\ \hline
\textbf{case14}                     & \multicolumn{1}{c|}{4}                                                                           & \multicolumn{1}{c|}{5}                                                                          & 1                      & \multicolumn{1}{c|}{0.00578}                                                                    & \multicolumn{1}{c|}{0.00578}                                                                    & 0.00062                \\ \hline
\textbf{case118}                    & \multicolumn{1}{c|}{4}                                                                           & \multicolumn{1}{c|}{6}                                                                          & 1                      & \multicolumn{1}{c|}{0.0035}                                                                     & \multicolumn{1}{c|}{0.0035}                                                                     & 0.00348                \\ \hline
\textbf{case2383wp}                 & \multicolumn{1}{c|}{\color{red}diverge}                                                                     & \multicolumn{1}{c|}{6}                                                                          & 1                      & \multicolumn{1}{c|}{\color{red}diverge}                                                                          & \multicolumn{1}{c|}{0.01317}                                                                    & 0.00139                \\ \hline
\textbf{case3375wp}                 & \multicolumn{1}{c|}{5}                                                                           & \multicolumn{1}{c|}{\color{red}diverge}                                                                    & 1                      & \multicolumn{1}{c|}{0.04929}                                                                    & \multicolumn{1}{c|}{\color{red}diverge}                                                                          & 0.00152                \\ \hline
\textbf{case6468rte}                & \multicolumn{1}{c|}{7}                                                                           & \multicolumn{1}{c|}{\color{red}diverge}                                                                    & 1                      & \multicolumn{1}{c|}{0.04304}                                                                    & \multicolumn{1}{c|}{\color{red}diverge}                                                                          & 0.00793                \\ \hline
\textbf{case9241pegase}             & \multicolumn{1}{c|}{11}                                                                          & \multicolumn{1}{c|}{\color{red}diverge}                                                                    & 1                      & \multicolumn{1}{c|}{0.06532}                                                                    & \multicolumn{1}{c|}{\color{red}diverge}                                                                          & 0.01248                \\ \hline
\textbf{ACTIVSg25k}                 & \multicolumn{1}{c|}{\color{red}diverge}                                                                     & \multicolumn{1}{c|}{\color{red}diverge}                                                                    & 1                      & \multicolumn{1}{c|}{\color{red}diverge}                                                                          & \multicolumn{1}{c|}{\color{red}diverge}                                                                          & 0.00371                \\ \hline
\end{tabular}
\end{table}

The divergences in the traditional WLS-based method reflect that traditional iterative method relies highly on good initial guess. Clearly, the different starting points lead to different solution trajectories that vary in number of iterations and convergence. As results show, starting from an initial guess from the input case file is better than flat start with 1 p.u. for all buses, but neither initialization assures convergence for all of our examples. Especially when the cases become larger, the traditional algorithm, without proper initialization, is very susceptible to divergence. 

Our circuit-based method, in contrast, doesn’t need any initial guess or iterative updates and obtains a guaranteed global optimal solution in 1 step, resolving the convergence issues successfully. Particularly for larger cases, the guaranteed convergence and reduced runtime is extremely beneficial. Also, by comparison of residual metric res, we can see that our circuit-based method better minimizes the objective function.

Furthermore, comparison of accuracy by $RMSE$ shows that our circuit-based estimator reaches more accurate estimates. Specifically, the lower $RMSE$ validates that our adding additional system knowledge in the form of zero-injection bus constraints produces a solution closer to the true operating point, enabling more reliable control actions to be made.

\chapter{Sparse optimization and intrinsic robustness in simulation and estimation}\label{ch: physics}
\section{Sparse optimization and robustness}

Sparse optimization is able to provide simple and structured vector or matrix values,  where most entries are zero, but only a few are non-zero.  This is usually achieved by either adding (L1) regularization or imposing additional constraints to encourage sparsity in the solution. The LASSO method \cite{sparsePFref10-lasso} has been popular in explaining a signal as a sparse weighted sum of basis features. And weighted least absolute value (WLAV) methods have also been widely used for robust estimation\cite{robust-estimation-Huber}.  
The idea was first  introduced by Peter J. Huber\cite{robust-estimation-Huber}\cite{robust-regression-Huber} to apply L1 norm-based method as alternatives to least squares estimators for better handling outliers and non-Gaussian data distributions.

Enforcing sparse structures can be helpful in engineering applications. 
For system control and optimization, we sometimes prefer easy actions taken at a few critical locations instead of all, which represents sparse recommendations. For signal processing, we make a selection of the most relevant features,  which represents a sparse recommendation or explanation. Moreover, when performing estimation or data processing problems, we are often encountered with a sparse distribution of outliers or errors, which corresponds to a sparse explanation.  
Such kind of sparse recommendations and sparse explanations are ubiquitous to a real-world system.  A famous problem solving principle called Occam’s Razor stated that the simplest explanation is usually the best one.

\begin{figure*}[h]
     \centering
     \begin{subfigure}[b]{0.49\linewidth}
         \centering         \includegraphics[width=0.8\textwidth]{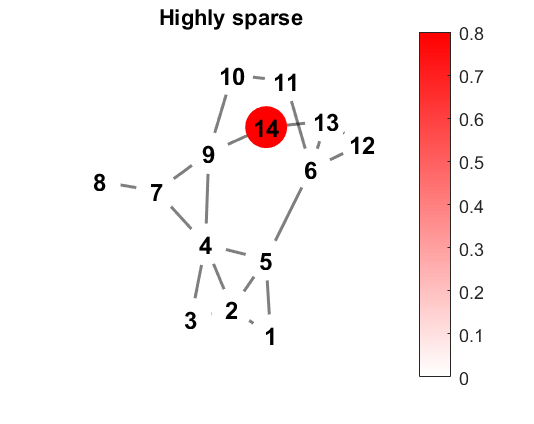}
         \caption{Simulation with sparse infeasibility indicators to pinpoint a dominant source of blackout: a sparse explanation of failure as well as sparse recommendation of fix.}
         \label{fig: sparsePF graph illu}
     \end{subfigure}
     \hfill
     \begin{subfigure}[b]{0.49\linewidth}
         \centering         \includegraphics[width=0.95\textwidth]{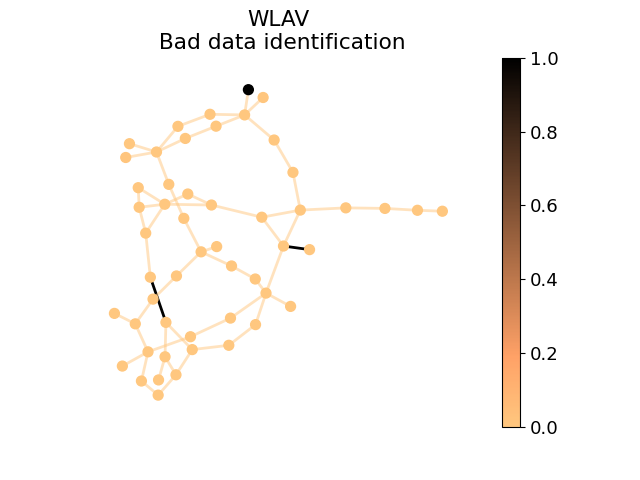}
         \caption{Estimation with sparse error indicators to pinpoint 1 bad data and 2 topology errors: a sparse explanation of anomalous data.}
         \label{fig: sparse bdd graph illu}
     \end{subfigure}
     \caption{Toy examples: sparse optimization adding robustness to simulation and estimation. Sparse threat indicators pinpoint dominant threat sources and suggest actionable corrections.}
        \label{fig: sparse threat indicators}
 \end{figure*}

In this Section, we earlier simulators and estimators to further advance their intrinsic robustness via sparse optimization. The key strategy is to exploit sparse structures in the "threat indicators" of circuit-based simulation and estimation problems. The resulting sparse threat indicators are able to pinpoint critical failure and error sources and make analytical tools immune to these threats. Figure \ref{fig: sparse threat indicators} gives illustrative examples of our work. 

In the remainder of this Section, we first develop robust simulator by enforcing sparse infeasibility indicators in Section \ref{sec: sparsePF}; and then develop robust estimator by encouraging sparse error indicators in Section \ref{sec: ckt-GSE}.

\section{Robust simulator with sparse infeasibility indicators}\label{sec: sparsePF}

Methods based on infeasibility-quantified simulation as defined in Problem Definition \ref{def: sugar simulation} can converge for a collapsed grid state. However, they do not provide a specific cause of power outage, nor do they identify localized locations that are disrupting system security and robustness. In most situations, it would be desirable to know the smallest possible set of dominant nodes that are causing system collapse with some quantifiable metric, so that fixing the system becomes easier without having to take actions everywhere. 

How to finding the best and dominant subset of nodes to fix the deficiency of power (real and reactive) at sparse critical locations? Usually, this question is answered in asset planning problems. For instance, consider reactive power planning (RPP) problems \cite{sparsePFref5-planningFACTS4congestion}\cite{sparsePFref6-planningSVCnTCSC}\cite{sparsePFref7-planningSVC}\cite{sparsePFref8-planningFACTS} that aim to find the optimal allocation of reactive power support through capacitor banks or FACTS devices such as static VAR compensators (SVC). Such problems correspond to finding the sparsest reactive power compensation vector that satisfies system power balance and operation limits in an optimization-based power flow study. However, convergence of such optimization-based studies becomes more difficult with increased system size and operating limits. Most state-of-the-art placement planning strategies \cite{sparsePFref5-planningFACTS4congestion}\cite{sparsePFref6-planningSVCnTCSC}\cite{sparsePFref7-planningSVC}\cite{sparsePFref8-planningFACTS} are only shown to handle small cases with hundreds of buses or less and are known to suffer from lack of robust convergence. Instead, a sparsity enforcement method is preferred. The objective of this method will be to provide a sparse set of nodes, that along with quantified infeasibility power, can be added to each of the corresponding set of nodes to make the model feasible. Figure \ref{fig: sparsePF result, case14} illustrates how a sparse solution can be helpful in understanding and fixing a blackout failure. 

\begin{figure}[h]
	\centering
\includegraphics[width=1\linewidth]{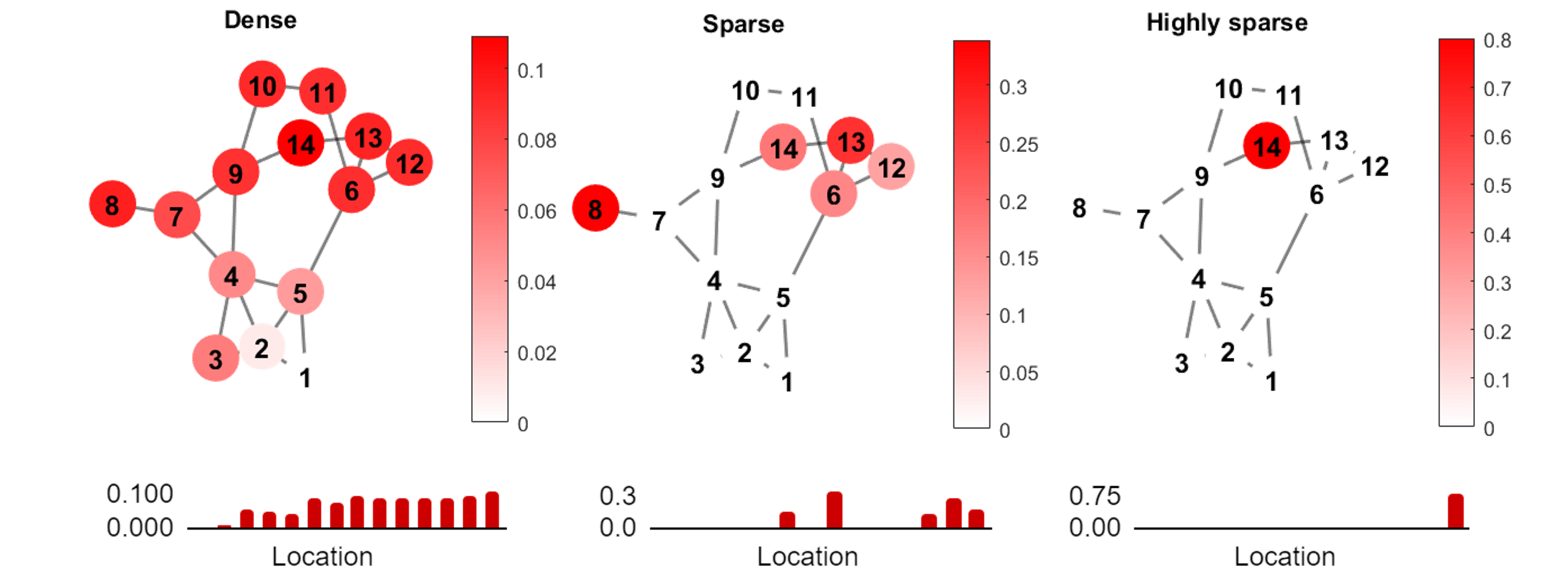}
	\caption{Sparse indicators pinpoint dominant sources of failure and focused corrective actions at key locations. This figure is a visualization of our experiment results on Case 14 in Table \ref{tab: sparsePF result, case14} in Section \ref{sec: sparsePF result}. Left: prior work of infeasibility-quantified simulation; middle: a weaker version of our work using L1 regularization; right: proposed robust actionable simulation using bus-wise sparsity enforcer. Result on the right figure indicates that bus 14 is the dominant source of failure and we can possibly fix the system by building a new power plant here producing 0.8 per unit current injection.}
	\label{fig: sparsePF result, case14}
\end{figure}

This thesis develops a novel method shed light on the dominant sources of blackout failure and potential easy actions to fix the system, not via planning, but from a robust actionable simulation method. 
And we achieve this by taking a further step from the infeasibility-quantified simulation we defined earlier in Problem Definition \ref{def: sugar simulation}. In reality, the most useful solution for expansion planning and corrective action is not the one with the smallest L2 norm and infeasible current values at multiple sources, but rather a solution that has non-zero "compensation terms" $n$ in the least number of locations. Therefore, extending Problem Definition \ref{def: sugar simulation} by optimizing a sparse $n$ vector is likely to give us the desired result.

\subsection{Formulating robust actionable simulation} 
To foster a sparse solution, we are inspired by LASSO \cite{sparsePFref9-fusedlasso}\cite{sparsePFref10-lasso}, a method that is used to enforce sparsity in feature selection of a model by L1-regularization. But instead of using a direct L1-regularization, we define a new approach of enforcing sparsity such that the $n_i$ values corresponding to geographically localized buses are correlated through a bus-wise sparsity enforcer. Mathematically, this problem is formulated as a non-convex optimization problem implemented based on an equivalent circuit formulation (ECF):

\begin{problem}[Robust actionable simulation with bus-wise sparsity enforcer]\label{def: sparsePF simulation} evaluates "what-if" scenarios while locating possible blackouts to a sparse set of dominant bus locations, by solving $x=[V, \theta]$ and sparse compensation terms $n$ from: 
\begin{align}
&\min_{x,n} \frac{1}{2}||n||_2^2 + \sum_i c_i(|n_i^{R}| + |n_i^{I}|)\notag\\
\text{s.t. }
   & F_i(x)+n_i=0 \text{ for }  i=1,2,...,N_{bus} 
\end{align}
\end{problem}
\noindent Here under equivalent circuit formulation of the power system, $n_i^{R}$ and $ n_i^{I} $ represent the real and imaginary parts of independent current sources injected at each bus $i$ in order to recover network (current) balance enforced by Kirchhoff's Current Law (KCL)  equations; and $c_i$ is the bus-wise sparsity enforcer that promotes sparse vector $n$ in the final solution. 

Why introduce a bus-wise sparsity enforcer instead of directly using L1-regularization? We propose this technique in order to reach a sufficiently sparse solution that a regular L1-regularization may not be able to reach. And in the next few subsections, we would like to illustrate our new sparsity enforcing mechanism by first providing insight into the ill-conditioning of L1-regularization, and then how a better trajectory of sparse solution can be established through bus-wise sparsity enforcer.

\subsection{Why L1 regularization is not ideal}

We start from analyzing L1-regularization and its limitations in reaching sparsity. Inspired by LASSO \cite{sparsePFref10-lasso}\cite{sparsePFref9-fusedlasso}, an immediate idea for enforcing sparsity can be applied by adding an L1-norm based penalty term in the objective function: 
\begin{problem}[Enforcing sparse infeasibility with classical L1-regularization]\label{def: sparsePF simulation - L1}
\begin{align}
&\min_{x,n} \frac{1}{2}||n||_2^2 + c||n||_1\notag\\
\text{s.t. }
   & F_i(x)+n_i=0 \text{ for }  i=1,2,...,N_{bus} 
\end{align}
\end{problem}

This formulation is not ideal for multiple reasons. It neglects the correlation of real and imaginary counterparts during sparsity enforcement, while in reality, the nonzero $n_i^{R}, n_i^{I} $ at the same bus are often coupled terms emerging concurrently. 

And moreover, the desired sparseness requires high values assigned to regularization parameter c, leading to ill-conditioning and convergence difficulties. Specifically, the inclusion of L1-norm leaves an non-differentiable objection function. To tackle this, we introduce slack variable t and convert the problem into the following constrained optimization form \cite{sparsePFref13-L1slack}:
\begin{align}
&\min_{x,n} \frac{1}{2}||n||_2^2 + c\cdot t\notag\\
\text{s.t. }
   & F_i(x)+n_i=0 \text{ for }  i=1,2,...,N_{bus} \\
   & n\preccurlyeq t \label{eq: l1 n<t}\\ 
   & -n\preccurlyeq t \label{eq: l1 -n<t}
\end{align}
\noindent where the $\preccurlyeq$
denotes componentwise inequality such that  $x_k\preccurlyeq y_k$ holds
 for every index $k$ given any two vectors $x, y$. Here the slack variable $t$ intuitively represents the upper bound on the infeasibility vector $n$. The magnitude of $n_i^{R}$ or $n_i^{I}$ will be bounded by $t_i^{R}$ or $t_i^{I}$ such that $|n_i^{R/I}|\leq t_i^{R/I}$.
 
And then we can write its Lagrangian function as:
\begin{equation}
   L(x,n,t,\lambda,\bar{\mu}, \underline{\mu}) = \frac{1}{2}||n||_2^2 + c\cdot t +\lambda^T(F(x)+n) +\bar{\mu}(n-t)+\underline{\mu}(-n-t)
\end{equation}

The dual variables $\bar{\mu}, \underline{\mu}$, which are Lagrangian multipliers corresponding to inequalities. The optimal solution of these variables satisfy the following equations enforced by (perturbed) KKT conditions:
\begin{align}
    & n^*=\lambda-\bar{\mu}+ \underline{\mu}\label{eq:sparsePF l1 dL/dn}\\
    &\bar{\mu}+ \underline{\mu}=c\cdot {\bf 1}\label{eq:sparsePF l1 dL/dt}\\
    &\bar{\mu}(n-t)=-\epsilon\label{eq:sparsePF l1 CS-U} \text{ (upper bound)}\\
    &\underline{\mu}(-n-t)=-\epsilon\label{eq:sparsePF l1 CS-L} \text{ (lower bound)}
\end{align}
\noindent with (\ref{eq:sparsePF l1 dL/dn}) and (\ref{eq:sparsePF l1 dL/dt}) being part of the stationarity conditions derived from $\frac{\partial L}{\partial n}=0$ and $ \frac{\partial L}{\partial t}=0$, respectively; and (\ref{eq:sparsePF l1 CS-U}) and (\ref{eq:sparsePF l1 CS-L}) being complementary slackness conditions related to the inequalities in (\ref{eq: l1 n<t})-(\ref{eq: l1 -n<t})

By further manipulation based on properties of Lagrangian multiplier, the optimal primal-dual pair ($n^*,\lambda^*$) should satisfy:
\begin{equation}
    |n^*| = |\lambda^*|-c
\end{equation}

\begin{figure}[h]
	\centering
\includegraphics[width=0.3\linewidth]{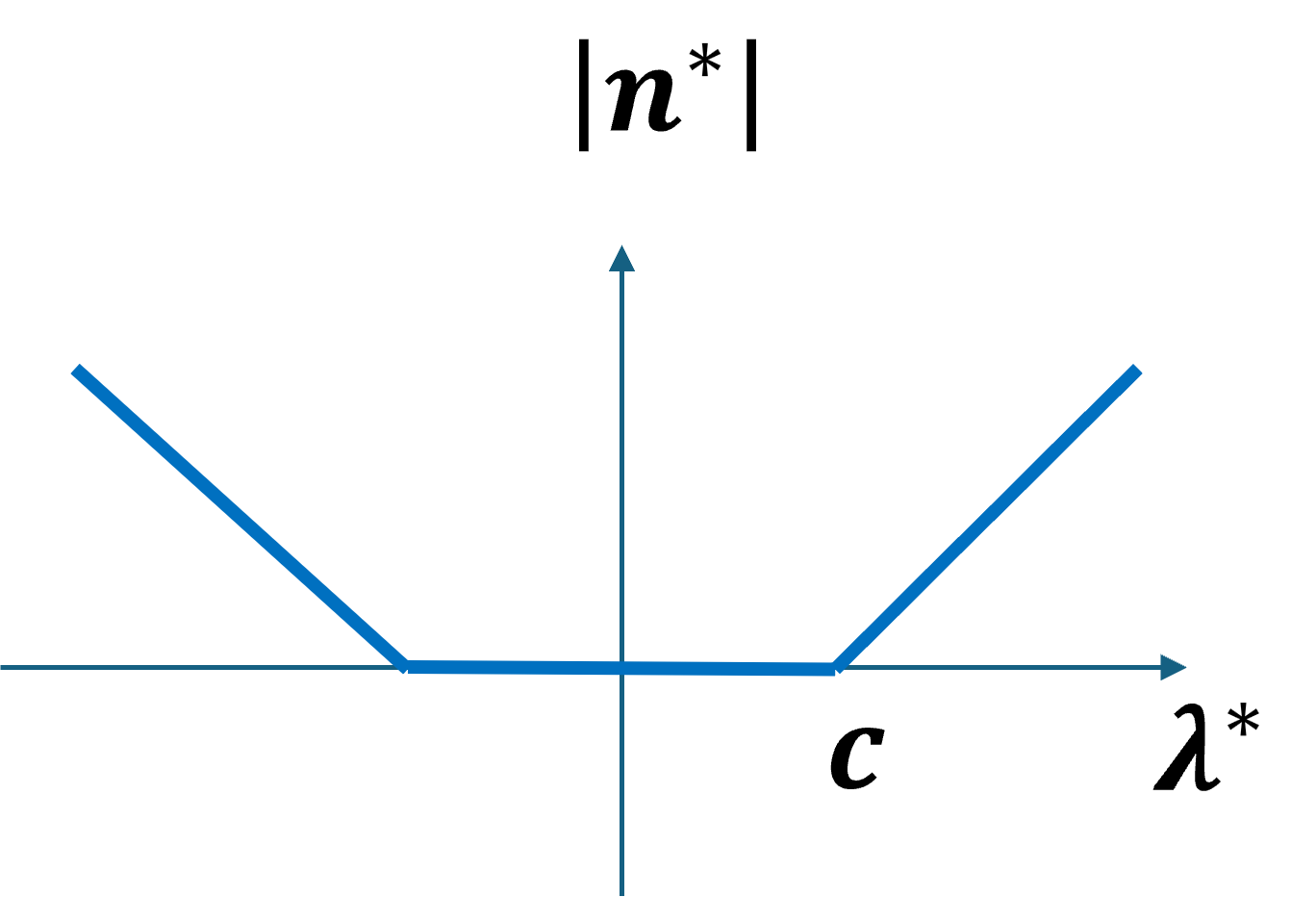}
	\caption{Relationship between $n^*$ and $c$: a blocking effect.}
	\label{fig: sparsePF, blocking effect}
\end{figure}

This primal-dual relationship can be clearly illustrated by Figure \ref{fig: sparsePF, blocking effect}, and inspires us to attach intuitive physical meanings:
\begin{itemize}
    \item Bus-wise Lagrangian multiplier $\lambda$ is a source of additional current flow into the network
    \item Scalar $c$ is a threshold such that any $\lambda$ below threshold are blocked out and only those above this threshold flows  into the system, appearing as non-zero $n$ sources.
\end{itemize}

This reveals a simple mechanism through which the threshold $c$ encourages a sparse solution by confining most $n$ entries to near zero value. Whenever threshold $c$ is added, the blocking effect reduces the number of non-zero infeasibility sources in the network. As the threshold $c$ is increased, the number of non-zero $n$ entries decreases, and any remaining non-zero $n$ entries adjust their value to make the network feasible. Therefore, with a high enough threshold value, only a few sources turn out to be above threshold and appear as nonzero elements in $n$.
In summary, our approach utilizes that: raising the value of $c$ encourages more near-zero $n$  elements by making the threshold hard-to-pass.

However, there exists a serious convergence problem with a single scalar $c$ as a tuning parameter for regularization. This challenge can be characterized by an unwanted trade-off and inflexibility. Let us illustrate this further.
With $t$ representing the upper bound of $n$, if there exists nonzero infeasibility (e.g. $n_i^{R}>0$) at bus $i$, due to the minimization of $t$ in the objective function, the upper bound tends to be very tight (i.e., $t_i^{R}-n_i^{R}\approx 0$). Hence, if we utilize a single large scalar $c$ value to achieve a sufficiently sparse solution, the tightness property of the algorithm results in convergence difficulties due to the steep and highly non-linear regions of the complementary slackness conditions given by (\ref{eq:sparsePF l1 CS-U})-(\ref{eq:sparsePF l1 CS-L}).

\begin{figure}[h]
	\centering
\includegraphics[width=0.5\linewidth]{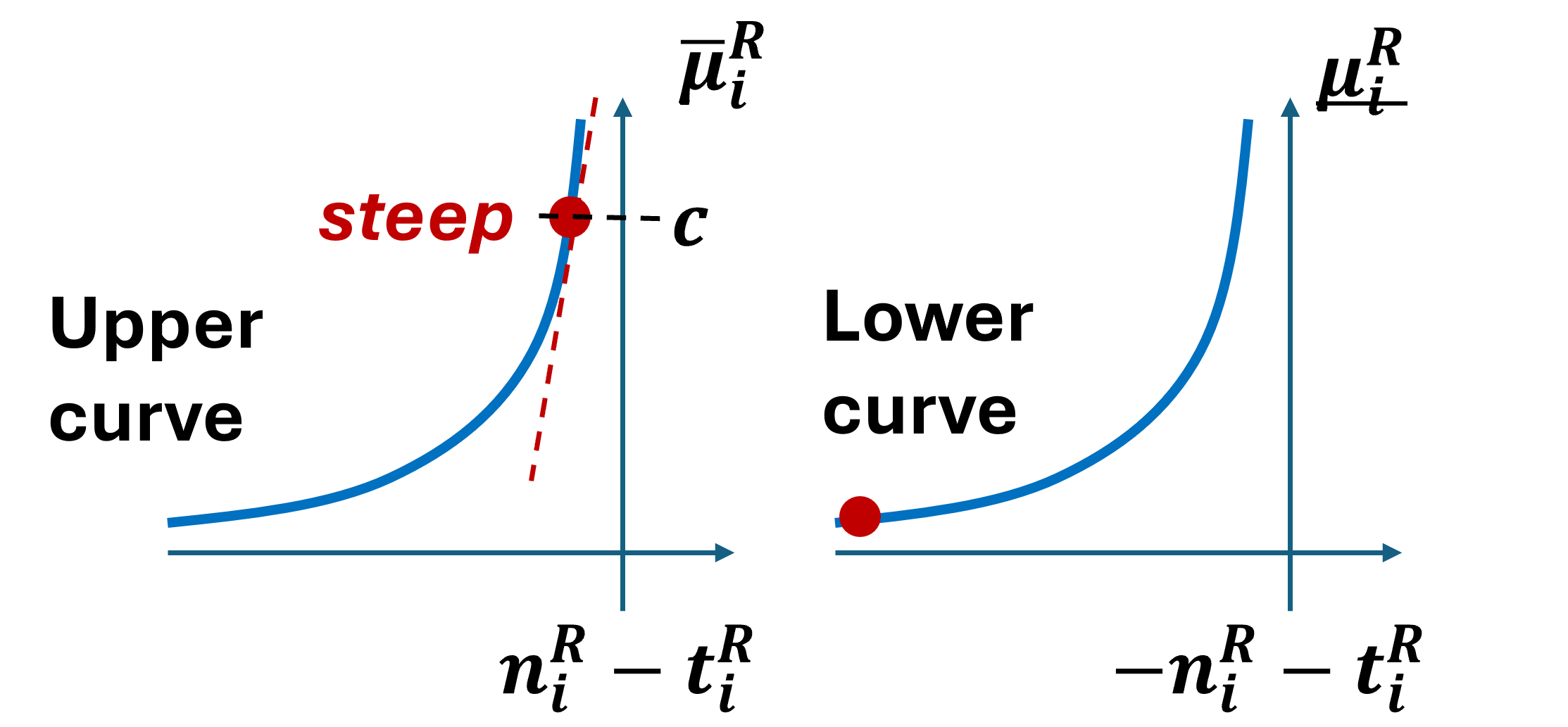}
	\caption{High $c$ value causes steep convergence region on the complementary slackness curve. Left: upper bound curve $\bar{\mu}_i^R(n_i^R-t_i^R)=-\epsilon$, right: lower bound curve $\underline{\mu}_i^R(-n_i^R-t_i^R)=-\epsilon$. When $n_i^R >0$, we have $\bar{\mu}_i^R\rightarrow c, \underline{\mu}_i^R\rightarrow 0$. $(n_i^R, t_i^R, \bar{\mu}_i^R) $ converges on a difficult region of the upper curve.}
	\label{fig: high c and steep CS}
\end{figure}

This problem can be illustrated in Figure \ref{fig: high c and steep CS}. As the number of "infeasible buses" increases, numerous buses encounter difficult steep regions of this kind, making it difficult for the algorithm to converge.
Thus, the selection of the value of the $c$ parameter is a trade-off between sufficient sparsity and robust convergence, both of which are essential for meeting our eventual goal. With $c$ being a single scalar value, there is little freedom for us to manipulate its value and achieve the desired performance.

\subsection{Bus-wise sparsity enforcement: a better pathway to sparse solutions}

To address the aforementioned challenges, we propose a new method that defines threshold $c_i$ for each bus $i$. This $c_i$ parameter is a "bus-wise sparsity enforcer" such that, according to the thresholding effect, raising $c_i$ encourages a zero $n_i$ at bus $i$ in the solution. As previously defined in Problem Definition \ref{def: sparsePF simulation}, the problem can be re-formulated as: 
\begin{align}
&\min_{x,n} \frac{1}{2}||n||_2^2 + \sum_i c_i(|n_i^{R}| + |n_i^{I}|)\notag\\
\text{s.t. }
   & F(x)+n=0  
\end{align}

In this approach, we convert a single scalar $c$ to a vector of bus-wise sparsity enforcers $c_i, \forall i=1,2,...,N_{bus}$. Then, to determine the values of $c_i$, we use the following assumptions that are based on the grid physics:
\begin{itemize}
    \item Uneven distribution of infeasibility sources: an infeasible system is caused by and can be characterized by failures on isolated locations, rather than outages of equal seriousness at each bus.
    \item There is a high probability that the dominant sources (locations) of failure in the system are reflected by the nodes with highest magnitude of the "infeasibility indicators" $n_i$ in the simulation.
\end{itemize}

Based on these assumptions, we can simply make flexible adjustments to $c_i$ at each bus, according to the qualitative classification of bus-wise infeasibility, as shown in Algorithm \ref{alg: bus-wise sparse enforcer}. For simplification and efficiency, we simply classify all buses into ‘major’ and ‘minor’ categories, according to their $n_i$ magnitude and the sparsity goal. For buses in the ‘major’ group, i.e. with high infeasibility quantities ($||n_i||>>0$), we assume that they are very likely the dominant sources of failure and assign a low value $c_L$. This encourages non-zero $n_i$ on those locations. For buses in the minor group, we assign a higher threshold $c_H$ such that we can force their $n_i$ values to zero or near zero values.
\begin{algorithm}
	\caption{Bus-wise sparsity enforcer assignment} 
	\label{alg: bus-wise sparse enforcer}
	\KwIn{sparse goal $k$, threshold ($c_H,c_L$), existing infeasibility indicators $n$}
	\KwOut{updated bus-wise sparsity enforcer $c_i, i=1,2,…,N_{bus}$}
{\bf Sort all buses} by $|n_i|$ values (the magnitude of $n_i, \forall i$) in descending order

{\bf Classify bus category: }
        buses with top $k$ largest $|n_i|$ are grouped into "major" bus set $\bm{B}_{major}$; and the remaining buses are grouped in "minor" bus set $\bm{B}_{minor}$  
        
{\bf Assign bus-wise enforcer: }
         $$c_i=c_L \text{ for } \forall i\in \bm{B}_{major}$$
         $$
         c_i=c_H \text{ for } \forall i\in \bm{B}_{minor}$$	
\end{algorithm}

\begin{algorithm}
	\caption{Robust actionable simulation with a (soft) $k-sparse$ goal to localize possible failure} 
	\label{alg: sparsePF}
	\KwIn{system testcase, initial guess $x_0, n_0, \lambda_0$ sparse goal $k$, threshold $(c_H,c_L)$}
	\KwOut{simulation output $[x, n]$ with sparse compensation terms $n$ (may not be strictly $k-sparse$)} 
 
{\bf 
Initialize} $t, \bar{\mu}, \underline{\mu}$

{\bf Bus-wise sparse enforcer assignment} using Algorithm \ref{alg: bus-wise sparse enforcer} to get $c_i$ for $i=1,2,...,N_{bus}$ 

{\bf Solve robust actionable simulation problem} as defined in Problem Definition \ref{def: sparsePF simulation}, using a circuit-theoretic primal-dual interior point solver
	
\end{algorithm}
Our infeasibility localization method is summarized in Algorithm \ref{alg: sparsePF}, where $k$ defines the number of locations where non-zero $n_i$ values might be allowed. From another intuitive viewpoint, this method unevenly penalizes infeasibility values at different buses. For any bus $i$ assigned with a high threshold $c_H$, we deliberately attach high penalty to any compensation $n_i$ made at those buses, thereby forcing compensations at other sets of buses to restore a feasible system. More importantly, this ($c_H,c_L$) configuration removes the need for high values of parameter $c_H$, as sparsity is dependent on the ratio of $c_H$ and $c_L$, not the absolute value of the threshold. 

Simple principles for selecting ($c_H,c_L$) are:
\begin{itemize}
    \item $c_H$ is chosen to be sufficiently larger than $\lambda_i^{R/I}$ such that ‘minor’ buses end up with zero or near zero $n_i$ values. This enables sufficiently sparse solutions with small  ($c_H,c_L$) values, thereby avoiding ill-conditioning and convergence difficulties.
    \item $c_L$ is chosen to be sufficiently lower than $c_H$ such that the threshold is ‘easy-to-pass’ for both $\lambda_i^{R}$ and $\lambda_i^{I}$, making nonzero $n_i^R, n_i^I$ coexist at 'major' buses. This is a necessary condition for practical applications. Due to the nature of the power flow equations, grid devices provide both real and imaginary currents. Therefore, for any corrective actions, it is preferable to achieve sparse solutions that have infeasibilities localized to the fewest number of bus locations.
\end{itemize}

Additionally, if the $k-sparse$ goal is not practical, $n_i{R/I}$ in the ‘minor’ group leaves room for infeasible sources on more than $k$ locations, and the final solution can be $(k+m)-sparse$.

\subsection{Extending the algorithm to large-scale systems}

Since the power system is highly nonlinear, it's important to facilitate time-efficient convergence on large-scale systems. For a practical large-scale power system, we do not have accurate knowledge in advance about the severity of the system collapse, and therefore, it is hard to define a reasonable guess of the k-sparse goal. Importantly, since infinite possible combinations of $\{n_i\}$ can make the system network balance equations correspond to a feasible network, the ‘major’ locations in a dense solution are likely to be the dominant sources with a high probability; however, we must note that this is not always true. 

With these considerations we extend our method to large-scale networks by iteratively adjusting sparse enforcers and gradually reaching sparser solutions from denser ones. For robust convergence, we start from a dense solution from infeasibility-quantified simulation (Problem Definition \ref{def: sugar simulation}) \cite{SUGAR-pf-marko}\cite{sparsePFref14-SUGAR-pf-txstepping} in all locations and gradually update the $k-sparse$ goal by some shrinkage rate. This is equivalent to splitting the original problem into a series of subproblems, where each subproblem uses a solution from the previous one as its initial guess, and easily reaches its optimal solution within a few iterations. Our method is shown in Algorithm \ref{alg: sparsePF-large}.

\begin{algorithm}
	\caption{
Robust actionable simulation for large-scale systems} 
	\label{alg: sparsePF-large}
	\KwIn{system testcase, shrinkage rate $r$}
	\KwOut{simulation output $[x, n]$ with sparse infeasibility indicators (i.e., compensation terms)} 

{\bf Initialize} $x_0, n_0, \lambda_0$ by solving infeasibility-quantified simulation (Problem Definition \ref{def: sugar simulation})

{\bf Initialize} thresholds ($c_H, c_L$) and sparsity goal $k=round(N_{bus}*r)$; default values for power systems are $c_H=10, c_L=0.1, r=0.75$
 
	\While {not sparse enough}{
 
{\bf 1. Bus-wise sparse enforcer assignment} as in Algorithm \ref{alg: bus-wise sparse enforcer}

{\bf 2. Robust actionable simulation  with a k-sparse goal} as in Algorithm \ref{alg: sparsePF}

{\bf 3. Check solution $n$ and update sparse goal:} $k=k*r$
            
{\bf 5. (Optional) adjustments:} adjust $c_H,c_L$ and shrinkage rate $r$ if needed}

\end{algorithm}

\subsection{Comparison and results: evaluation on Eastern Interconnection system}\label{sec: sparsePF result}

This method presents experimental results to compare different methods, and also results on large cases that include the U.S. Eastern Interconnection sized 80k+ bus network. 

To prove the efficacy and scalability of our proposed method under blackout failures, we create infeasible scenarios (past the nose curve) on power system cases, by increasing their load factors. And parameters of our proposed method are set to default values $c_H=10, c_L=0.1, r=0.75$.

We first tested standard Case 14 which is infeasible by increasing load factor to $4.5$. Table \ref{tab: sparsePF result, case14} presents the values of compensation terms (or infeasibility indicators) at all buses. We compare the 3 different methods 1) non-sparse feasibility-quantified simulation method \cite{SUGAR-pf-marko} (Problem \ref{def: sugar simulation}), 2) enforcing sparse infeasibility indicators using L1-regularization (Problem \ref{def: sparsePF simulation - L1}), and 3) our proposed method using bus-wise sparsity enforcer (Problem \ref{def: sparsePF simulation}). 
Comparison shows that our method reaches 1-sparse solution and localizes infeasibility to bus 14, indicating easy action-taking at the minimum number of places to fix the system. Whereas, L1-regularization is less actionable with more locations being the sources of blackout failure; and the regular non-sparse method \cite{SUGAR-pf-marko} indicates infeasibility at almost all buses, making the approach impractical for expansion planning or applying corrective action. 

\begin{table}[htbp]
\centering
\caption{Simulation of Case 14 with blackout (load factor 4.5). Results in this table are also visualized in Figure \ref{fig: sparsePF result, case14} to better illustrate the advantage of sparse $n$ in understanding and fixing the system.} 
\label{tab: sparsePF result, case14}
\begin{tabular}{c|ccc}
\hline
                                  & \multicolumn{3}{c}{\textbf{Magnitude of infeasibility indicators / compensation terms}}                                                      \\ \cline{2-4} 
\multirow{-2}{*}{\textbf{Bus ID}} & \multicolumn{1}{c|}{\textbf{Non-sparse}} & \multicolumn{1}{c|}{\textbf{L1-regularization}} & {\color[HTML]{C00000} \textbf{Proposed bus-wise sparsity enforcer}} \\ \hline
1                                 & \multicolumn{1}{c|}{0}                   & \multicolumn{1}{c|}{0}                          & 0                                               \\
2                                 & \multicolumn{1}{c|}{0.00858402}          & \multicolumn{1}{c|}{0}                          & 0                                               \\
3                                 & \multicolumn{1}{c|}{0.0561223}           & \multicolumn{1}{c|}{0}                          & 0                                               \\
4                                 & \multicolumn{1}{c|}{0.05097014}          & \multicolumn{1}{c|}{0}                          & 0                                               \\
5                                 & \multicolumn{1}{c|}{0.04278203}          & \multicolumn{1}{c|}{0}                          & 0                                               \\
6                                 & \multicolumn{1}{c|}{0.08877886}          & \multicolumn{1}{c|}{0.16111856}                 & 0                                               \\
7                                 & \multicolumn{1}{c|}{0.07740694}          & \multicolumn{1}{c|}{0}                          & 0                                               \\
8                                 & \multicolumn{1}{c|}{0.09593462}          & \multicolumn{1}{c|}{0.33915759}                 & 0                                               \\
9                                 & \multicolumn{1}{c|}{0.08860328}          & \multicolumn{1}{c|}{0}                          & 0                                               \\
10                                & \multicolumn{1}{c|}{0.09134275}          & \multicolumn{1}{c|}{0}                          & 0                                               \\
11                                & \multicolumn{1}{c|}{0.08889756}          & \multicolumn{1}{c|}{0}                          & 0                                               \\
12                                & \multicolumn{1}{c|}{0.09065051}          & \multicolumn{1}{c|}{0.1244972}                  & 0                                               \\
13                                & \multicolumn{1}{c|}{0.09368859}          & \multicolumn{1}{c|}{0.27381069}                 & 0                                               \\
14                                & \multicolumn{1}{c|}{0.10908567}          & \multicolumn{1}{c|}{0.1824952}                  & {\color[HTML]{C00000} 0.80006182}               \\ \hline
\end{tabular}
\end{table}
 
Next, we also test on 5 large-scale systems. Table \ref{tab: sparsePF result, large} shows our method efficiently localizes dominant sources of system blackout to sparse distributions.  

\begin{table}[htbp]
\centering 
\caption{Robust actionable simulation results on large systems.}
\label{tab: sparsePF result, large}
\begin{tabular}{c|c|c|c}
\hline
\textbf{Case Name} & \textbf{Load factor} & {\color[HTML]{C00000} \textbf{k-sparse solution}} & \textbf{Dominant infeasible buses name} \\ \hline
MMWG80K            & 1.07                 & {\color[HTML]{C00000} 1}                          & ‘155753’                                \\ \hline
ACTIVSg25K         & 1.8                  & {\color[HTML]{C00000} 42}                         & Not listed here                         \\ \hline
CASE9241pegase     & 1.15                 & {\color[HTML]{C00000} 1}                          & ‘2159’                                  \\ \hline
CASE6515rte        & 1.15                 & {\color[HTML]{C00000} 2}                          & ‘3576’,’ 4356’                          \\ \hline
CASE6468rte        & 1.29                 & {\color[HTML]{C00000} 1}                          & ‘3718’                                  \\ \hline
\end{tabular}
\end{table}

\section{Robust estimator with sparse error indicators}\label{sec: ckt-GSE}

Methods based on (error-quantified) circuit-based estimation as defined in Problem \ref{def: sugar-SE} can solve steady-state estimation with closed form solution, while capturing the amount of measurement noises at each element using indicators $n_i$. However, they do not provide the necessary robustness to retain accurate voltage solutions when bad data and topology errors exist. And also data errors need to be identified using extra hypothesis testing. 

\begin{figure*}[h]
     \centering
     \begin{subfigure}[b]{0.49\linewidth}
         \centering         \includegraphics[width=0.95\textwidth]{figures/ckt-se/wlav_bdd_graph.png}
         \caption{Sparse error indicators: pinpointing 1 random bad data and 2 topology errors}
         \label{fig: wlav bdd graph}
     \end{subfigure}
     \hfill
     \begin{subfigure}[b]{0.49\linewidth}
         \centering         \includegraphics[width=0.95\textwidth]{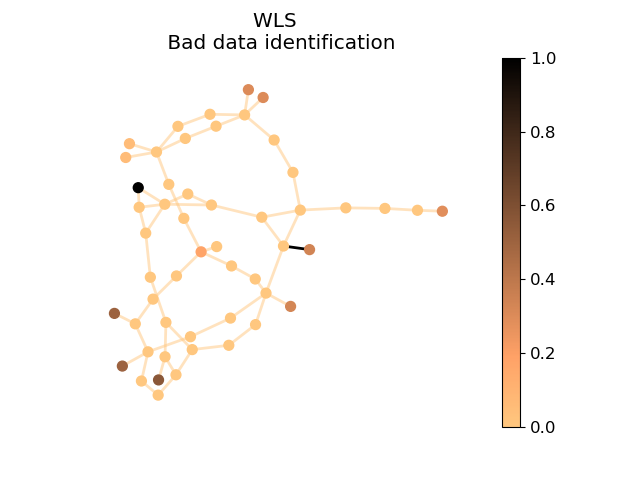}
         \caption{Dense error indicators: unable to locate data errors accurately}
         \label{fig: wls bdd graph}
     \end{subfigure}
     \caption{Identifying data errors on Case 118}
        \label{fig: sparse error indicators}
 \end{figure*}

In the estimation task, it would be desirable to have intrinsic robustness to directly identify  data errors and automatically reject them so that an accurate solution of voltages can be obtained. To this end, a sparsity enforcement method is preferred. The objective of the method will be to provide a sparse set of error indicators. Figure \ref{fig: sparse error indicators} illustrates how a sparse solution can be
helpful in pinpointing a mixture of random bad data and topology errors.

\subsection{Assuming a possibly wrong topology}\label{sec: BB to NB}

To consider the possibly wrong switching statuses which contribute to topology errors, we need node-breaker models of the power grid which contain nodes, switches, and branches. 
However, a node-breaker model is not directly available sometimes, because the majority of estimation, simulation and optimization processes and studies on today's power grid still rely on the bus-branch model as the direct input. Nevertheless, even if the node-breaker model is obtainable, it might be sometimes undesirable to use, since the network is too large-scale and complicated, whereas we are only interested in analyzing a subset instead of a complete set of all connection statuses. 

\begin{figure}[t]
     \centering
     \begin{subfigure}[b]{0.24\linewidth}
         \centering         \includegraphics[width=01\linewidth]{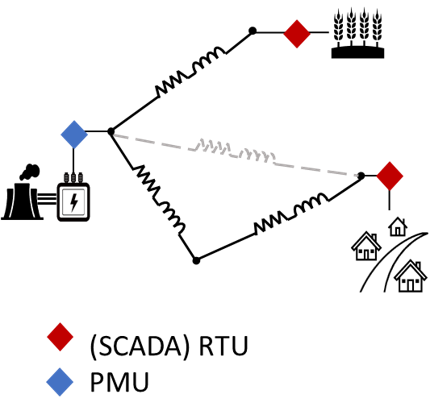}
         \caption{}
         \label{fig:case4}
     \end{subfigure}
     \hfill
     \begin{subfigure}[b]{0.24\linewidth}
         \centering         \includegraphics[width=01\linewidth]{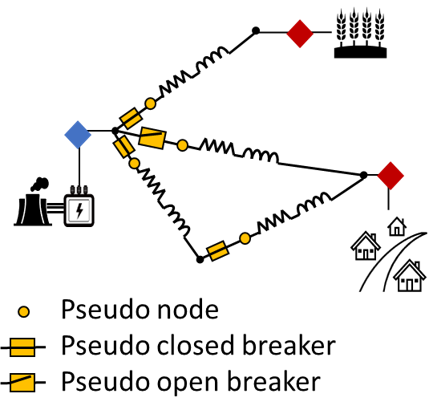}
         \caption{}
         \label{fig:case4 circuit model}
     \end{subfigure}
     \hfill
     \begin{subfigure}[b]{0.49\linewidth}
         \centering         \includegraphics[width=\linewidth]{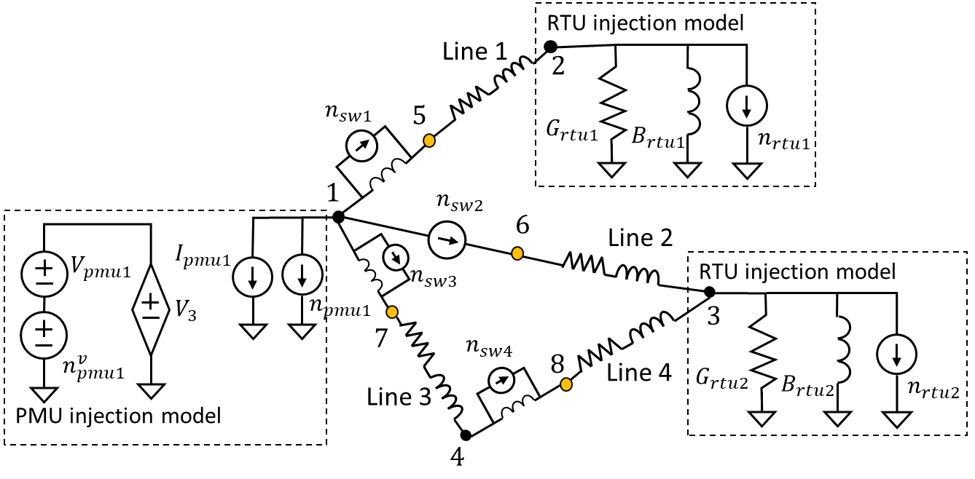}
         \caption{}
         \label{fig:case4 circuit model}
     \end{subfigure}
        \caption{Extending a bus-branch model with pseudo nodes and switches.}
        \label{fig: BB to NB}
\end{figure}

Therefore, whenever we are given a bus-branch model but need to consider topology errors on it, we extend the bus-branch model to include the switching devices of interest. Figure \ref{fig: BB to NB} illustrates the high-level idea. Specifically, at any time $t$, given a power grid Figure \ref{fig: BB to NB}(a) that is measured by SCADA and modern PMUs, the gray dashed transmission line is inactive. If it is a bus-branch model, we add pseudo buses and pseudo circuit breakers, in order to estimate topology, as in Figure \ref{fig: BB to NB}(b). The breaker statuses represent the associated line statuses, i.e., an active line connects to a closed breaker and an inactive line connects to an open breaker. Then the measured elements are replaced with equivalent circuit models (PMU model, RTU model, and open/closed switch model) to transform the power grid into an aggregated linear equivalent circuit on which a convex constrained optimization problem can be defined for estimation purpose, as in Figure \ref{fig: BB to NB}(c).

\subsection{Formulating robust circuit-based state estimator}

In the presence of bad data, the WLS formulation is not robust: it does not produce accurate estimates and requires post-processing to isolate suspicious measurements, followed by iteratively re-running the algorithm to obtain reliable estimates. To enhance the intrinsic robustness of the solution against wrong status and bad (continuous) data, we are using WLAV to develop robust estimators. A problem definition of our approach is given as below:

\begin{problem}[Robust circuit-based (generalized) state estimator using WLAV]\label{def: ckt-GSE} Given a real-time set of continues measurements $z_{cont}$, status measurements $z_{sw}$ on a set of switches $\{sw_1, sw_2,...sw_K\}$, the robust circuit-based state estimator solves state $x$ and sparse error indicators $n=[n_{pmu}, n_{rtu}, n_{sw}]$ to pinpoint bad data and topology errors. This is by solving a weighted least absolute value (WLAV) estimation problem subject to AC-network constraints at all nodes:
\begin{equation}
    \min_{x,n}\sum_k w_k|n_{swk}|+\sum_i\alpha_i|n_{rtui}|\notag\\
    + \sum_j\beta_{j}|n_{pmuj}| 
    + \gamma_{j}|n_{pmuj}^{v}|   
\end{equation}
    \text{s.t. (linear) KCL equations at all nodes: } 
    \begin{equation}
        [Y,B]\begin{bmatrix}
    x\\n
\end{bmatrix} = J 
\label{eq: ckt-GSE}
    \end{equation}
Notably, this problem definition applies to both robust state estimation and generalized state estimation. If we assume an accurate topology is available, we just use the bus-branch model as the input, and the switch set is empty.
\end{problem}

\noindent Section \ref{sec: linear circuit and KCL} has taken the example in Fig \ref{fig: BB to NB}(c) to show how the KCL equations can be written linearly as in (\ref{eq: kcl example start})-(\ref{eq: kcl example end}).

Here the state vector $x=[V^R_1,V^I_1,...,V^R_N,V^I_N]$ contains real and imaginary bus voltages. $n_{rtu}, n_{pmu},$ and $n_{sw}$ represent the noise/error terms for RTUs, PMUs, and switches, respectively. Also,  ${w,\alpha,\beta,\gamma}$ are weights on each measurement model to represent a level of uncertainty, the selection of which will be discussed in Section \ref{sec: sensitivity}.

The use of the WLAV objective is inspired by the assumption that the data errors are sparsely distributed amongst the total measurement set since anomalies are rare in reality. 
As it minimizes the L1-norm objective, the WLAV estimator enforces a sparse vector of 'noise terms' that matches the sparse population of measurement errors. Large non-zero values only appear on locations with bad continuous data and wrong switch statuses, whereas the solution fits other high-quality measurements, providing robust estimates. 

\subsection{Fast solution method for large systems}

Mathematically, the formulation in (\ref{eq: ckt-GSE}) is a \textit{linear programming (LP)} problem, which is guaranteed to reach a global optimum under the hold of certain conditions. 
We aim to develop a  primal-dual interior point (PDIP) algorithm \cite{PDIP-L1norm} with novel problem-specific heuristics to reach the optimal solution with fast speed. The specific goal is to give superior convergence properties over simplex-based \cite{simplex} algorithms for large-scale problems. Simplex method is generally better suited for smaller problems since it traverses through a set of vertices of the feasible space until the optimal solution is found. However, as the problem size increases, the number of vertices grows exponentially, making the method impractical for larger networks. Compared with the exponential computational complexity of the Simplex method, solving a linear programming problem with PDIP has proven to be effective on large-scale problems \cite{PDIP-L1norm}, with its worst-case complexity being polynomial to problem dimension. 

The practical challenge stems from the non-differentiable L1 terms in the objective. To efficiently deal with the problem-solving, we first converted the objective function to a differential form:
\begin{subequations}
\label{prob: differentiable ckt-GSE}
\begin{equation}
\min_{x,n,t} c^Tt
\end{equation}
$$ s.t.\text{ network equations as in } (\ref{eq: ckt-GSE}) $$
\begin{equation}
    |n|\preceq t
    \label{con neq}
\end{equation}
\end{subequations}
\noindent with $c=[w,\alpha, \beta, \gamma]$, {  and the $t$ variable physically corresponds to the upper bound of the slack sources $n$.}

Then we adopt a circuit-theoretic LP solver by augmenting the standard primal-dual interior point (PDIP) algorithm with circuit-theoretic heuristics to speed up convergence. 
Specifically, the PDIP method solves the differentiable problem in (\ref{prob: differentiable ckt-GSE}) by iteratively solving the nonlinear perturbed KKT conditions {   as follows:

\begin{subequations}
\noindent \text{Primal feasibility:}
\begin{align}
        Y\begin{bmatrix}
    x\\n
\end{bmatrix} = J\\
    |n|\preceq t
\end{align}
Complementary slackness:
\begin{align}
    \overline{\mu} (n-t) = -\epsilon\\
    \underline{\mu} (-n-t) = -\epsilon
\end{align}
Dual feasibility:
\begin{align}
   \mu \succeq 0, \mu = [\overline{\mu},\underline{\mu}]
\end{align}
\noindent Stationarity:
\begin{align}
    Y^T\lambda = 0\\
    \overline{\mu}-\underline{\mu} + B^T\lambda = 0\\
    \overline{\mu}+\underline{\mu} = c
\end{align}
\end{subequations}
}
where $\lambda$ denotes a vector of Lagrangian multipliers associated with the linear constraints.

Standard toolboxes can be used to solve the linear programming problem. These include CVXOPT 
\cite{CVXOPT}, SciPy \cite{scipy}, etc. Yet the (speed) performance of these solvers is limited for large grid cases. Therefore, to further improve the efficiency, we solve the perturbed KKT conditions with problem-specific limiting heuristics. Taking into account that the problem is convex and only local nonlinearity exists in the complementary slackness component of the perturbed KKT conditions, we apply simple step-limiting only on dual variables $\mu$ (corresponding to inequalities) and $t$ to make each iteration update faster and more efficient. This approach moves away from standard filter line-search algorithms \cite{CVXOPT} used by other generic tools. 

Algorithm \ref{alg:LP solver} illustrates our circuit-theoretic variable limiting heuristics.
{In this algorithm, step 1 adjusts the update of $\mu$ based on the limits defined in the dual feasibility $\mu \succeq 0$ and the stationarity $\overline{\mu}+\underline{\mu} = c$. And step 2 adjusts $t_j$ to guarantee the satisfaction of the primal feasibility $|n|\preceq t$.}
\begin{algorithm}
	\caption{Variable limiting heuristics to solve LP problem}
	\label{alg:LP solver}
	\KwIn{previous solution $\mu_{old}$,
	new solution $\mu,t,n$, step limit $d$}
	\KwOut{new solution $\mu,t$ after limiting}
	For each element $\mu_j$ in $\mu$:
	$$\Delta\mu_j=\mu_j-\mu_{old,j}$$
	$$dir = sign(\Delta\mu_j)$$
	$$h = \begin{cases} c_j-\mu_{old,j} & dir \ge 0 \\ \mu_{old,j} & dir < 0 \end{cases}$$
    $$\mu_j=dir*\min(d,h)$$
	
	For each element $t_j$: 
	$$t_j = \begin{cases} 2|n_j| & |n_j|>t_j \\ t_j & else \end{cases}$$
\end{algorithm}

As (\ref{prob: differentiable ckt-GSE}) is convex and applicable to realistic settings of meters (both SCADA meters and modern PMUs), the proposed method improves existings works of WLAV based robust state estimation\cite{abur-robustSE-PMU} and robust generalized state estimation \cite{TESE-GSE-PMUabur}\cite{convexTESE-SDP-weng} which were either limited to only PMUs \cite{TESE-GSE-PMUabur} or applied non-scalable relaxation techniques \cite{convexTESE-SDP-weng} to convexify the problem.

\subsection{Hypothesis test to validate wrong switch status} \label{sec:hypothesis_test}
By using WLAV formulation on the node-breaker model, the proposed estimation algorithm above provides a robust solution that implicitly rejects any data errors. While the sparsity of the noise vector is already indicative of the location of suspicious data samples, we propose the use of a hypothesis test to formally identify wrong switch statuses (i.e., topology errors). It follows from grid physics that an open switch should have zero current flow, whereas a closed switch should have nearly zero voltage across it, see Table \ref{tab:hypothesis test}. In this work, thresholds $\tau_I$ and $\tau_V$ are chosen from empirical values $\tau_I=0.01, \tau_V=0.01$. 
\begin{table}[htbp]
\centering
\caption{Hypothesis test to detect wrong switch status}
\label{tab:hypothesis test}
\begin{tabular}{lll}
\hline
\bf Measured status & \bf Hypothesis test & \bf Conclusion\\
\hline
Open & \begin{tabular}[c]{@{}l@{}}  $|I_{sw}|>\tau_I?$ \end{tabular} & \begin{tabular}[c]{@{}l@{}}If YES, switch should be closed\end{tabular}\\
\hline
Closed & \begin{tabular}[c]{@{}l@{}}  $|V_{sw}|>\tau_V?$ \end{tabular} & \begin{tabular}[c]{@{}l@{}}If YES, switch should be open\end{tabular}\\
\hline
\end{tabular}
\end{table}

\subsection{Hyperparameter tuning: trade-off in weight selection}\label{sec: sensitivity}

As formulated in the estimation problem in (\ref{eq: ckt-GSE}), each measurement device is assigned a weight in the objective function. This weight represents a level of confidence in each measurement and determines the algorithm’s sensitivity to different data errors. As our proposed method detects and localizes erroneous data by the sparse vector of noise terms $n$, the algorithm's sensitivity to data errors can be mathematically defined as the sensitivity of $n$ for any perturbation on the data (i.e., true data errors). 
Specifically, a lower weight $c_j$ for a particular measurement $j$ indicates the measurement is less trustworthy, and while minimizing $c_j|n_j|$ in the WLAV objective, a low $c_j$ tends to push the corresponding $n_j$ to a larger value, making the corresponding data error, if any, easily detectable.
Therefore, a lower weight makes the algorithm more sensitive to data errors at this location. This is a desirable feature as we expect less trustworthy meters to be more prone to gross data errors.

The selection of weights for continuous measurements $(\alpha, \beta, \gamma$ in objective function of \ref{eq: ckt-GSE}) is statistically related to the variance (or dispersion) of the measurement tolerance (especially when assuming noise $n$ as Gaussian). Most existing works \cite{WLS-SE}\cite{TE-WLSE-BDI} set weights as $\frac{1}{\sigma^2}$ which is the reciprocal of the variance of the noise. This results in a statistical property wherein minimizing weighted least squares of the noise in the objective is equivalent to a maximal likelihood estimation (MLE) if we assume that noise $n$ follows Gaussian distribution. However, this can also lead to numerical issues as a high-quality measurement device (which has a very low noise variance) corresponds to an extremely high weight value which can cause ill-conditioning issues.
In this paper, to avoid extremely high weights, we scale the reciprocal of variance such that any RTU device with noise $\sigma=0.001$ has weight=1.

In contrast, the selection of weights for switches ($w$ in (\ref{eq: ckt-GSE})) requires additional tuning. 
Unlike continuous measurements, the switch statuses are discrete data, and the assumption of Gaussian noise no longer holds, making the statistical variance inapplicable. Instead, this paper’s selection of switch weights is based on considering a trade-off between convergence stability and the algorithm's sensitivity to topology error.
Specifically, when the weights of switches are high, the resulting low sensitivity to topology errors can cause a wrong switch status to be falsely identified as multiple bad continuous data and degrade solution quality. While very low switch weights will result in a high sensitivity to topology errors and allow easy detection of wrong switch status data, low weights can cause numerical difficulties, which will deteriorate convergence efficiency. This paper applies hyper-parameter tuning to select the weights that give the lowest misclassification rate. Based on our empirical findings, the weights of switching devices should be lower than continuous meters to provide the necessary sensitivity for the wrong switch status. In this paper, we set all switch weights to $0.001$ for the 8 substation case, and $0.01$ for the 300-substation case and Taxas CP-2000 case.

\subsection{Experiments to evaluate robust estimator: robustness and scalability }

To validate the efficacy of the proposed models and method, we design experiments to answer the following questions:
\begin{enumerate}
    \item \textbf{Robustness:} Is the method robust against random bad data and topology error? We experiment on generalized state estimation on node-breaker model to account for a mixture of these two types of random errors.
    \item \textbf{When does it fail:} 
    How does the (solution accuracy) performance change as the number of data errors increase?
    \item \textbf{Scalability:} Is the method applicable to large networks?
\end{enumerate}
\noindent 

{ 
\textbf{Reproducibility:} All test cases are from the CyPRES public dataset available at {\color{blue} \url{https://cypres.engr.tamu.edu/test-cases}}. 
And all experiments are run on a laptop computer with 11th Gen Intel(R) Core(TM) i7-1185G7 @ 3.00GHz   1.80 GHz processor and 32 GB RAM. 

\textbf{Assumption of meter placement: }Today's industrial practices and guidelines \cite{PMU_install_guideline} suggest the installation of PMUs at plants generating more than 100 MVA, large load buses, and grid control devices. Thus, in this paper, we assume the installation of PMUs on every generation bus and traditional SCADA meters (RTUs) on other injection buses without generators. We further assume line power flow measurements at randomly selected transmission lines that have an RTU located at either the from or to node.
}

\subsubsection{Robustness: WLAV outperforms WLS}
 \begin{figure}[b]
	\centering	\includegraphics[width=0.45\linewidth]{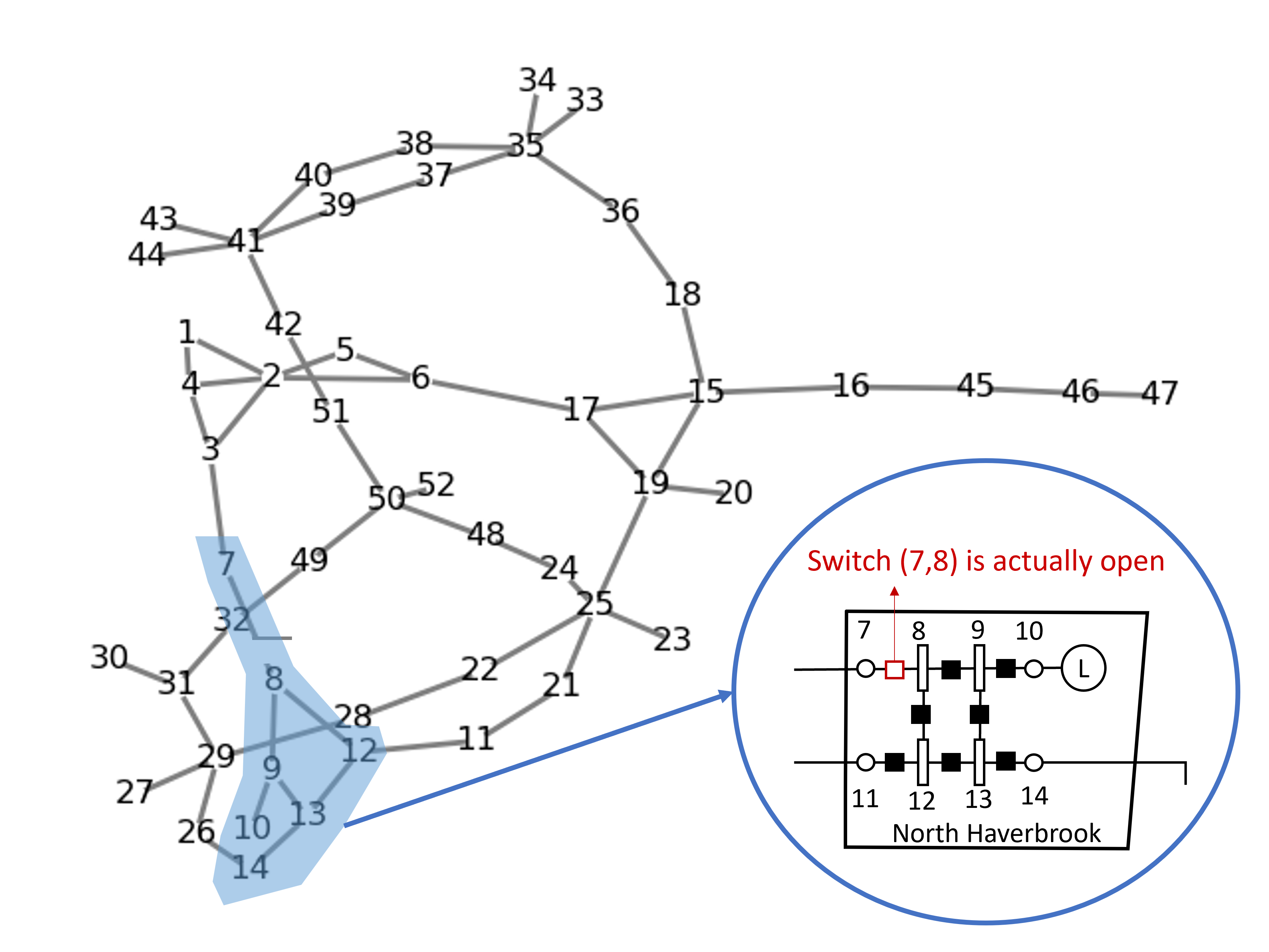}
	\caption{{  CyPRES 8 substation network. (The case is modified by opening the switch (7,8))}}
	\label{fig:8-substation case}
\end{figure}

{ Here, we evaluate the robustness of our robust estimator. Here the weighted least absolute value (WLAV) based method is expected to have two desirable properties that a weighted least square (WLS) method does not have:

\begin{itemize}
    \item  \textbf{automatically reject} data errors: the state solution is still accurate when data errors exist. In this paper, the evaluation metrics for solution accuracy include: 
    \begin{enumerate}
        \item root mean squared error (RMSE) which evaluates the overall deviation from the true states:
        \begin{equation}
            RMSE = \sqrt{||x_{est}-x_{true}||_2^2}
            \label{eq:rmse}
        \end{equation}
    \item number of inaccurate bus estimates: which is the number of buses whose estimated states have $>0.02 pu$ $|V|$ error or $>2^{\circ}$ phase angle ($\theta$) error, i.e.,
    \begin{align}
        &\text{Number of inaccurate bus estimates}\notag\\
        =&\sum_{busi} I\{|\Delta|V_i||>0.02, or|\Delta\theta_i|>2^\circ\}\label{eq:num of inaccurate bus}\\
        &\text{with } \Delta|V_i|=|V_i|_{est}-|V_i|_{true},\notag \\ &\Delta|\theta_i|=|\theta_i|_{est}-|\theta_i|_{true} \notag
    \end{align}
    A small value means that solution inaccuracy only exists
    regionally on a subset of buses.
    \end{enumerate}
    \item \textbf{identify} data errors: multiple types of data errors (even when they co-exist) which affect state estimation can be detected and localized:
    \begin{align}
        &\text{For PMU i, create alarm if } |n_{pmui}|>0.1\notag\\
        &\text{For RTU j, create alarm if } |n_{rtuj}|>0.1 \label{eq:bdd criteria}\\
        &\text{For sw k, raise suspicion if } |n_{swk}|>0.05,\notag\\
        &\text{ and create alarm by } \text{hypothesis test (Table \ref{tab:hypothesis test})}\notag
    \end{align}
     These bad data identification thresholds are empirically learned from our synthetic data, specifically by observing the data and finding a threshold value that effectively separates bad data points from normal ones, and they work well in our experiments. In real-world applications, the grid operators may need to learn their own optimal threshold from their real data, by observation, experience, or checking the area under curve (AUC) metric.
    However, due to the redundancy in realistic switch installation, some wrong switches will not affect the state estimation, and they are undetectable, as discussed later in Section \ref{sec:detectability}.  Thus in this work, we do not adopt any performance metric since they may not reflect the quality of estimation.  
\end{itemize}
}

Here, we consider the following types of data errors that can realistically occur and disrupt estimation:
\begin{enumerate}
    \item \textbf{topology error}: either 1) a switch is actually \textit{open} but reported as \textit{closed},
    or 2) a switch is actually \textit{closed} but reported as \textit{open}
    \item \textbf{bad (continuous) data} from RTU or PMU, also known as \textit{(traditional) bad data}, which appears as a large deviation (1 p.u. in this paper) from the true value 
\end{enumerate}

We conduct experiments on an 8 substation node-breaker case.
Table \ref{tab:case info} and Figure \ref{fig:8-substation case} shows the case information and experiment settings.

\begin{table}[htbp]
\caption{Experiment settings on 8-substation case}
\label{tab:case info}
\begin{tabular}{m{0.16\linewidth}m{0.73\linewidth}}
\hline
{\bf Case name} & CyPRES 8-substation cyber-physical power system case \\
\hline
{\bf Case info} & 
\begin{compactitem}
    \item 52 nodes, 49 breakers (switches)
    \item 5 generators (4 of them are active), 6 loads, 1 shunt 
    \item 1 transformer, 11 transmission lines
\end{compactitem}\\
\hline
{\bf Synthetic meters: Location and Type} & 
\begin{compactitem}
    \item Status data created on 49 breakers
    { 
    \item 5 PMU buses: each generator bus has a PMU installed to collect voltage and current phasors
    \item 7 RTU buses: each load bus has an RTU installed to collect $P_{rtu}, Q_{rtu}, |V|_{rtu}$ data
    }
    \item 22 RTU line meters (measurements include $P_{ij,rtu}, Q_{ij,rtu}$ and voltage magnitude at one end $|V_{i,rtu}|$ ) on selected lines
    \item Data generated by adding Gaussian noise (std=0.001) to power flow solution
\end{compactitem}\\
\hline
{  {\bf Data error generation}} &
{  Bad RTU and topology errors are created (randomly):
 \begin{compactitem}
     \item sw (7,8): actually open but measured as closed
     \item sw (20,19): actually closed but measured as open   
     \item Bad RTU meter on bus 34: measurement of load values are perturbed by large random noise
 \end{compactitem}
 }
\\
\hline
{\bf Hyper-parameters} & 
\begin{compactitem}
    \item{  RTU weights $= 1$, PMU weights $=1$}
    \item Switch weights $=0.001$
\end{compactitem}
(See Section \ref{sec: sensitivity} for details of weight selection.)
\\
\hline
\end{tabular}
\end{table}

\begin{figure*}[h]
     \centering
     \begin{subfigure}[b]{0.49\linewidth}
         \centering         \includegraphics[width=0.95\textwidth]{figures/ckt-se/wlav_bdd_graph.png}
         \caption{The WLAV-based estimator pinpoints 2 topology errors and 1 bad RTU. (Node values are $|n_{rtu}|, |n_{pmu}|$ scaled to $[0,1]$ by $\frac{|n|}{max(n)}$; edge values are switch alarms (0 or 1).}
         \label{fig: wlav bdd graph}
     \end{subfigure}
     \hfill
     \begin{subfigure}[b]{0.49\linewidth}
         \centering         \includegraphics[width=0.95\textwidth]{figures/ckt-se/wls_bdd_graph.png}
         \caption{The WLS counterpart gives dense error indicators, unable to recognize all data errors correctly.}
         \label{fig: wls bdd graph}
     \end{subfigure}
     \begin{subfigure}[b]{0.49\linewidth}
         \centering         \includegraphics[width=0.9\textwidth]{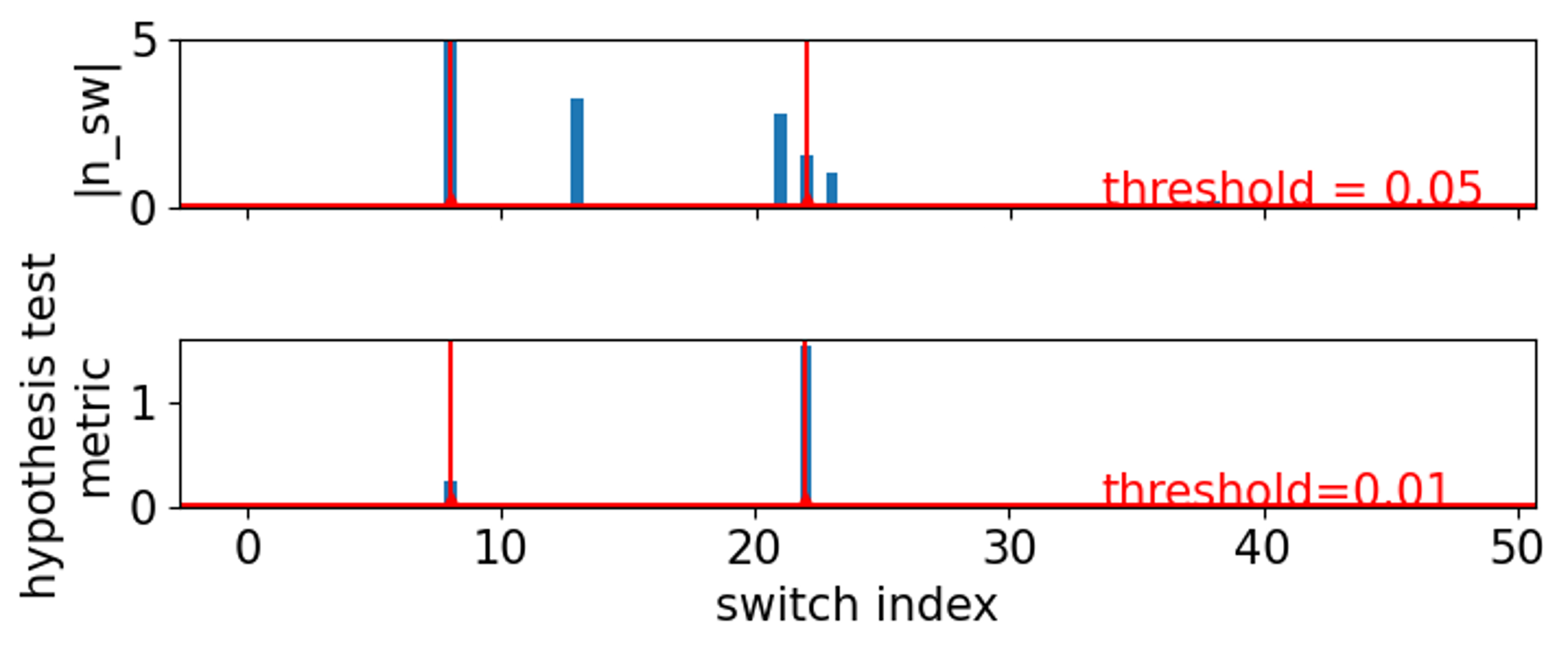}
         \caption{WLAV: sparse $n_{sw}$ 
         filters out suspicious switches and hypothesis test verifies the suspicion. Red vertical lines mark the true locations of topology errors.
          }
         \label{fig: wlav switch bdd}
     \end{subfigure}
     \hfill
     \begin{subfigure}[b]{0.49\linewidth}
         \centering         \includegraphics[width=0.9\textwidth]{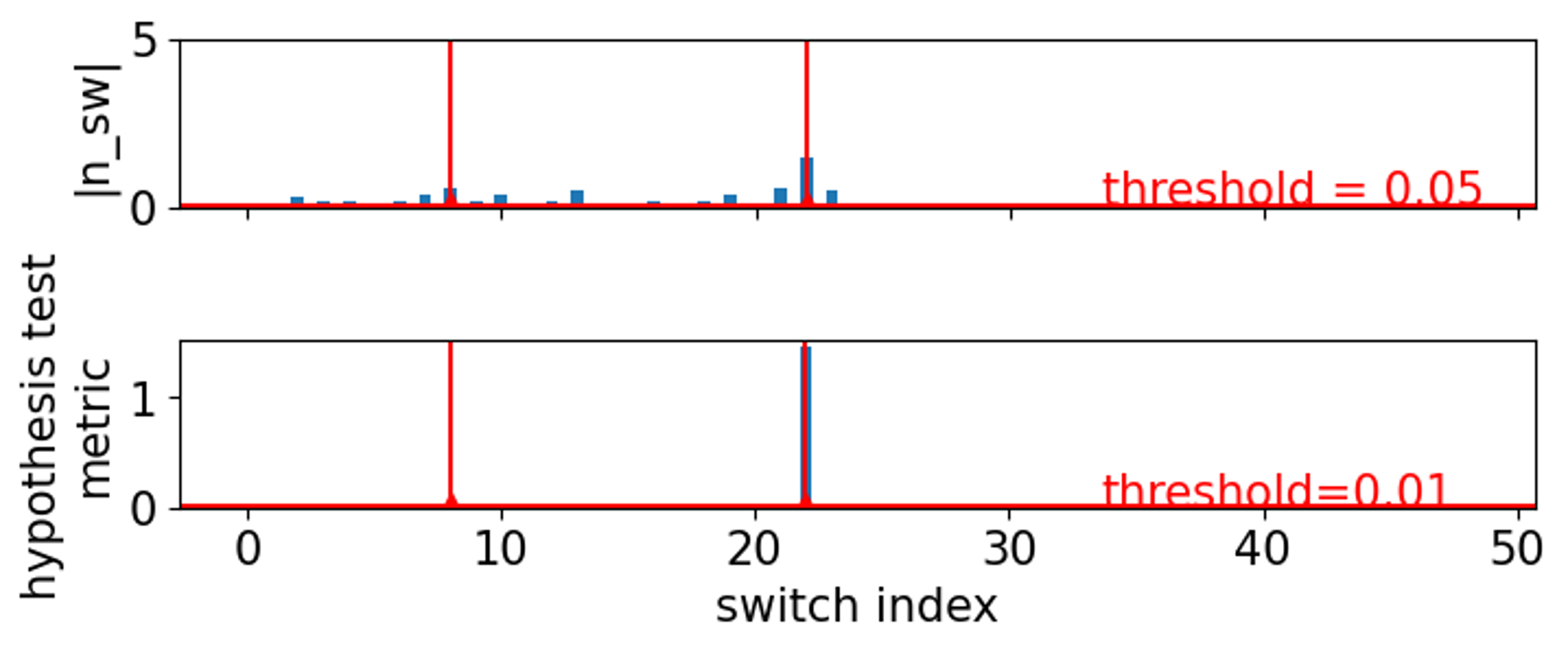}
         \caption{The WLS counterpart results in denser estimates of $n_{sw}$, and fails to recognize all topology errors after hypothesis test.}
         \label{fig: wls switch bdd}
     \end{subfigure}
     \begin{subfigure}[b]{0.49\linewidth}
         \centering         \includegraphics[width=0.95\textwidth]{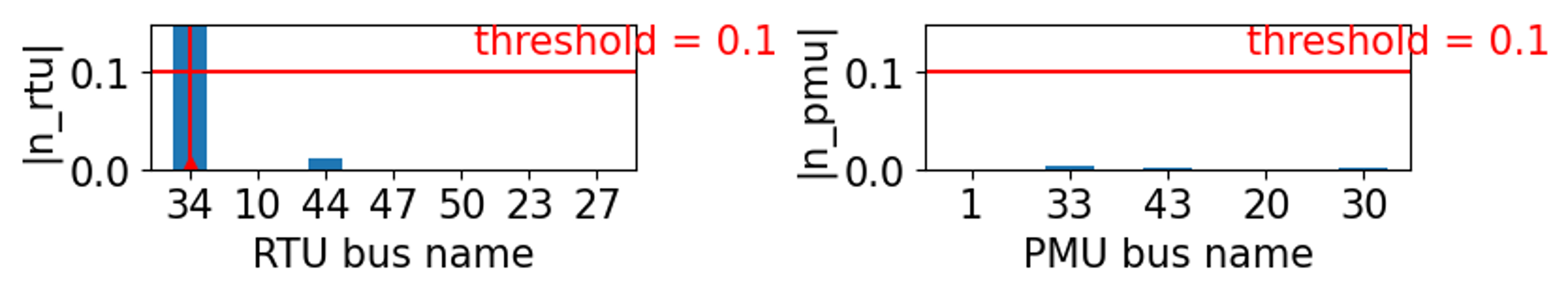}
         \caption{WLAV: sparse $n_{rtu}$ clearly identifies the bad RTU at bus 34. }
         \label{fig: wlav bus bdd}
     \end{subfigure}
     \hfill
     \begin{subfigure}[b]{0.49\linewidth}
         \centering         \includegraphics[width=0.95\textwidth]{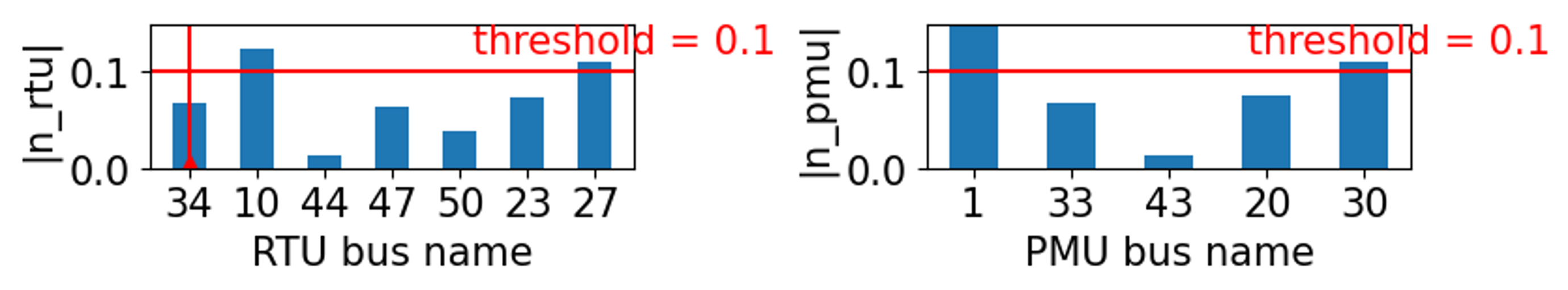}
         \caption{The WLS counterpart results in dense $n_{rtu}$ and $n_{pmu}$, leading to false alarms.}
         \label{fig: wls bus bdd}
     \end{subfigure}
     \begin{subfigure}[b]{0.49\linewidth}
         \centering         \includegraphics[width=0.9\textwidth]{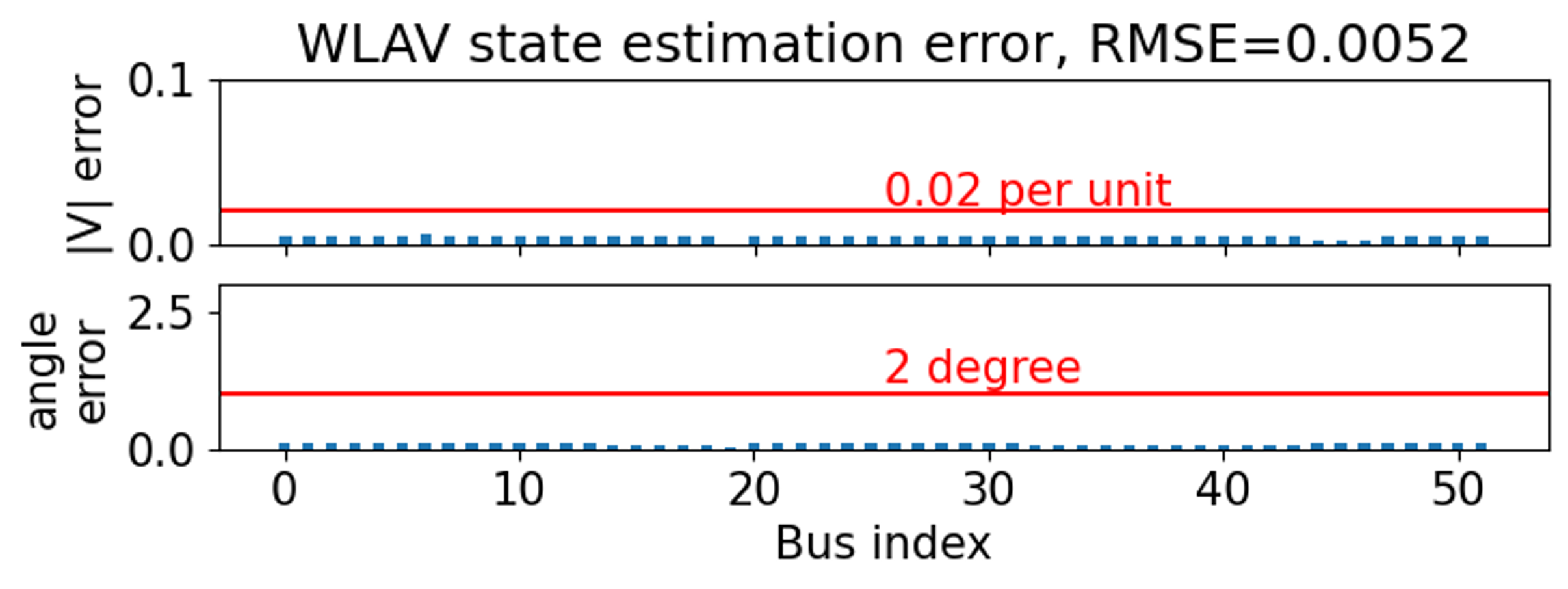}
         \caption{WLAV: Solution of bus voltages remains accurate.}
         \label{fig: wls voltage error}
     \end{subfigure}
     \hfill
     \begin{subfigure}[b]{0.49\linewidth}
         \centering         \includegraphics[width=0.9\textwidth]{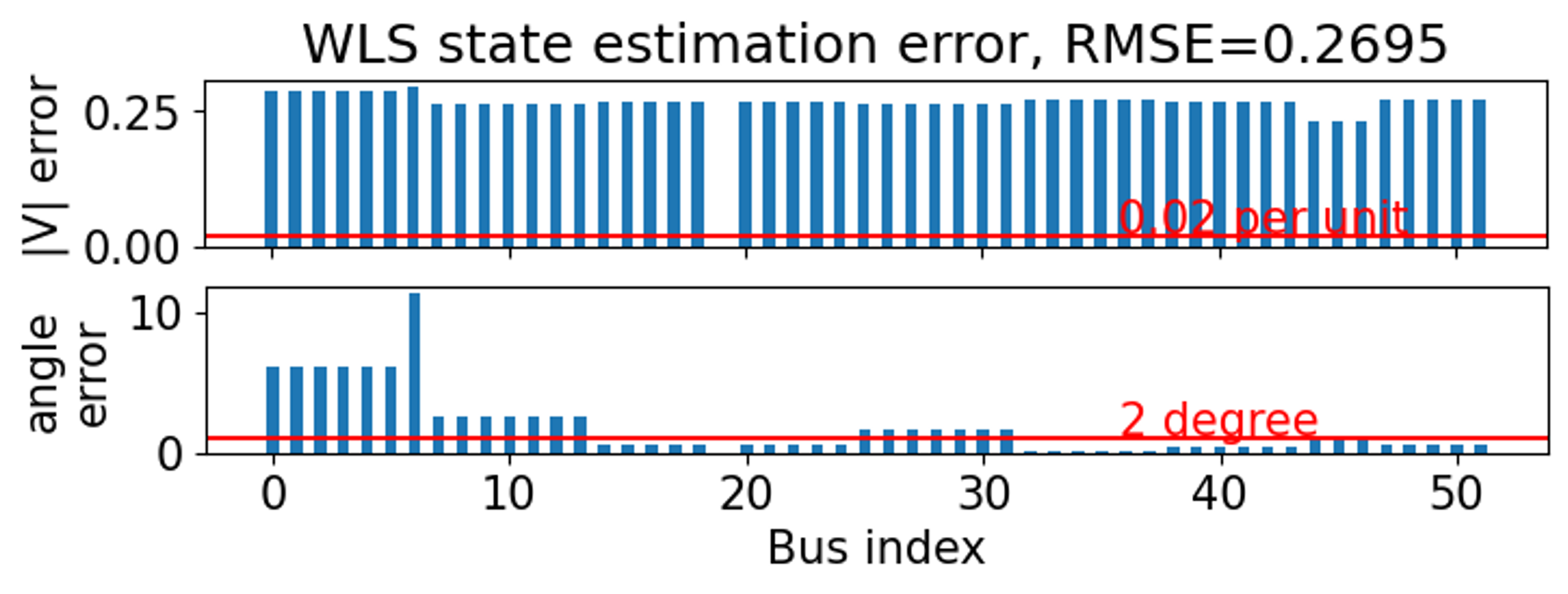}
         \caption{WLS: solution has widespread  $|V|$ and angle errors. }
         \label{fig: wls voltage error}
     \end{subfigure}
        \caption{{  Robustness of WLAV (left) vs WLS (right): WLAV better identifies data errors and obtains accurate estimates. }}
        \label{fig: wlav vs wls}
        
\end{figure*}

{ 
Results in Figure \ref{fig: wlav vs wls} demonstrate the robustness of the proposed WLAV-based robust estimator (Problem Definition \ref{def: ckt-GSE}) by comparing it against its WLS-based non-robust counterpart \ref{def: sugar-SE}. 
In terms of data error identification, Fig. (\ref{fig: wlav bdd graph}) and (\ref{fig: wls bdd graph}) demonstrate that the proposed WLAV model can provide sparse error indicators to precisely identify the topology errors and bad RTU bus; however, the WLS method fails to identify all topology errors and instead results in false alarms at many bus locations. 
Fig. (\ref{fig: wlav switch bdd}) - (\ref{fig: wls bus bdd}) further illustrates how the values of $n_{sw}, n_{rtu}, n_{pmu}$ along with hypothesis test can effectively identify different data errors. 
Further, in terms of the accuracy of state estimates, the WLAV model provides accurate solution with significantly smaller $|V|$ error, angle error, and RMSE. In contrast, the WLS solution  is significantly perturbed by data errors.

\subsubsection{The boundary of robustness: when does it fail?}

{ 
As the WLAV-based robust estimator achieves its desired robustness by enforcing sparsity, it relies on a basic assumption that the data errors are sparse. However, this property does not hold under higher penetration of data errors. In this Section, we explore how the growing percentage of topology errors will affect robustness.

\begin{figure*}[h]
     \centering
     \begin{subfigure}[b]{0.32\linewidth}
         \centering         \includegraphics[width=0.99\textwidth]{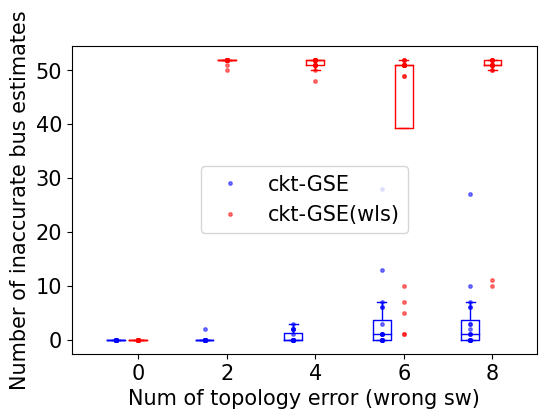}
     \end{subfigure}
     \hfill
     \begin{subfigure}[b]{0.32\linewidth}
         \centering         \includegraphics[width=0.99\textwidth]{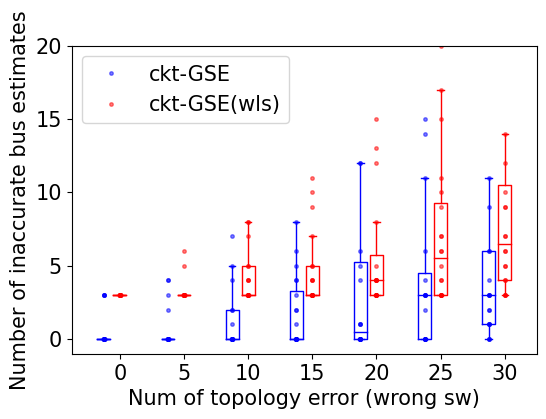}
     \end{subfigure}
     \begin{subfigure}[b]{0.32\linewidth}
         \centering         \includegraphics[width=0.99\textwidth]{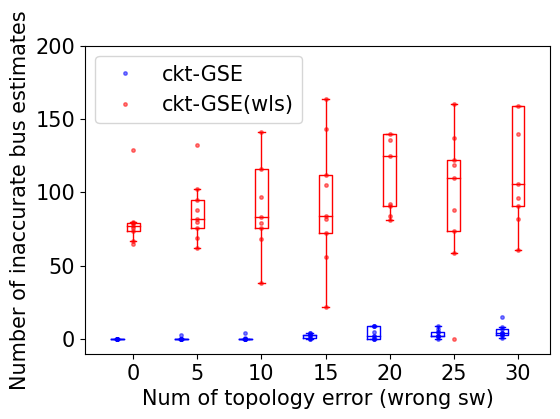}
     \end{subfigure}
     \hfill
     \begin{subfigure}[b]{0.32\linewidth}
         \centering         \includegraphics[width=0.99\textwidth]{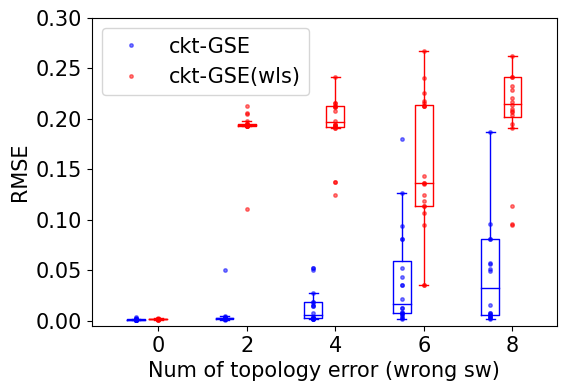}
         \caption{8-substation case: 51 nodes, 49 sw(itches)}
     \end{subfigure}
     \hfill
     \begin{subfigure}[b]{0.32\linewidth}
         \centering         \includegraphics[width=0.99\textwidth]{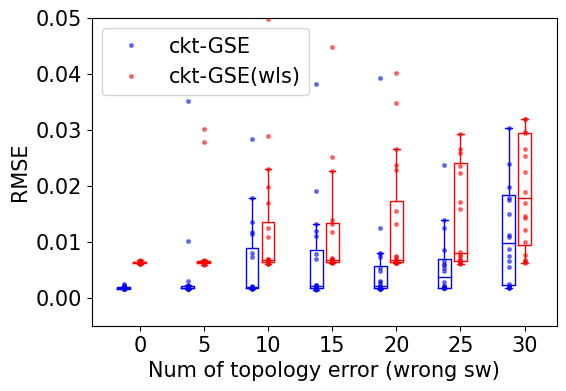}
         \caption{300-substation case: 1.6k nodes, 1.8k sw}
     \end{subfigure}
     \hfill
     \begin{subfigure}[b]{0.32\linewidth}
         \centering         \includegraphics[width=0.99\textwidth]{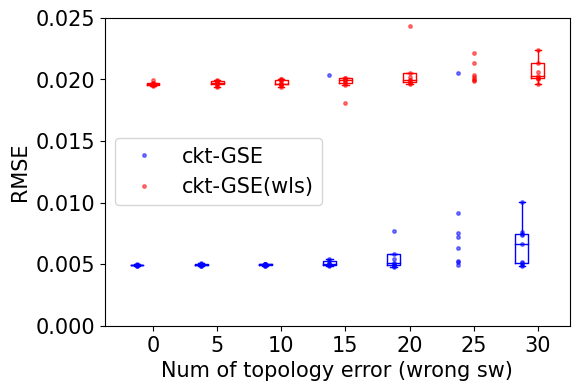}
        \caption{2000-substation case: 24k nodes, 23k sw}
     \end{subfigure}
    \caption{{  
    Robustness on different sized networks: The top row shows the number of inaccurate bus estimates defined in (\ref{eq:num of inaccurate bus}) where a small value means inaccurate estimates only exist on a subset of nodes, and the bottom row shows RMSE defined in (\ref{eq:rmse}) which reflects the overall inaccuracy of solution. Our robust estimator remains very robust under low (sparse) penetration of topology errors and degrades robustly as topology errors grow, i.e., inaccuracy gradually appears on a larger subset of nodes. In contrast, the WLS model is not robust: even a few topology errors result in widespread and significant state estimate inaccuracies. }}
        \label{fig: robustness boundary}
        
\end{figure*}
Table \ref{tab: exp setting, 3 cases} shows the experiment settings for different cases and Figure \ref{fig: robustness boundary} shows the results. We evaluate solution quality under a growing number of topology errors. 
Results show that our robust estimator has nearly zero inaccurate bus estimates and nearly zero RMSE under a small number of topology errors. This means it remains very robust under low (sparse) penetration of topology errors. As we have more topology errors, performance degrades in a robust way: it makes mistakes at a subset of locations (regionally in subsets around the wrong switches), whereas the remaining bus locations still obtain accurate estimates.  As there are more topology errors, inaccuracy gradually spreads out. Whereas for the WLS non-robust counterpart, there is always a larger number of buses whose state estimation is inaccurate, and a larger RMSE. This means the WLS solution is not robust and inaccuracy is wide-spread  even with a few topology errors.

Thus the main finding is that the robustness of our robust circuit-based state estimator degrades when the population of data errors becomes large. This holds for all WLAV-based models in general. The limitation is due to the violation of the sparse-data-error assumption and the algorithm's sensitivity to topology errors. Section \ref{sec: sensitivity} includes more discussions on the algorithm's sensitivity to different types of data errors.
}

\begin{table}[htbp]
\caption{ Experiment settings on different cases}
\label{tab: exp setting, 3 cases}
\begin{tabularx}{\linewidth}{ll}
\hline
\bf CASE &\bf \begin{tabular}[c]{@{}l@{}} Settings\\(sec) \end{tabular} \\
\hline
\begin{tabular}[c]{@{}l@{}}
\textbf{8-substation case}\\
- 52 nodes\\ - 49 switches\\
\end{tabular}& \begin{tabular}[c]{@{}l@{}}
- 5 PMUs, 7 RTUs, 22 RTU line meters \\
- experiments repeated 20 times with different random data error locations
\end{tabular}\\
\hline
\begin{tabular}[c]{@{}l@{}}
\textbf{300-substation case}\\
- 1598 nodes\\ - 1816 switches\\
\end{tabular} & \begin{tabular}[c]{@{}l@{}}- 69 PMUs, 224 RTUs, 608 RTU line meters \\
- experiments repeated 20 times with different random data error locations\end{tabular}\\
\hline
\begin{tabular}[c]{@{}l@{}}
\textbf{Texas CP-2000 case}\\
\textbf{(2000-substation case)}\\
- 24360 nodes \\- 22632 switches\\
\end{tabular}  & \begin{tabular}[c]{@{}l@{}}- 522 PMUs, 1524 RTUs, 4690 RTU line meters \\
- experiments repeated 10 times with different random data error locations\end{tabular}\\
\hline
\end{tabularx}
\footnotesize{
*The Texas CP-2000 case is a cyber-physical model built from the footprint of the Texas grid.\\
*PMUs are placed on generation buses, RTUs are placed on load buses, and RTU line meters are placed on random lines.
}
\end{table}

\subsubsection{Scalability}
The estimator needs to be time-efficient on large-scale networks to be applicable in real-world control rooms.
{ Here we evaluate the speed of our proposed circuit-based (ckt) solvers by comparing with standard LP solvers:
\begin{itemize}
    \item interior-point (IP) solver in python CVXOPT toolbox
    \item Simplex method in SciPy which solves min-max model
\end{itemize}
Figure \ref{fig:scalability GSE} shows the speed performance on different sized networks. By comparison, our ckt solver is significantly faster than the standard toolbox on large scale cases. 
}

\begin{figure}[h]
	\centering
	\includegraphics[width=0.4\linewidth]{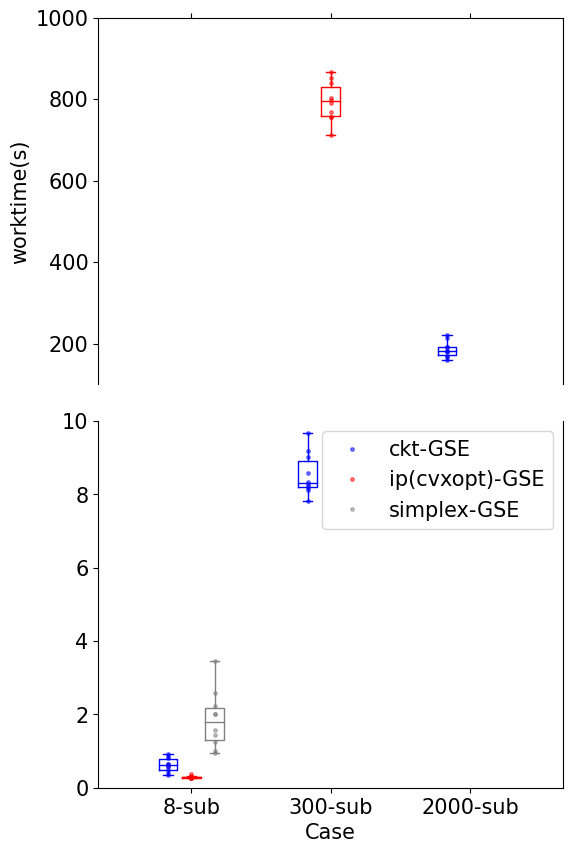}
	\caption{  Speed and scalability: generalized state estimation (considering topology errors). We compare work time of circuit-based (ckt) solver VS a standard interior-point (IP) solver in CVXOPT toolbox and a Simplex solver in Scipy toolbox. 
    Our ckt solver is efficient on different sized networks. CVXOPT is comparable with ckt solver only on small 8-substation case (a 52-node network), whereas it becomes significantly slower on larger cases and even fails on 2000-substation case (a >24k-node network)).
    Simplex solver is the slowest. It only works on the smallest case and fails on others (thus is not shown). } 
	\label{fig:scalability GSE}
\end{figure}

\subsection{Discussion: when can topology errors become undetectable on node-breaker model?}\label{sec:detectability}

Although the use of the node-breaker model enables considering all switching devices and detecting topology errors, not all wrong switch statuses are detectable (i.e., undetectability) using the observed data. In the real world, there exist cases where different grid configurations and anomaly scenarios have the same physical effect, and thus at times, one cannot accurately localize the source of an anomaly (i.e., mis-localization). These issues are limitations for all node-breaker based estimation methods such that some data errors can be \textit{undetectable} or \textit{mis-localized}, unless additional sources of information are included. The major cause is the redundancy of power system components. Figure \ref{fig:detectability} illustrates 3 realistic scenarios where we may observe these issues, and Table \ref{tab:detectability issues} further describes the causes of potential \textit{undetectability} and \textit{mis-localization} of wrong switch status in these scenarios, as well as how these limitations affect the solution quality.

\begin{figure*}[ht]
	\centering\includegraphics[width=0.9\linewidth]{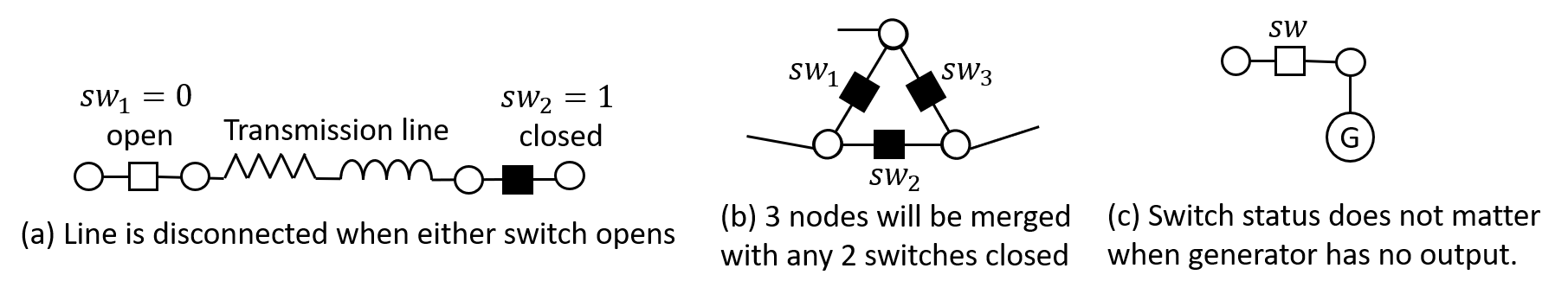}
	\caption{Detectability issues: three examples where the robust circuit-based generalized state estimator will have problems identifying wrong switch statuses. See Table \ref{tab:detectability issues} for illustration.} 
	\label{fig:detectability}
\end{figure*}

\begin{table*}[b]
\caption{Detectability issues of wrong status data on node breaker model}
\label{tab:detectability issues}
\small
\begin{tabularx}{\linewidth}{m{0.19\linewidth}m{0.41\linewidth}m{0.33\linewidth}}
\hline
\bf Realistic condition & \bf Detectability issues & \bf Impact of the issue\\
\hline
\textbf{Equivalent line switches:} a realistic transmission line usually has switching devices at both ends of it, see Figure \ref{fig:detectability}(a).
& \textit{Undetectability:} When one switch is open, any wrong status on the other switch is undetectable. This is because the transmission line is disconnected with either switch open, and the other switch, whether open or closed, does not impact the true grid states.
& Such instances of undetectability do not affect the quality of solution as it has no impact on other grid states. From the viewpoint of  bus-branch model, such wrong switch status has no impact on grid topology.\\
& \textit{mis-localization}: When one switch is open and the other is closed, any wrong status on the open switch can be mis-localized at the wrong position. E.g., let $0$ denote open, and $1$ denote closed, when the true status is $[sw_1,sw_2]=[0,1]$, measured status is $[1,1]$, the wrong status localization may estimate the status to be $[1,0]$.
& Such mis-localization has no impact on the solution since the true state and the mis-localized state are equivalent, with the same system topology in the bus-branch model.\\
\hline
\textbf{Cyclic connection of switches:} due to system redundancy some closed switches can form a cyclic graph, see Figure \ref{fig:detectability}(b). 
&  \textit{Undetectability:} In Figure \ref{fig:detectability}(b), when any 2 switches are closed, the status of the third switch, whether closed or open, has no impact on the grid operation. Therefore, we cannot detect the wrong status indication for the third switch. 
&Such an instance of undetectability does not affect the quality of the solution as it has no impact on other grid states. From the viewpoint of the bus-branch model, topology remains unchanged independent of the status of the third switch. \\
& \textit{mis-localization}: When 2 switches are closed and one open, wrong status on any closed switch can be mis-localized. E.g., the true status is $[sw_1,sw_2,sw_3]=[1,1,0]$, the measured status is $[1,0,0]$, then a wrong estimation could be $[1,0,1]$
& Such mis-localization has no impact on the solution as the mis-localization does not change the bus configuration in bus-branch model.\\
\hline
\textbf{Switching device connected to a node of degree one}, see Figure \ref{fig:detectability}(c). 
&\textit{Undetectability:} In Figure \ref{fig:detectability}(c), when a generator has no output (produces no power), the switch status has no impact on the grid operating state and its bus-branch model. Therefore, the wrong status  on this switch is undetectable.
&The undetectability does not affect the quality of our estimator as it has no impact on the grid operating state.\\
\hline
\end{tabularx}
\end{table*}

\chapter{Sparse structures and lightweight ML}\label{ch: ML}
\section{Temporal and spatial sparse structures for ML}

The previous Chapter \ref{ch: physics} has discussed the use of sparse optimization to bring intrinsic robustness into the simulation and estimation tools, producing sparse threat indicators to pinpoint failure sources and anomalies of interest. However, the modern threats are still not well handled in these physics-based simulation and estimation methods, leading to remaining gaps to be fulfilled. Below are two examples of remaining gaps:
\begin{itemize}
    \item Remaining efficiency gap: large-scale systems are hard to simulate due to the nonlinear nature of the power system simulation problem.
    \item Remaining robustness gap: existing analytical tools are not good at cyberthreats: estimation problems can be compromised by the anomalous injected by false data injection attack (FDIA)\cite{fdia-acse}\cite{fdia-review}; and simulation problems are not good at multiple-location disturbances like those induced by MadIoT attack \cite{madiot} because the simulator hardly has a initial condition close to the solution point when system changes significantly (this is also an efficiency gap)
\end{itemize}

These remaining gaps motivate the use of ML to predict and detect under modern threats. 

In this Chapter, we highlight that the exploitation of sparse structures is helpful to develop simple-structured ML models with high-performance generalization and scalability. These ML models will contribute to situational awareness with advanced robustness and efficiency. 

We start from the discussing a temporal sparse structure. An important question ML often faces is: how to make a ML model generalize to (big) data that unavoidably come from different distributions? ML research is always struggling with bias and variance issues. Learning from too many irrelevant data can lead to high bias and even erroneous predictions, whereas not making use of the large data set can potentially cause high-variance (i.e., overfitting). 

\begin{figure}[h]
	\centering
\includegraphics[width=0.8\linewidth]{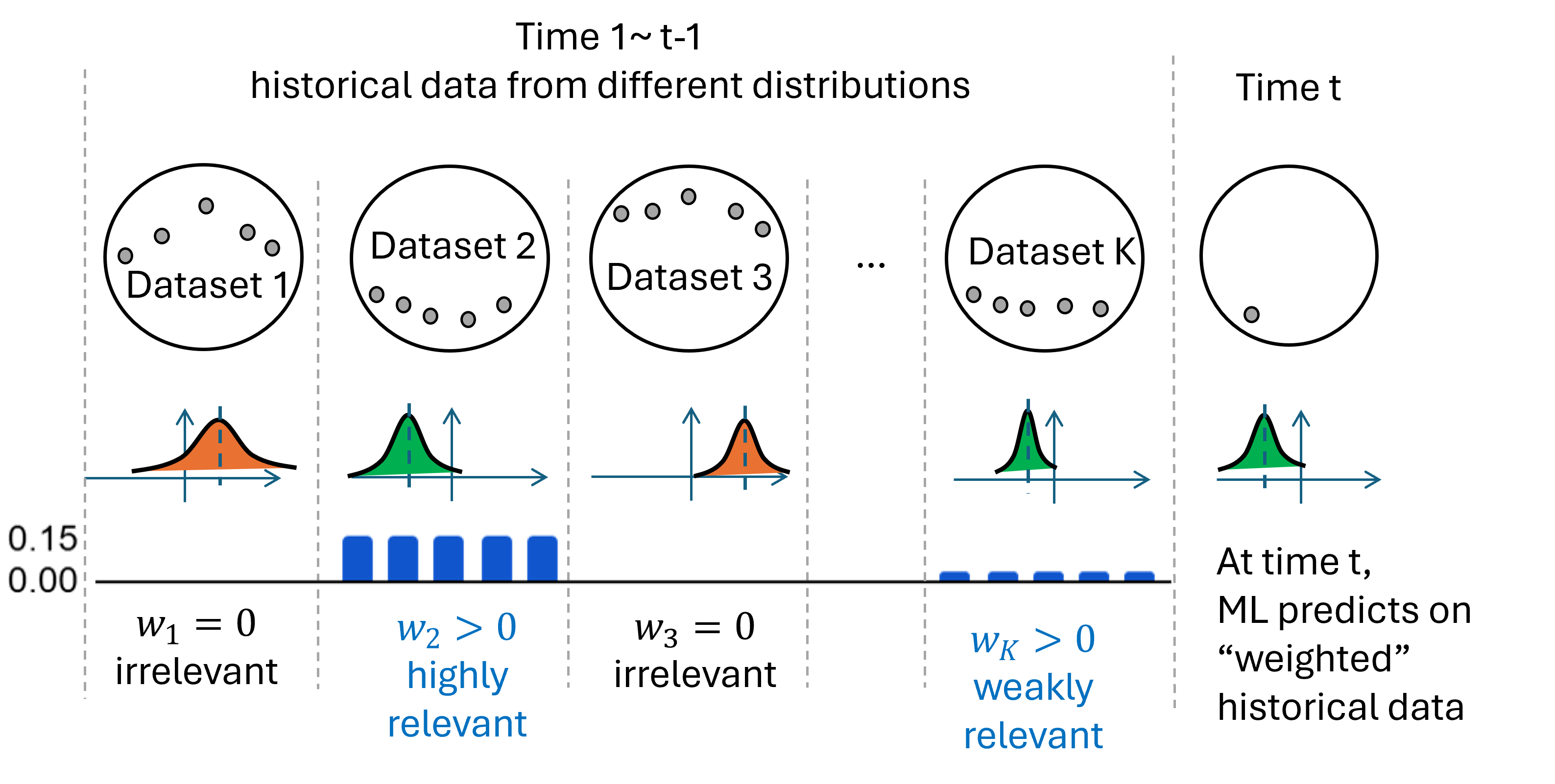}
	\caption{Temporally sparse weights pinpoint the relevant historical data: a sparse recommendation.}
	\label{fig: temporal sparsity}
\end{figure}

Assigning sparse temporal weighting provides a solution to this issue, by giving sparse recommendations of relevant historical data. As shown in Figure \ref{fig: temporal sparsity}, suppose we need to make a data-driven prediction at time $t$, from historical data coming from $K$ different distributions. Grouping these data into $K$ datasets, we can assume that each dataset is drawn from its corresponding distribution $N_k, \forall k=1,2,...,K$. Suppose the true distribution takes the form as shown in the figure. 
Dataset 1 and 3 are assigned with zero weights, since they are irrelevant data too far from the distribution at time $t$. Whereas, the remaining relevant datasets are assigned with non-zero weights.
Suppose we are able to optimally assign sparse weights reflecting the relevance of data; and make time $t$ predictions using weighted historical data. Then these sparse weights are able to pinpoint the level of contribution each dataset should make in our final prediction for time $t$, and remove the irrelevant data that might cause a biased output. Section \ref{sec: dynwatch} will further elaborate on this idea, by developing \method, a distance-based method to obtain the sparse weight vector from an optimal bias-variance trade-off.

On the other hand, a large power system is typically spatially sparse due to its sparse edge connections, i.e., each bus (node) is only connected to a few adjacent lines, instead of fully connected with all other nodes. This leads to a highly sparse admittance matrix. \cite{Alvarado1976computation} shows that for typical power systems, a reasonable complexity of solving linear $Y_{bus}x=b$ is $n^{1.2}$. In Section \ref{sec: gridwarm}, we will exploit sparse graphical structures to design a "lightweight" ML predictor, which is simple-structured, but generalizable, scalable and interpretable.

\section{Application 1 -  \method: exploiting temporal sparsity for generalization on dynamic graphs}\label{sec: dynwatch}

In this section, we introduce \method which exploits sparse temporal weighting to make ML prediction and detection account for the different distributions of historical data. We consider a more generalized scenario where the true distributions of historical data are not known, but we can still manage to quantify the relevance of historical data via a distance-based method. \method can be useful in aiding the estimation tool via a Physics-ML Synergy design later in Chapter \ref{ch: synergy}.

Specifically, we focus on scenarios where distribution differences are caused by changes in network topology.
\begin{figure}[h]
	\centering
	\includegraphics[width=0.6\linewidth]{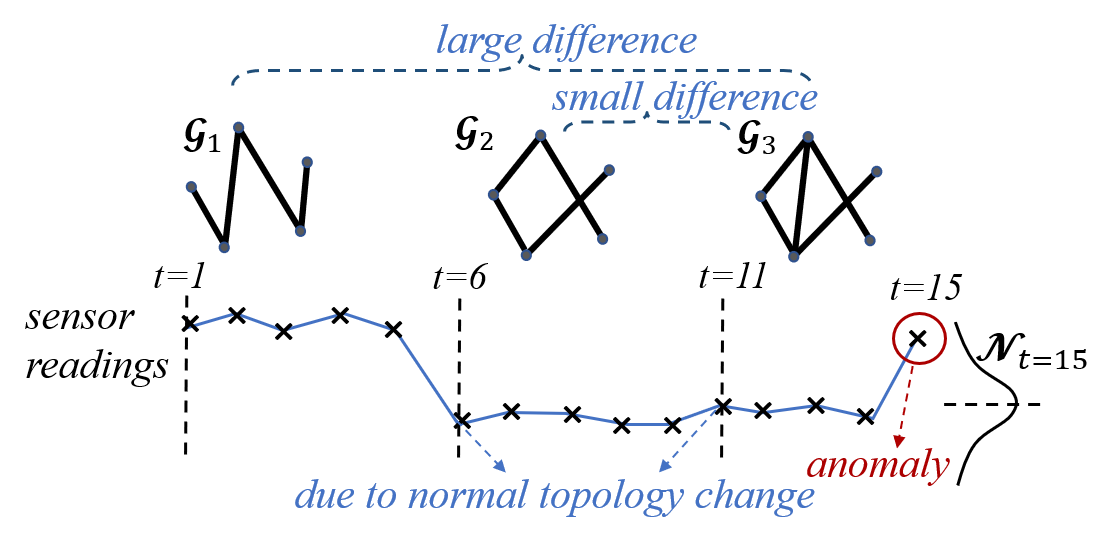}
	\caption{Toy example of \method: different distributions are caused by topology changes.}
\label{fig:toyexample_nonlocal}
\end{figure}

Consider the simple power grid shown in Figure \ref{fig:toyexample_nonlocal}, which evolves over time from $\G_1$ to $\G_2$ to $\G_3$. For simplicity, assume that we have a single sensor, from which we want to detect anomalous events. How do we evaluate whether the current time point ($t=15$) is an anomaly? If the graph had not been changing, we could simply combine all past sensor values to learn a distribution of normal behavior (e.g. fitting a Gaussian distribution as in $\mathcal{N}_{t=15}$), then evaluate the current time point using this Gaussian distribution (e.g. in terms of the number of standard deviations away from the mean).

In the changing graph setting, we still want to learn a model of normal behavior ($\mathcal{N}_{t=15}$), but while taking the graph changes into account. Note that $\G_2$ and $\G_3$ are only slightly different, while $\G_1$ and $\G_3$ are very different. Hence, the sensor values coming from $\G_2$ (i.e. time $6$ to $10$) should be taken into account more highly when constructing $\mathcal{N}_{t=15}$, as compared to those from $\G_1$. Intuitively, the sensor values from $\G_1$ are drawn from a very different distribution from the current graph, and thus should not influence our learned model $\mathcal{N}_{t=15}$. In general, the more similar a graph is to the current graph, the more we should take its sensor values into account when learning our current model. This motivates the 3-step process we use:
\begin{enumerate}
\item {\bf Graph Distances:} Measure the distance between each past graph and the current graph.
\item {\bf Temporal Weighting:} Weight the past sensor data, where data from graphs that are similar to the current one are given higher weight.
\item {\bf Prediction and Anomaly Detection:} Learn a distribution of normal behavior ($\mathcal{N}_{t}$) from the weighted sensor values, and measure the anomalousness at the current time based on its deviation from this distribution.
\end{enumerate}

For electric power grids, we assume that the topology $\G(t)$ at any time $t$ can be available; and that the  measurements should be consistent with distribution for this topology. In practice, $\G(t)$ can be the output from topology estimation (NTP) in the grid energy management system (EMS). It is the best estimate of the grid topology at any time. Notably, this topology is not assumed to be perfect and accurate, as anomalies like an unknown line outage can cause us to have a wrong topology that does not reflect the disconnection of this line. The broad variety of anomalies detectable by \method includes but is not limited to  line/generator/load contingencies, and anomalous data.

In the remainder of this Section, we first introduce our power-grid-specific graph distance measure based on Line Outage Distribution Factors (LODF)~\cite{wood2013power}. Then, we describe our temporal weighting and anomaly detection framework, which flexibly allows for any given graph distance measure. Finally, we present an alternate distance measure that is locally sensitive, i.e., it accounts for the local neighborhood around a given sensor.

\subsection{Graph Distance Measure for Power Grid} \label{sec:measure}

In this section, we describe our proposed graph distance measure to calculate the distance $D(\G_i, \G_j)$ between any pair of graphs. For ease of understanding, the rest of this Section uses the example of anomaly detection at $t=15$ in Figure \ref{fig:toyexample_nonlocal} as an extended case study, but our approach can be easily extended to the general case. 

For an anomaly detection algorithm to work well on the power grid applications, the choice of graph distance needs to consider {\bf problem-specific challenges} along with desirable properties for an anomaly detection algorithm (scalability, sensitivity to change, and {\bf ‘importance-of-change awareness’}). One grid-specific challenge is that the ideal graph distance should capture ‘grid physics’ rather than only graph structural changes. Specifically, the distance should be sensitive to the redistribution of power flow, not only the addition/deletion of nodes/edges, since the anomaly information is extracted from power flow measurements. Meanwhile, as the ‘importance-of-change awareness’ indicates, grid changes that cause big shifts in power flow (measurements) should result in larger graph distances, than changes that cause minor power impact. Unfortunately, none of the classical graph distances can capture the physics of the power flow and quantify the impact of graph change in terms of power. To handle this, this work proposes a novel design of graph distance by making use of the power sensitivity factor.

Intuitively, the goal is for our graph distance to represent {\bf redistribution of line power flow}. Critical changes in topology result in large redistributions of power. Thus, the graph distance arising from a topology change should be large if the changed edges can potentially cause large amounts of power redistribution.

Hence, given two graphs $\G_i(\V(i),\E(i))$ and $\G_j(\V(j),\E(j)$ with different topology, we first define a transition state which takes the union of the two graphs:
$$\G_{trans}=(\V(i)\cup\V(j),\E(i)\cup\E(j))$$ 

\begin{figure}[h]
	\centering
	\includegraphics[width=0.3\linewidth]{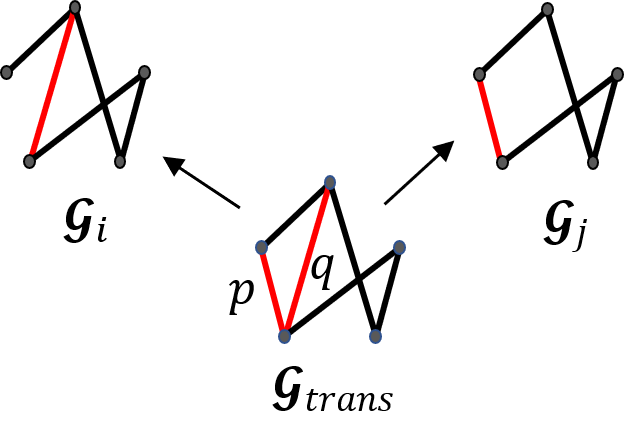}
	\caption{Transition state of two graphs: the union of two graphs.}
	\label{fig:graphdistance}
\end{figure}

Then the topology changes from $\G_i$ to $\G_j$ can be considered as different line deletions from their base graph $\G_{trans}$. For each single line deletion, e.g. line $p$, we define its contribution $x_p$ to graph distance by taking the average of its power impacts on all other lines as measured by LODF:
$$x_p = \frac{1}{|\E(i)\cup\E(j)|} \sum_{l\in \E(i)\cup\E(j)\backslash\{p\}}(|d_l^p|)$$
where $|\E(i)\cup\E(j)|$ denotes the cardinality of the set $\E(i)\cup\E(j)$, $d_l^p$ denotes the LODF coefficient with $p$ as outage line and $l$ as observed line. 

Then graph distance $D(\G_i, \G_j)$ is given by summing up the contributions of different line deletions from the base graph:
$$D(\G_i, \G_j) = \sum_{p\in (\E(i)-\E(j)) \cup (\E(j)-\E(i))}x_p$$ 
where $\E(i)-\E(j)=\{p|p\in \E(i),p\notin \E(j)\}$ and accordingly, $(\E(i)-\E(j)) \cup (\E(j)-\E(i))$ denotes all the edge changes between the two graphs.

This definition uses LODF as a measure of the impact on power flow of the removal of line $p$. Hence, edges with high LODF to many other edges can potentially cause greater changes in power flow, and thus our graph distance measure places greater importance on these edges.  Section \ref{apdx: compare distances} demonstrates the effectiveness of the LODF-based graph distance by comparing it against traditional distance measures for anomaly detection.

\subsection{Proposed Temporal Weighting Framework} \label{sec:framework}

In this section, we assume that we are given any distance measurements $D(\G_i, \G_j)$ between any pair of graphs $\G_i$ and $\G_j$, and explain how to use them to assign weights to each previous sensor data. This procedure can take the LODF-based distance defined in the previous subsection as input, but also allows us to flexibly use any given graph distance measure. The proposed Temporal Weighting is given in Algorithm \ref{alg:framework}.

\begin{algorithm}
	\caption{Temporal Weighting Framework at time $t=15$ (see toy example in Figure \ref{fig:toyexample_nonlocal})} 
	\label{alg:framework}
	\KwIn{Graph distance $D(\G_1, \G_3)$, $D(\G_2, \G_3)$, $D(\G_3, \G_3)$; sensor data $s_i(t)$ with $t=1,2,...15$, $ i=1,2,..., N_{sensor}$.}
	\KwOut{Anomaly score $A(15)$.}
	{\bf Extend graph distance to tick-wise distance.} Each previous time tick is given a distance $d_t$ according to the graph it comes from:
	$$d_t = \begin{cases}
	D(\G_1, \G_3) & \text{for } t = 1,2,..., 5 \\
	D(\G_2, \G_3) & \text{for } t = 6,7,..., 10 \\
	D(\G_3, \G_3) & \text{for } t = 11,..., 14
	\end{cases}$$
	
	{\bf Temporal Weighting:} Use $d_1,...d_{14}$ to assign weights $w_1$, ..., $w_{14}$ to the past sensor data using Algorithm \ref{alg:weighting}.
\end{algorithm}

For the purpose of utilizing previous data from a series of dynamic graphs, {\bf Temporal Weighting} plays an important role. The resulting weights directly determine how much information to extract from each previous record, thus requiring special care. Intuitively, the weights should satisfy the following principles:
\bit
\item The larger the distance $d_t$, the lower the weight $w_t$. This is because high $d_t$ indicates that time tick $t$ is drawn from a very different graph from the current one, and thus should not be given high weight when estimating the expected distribution at the current time
\item Positivity and Normalization: $\sum_{t}w_t = 1, w_t\geq0$
\eit
To satisfy these conditions, we use a principled optimization approach based on {\bf bias-variance trade-off}. Intuitively, the problem with using data with high $d_t$ is {\bf bias}: it is drawn from a distribution that is very different from the current one, and that can be considered a biased sample. We treat $d_t$ as a measure of the amount of bias. Hence, given weights $w_1, \cdots, w_{14}$ on previous data (in Figure \ref{fig:toyexample_nonlocal} example), the total bias we incur can be defined as $\sum_{t\in\{1,...,14\}}w_t d_t$. 

We could make the bias low simply by assigning positive weights to only time points from the most recent graph. However, this is still unsatisfactory as it results in a huge amount of {\bf variance}: since very little data is used to learn $\mathcal{N}_{t=15}$, the resulting estimate has high variance. Multiplying a fixed random variable by a weight $w_t$ scales its variance proportionally to $w_t^2$. Hence, given weights $w=[w_1,w_2,\cdots]$, the total amount of variance is proportional to $\frac{1}{2}w^Tw$, which we define as our variance term.

We thus formulate the following optimization problem as minimizing the sum of bias and variance, thereby balancing the goals of low bias (i.e. using data from similar graphs) and low variance (using sufficient data to form our estimates). We formulate the problem as:
\begin{problem}[DynWatch: temporal weighting]\label{def: dynwatch}
a bias-variance trade-off optimization is used to return a sparse vector of weights assigned to historical data from different topology (distributions):
\begin{align}
\min_{w}&{\sum_{t}w_t d_t+\frac{1}{2}w^Tw}\\
\text{s.t. }
&\sum_{t}w_t = 1\\
&w_t\geq 0,\forall t
\end{align}
\end{problem}

By writing out its Lagrangian function:
\begin{align}
L(w,\lambda,u)=d^Tw+\frac{1}{2}w^Tw+\lambda(1-\sum_t w_t)-u^Tw
\end{align}
 and applying KKT conditions, we can see the optimal primal-dual solution $(w,\lambda^*,u^*)$ must satisfy:
\begin{align}
d_t+w_t-\lambda^*-u_t^*=0
\end{align}

Since we have $d_t\geq0$, by further manipulation we have:
\begin{align}
w_t = \max\{\lambda^*-d_t,0\}
\end{align}
Moreover, there is a unique choice of $\lambda^*$ such that the resulting weights $w_t$ sum up to $1$. 
This $w_t$ against $d_t$ relationship is shown in Figure \ref{fig:w_and_d}. This result is intuitive: as $d_t$ increases, the resulting weight we assign $w_t$ decreases, and if $d_t$ passes a certain threshold, it becomes large enough so that any reduction in variance it could provide is more than offset by its large bias, in which case we assign it a weight of $0$. 
 \begin{figure}[h]
 	\centering
 	\includegraphics[width=0.3\linewidth]{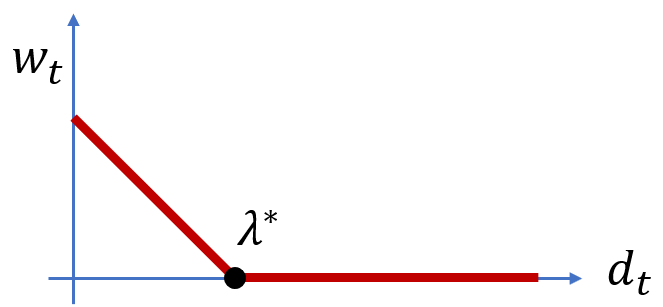}
 	\caption{$w_t-d_t$ relationship.}
 	\label{fig:w_and_d}
 \end{figure}

Our Temporal Weighting algorithm is in Algorithm \ref{alg:weighting}. During implementation, we adjust the relative importance of bias and variance by normalizing and scaling the graph distances (scaling factor 0.005 works well based on our empirical observation).
\begin{algorithm}
	\caption{Computing Temporal Weights $w_t$}
	\label{alg:weighting}
	\KwIn{distance $d_t$, with $t = 1,2,\cdots, N$}
	\KwOut{weights $w_t$, with $t = 1,2,\cdots, N$}
	Compute the unique $\lambda^*$ that satisfies:
	$$\sum_{t\in\{1,2,...N\}} \max\{\lambda^*-d_t,0\}=1$$
	
	Get weights $w_t$: 
	$$w_t = \max\{\lambda^*-d_t,0\}$$
\end{algorithm}

\subsection{Proposed Anomaly Detection Algorithm} \label{sec:anomaly}
Having obtained our weights $w_t$, the remaining step is to compute our anomaly score, as shown in Algorithm \ref{alg:anomaly}. 

We focus on 3 metrics from sensor data as indications of power system anomalies. These metrics were studied in \cite{hooi2018gridwatch} and found to be effective for detecting anomalies in power grid sensor data. In our setting, recall that for each sensor, we can obtain $\Delta s_i$ that contains changes of real and reactive power on the adjacent lines, over time. The 3 metrics are:
\bit
\item \emph{Edge anomaly metric}: $X_{edge,i}(t) = \max_{l\in{E_{adj}}}\Delta s_{i,l}$ which measures the maximum line flow change among lines connected to the sensor. Let $E_{adj}$ denote the set of lines connected to sensor $i$:
\item \emph{Average anomaly metric}: $X_{ave,i}(t) = \text{mean}\{\Delta s_{i,l}\ |\ l\in{E_{adj}}\},$ which measures the average line flow change on the lines connected to the sensor:
\item \emph{Diversion anomaly metric}: $X_{div,i}(t) = \text{std}\{\Delta s_{i,l}\ |\ l\in{E_{adj}}\},$ which measures the standard deviation of line flow change over all lines connected to the sensor:
\eit

Intuitively, for each metric, we want to estimate a model of its normal behavior. To do this, we compute the weighted median and interquartile range (IQR)\footnote{IQR is the difference between 1st and 3rd quartiles of the distribution, and is commonly used as a robust measure of spread.} of the detection metric, weighting the time points using our temporal weights $w_1, \cdots, w_t$. (Weighted) median and IQR are preferred choice of distribution parameters over the mean and variance for anomaly detection since they are robust measures of central tendency and statistical dispersion (i.e. they are less likely to be impacted by outliers)\cite{robust_statistics}. We can then estimate the anomalousness of the current time tick by computing the current value of a metric, then subtracting its weighted median and dividing by its IQR. The exact steps are given in Algorithm \ref{alg:anomaly}.

\begin{algorithm}
	\caption{Anomaly Detection (see Figure \ref{fig:toyexample_nonlocal})}
	\label{alg:anomaly}
	\KwIn{Temporal weights $w_t$; sensor data $s_i(t)$ with $t=1,2,...15$, $ i=1,2,...N_{sensor}$}
	\KwOut{anomaly score $A(15)$}
	\For{{$i \gets 1$ to $N_{sensor}$} }{
	
	
	
	{\bf Compute weighted median and IQR:}
	\begin{align} 
	\mu_{edge}=\text{Weighted Median}\{X_{edge, i}(t)\ |\ t=1,...,14\}\\
	IQR_{edge}=\text{Weighted IQR}\{X_{edge, i}(t)\ |\ t=1,...,14\}
	\end{align}
	\ \ weighted by $w_1,...w_{14}$ (similarly for $X_{ave}, X_{div}$). 
	
	{\bf Calculate sensor-wise anomaly score at t=15:} 
	\begin{align}
	a_i(15) = \max_{metric\in\{edge, ave, div\}}\frac{X_{metric, i}(15)-\mu_{metric}}{IQR_{metric}}
	\end{align}
	}

{\bf Calculate anomaly score for target time tick}, as the max score over sensors:
\begin{align}
A(15) = \max_{i\in\{1,...,N_{sensor}\}}{ a_i(15)}
\end{align}
\end{algorithm} 


\subsection{Extension to large-scale system: locally sensitive distance measure} \label{sec:local}

\begin{figure}[h]
	\centering
	\includegraphics[width=0.4\linewidth]{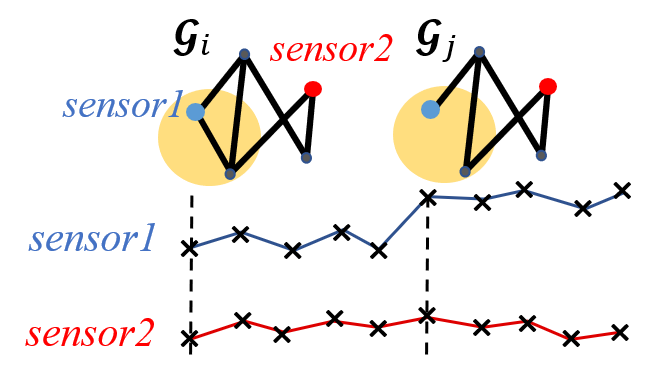}
	\caption{Simple motivating example: \methodls. The two graphs are very different within the yellow localized region. However, sensor 2 is far away from the yellow region and thus experiences no changes.} 
	\label{fig:toyexample_local}
\end{figure}
In the previous section, we computed a single distance value $D(\G_i, \G_j)$ between any pair of graphs. However, consider two graphs $\G_i$ and $\G_j$ in Figure \ref{fig:toyexample_local} that are very different due to a small yellow localized region (e.g. in a single building that underwent heavy renovation). Hence, $D(\G_i, \G_j)$ is large, indicating not to use data from $\G_i$ when we analyse a time tick under $\G_j$. However, from the perspective of a single sensor $s$ (sensor 2) far away from the localized region, this sensor may experience little or no changes in the power system's behavior, so that data from graph $\G_i$ may have a similar distribution as data from graph $\G_j$, and so for this sensor (sensor 2) we can still use data from $\G_i$ to improve anomaly detection performance. Hence, rather than computing a single distance $D(\G_i, \G_j)$, we compute a separate {\bf locally-sensitive} distance $D_s(\G_i, \G_j)$ specific to each sensor, which measures the amount of change between graphs $\G_i$ and $\G_j$ in the `local' region to sensor $s$. Clearly, the notion of `local regions' must be carefully defined: we will define them based on LODF, recalling that LODF measures how much changes on one edge affect each other edge.

\begin{figure}[h]
	\centering
	\includegraphics[width=0.3\linewidth]{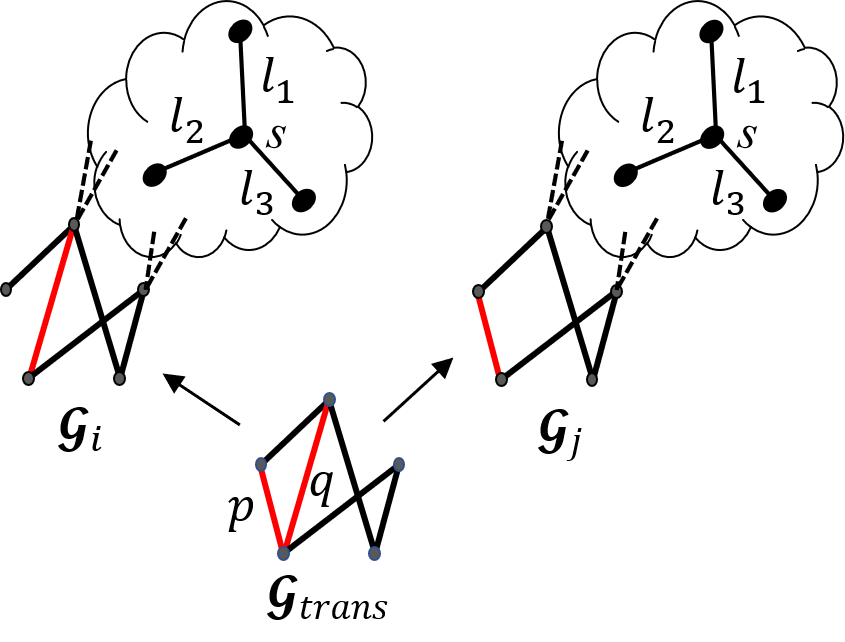}
	\caption{Local graph distance: the adjacent lines connected to each sensor $s$ are considered.}
	\label{fig:localgraphdistance}
\end{figure}

Intuitively, the local distance between two graphs with respect to sensor $s$ is large if the changed edges can potentially cause large power change nearby the sensor. Hence, given two graphs $\G_i(\V(i),\E(i))$ and $\G_j(\V(j),\E(j))$ with their transition state $\G_{trans}$ and a sensor $s$ of interest, the local graph distance contribution $y_p$ of line $p$ with respect to sensor $s$ can be calculated by multiplying the whole-grid-wide contribution $x_p$ with a weighing factor $c_p^s$. This $c_p^s$ coefficient filters the power impact for sensor $s$ using the maximum power impact of line deletion on lines around this sensor:
\begin{align}
c_p^s &= \max_{l\in\Exp_{sensor}(s)}|d_l^p|\\
y_p &= x_pc_p^s
\end{align}
where $\Exp_{sensor}(s)$ denotes the set of edges around sensor $s$ (e.g. in Figure \ref{fig:localgraphdistance}, $\Exp_{sensor}(s)=\{l_1,l_2,l_3\}$), and $d_l^p$ denotes the LODF with $p$ as outage line and $l$ as observed line.)

Then, as before, the local graph distance with respect to sensor s is defined by summing up the local graph distance contributions of different line deletions from the graph:
$$D_s(\G_i, \G_j) = \sum_{p\in (\E(i)-\E(j)) \cup (\E(j)-\E(i))}y_p$$

\subsection{Statistical error analysis}\label{sec: statistical error analysis}

This section focuses on a quantitative analysis of the performance of our method, through statistical error analysis.

Let $T$ denote the width of time window for analysis, then for $\forall$ sensor $s$, the anomalousness of its observation $x_{T+1}$ is evaluated based on its previous data $x_1, x_2,...,x_T$. The anomaly detection method works by assigning weights $w_1, w_2, ..., w_T$ ($w_t\geq 0, \forall t, \sum_{t=1}^{T}w_t=1$) to all the previous observations and an alarm is created if $x_{T+1}$ deviates $\sum_{t=1}^{T}w_tx_t$ by a certain threshold. 

Here we investigate the properties of statistical error based on the following definitions and assumptions:

\begin{assumption}[Temporal independence]  
For any sensor $s$ and time $t$, its measured data $x_t$ is drawn from a Gaussian distribution $P(x_t)=N(\mu_t,\sigma^2)$ independently from other time ticks, where $\sigma^2$ accounts for all  uncertainties caused by measurement noise, load/generation variation, weather uncertainty, etc.
\label{assumption: temporal independence}
\end{assumption}

\begin{assumption}[identical distribution conditioned on topology]
Given a certain topology $G$ and $\forall$ sensor $s$, all data of $s$ under the same topology $G$ are drawn independently from the same distribution $P(x_t|G)=N(\mu_G,\sigma^2)$ (i.e., for any two time ticks $t_1,t_2$ with the same topology $G$, we have $\mu_{t_1}=\mu_{t_2}=\mu_G$.)
\label{assumption: identical distribution conditioned on G}
\end{assumption}

\begin{definition}[Optimal graph distance]
For any time-series data $x_1,x_2,...,x_T$ of a sensor $s$ and its latest observation $x_{T+1}$, $d_t$ denotes the graph distance between the graph at time $t$ and the graph at $T+1$, i.e., $d_t=D(G_t,G_{T+1}),d_t\geq 0,\forall t$. Then the optimal graph distance $d^*_{t}$ for $\forall t$ satisfies $|\mu_t-\mu_{T+1}|=|\mu_{G_t}-\mu_{G_{T+1}}|\propto d_t^*$, or equivalently, $\exists$ constant $c$ such that $|\mu_t-\mu_{T+1}|=c\cdot d_t^*$.
\label{def: optimal graph distance}
\end{definition}

We first demonstrate that the statistical error can be bounded:

\begin{mytheorem}[Error bound]
\label{theorem: error bound}
Based on Assumption \ref{assumption: temporal independence}, \ref{assumption: identical distribution conditioned on G} and Definition \ref{def: optimal graph distance}, the statistical error $\Exp_{x_1,x_2,...,x_T,x_{T+1}}[(\sum_{t=1}^{T}w_tx_t-x_{T+1})^2]$ with $w_t\geq0, \forall t$ and $\sum_{t=1}^{T}w_t=1$, satisfies:
\begin{equation}
    \sigma^2\leq \Exp[(\sum_{t=1}^{T}w_tx_t-x_{T+1})^2]\leq (1+\max_t w_t)\sigma^2+c\max_t d_t^*
\end{equation}
\end{mytheorem}
Upon obtaining the error bound, here are some intuitive explanations of the upper bound being dependent on $\max_t w_t$ and $\max_t d^*_t$:
\begin{itemize}
    \item $\max_t w_t$: large value for this term indicates that the estimation method depends heavily on a particular prior data point with weight $w_t$. This can lead to overfitting and as a result higher error (upper bound) due to high variance.
    \item $\max_t d^*_t$: large value for this term indicates that a prior data point from a very different distribution has been used for estimation, which can lead to higher error (upper bound) due to high bias.
\end{itemize}

\noindent Another question of interest to us is the properties in the limit of infinite data:
\begin{mytheorem}[Unbiased estimation under infinite data]
\label{theorem: unbiased est}
In the limit of infinite data, the statistical error limits at the lower bound:
$$\Exp_{x_1,x_2,...,x_T,x_{T+1}}[(\sum_{t=1}^{T}w_tx_t-x_{T+1})^2]=\sigma^2$$
and an unbiased estimate of the true distribution $x_{T+1}\sim N(\mu_{T+1},\sigma^2)$ is obtainable using the previous samples, i.e.,
$$\Exp[\sum_{t=1}^{T}w_tx_t]=\mu_{T+1}$$
$$\Exp[\frac{1}{T-1}\sum_t(x_t-\frac{\sum_t x_t}{T})^2]=\sigma^2$$
\end{mytheorem}

\noindent Detailed proofs of the two theorems are shown below:
\begin{derivation}[Proof of error bound] we derive the statistical error bound demonstrated in Theorem \ref{theorem: error bound}: 
\begin{equation}
    \sigma^2\leq \Exp[(\sum_{t=1}^{T}w_tx_t-x_{T+1})^2]\leq (1+\max_t w_t)\sigma^2+c\max_t d_t^*
    \notag
\end{equation}
Starting from the statistical error $\Exp_{x_1,x_2,...x_T}[(\sum_{t=1}^{T}w_tx_t-x_{T+1})^2]$, we have:
\begin{align}
    &\Exp_{x_1,x_2,...x_T}[(\sum_{t=1}^{T}w_tx_t-x_{T+1})^2] \\
    =& \Exp[(\sum_{t=1}^{T}w_tx_t-\mu_{T+1}+\mu_{T+1}-x_{T+1})^2]\\
=&\Exp[(\sum_{t=1}^{T}w_tx_t-\mu_{T+1})^2] + \Exp[(\mu_{T+1}-x_{T+1})^2] \\
    &+ 2\Exp[(\sum_{t=1}^{T}w_tx_t-\mu_{T+1})(\mu_{T+1}-x_{T+1})]
\end{align}

Based on Assumption \ref{assumption: temporal independence}, we have $\Exp[\mu_{T+1}-x_{T+1}]=0$, and thus
\begin{align}
    &\Exp[(\sum_{t=1}^{T}w_tx_t-\mu_{T+1})(\mu_{T+1}-x_{T+1})\\
    =&\Exp[\sum_{t=1}^{T}w_tx_t(\mu_{T+1}-x_{T+1})]-\mu_{T+1}^2+\mu_{T+1}\Exp[x_{T+1}]\\
    =&\Exp[\sum_{t=1}^{T}w_tx_t]\Exp[\mu_{T+1}-x_{T+1}]-\mu_{T+1}^2+\mu_{T+1}^2\\
    =&0
\end{align}
Therefore we have
\begin{align}
    &\Exp_{x_1,x_2,...x_T,x_{T+1}}[(\sum_{t=1}^{T}w_tx_t-x_{T+1})^2] \\
    =&\Exp[(\sum_{t=1}^{T}w_tx_t-\mu_{T+1})^2] + \Exp[(\mu_{T+1}-x_{T+1})^2] \\
    =&\Exp[(\sum_{t=1}^{T}w_tx_t-\sum_{t=1}^{T}w_t\mu_t+\sum_{t=1}^{T}w_t\mu_t-\mu_{T+1})^2] \\
    &+ \Exp[(\mu_{T+1}-x_{T+1})^2] \\
    =&\Exp[(\sum_{t=1}^{T}w_tx_t-\sum_{t=1}^{T}w_t\mu_t)^2] +\Exp[(\sum_{t=1}^{T}w_t\mu_t-\mu_{T+1})^2]\\
    &+2\Exp[(\sum_{t=1}^{T}w_tx_t-\sum_{t=1}^{T}w_t\mu_t)(\sum_{t=1}^{T}w_t\mu_t-\mu_{T+1})]\\
    & +\Exp[(\mu_{T+1}-x_{T+1})^2] 
\end{align}
Similarly based on Assumption \ref{assumption: temporal independence}, it is easy to show that
\begin{align}
    \Exp[(\sum_{t=1}^{T}w_tx_t-\sum_{t=1}^{T}w_t\mu_t)(\sum_{t=1}^{T}w_t\mu_t-\mu_{T+1})]=0
\end{align}
Thus we have
\begin{align}
    &\Exp_{x_1,x_2,...x_T,x_{T+1}}[(\sum_{t=1}^{T}w_tx_t-x_{T+1})^2]\\
    =&\underbrace{\Exp[(\sum_{t=1}^{T}w_tx_t-\sum_{t=1}^{T}w_t\mu_t)^2]}_\text{Variance}+
    \underbrace{E[(\sum_{t=1}^{T}w_t\mu_t-\mu_{T+1})^2]}_\text{Bias$^2$}\\
    &+\underbrace{\Exp[(\mu_{T+1}-x_{T+1})^2] }_\text{irreducible error}
\end{align}
Based on Assumption \ref{assumption: temporal independence} and $w_t\geq0$ for $\forall t, \sum_{t=1}^{T}w_t=1$, the variance term can be upper bounded by:
\begin{align}
    \Exp[(\sum_{t=1}^{T}w_tx_t-\sum_{t=1}^{T}w_t\mu_t)^2]&=\sum_{t=1}^{T}w_t^2\Exp[(x_t-\mu_t)^2]\\
    &\leq (\max_tw_t)\sigma^2
\end{align}
Further making use of Assumption \ref{assumption: identical distribution conditioned on G}, it is easy to get an upper bound for the bias$^2$ term:
\begin{align}
    E[(\sum_{t=1}^{T}w_t\mu_t-\mu_{T+1})^2]&=\Exp[(\sum_{t=1}^{T}w_t|\mu_t-\mu_{T+1}|)^2]\\
    &=\sum_{t=1}^{T}w_tcd^*_t\\
    &\leq c\max_td^*_t
\end{align}
Finally, as $\Exp[(\mu_{T+1}-x_{T+1})^2]=\sigma^2$ based on the assumption that $x_{T+1}\sim N(\mu_{T+1},\sigma^2)$, we are able to \textbf{upper bound} the statistical error as:
\begin{align}
    &\Exp_{x_1,x_2,...x_T,x_{T+1}}[(\sum_{t=1}^{T}w_tx_t-x_{T+1})^2]\\
    &\leq (\max_t w_t)\sigma^2+\max_t cd_t^*+\sigma^2\\
    &= (1+\max_t w_t)\sigma^2+c\max_td_t^* 
\end{align}
Meanwhile the \textbf{lower bound} is also obvious:
\begin{align}
     \Exp_{x_1,x_2,...x_T,x_{T+1}}[(\sum_{t=1}^{T}w_tx_t-x_{T+1})^2]\geq \sigma^2
\end{align}
\end{derivation}

\begin{derivation}[Proof of unbiased estimation under infinite data] we prove the result in Theorem \ref{theorem: unbiased est}
In the limit of infinite data, there exist infinite data with the same topology as $x_{T+1}$, thus it is possible to find the time series data such that $x_1,x_2,...,x_T$ are drawn independently from the same distribution, s.t, $x_t\sim N(\mu_{T+1},\sigma^2), \forall t$ and $T\longrightarrow \infty$ . 

From Theorem \ref{theorem: error bound}, we have:
\begin{equation}
    \sigma^2\leq \Exp_{x_1,x_2,...,x_T,x_{T+1}}[(\sum_{t=1}^{T}w_tx_t-x_{T+1})^2]\leq  \sigma^2
\end{equation}
Now given that $w_t=\frac{1}{T}\longrightarrow 0, d^*_t=0$ for $\forall t\in{0,1,...,T}$, it is easy to show:
\begin{align}
\Exp_{x_1,x_2,...,x_T,x_{T+1}}[(\sum_{t=1}^{T}w_tx_t-x_{T+1})^2]= \sigma^2\\
\Exp[\sum_{t=1}^{T}w_tx_t]=\sum_{t=1}^{T}w_t\Exp[x_t]=\mu_{T+1}
\end{align}
And based on Bessel's correction, it is easy to get
\begin{equation}
    \Exp[\frac{1}{T-1}\sum_t(x_t-\frac{\sum_t x_t}{T})^2]=\sigma^2
\end{equation}
\end{derivation}

\subsection{Experiments to evaluate \method: scalable detection on dynamic graphs}
We design experiments to answer the following questions:
\begin{itemize}
    \item \textbf{Q1. Anomaly Detection Performance:} how accurate is the anomaly detection from our method compared to other ML baselines?
    \item \textbf{Q2. Scalability:} how do our algorithms scale with the graph size?
    \item\textbf{Q3.  Robustness against cyberattack of false data:} how can our algorithm enhance the standard practices of (bad data) detection in today's grid operator, when considering modern cyberthreat?
\end{itemize}

Our code and data are publicly available at \codeurl. 
Experiments were done on a 1.9 GHz Intel Core i7 laptop, 16 GB RAM running Microsoft Windows 10 Pro. 

{\bf Case Data:} We use 2 test cases: \datasmal is an accurate reconstruction of part of the European high voltage network, and \dataxlarge is a synthetic network that mimics the Texas high-voltage grid in the U.S. The \dataxlarge represents a similarly sized system as the PJM (the largest independent service operator (ISO) in the U.S.) grid, which contains around 25 to 30k buses~\cite{zimmerman2011matpower}. 



{\bf Selection of sensors and network observability:}  Due to the spatial impact of grid anomalies and the efficacy of anomaly metrics, full observability\cite{SE_observability}\cite{grid_observability_analysis} and optimal sensor placement for observability\cite{OPP_observability} are not necessary for Dynwatch to perform. However, access to more sensors as inputs alongside optimal sensor selection\cite{hooi2018gridwatch} can improve the detection performance and help with localization of anomaly. To be conservative, for these experiments, we select random subsets of sensors (of varying sizes) as input. The good performance even with randomly selected sensor measurements validates the effectiveness of our method in selecting relevant time frames from historical data.

{\bf Threshold tuning:} For a fair comparison of different methods, our experiment section, without using any threshold, compares the top K anomalies scored by each algorithm, where K is the number of anomalies simulated. However, in practical use, a detection threshold is required for the algorithm to identify an anomaly. A proper threshold can be either a fixed threshold from an empirical value or domain-knowledge or a learned threshold to facilitate optimal decision making. In particular, optimal threshold tuning needs to take {\bf class imbalance and asymmetric error} into account. Since only a minority of instances are expected to be abnormal, there is an unbalanced nature of data. Moreover, as grid applications are safety-critical, mislabeling an anomaly as normal, i.e., false negatives (FN), could cause fatal consequences, while false positives (FP) which cause loss of ‘fidelity’ are less serious. Proper techniques for tuning a threshold {\bf offline} include: 
\begin{enumerate}
    \item Calculating the evaluation metric (e.g. F-measure which quantifies the balance of precision and recall) for each threshold to select the one that maximizes the metric. 
    \item Plotting the ROC curve or precision-recall curve to select the threshold that gives the optimal balance.
    \item A cost approach that, when the cost of FP, FN, TP, TN are available, minimizes the average overall cost of a diagnostic test, yet domain-specific knowledge is needed for reasonable quantification of the costs.
\end{enumerate}

\subsubsection{Q1. Anomaly Detection Performance}

\begin{figure*}[htp]
	\centering
	\includegraphics[width=0.9\linewidth]{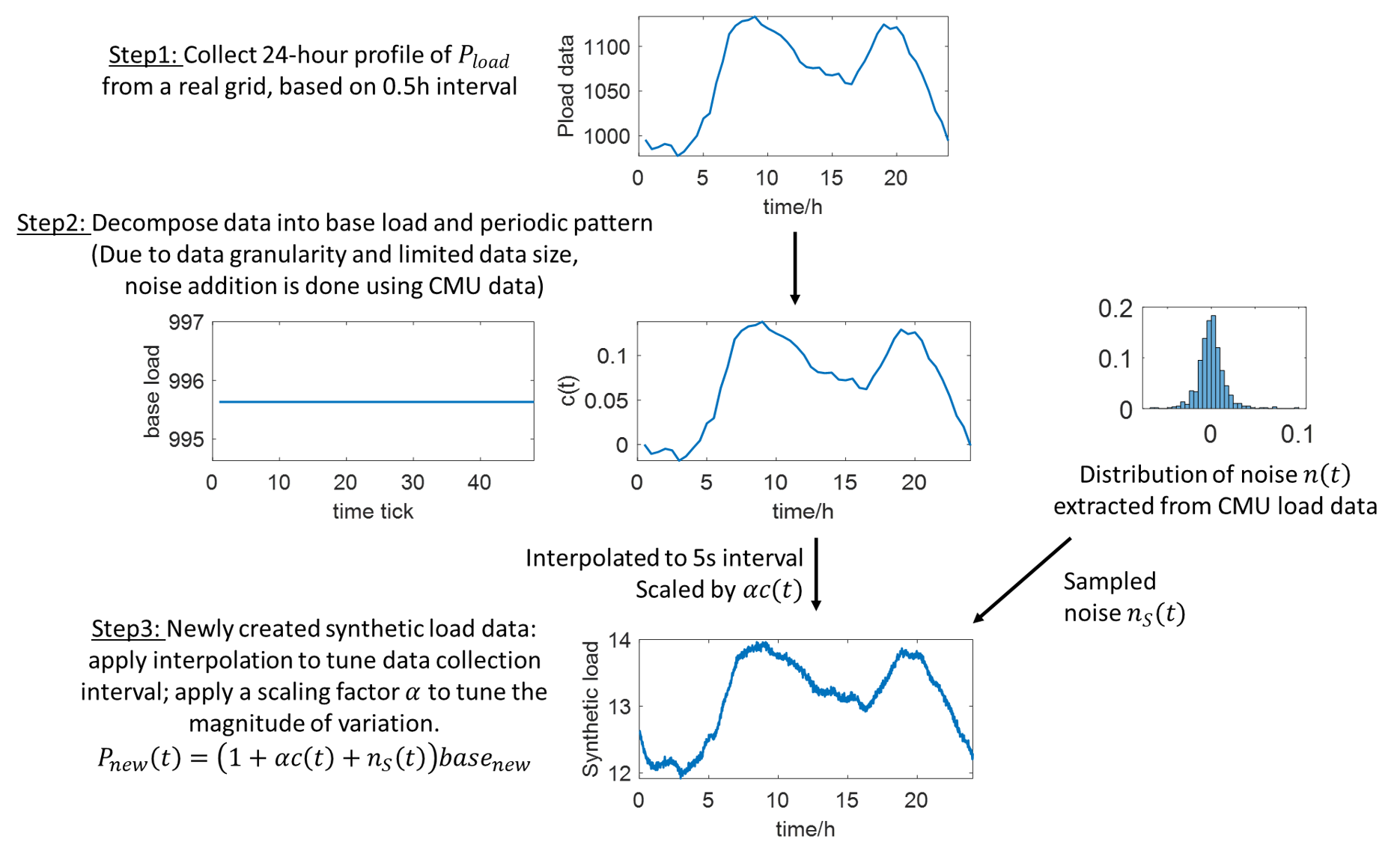}
	\caption{Synthetic data generation for evaluating \method: 1200 time-ticks are created to model the time-series data of grid operation over 1h 40min, using real utility-provided load data.}
	\label{fig:synthetic data TVA}
\end{figure*}

In this section, we compare \method against baseline anomaly detection approaches, while varying the number of sensors in the grid. 

{\bf Experimental Settings:} 
Starting with a particular test case as a base graph $G$, we first create $20$ different topology scenarios where each of them deactivates a randomly chosen branch in the base graph. These subsequent $20$ network topologies represent the dynamic grid with topology changes due to operation and control. Then for each topology scenario, we use MatPower~\cite{zimmerman2011matpower}, a standard power grid simulator, to generate $60$ sets of synthetic measurements based on the load characteristics described in the following paragraph. As a result, the multivariate time series with $20\times60=1200$ time ticks mimics the real-world data setting where sensors receive measurements at each time tick $t$, and the grid topology changes every $60$ time ticks. Finally, we sample $50$ random ticks out of $1200$ as times when anomalies occur. Each of these anomalies is added by randomly deleting an edge on the corresponding topology.


Following \cite{hooi2018gridwatch}, to generate an input time series of loads (i.e. real and reactive power at each node), we use the patterns estimated from two real datasets:
\begin{itemize}
    \item Carnegie Mellon University (CMU) campus load data recorded for $20$ days from July 29 to August 17, 2016;
    \item Utility-provided 24 hour dataset of a real U.S. grid. 
\end{itemize}
to create synthetic time-series load based on a 5s interval, with the magnitude of daily load variation scaled to a predefined level, and with added Gaussian noise sampled from the extracted noise distribution~\cite{song2017powercast}. The detailed data generation process is shown in Figure \ref{fig:synthetic data TVA}.

Given this input, each algorithm then returns a ranking of the anomalies. We evaluate this using standard metrics, AUC\footnote{AUC is the probability of correct ranking of a random “positive”-“negative” pair.} (area under the ROC curve) and F-measure\footnote{F-measure is a trade-off between precision and recall.} ($\frac{2\cdot \text{precision} \cdot \text{recall}}{\text{precision} + \text{recall}}$), the latter computed on the top $50$ anomalies output by each algorithm. 

{\bf Baselines:} Dynamic graph anomaly detection approaches \cite{akoglu2010oddball,chen2012community,araujo2014com2,shah2015timecrunch} cannot be used as they consider graph structure only, but not sensor data. \cite{mongiovi2013netspot} allows sensor data but requires graphs with fully observed edge weights, which is inapplicable as detecting failed power lines with all sensors present reduces to checking if any edge has current equal to $0$. Hence, instead, we compare \method to GridWatch~\cite{hooi2018gridwatch}, an anomaly detection approach for sensors on a static graph, and the following multidimensional time-series based anomaly detection methods: Isolation Forests~\cite{liu2008isolation}, Vector Autoregression (VAR)~\cite{hamilton1994time}, Local Outlier Factor (LOF)~\cite{breunig2000lof}, and Parzen Window~\cite{parzen1962estimation}. Each uses the currents and voltages at the given sensors as features. For VAR, the norms of the residuals are used as anomaly scores; the remaining methods return anomaly scores directly.
For Isolation Forests, we use $100$ trees (following the defaults in scikit-learn\cite{scikit-learn}). For VAR, following standard practice, we select the order by maximizing AIC. For LOF we use $20$ neighbors (following the default in scikit-learn), and we use $20$ neighbors for Parzen Window.

As shown in Figure \ref{fig:result_all}, \method clearly outperforms the baselines on both metrics, having an F-measure of $>20\%$ higher than the best baseline. 
\begin{figure}[h]
	\centering
	\includegraphics[width=1\linewidth]{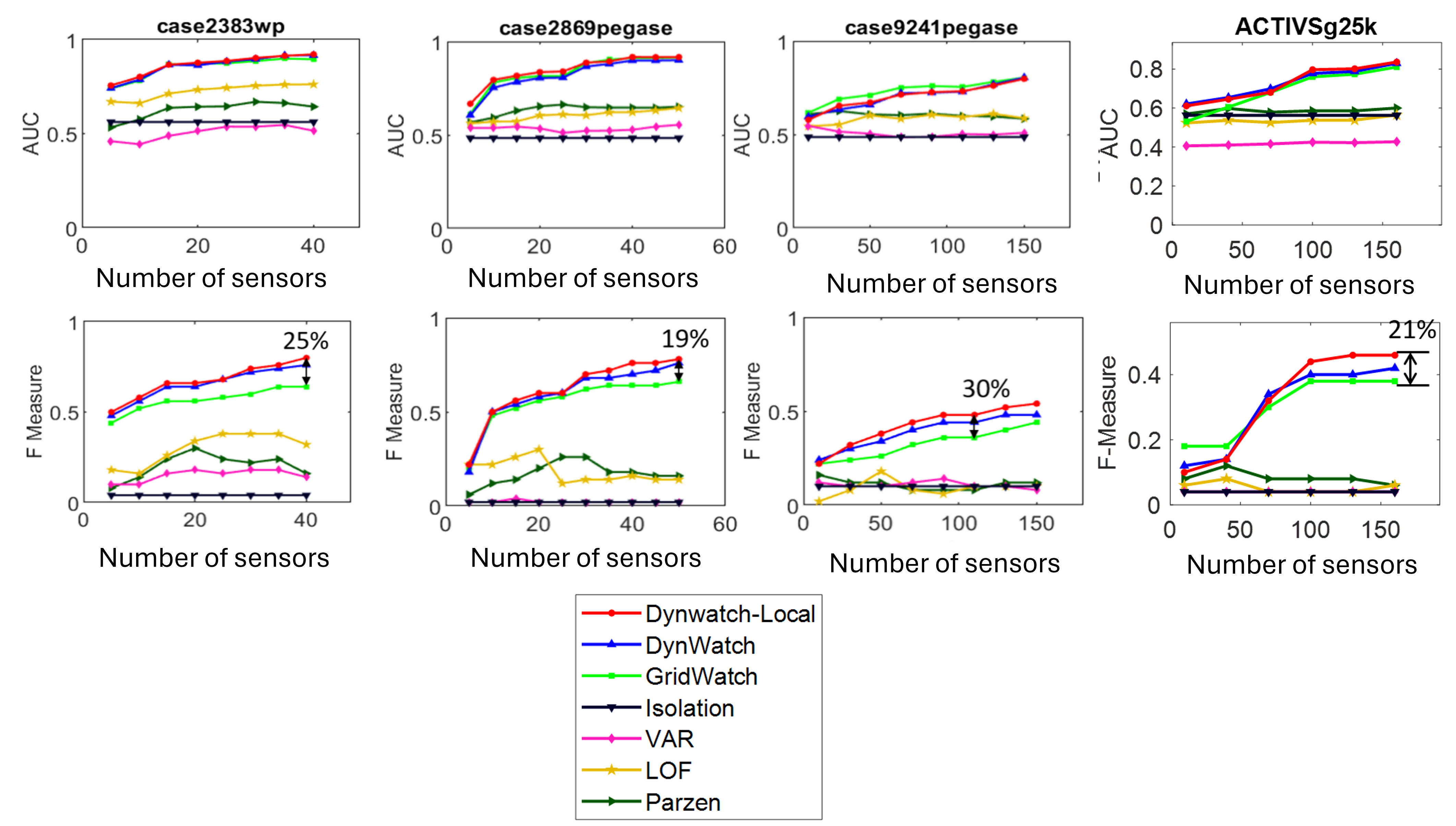}
	\caption{Performance of \method: AUC and F-measure.}
	\label{fig:result_all}
\end{figure}

\subsection{Q2. Scalability}
In this subsection, we seek to analyze the scalability of our \method and \methodls. In reality, PJM, the largest ISO in the U.S., runs ACSE on a 28k bus model, performed every 1 min\cite{PJM-manual-12}, thus any anomaly detection algorithm that takes significantly less than 1 min may provide valuable information to prevent wrong control decisions in real-time. The following results demonstrate the proposed method's capability to achieve this.


Here, 
we generate test cases of different sizes by starting with the \datasmal case and duplicating it $3,4,5,\cdots,12$ times. After each duplication, edges are added to connect each node with its counterpart in the last duplication, so that the whole grid is connected. Then for each testcase, we generate 20 dynamic grids and sensor data with 1 randomly chosen sensor and 1200 time ticks, following the same settings as the previous sub-section. Finally, we measure 
the graph distance calculation time, as well as the time is taken for the whole \method and \methodls methods. 
\begin{figure}[h]
	\centering
	\includegraphics[width=0.75\linewidth]{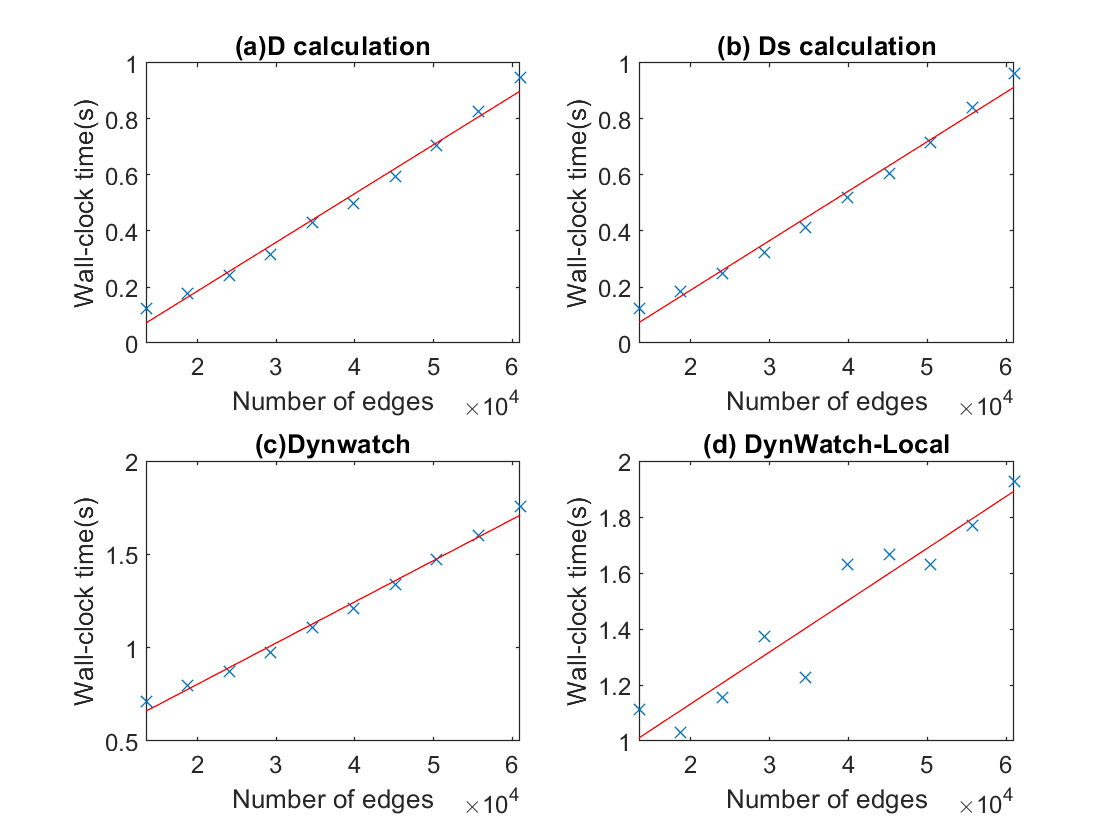}
	\caption{\label{fig:scability_all_new}\method scales linearly with the number of edges, when detecting all 1200 time ticks. The red lines are best-fit regression lines. }
\end{figure}

Figure \ref{fig:scability_all_new} shows that our method is fast: even on a large case with 60k+ branches, both methods took less than 2s to apply anomaly detection on all 1200 time ticks of the sensor, corresponding to an average of less than 1.7ms per time tick per sensor. The \dataxlarge (realistic power system case with 25K buses) has 32k+ branches, and thus run-time of anomaly detection at each time $t$ will be significantly less than 1 min. The figure also shows that our methods scale close to linearly with the grid size. 



\subsection{Q3. Robustness against cyberattack of false data}\label{sec:experiment on FDIA}

In this section, we explore how the proposed Dynwatch algorithm can improve the performance of the standard residual-based ACSE bad-data detection (BDD) method, by testing a type of grid-specific anomaly that SE BDD is known to fail against False Data Injection Attack (FDIA).

False Data Injection Attack (FDIA)\cite{fdia-review} is a cyber-attack in which attackers manipulate the value of measurements according to the grid physical model such that the SE outputs incorrect grid estimates while ensuring that its residual does not change by much (ideally remains unchanged). 

In this experiment, we construct an attack on a 14 bus network to mislead the operator into thinking that the load reduces by 20\%. For any anomalous time tick $t$, this is implemented by simulating power flow with the reduced load and generating measurements based on that. The time-series measurement data and a comparison of anomaly scores are shown in Figure \ref{fig: dynwatch on FDIA}. 
\begin{figure}[htp]
\centering
\includegraphics[width=0.7\linewidth]{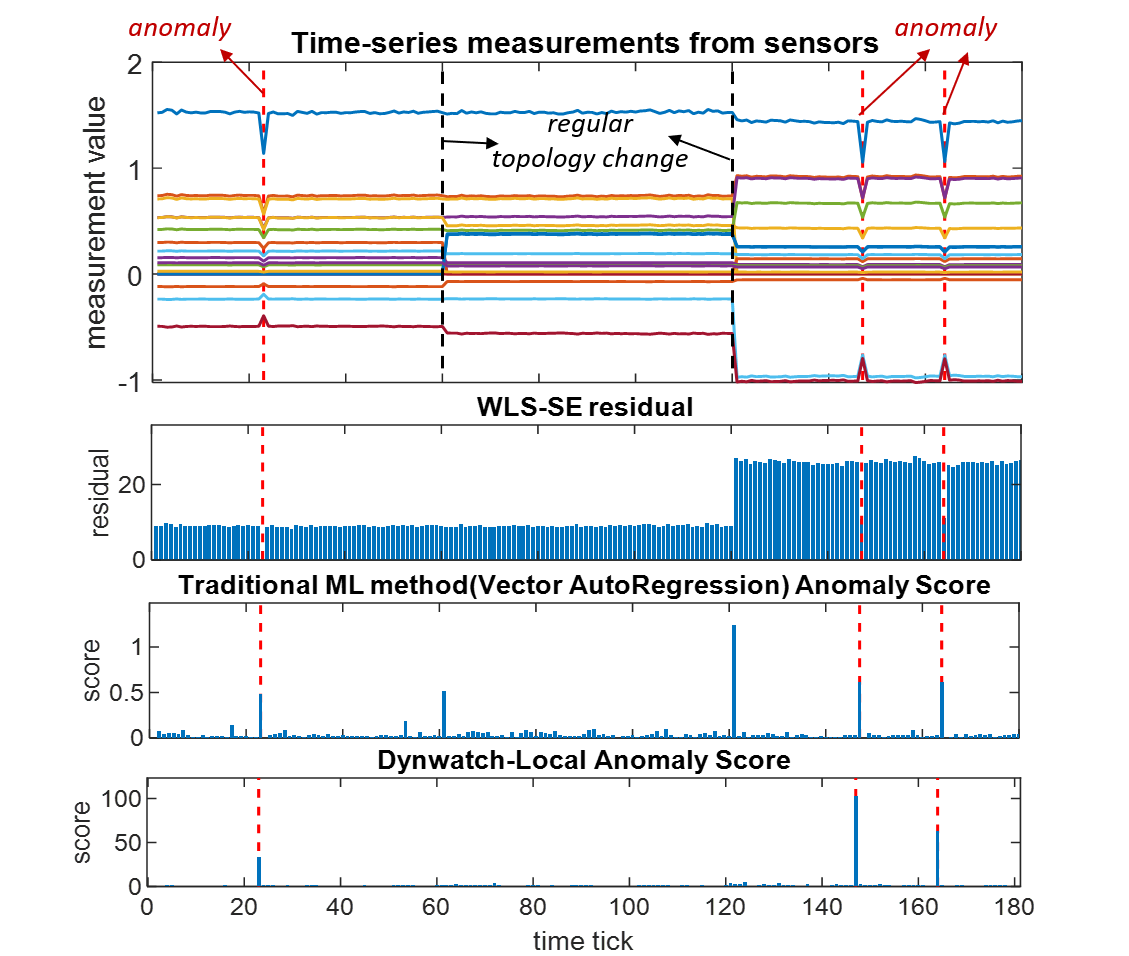}
\caption{Change of line status, total load and generation on for a real-world load dataset.}
\label{fig: dynwatch on FDIA}    
\end{figure}

Results show that the ACSE residual, which is metric for BDD reduced in value when anomalies occur (see the residual decrease in Figure \ref{fig: dynwatch on FDIA} during anomalous operation shown by the dotted red line), implying that the standard SE BDD, along with any other residual-based method, will not be able to detect this coordinated attack.

In addition to standard ACSE BDD, we also implemented the auto-regression (VAR) method to detect grid anomalies. As the VAR algorithm does not consider dynamic graph properties of the power grid, it tends to create alarms on all abrupt measurement changes. This will easily lead to false positives since regular topological changes also result in sudden temporal change. This can be seen in the Figure \ref{fig: dynwatch on FDIA} wherein during regular topology changes (shown in black dotted line) sudden spikes in VAR anomaly score are observed.

In comparison, our proposed method is able to detect all anomalies without False positives (FP). This indicates that the proposed algorithm is more likely to detect anomalies due to complex attack scenarios while being able to reduce the occurrences of false positives. 

\subsection{Comparison of graph distance measures}\label{apdx: compare distances}
To quantitatively justify the effectiveness of our proposed graph distance, we compared the proposed distance with other traditional measures applicable to power grids:
\begin{itemize}
    \item Simple GED\cite{GED_MCS_shoubridge2002}: the distance between two graphs is equal to the number of changed edges.
    \item Variant of GED with line admittance used as weights assigned to the changed edges. Admittance is used here because the larger the admittance, the more likely the edge has large power flows on it, meaning it is important to the grid.
    \item Equivalent conductance-based measure: the distance between two graphs is equal to the sum of the equivalent conductance of all changed edges. Equivalent conductance is able to take more consideration of the system-wise impact of each edge.
\end{itemize}

Result in Figure \ref{fig:compare different distances} shows our proposed measure outperforms the baselines above. Here the time series data is generated using the pattern from the utility-provided data set, following the process described in Figure \ref{fig:synthetic data TVA}.
\begin{figure}[h]
	\centering
	\includegraphics[width=0.6\linewidth]{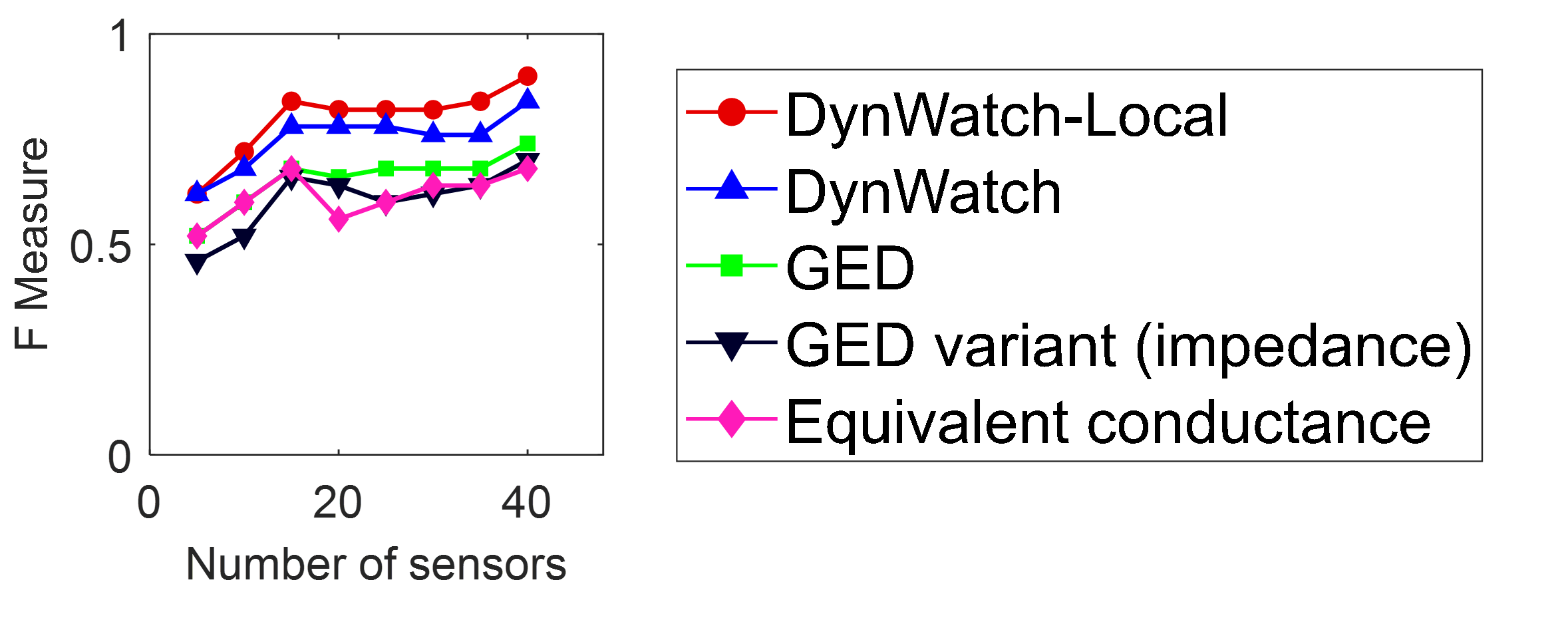}
	\caption{Result of F-measure on case2383wp, with 40 sensors installed: the proposed LODF-based graph distance outperforms other distance measures.}
	\label{fig:compare different distances}
\end{figure}

\section{Application 2 - Gridwarm: exploiting spatial sparsity for generalizable, scalable, and interpretable prediction}\label{sec: gridwarm}

In this Section, we will develop a ML predictor that exploits the sparse graphical structure. The ML model is able to predict the impact of disturbances on power systems, which can be useful in aiding the simulation tool via a Physics-ML Synergy design later in Chapter \ref{ch: synergy}.

We develop a ML model, Gridwarm\cite{gridwarm}, to predict the impact of a disturbance with nice properties of being: 
\begin{itemize}
    \item \textbf{generalizable} to topology change by using a graphical model
    \item \textbf{lightweight and scalable} by exploiting the sparse graphical structures and applying regularization techniques
    \item \textbf{physically interpretable} by forming a linear system proxy
\end{itemize}

\subsubsection{Defining a probabilistic graphical model}

Let us start with the task definition of Gridwarm with Table \ref{tab:notations} showing the symbols used in the method.
\begin{problem}[{Gridwarm predictor}]\label{def: gridwarm}
given an input $\bm{x}$ which contains disturbance information $c$ and (pre-disturbance) system information $G$, a warm starter makes prediction $\bm{y}$ which is an estimate of the post-disturbance bus voltages $\bm{v}^{post}$. The model is a function mapping, which is learned from training dataset  $Data=\{(\bm{x^{(j)},\bm{y^{(j)}}})\}$, where $(j)$ denotes the $j$-th sample. 
\end{problem}

\begin{table}[htbp]
\small
\centering
	\caption{Symbols and definitions \label{tab:notations}}
	\begin{tabular}{ @{}rl@{} }  
	\toprule
	\textbf{Symbol} & \textbf{Interpretation} \\ \midrule
		$G$    
     &case data before disturbance
     containing topology, generation, and load settings\\
     \midrule
     $\bm{v_i}$   & the voltage at bus $i$,
     $\bm{v_i}=[v_i^{real},v_i^{imag}]^T$\\
     \midrule
     $\bm{v}^{pre/post}$ & $\bm{v}^{pre/post}=[\bm{v}^{pre/post}_1,\bm{v}^{pre/post}_2,...,\bm{v}^{pre/post}_n]^T$ the pre/post-disturbance voltages at all buses\\
     \midrule
     $c$ & disturbance setting $\textit{(type, location, parameter)}$\\
     &e.g. $(\textit{MadIoT}, [1,3], 150\%)$: increasing loads at bus 1\\
     & and 3 to $150\%$ of the original value via MadIoT attack.\\
     \midrule
     $i,n$ & bus/node index; total number of nodes\\
     $(s,t)$& a branch/edge connecting node $s$ and node $t$\\
     $\mathcal{V},\mathcal{E}$ & set of all nodes and edges: $i\in\mathcal{V},\forall i; (s,t)\in\mathcal{E}$ \\
     $j,N$ & data sample index; total number of {  (training)} samples\\
     \midrule
     $(\bm{x},\bm{y})$ & a sample with feature $\bm{x}$ and output $\bm{y}$, 
     $\bm{y}=[\bm{y_1,...,y_n}]^T=[\bm{v}^{post}_1,...,\bm{v}^{post}_n]^T$ \\
	\bottomrule
	\end{tabular} 
\end{table}

As the power grid's network topology can change over time and experience line outages, we need a model that can be generalizable to different topologies. To achieve this goal, we define the model as a probabilistic graphical model in Figure \ref{fig: graphical model}, as the power grid can be naturally represented as a graph. 

\begin{figure}[h]
	\centering
	\includegraphics[width=0.8\linewidth]{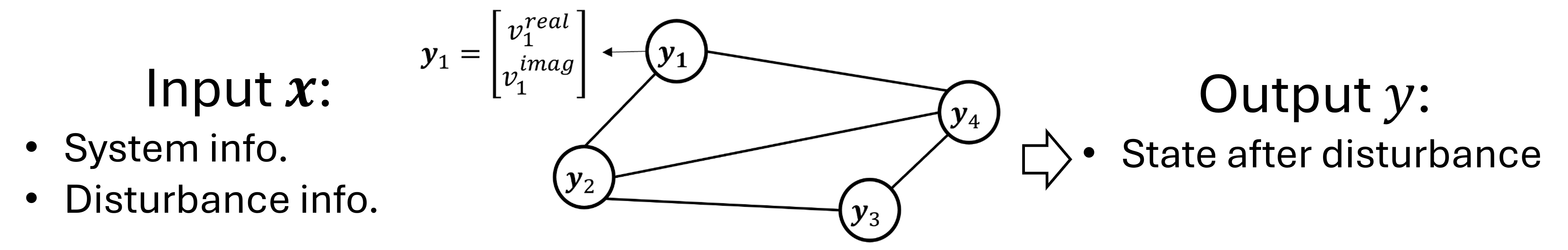}
	\caption{ A power grid can be naturally represented as a graphical model. Each node represents the bus voltage after disturbance, each edge represents a branch status after disturbance. Now conditioned on an original power grid $G$ and a disturbance $c$ that happens on it, we want to know the bus voltages after disturbance. }
	\label{fig: graphical model}
\end{figure}

On this graphical model, nodes and edges on the graph correspond to power grid buses and branches (lines and transformers), respectively. Each node represents a variable $\bm{y_i}$ which denotes voltage phasor at bus $i$, whereas each edge represents a direct inter-dependency between adjacent nodes. { Such an undirected graphical model that models the dependency relationship among the random variables can be defined as a Markov Random Field (MRF), if Markov Properties hold.} On an electrical power grid, Markov Properties (which enforce conditional independence relationships between variables) naturally hold true with real meanings. Specifically, for $\forall$ node $i$, the set of all immediate neighbors forms its Markov blanket, and thus the Local Markov property holds that, given the states on all its neighboring nodes, the state at node $i$ is conditionally independent of the rest of the nodes, as it can be determined from the balance equation at node $i$. And one can easily show that the global Markov property and pairwise Markov property also hold.

The use of MRF enables a compact way of parameterizing the joint distribution and performing inference thereafter, using observed data. According to the Hammersley-Clifford theorem \cite{Hammersley-Clifford-1971}\cite{Hammersley-Clifford-1990}, any strictly positive distribution factorizes with respect to its (maximal) cliques. Although the theoretical proof \cite{Hammersley-Clifford-proof} was established for decomposing the graph into its maximal cliques, it is noted (e.g., in \cite{MRF}) that we are free to restrict the parameterization to the edges of the graph, instead of maximal cliques, and this is called a pairwise MRF (widely used due to its simplicity although not as general). 

Therefore in this work, we consider the setting where functions with cliques are non-trivial only for the nodes and edges. Thus, the joint distribution of variables $\bm{y}$ conditioned on input features $\bm{x}$ can be factorized in a pairwise manner, {leading to a pairwise Markov Random Field \cite{MRF} as shown in (\ref{eq:pairwise parameterization}):}
\begin{align}
    &p(\bm{y}|\bm{x},\bm{\theta})=\frac{1}{Z(\bm{\theta},\bm{x})}\prod_{i=1}^n\psi_i(\bm{y_i}) \prod_{(s,t)\in\bm{E}}\psi_{st}(\bm{y_s},\bm{y_t})
    \label{eq:pairwise parameterization}
\end{align}
where $\bm{\theta}$ denotes model parameters that maps $\bm{x}$ to $\bm{y}$; $\psi_i(\bm{y_i}),\psi_{st}(\bm{y_s},\bm{y_t})$ are node and edge potentials conditioned on $\bm{\theta}$ and $\bm{x}$; and $Z(\bm{\theta},\bm{x})$ is called the partition function that normalizes the probability values such that they sum to the value of 1.

The factorization model in (\ref{eq:pairwise parameterization}) is inspired by pairwise continuous MRF \cite{MRF} and has an intuitive form: every edge potential encodes the mutual correlation between two adjacent nodes; both node and edge potentials represent the local contributions of nodes/edges to the joint distribution. In the task of disturbance analysis, each potential function intuitively represents how the status of each bus and branch 'independently' impacts the bus voltages.

Given a training dataset of $N$ samples $\{(\bm{x}^{(j)},\bm{y}^{(j)})\}$, the training and inference can be described briefly as:
\begin{itemize}
    \item \textbf{Training:} With proper definition of the potential functions $\psi_i(\bm{y_i}|\bm{x,\theta})$, $\psi_{st}(\bm{y_s},\bm{y_t}|\bm{x,\theta})$ (see Section \ref{sec:cGRF} and \ref{sec: NN-node and NN-edge}), and the parameter $\bm{\theta}$ can be  learned by maximizing log-likelihood
   \begin{equation}
       \bm{\hat{\theta}}=arg\max_{\bm{\theta}} \sum_{j=1}^N log l(\theta)^{(j)}
   \end{equation}
   where $l(\theta)^{(j)}$ denotes the likelihood of the j-th sample:
   \begin{equation}
       l(\theta)^{(j)}=p(\bm{y}^{(j)}|\bm{x}^{(j)},\bm{\theta})
   \end{equation}
    \item \textbf{Inference:} For any new input $\bm{x}$, we make use of the estimated parameter $\hat{\bm{\theta}}$ to make a single-point prediction 
    \begin{equation}
        \bm{\hat{y}_{test}}=arg\max_{\bm{y}} p(\bm{y|x,\hat{\theta}})
        \label{eq: general inference}
    \end{equation}
\end{itemize}

The use of a probabilistic graphical setting naturally integrates the domain knowledge from grid topology into the method:  
\begin{knowledge}[Grid topology] Power flow result is conditioned on the grid topology. Bus voltages of two adjacent buses connected and directed by a physical linkage (line or transformer) have direct interactions.  
\end{knowledge}

Each sample in this method can have its topology and each output is conditioned on its input topology. The following sections will discuss how the graphical model and domain knowledge enable an efficient and physically interpretable model design.

\subsubsection{Defining location-specific parameters to learn}
\label{sec:cGRF}

Upon representing the power grid and its disturbance as a conditional pairwise MRF factorized in the form of (\ref{eq:pairwise parameterization}), we obtain the benefits that the learnable parameters in the model become sparse and location-specific, i.e., specific to each node and edge in the network. This can be shown more clearly as we define the potential functions $\psi_i(\bm{y_i}),\psi_{st}(\bm{y_s},\bm{y_t})$. 

This paper builds a Gaussian random field which equivalently assumes that the output variable $(\bm{y})$ satisfies multivariate Gaussian distribution, i.e., $P(\bm{y}|\bm{x},\bm{\theta})$ is Gaussian. The justification and corresponding benefits of using Gaussian Random Field are:
\begin{itemize}
    \item partition function $Z(\bm{\theta},\bm{x})$ is easier to compute due to nice properties of Gaussian distribution. {  Specifically in the case of Gaussian, the normalization constraint can be computed easily by calculating matrix determinant $|\Lambda|$, whereas the use of other potential functions might lead to computation difficulties, potentially NP-hard \cite{MRF}.}
    \item high physical interpretability due to a physically meaningful inference model. We will discuss this later.
\end{itemize}

\noindent The potential functions for Gaussian random field \cite{MRF} are defined as follows:
\begin{align}
    &\psi_i(\bm{y_i})=exp(-\frac{1}{2}\bm{y_i}^T\bm{\Lambda_{i}}\bm{y_i}+\bm{\eta_i}^T\bm{y_i})\label{eq:gaussian_node_potential}\\
    &\psi_{st}(\bm{y_s},\bm{y_t})=exp(-\frac{1}{2}\bm{y_s}^T\bm{\Lambda_{st}}\bm{y_t})
    \label{eq:gaussian_edge_potential}
\end{align}
where $\bm{\Lambda_{i}}, \eta_i$ represents node-specific parameters with $\bm{\Lambda_{i}}$ in the form of a $2\times 2$ matrix, and  $\eta_i$ in the form of a $2\times 1$  vector. $\bm{\Lambda_{st}}$ represents edge-specific parameters in the form of a $2\times 2$ matrix.

By plugging \eqref{eq:gaussian_node_potential} and \eqref{eq:gaussian_edge_potential} into \eqref{eq:pairwise parameterization}, we have:
\begin{equation}
    p(\bm{y}|\bm{x},\bm{\theta})\propto exp(\bm{\eta}^T\bm{y}-\frac{1}{2}\bm{y}^T\Lambda\bm{y})
    \label{eq:gaussianMRF}
\end{equation}
where $\bm{\Lambda_{i}}$ and $\bm{\Lambda_{st}}$ parameters are the building blocks of matrix $\bm{\Lambda}$, and $\bm{\eta}$ is a column vector composed of all $\bm{\eta_i}$.

To further illustrate, consider a post-disturbance grid structure in Figure \ref{fig: graphical model}. The $\bm{\eta}$ and $\bm{\Lambda}$ variables for this grid structure will be shown later (in Figure \ref{fig: training process} where the $\bm{0}$ blocks in $\bm{\Lambda}$ matrix are structural zeros representing no \textit{active} edges at the corresponding locations).

In the model (\ref{eq:gaussianMRF}),   both $\bm{\eta}$ and $\bm{\Lambda}$ are functions of $\bm{x},\bm{\theta}$, i.e.,
\begin{equation}
    {\bm{\eta}}=f_\eta(\bm{x},\bm{{\theta_{\eta}}}), {\bm{\Lambda}}=f_\Lambda(\bm{x},\bm{{\theta_{\Lambda}}})
\end{equation}
and $P(\bm{y}|\bm{x},\bm{\theta})$ takes an equivalent form of a multivariate Gaussian distribution $N(\bm{\mu},\bm{\Sigma})$ ($\bm{\mu}$ is the mean and $\bm{\Sigma}$ is the covariance matrix) with
\begin{equation}
\bm{\eta}=\bm{\Lambda}\bm{\mu}, \bm{\Lambda}=\bm{\Sigma}^{-1}    
\end{equation}

Now based on these defined models, we seek to learn the parameter $\bm{\theta}$ through maximum likelihood estimation (MLE). The log-likelihood of each data sample can be calculated by:
\begin{equation}
    l(\bm{\theta})= logP(\bm{y}|\bm{x},\bm{\theta})= -\frac{1}{2}\bm{y}^T\bm{\Lambda}\bm{y}+\bm{\eta}^T\bm{y}-log Z(\bm{\theta},\bm{x}) \label{eq: loglikelihood}
\end{equation}
and the MLE can be written equivalently as an optimization problem that minimizes the negative log-likelihood loss on the data set of $N$ training samples:
\begin{equation}
    \min_{\bm{\theta}} -\sum_{j=1}^{N}l(\bm{\theta})^{(j)}
    \label{eq:MLE simplified}
\end{equation}

\textbf{Inference and Interpretation:}
Upon obtaining the solution of $\hat{\bm{\theta}}=[\hat{\bm{\theta_{\eta}}},\hat{\bm{\theta_{\Lambda}}}]^T$, parameters $\hat{\bm{\Lambda}}=f_\Lambda(\bm{x_{test}},\bm{\hat{\theta}})$, $\hat{\bm{\eta}}=f_\eta(\bm{x_{test}},\bm{\hat{\theta}})$ can be estimated thereafter. Then for any test disturbance sample $\bm{x_{test}}$, the inference model in (\ref{eq: general inference}) is equivalent to solving $\bm{\hat{y}_{test}}$ by:
\begin{align}
    \hat{\bm{\Lambda}}\bm{\hat{y}_{test}}=\hat{\bm{\eta}}
    \label{eq: linear approx}
\end{align}

Notably, the model in (\ref{eq: linear approx}) can be seen as a \textbf{linear system proxy} of the post-disturbance grid, providing a physical interpretation of the method. $\bm{\Lambda}$ is a sparse matrix with a structure similar to the bus admittance matrix where the zero entries are 'structural zeros' representing no branch connecting buses. $\bm{\eta}$ behaves like the net injection to the network. 

\subsubsection{Sparse graphical structures giving local lightweight learning: \textit{NN-node} and \textit{NN-edge}}\label{sec: NN-node and NN-edge} 


Next, we need to specify the functions of  $f_\eta(\bm{x},\bm{{\theta_{\eta}}}),f_\Lambda(\bm{x},\bm{{\theta_{\Lambda}}})$. 
Taking advantage of the sparsity of $\bm{\Lambda}$, the task here is to learn a function mapping from input $\bm{x}$ to only some edge-wise parameters $\bm{\Lambda_{st}}$ and node-wise parameters $\bm{\Lambda_{i}},\bm{\eta_i}$. Yet the number of $\bm{\Lambda_{i}},\bm{\Lambda_{st}}, \bm{\eta_i}$ parameters still increases with grid size, meaning that the input and output size of the model will explode for a large-scale system, requiring a much more complicated model to learn a high-dimensional input-output map. 

To efficiently reduce the model size, this paper implements the mapping functions using local Neural Networks: each node has a \textit{NN-node} to predict $\bm{\Lambda_{i}},\bm{\eta_i}$ using local inputs; each edge has a \textit{NN-edge} to output $\bm{\Lambda_{st}}$ in a similar way, as shown in Figure \ref{fig: NN-node and NN-edge}. This is inspired by our interpretation that $\bm{\Lambda}$ is a proxy of the bus admittance matrix whose elements represent some local system characteristics regarding each node and edge.

\begin{figure}[h]
	\centering
	\includegraphics[width=0.8\linewidth]{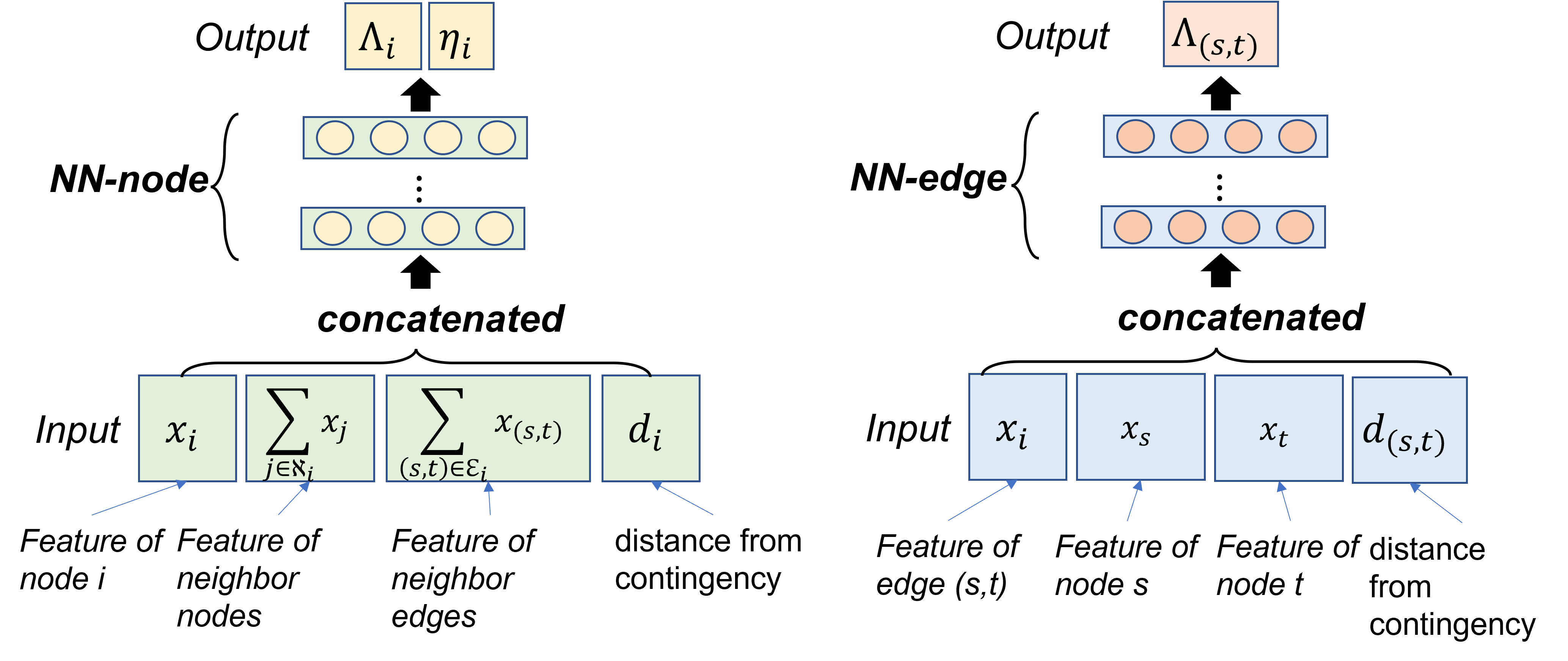}
	\caption{Local lightweight NNs: each node has a \textit{NN-node} and each edge has a \textit{NN-edge}, to map the input features to the post-disturbance system characteristics.}
	\label{fig: NN-node and NN-edge}
\end{figure}

Meanwhile, to effectively learn the mapping, we must answer the following question: \textbf{how to select the input features to the NN models optimally?} We apply \textit{domain knowledge} to design the input space that feeds most relevant features into the model:\\

\begin{knowledge}[Decisive features]
    The impact of disturbance depends heavily on the importance of disturbance components which can be quantified by the amount of its generation, load or power delivery.
\end{knowledge}

\begin{knowledge}[Taylor Expansion on system physics] Let $v=h(G)$ denote any power flow simulation that maps the case information to the voltage profile solution. By Taylor Expansion, the post-disturbance voltage can be expressed as a function depending on pre-disturbance system $G_{pre}$ and the system change $\Delta G$ caused by disturbance:
$$\bm{v^{post}} = h(G_{pre}) + h' (G_{pre})\Delta G + \frac{1}{2}h'' (G_{pre})\Delta G^2 + ...$$
\end{knowledge}

Therefore, the key features of the pre-disturbance system $(G_{pre})$ and system change $(\Delta G)$ are selected as node features to feed into the NN mappings, and include:
\begin{itemize}
    \item node feature $\bm{x}_i$: real and imaginary voltages ($v_i^{real}, v_i^{imag}$), power and current injections ($ P_i, Q_i, I_i^{real}, I_i^{imag}$), and shunt injections ($Q_{shunt,i}$) before a disturbance, and change in power injections ($\Delta P_{gen,i},\Delta P_{load,i},\Delta Q_{load,i}$) post disturbance
    \item edge feature $\bm{x}_{s,t}$: line admittance and shunt parameters of the transmission line, $\bm{x}_{s,t}=[G,B,B_{sh}]$
\end{itemize}

\subsubsection{Training the model with a surrogate loss}\label{sec:training}

With the models defined above, the training process is illustrated in Figure \ref{fig: training process}, where the forward pass of NN-node and NN-edge gives $\bm{\Lambda,\eta}$, and then the loss defined from the cGRF can be calculated to further enable a backward pass that updates the parameter $\bm{\theta}$.

\begin{figure}[h]
	\centering
\includegraphics[width=0.5\linewidth]{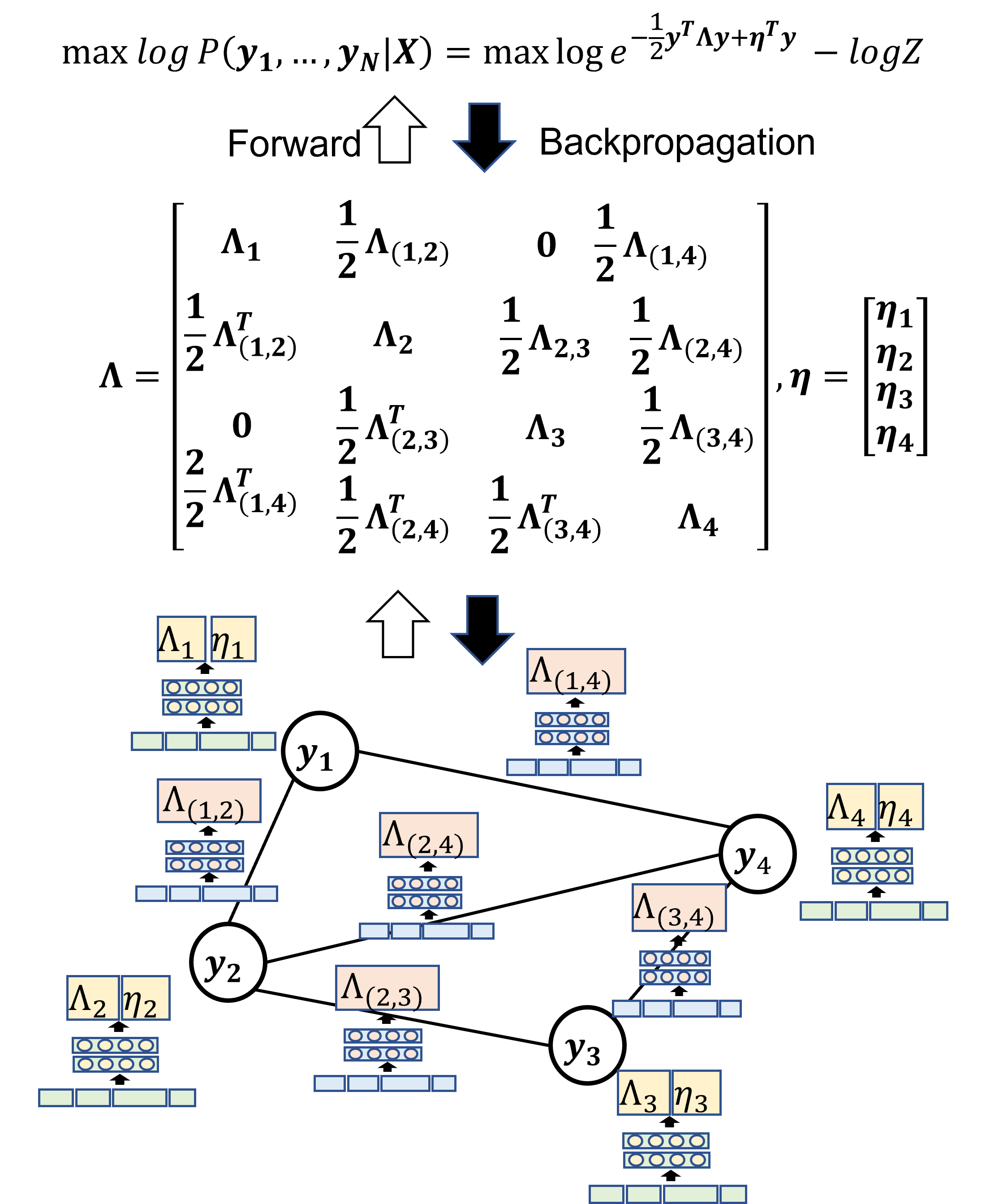}
	\caption{Training of Gridwarm: forward pass and back-propagation. Local lightweight NNs combine to learn system characteristics $\Lambda,\eta$. On large real-world power grids, $\Lambda$ is a very sparse matrix, enabling an overall lightweight ML model.}
	\label{fig: training process}
\end{figure}

As described in (\ref{eq: loglikelihood})-(\ref{eq:MLE simplified}), the loss function is the negative log-likelihood loss over the training data. Making use of the nice properties of Gaussian distribution, the partition function $Z(\bm{x,\theta})$ in the loss can be calculated analytically:
\begin{align}
    Z &= \int_{\bm{y}} exp(\bm{\eta}^T\bm{y}-\frac{1}{2}\bm{y}^T\Lambda\bm{y}) d\bm{y}\notag\\
&=\sqrt{\frac{2\pi}{|\Lambda|}} exp({\frac{\mu^T\Lambda\mu}{2}})
=\sqrt{\frac{2\pi}{|\Lambda|}} exp({\frac{\eta^T\Lambda^{-1}\eta}{2}})
    \label{eq: Z}
\end{align}

\noindent The detailed derivation is shown below:

\begin{derivation}[{Calculate partition function  $Z$}]\label{appendix: Z} 
For a multivariate Gaussian distribution $\bm{y}\sim N(\bm{\mu},\bm{\Sigma})$ where $\bm{\mu}$ denotes the mean and $\bm{\Sigma}$ denotes the covariance matrix, let $\bm{\Lambda}=\bm{\Sigma}^{-1}$, we have:
\begin{equation}
   \int_{\bm{y}} \sqrt{\frac{|\bm{\Lambda}|}{2\pi}}exp({-\frac{1}{2}(\bm{y-\mu})^T\bm{\Lambda}(\bm{y-\mu})}) d\bm{y}= 1
   \label{eq: standard Gaussian pdf integral}
\end{equation}

As mentioned earlier, the Gaussian CRF model $P(\bm{y}|\bm{x},\bm{\theta})=\frac{1}{Z(\bm{x,\theta})} exp(\bm{\eta}^T\bm{y}-\frac{1}{2}\bm{y}^T\Lambda\bm{y})$
is equivalent to a multivariate Gaussian distribution $N(\bm{\mu},\bm{\Sigma})$ with
$\bm{\eta}=\bm{\Lambda}\bm{\mu}, \bm{\Lambda}=\bm{\Sigma}^{-1}$. Thus (\ref{eq: standard Gaussian pdf integral}) can be rewritten as:
\begin{equation}
    \sqrt{\frac{|\bm{\Lambda}|}{2\pi}}exp({-\frac{\mu^T\Lambda\mu}{2}})\int_{\bm{y}} exp(\bm{\eta}^T\bm{y}-\frac{1}{2}\bm{y}^T\Lambda\bm{y}) d\bm{y}= 1
\end{equation}

Taking the nice properties of Gaussian distribution, the partition function $Z(\bm{x,\theta})$ can be calculated as:
\begin{equation}
   Z(\bm{x,\theta}) = \int_{\bm{y}} exp(\bm{\eta}^T\bm{y}-\frac{1}{2}\bm{y}^T\Lambda\bm{y}) d\bm{y}
=\sqrt{\frac{2\pi}{|\Lambda|}} exp({\frac{\mu^T\Lambda\mu}{2}})
\end{equation}
\end{derivation}

Further, we derive a surrogate loss function that mathematically approximates the original objective function. In this way, we removed the need to maintain positive-definiteness of the $\bm{\Lambda}$ in the learning process, whereas the prediction is made using a $\bm{\Lambda}$ computed from the forward pass, and thus it still considers the power grid structure enforced by the graphical model.

\begin{derivation}[{Surrogate loss}]\label{appendix: surrogate loss}
    Mathematically, due to the Gaussian distribution properties, the original optimization problem in (\ref{eq: actual optimization problem}) is equivalently:
\begin{align}
    &\min_{\bm{\theta}}\sum_{j=1}^{N}\frac{1}{2}\bm{(y^{(j)}-\mu^{(j)})}^T\bm{\Lambda^{(j)}}(\bm{y^{(j)}-\mu^{(j)}})
    -\frac{1}{2}log|\bm{\Lambda^{(j)}}| \\
    s.t. &\notag\\
    &\text{(forward pass) }\bm{\Lambda^{(j)}}=f_\Lambda(\bm{x^{(j)}},\bm{{\theta_{\Lambda}}}),\forall j\\
    &\text{(forward pass) } \bm{\eta^{(j)}}= f_\eta(\bm{x^{(j)}},\bm{{\theta_{\eta}}}), \forall j\\
    &\text{(positive definiteness) }\bm{\Lambda^{(j)}}\succ \bm{0}, \forall j\\
     &\text{(inference) } \bm{\mu^{(j)}=\Lambda^{-1(j)}\eta^{(j)}}, \forall j
\end{align}

To design a surrogate loss, we make an approximation $\bm{\Lambda^{(j)}=I}$ ($\bm{I}$ is identity matrix) only in the objective function, so that $log|\bm{\Lambda}|=0$ becomes negligible. 

Furthermore, to enable a valid distribution $P(\bm{y}|\bm{x},\bm{\theta})$ and unique solution during inference, it is required that the $\bm{\Lambda}$ matrix is positive definite (PD), i.e., $\bm{\Lambda}\succ \bm{0}$. Therefore, adding this constraint and substituting (\ref{eq: Z}) into the loss function, the optimization problem of the proposed method can be written as:
\begin{align}
    &\min_{\bm{\theta}}\sum_{j=1}^{N}
    \frac{1}{2}\bm{y^{(j)T}}\bm{\Lambda^{(j)}}\bm{y^{(j)}}\notag\\
    &-\bm{\eta^{(j)T}}\bm{y^{(j)}}
    -\frac{1}{2}log|\bm{\Lambda^{(j)}}| 
    + \frac{1}{2}\bm{\eta^{(j)T}\Lambda^{-1(j)}\eta^{(j)}}\\
    s.t. &\notag\\
    &\text{(forward pass) }\bm{\Lambda^{(j)}}=f_\Lambda(\bm{x^{(j)}},\bm{{\theta_{\Lambda}}}),\forall j\\
    &\text{(forward pass) } \bm{\eta^{(j)}}= f_\eta(\bm{x^{(j)}},\bm{{\theta_{\eta}}}), \forall j\\
    &\text{(positive definiteness) }\bm{\Lambda^{(j)}}\succ \bm{0}, \forall j
    \label{eq: actual optimization problem}
\end{align}

In this problem, maintaining the positive definiteness of matrix $\bm{\Lambda}$ for every sample is required, not only in the final solution (i.e., throughout the training process due to the $log|\bm{\Lambda}|$ in the loss). This can be computationally challenging, especially when the network size grows with large matrix dimensions. 

To address this issue, we design a \textbf{surrogate loss} $\sum_{j=1}^{N}\frac{1}{2}\bm{(y^{(j)}-\mu^{(j)})}^T(\bm{y^{(j)}-\mu^{(j)}})$, which acts as a proxy for the actual loss we want to minimize. With the use of surrogate loss function, the  overall problem converts into the following form:
\begin{align}
    &\min_{\bm{\theta}}\sum_{j=1}^{N}\frac{1}{2}\bm{(y^{(j)}-\mu^{(j)})}^T(\bm{y^{(j)}-\mu^{(j)}})\\
    s.t. &\notag\\
    &\text{(forward pass) }\bm{\Lambda^{(j)}}=f_\Lambda(\bm{x^{(j)}},\bm{{\theta_{\Lambda}}}),\forall j\\
    &\text{(forward pass) } \bm{\eta^{(j)}}= f_\eta(\bm{x^{(j)}},\bm{{\theta_{\eta}}}), \forall j\\
    &\text{(inference) } \bm{\mu^{(j)}=\Lambda^{-1(j)}\eta^{(j)}}, \forall j
    \label{eq: surrogate optimization}
\end{align}
\end{derivation}

From decision theory, both the original and the surrogate loss aim to return an optimized model whose prediction $\bm{\hat{y}}$ approximates the ground truth $\bm{y}$, and both make predictions by finding out the linear approximation of the post-disturbance system $\hat{\bm{\Lambda}}\bm{\hat{y}}=\hat{\bm{\eta}}$. 

Additionally, this surrogate optimization model can be considered as minimizing the mean squared error (MSE) loss $\frac{1}{2}||\bm{y-\hat{y}}||^2$ over the training data, where $\bm{\hat{y}}$ is the prediction (inference) made after a forward pass. 

\subsubsection{More regularization techniques enabling lightweight ML}

With each node having its own \textit{NN-node} and each edge having its \textit{NN-edge}, the number of parameters grows approximately linearly with grid size (more specifically, the number of nodes and edges). Can we reduce the model size further? The answer is yes! One option is to make all nodes share the same \textit{NN-node} and all edges share the same \textit{NN-edge}, so there are only two NNs in total.

Why does this work? Such sharing of \textit{NN-node} and \textit{NN-edge} is an extensive use of {parameter sharing} to incorporate domain knowledge into the network. Especially, from the physical perspective:

\begin{knowledge}
    [{Location-invariant (LI) properties}] the power grid and the impact of its disturbances have properties that are invariant to change of locations: 1) any location far enough from the disturbance location will experience little local change. 2) change in any location will be governed by the same mechanism, i.e., the system equations and Kirchhoff's laws.
\end{knowledge}

Parameter sharing across the grid network significantly lowers the number of unique model parameters. Also, it reduces the need for a large amount of training data to adequately learn the system mapping for larger grid sizes (like the networks representing continental U.S. networks with $>80k$ nodes).

Moreover, leveraging the zero-injection buses can also make $\eta$ vector more realistic:

\begin{knowledge}
    [{Zero-injection (ZI) buses}] A bus with no connected generation or load is called a zero-injection (ZI) bus. These buses neither consume nor produce power, and thus, injections at these buses are zero.
\end{knowledge}

In the proposed approach, the model parameter $\eta$ serves as a proxy to bus injections; therefore, we can integrate domain knowledge about zero-injection nodes into the method by setting $\bm{\eta}_i=0$ at any ZI node $i$.

\chapter{Physics-ML Synergy towards more efficient and robust situational awareness}\label{ch: synergy}
\section{High-level idea of Physics-ML Synergy}

Despite the advances in existing approaches, there remain some tasks that are challenging and even impossible for state-of-the-art physics-based and data-driven approaches. 
A key challenge regarding efficiency is to make simulation both fast and accurate for a large-amount of scenarios and for hard-to-solve scenarios. 
A significant challenge regarding robustness is the mitigation of modern threats in the form of cyberattacks leveraging system physics. 
State-of-the-art ML tools are successful in alarming these modern attacks, but they sometimes require human intervention to further mitigate some complex anomalies. This is because without the detailed physical model, even an advanced ML anomaly detector can only provide the alarm and a rough neighborhood indicating the suspicious anomalous region. Given that (state) estimation has been compromised, the true system state is hardly known, sometimes leaving it unknown what to follow from the current situation. This often necessitates human intervention in the anomaly mitigation process.


\begin{figure}[h]
     \centering
            \includegraphics[width=0.7\linewidth]{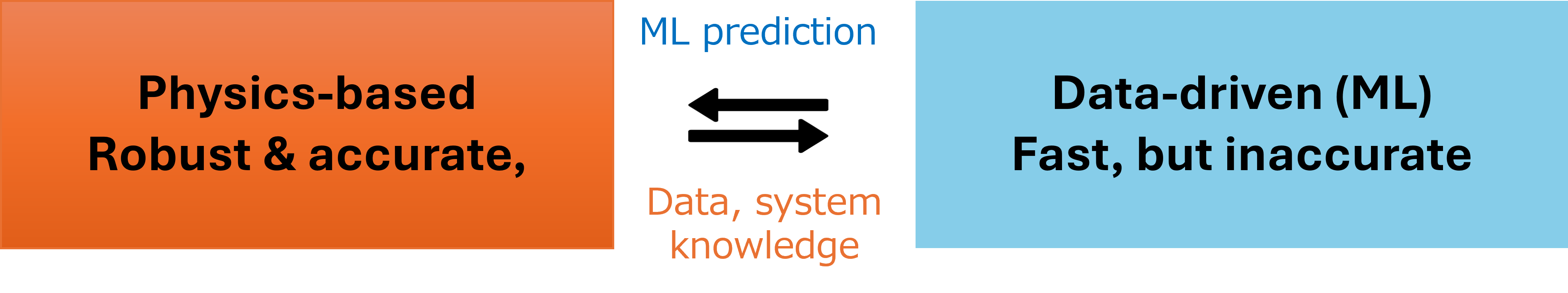}
    \caption{We advocate for a Physics-ML synergy: physics-based and data-driven models are interconnected to augment each other.
    }
    \label{fig: synergy concept}  
\end{figure}

Physics-based and data-driven approaches can be mutually complementary, motivating  Physics-ML Synergy framework in this thesis. In the remainder of this Chapter, we apply this high-level idea of synergy design in both simulation and estimation applications, advancing situational awareness through a collaborative architecture:
\begin{itemize}
    \item Section \ref{sec: synergy for simulation} develops Physics-ML Synergy for simulation via an interconnection of GridWarm ML predictor (Problem \ref{def: gridwarm} in Section \ref{sec: gridwarm}) and robust circuit-based simulator (Problem \ref{def: sparsePF simulation} in Section \ref{sec: sparsePF}. In this synergy, ML prediction is fed into a physics-based simulator as a warm starting point, while system knowledge and data are given to ML for learning of threat impact. The ultimate goal is to achieve both fast speed from ML and accuracy from physics-based simulators.
    \item Section \ref{sec: synergy for estimation} develops Physics-ML Synergy for estimation by interconnecting time-series ML \method (Problem \ref{def: dynwatch} in Section \ref{sec: dynwatch}) and robust circuit-based estimator (Problem \ref{def: ckt-GSE} in Section \ref{sec: ckt-GSE}). In this synergy, prior knowledge from ML comes in the form of regularization to augment estimation problems, and system knowledge is integrated into ML for predicting normal system behavior. The ultimate goal is to achieve a higher level of robustness to a mixture of traditional and modern threats.
\end{itemize}

\section{Application 1: Physics-ML Synergy for simulation: high speed for simulating modern threats}\label{sec: synergy for simulation}

\subsection{Identifying remaining gap of efficiency}
Now that our robust actionable simulation in Section \ref{sec: sparsePF} has  advanced steady-state simulation via sparse optimization. We have established simulation robustness to blackout failures such that the traditionally unsolvable cases have been made not only solvable but also actionable for making localized fixes. 

A remaining gap is related to efficiency: due to the nonlinear nature of simulation problems, large systems are slow to converge when good initial conditions are unavailable. Modern threats are also making good initial points harder to obtain. The system's state before a disturbance is typically the best possible starting point we can have, assuming no additional tools are used to refine it. However, facing modern threats which can cause disturbances at multiple locations and significant perturbations at each, the system  experiences substantial changes and the 'best possible' starting point could be far from the actual solution. \cite{madiot} proposed an attack model, namely \textit{BlackIoT} or \textit{MadIoT}, is a modern threat of this kind. Figure \ref{fig:madiot} provides a brief introduction to MadIoT. \cite{madiot} evaluates the impact of this attack using a steady-state simulation while considering the droop control, protective relaying, and thermal and voltage limits for various components. \cite{grid-cybersecurity-Vyas} evaluate the possible reduced impact of MadIoT by simulating with imperfect knowledge at the attacker's side.

\begin{figure}[h]
	\centering
	\includegraphics[width=0.7
	\linewidth]{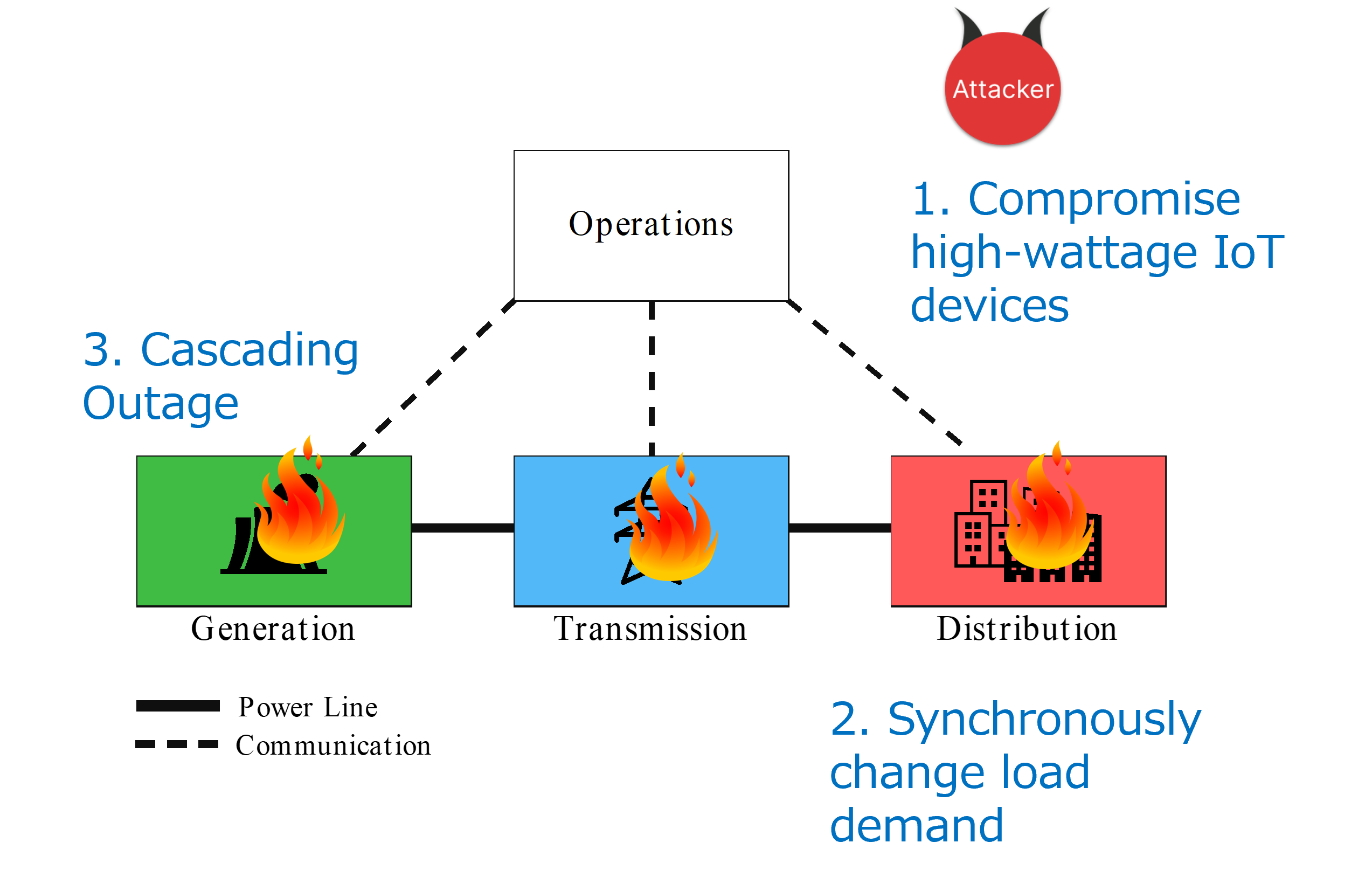}
	\caption{The proliferation of IoT devices has raised concerns about IoT-induced cyber attacks, where an attacker can manipulate multiple locations by synchronously turning on/off or scaling up/down some high-wattage IoT-controlled loads on the grid. This can lead to grid instability or grid collapse by further causing cascading outages.}
	\label{fig:madiot}
\end{figure}

These limited time-efficiency of simulation make it computationally prohibitive to evaluate many  scenarios of component failures and modern cyberthreats within a limited time during real-time operation. And this limits the operator's situational awareness.

To address the no-good-initial-point problem, we are motivated to leverage ML to fulfill this gap. However, replacing physical solvers with purely data-driven techniques makes it fast but has severe limitations in generalization, scalability and interpretability, as discussed in \cite{gridwarm}\cite{grid-reality-Li}. 

This thesis presents a Physics-ML Synergy approach that interconnects simulation with a ML-based predictor. The ML model is the GridWarm method developed in Section \ref{sec: gridwarm} which is scalable, interpretable and generalizable to topology differences.
Gridwarm predicts the post-disturbance system states before simulation starts and then the prediction is used to warm-start the physics-based simulation.

\subsection{Evaluating a synergy of Gridwarm and simulator on MadIoT attack}

This section runs experiments for disturbance events in the context of the MadIoT attack. The experiments are to validate that the proposed warm starter provides good initial states for \textit{hard-to-solve} N-k disturbances and enables faster convergence when compared to traditional initialization techniques.

We test three versions of the proposed method. These versions differ in the level of domain-knowledge that they incorporate within their model. Table \ref{tab: gridwarm variants} summarizes the domain-knowledge in these versions.

\setlength{\tabcolsep}{6pt}
\begin{table}[htbp]
\small
\centering
	\caption{Summary of domain knowledge \label{tab: gridwarm variants}}
	\begin{tabular}{ @{}rllccc@{} }  
	\toprule
	\textbf{Knowledge} & \textbf{Technique} &\textbf{Benefits} &\textbf{cGRF} &\textbf{cGRF-PS} & \textbf{cGRF-PS-ZI} \\ \midrule
		\textbf{Grid topology}  
     & graphical model & - physical interpretability & \checkmark& \checkmark& \checkmark \\
     &  &- generalization (to topology) & & & \\\\
     \midrule
     \textbf{Decisive features}
     & feature selection  & - accuracy & \checkmark& \checkmark& \checkmark\\
     & & - physical interpretability &&&\\
     & & - generalization (to load\&gen)\\
     \midrule
     \textbf{Taylor expansion }
     &feature selection & - accuracy & \checkmark& \checkmark& \checkmark\\
     \textbf{on system physics}& & - physical interpretability &&&\\
     \midrule
     \textbf{Location-invariant properties}
     &parameter & - trainability, scalability & & \checkmark& \checkmark\\
     &sharing (PS) &- generalization ($\downarrow$ overfitting) &&& \\
     \midrule
     \textbf{Zero-injection (ZI) bus}
     &enforce $\eta_i=0$ & - physical interpretability & & & \checkmark\\
     &at ZI buses & - generalization \\
	\bottomrule
	\end{tabular} 
\end{table}

\subsubsection{Data generation and experiment settings}
We generate synthetic MadIoT disturbances for the following two networks: i) IEEE 118 bus network \cite{IEEE118} ii) synthetic Texas ACTIVSg2000 network \cite{ACTIVSg2000}. { For each network, we generate $N_{data}$ disturbance samples $\{(x^{(j)},y^{(j))}\}_{N_{data}}$ where feature the notation of $x$ and $y$ has been illustrated in Figure \ref{fig: graphical model} and Table \ref{tab:notations}. } The algorithm to generate the synthetic disturbance data is given in Algorithm \ref{alg: main generation}.
\begin{algorithm}
	\caption{3-Step Data Generation Process} 
	\label{alg: main generation}
	\KwIn{Base case $G_{base}$, type of disturbance $t_c$, number of data samples $N_{data}$}
	\KwOut{Generated dataset $\{(x^{(j)},y^{(j))}\}_{N_{data}}$}
	\For{{$j \gets 1$ to $N_{data}$} }{
	{\bf 1. Create a random feasible pre-disturbance case $G_{pre}^{(j)}$:} each sample has random topology, generation and load level. 
	
	{\bf 2. Create disturbance $c^{(j)}$ on $G_{pre}^{(j)}$: } which has attributes \textit{type, location, parameter}.
	
	{\bf 3. Simulate with droop control:} run power flow to obtain the post-disturbance voltages $\bm{v}^{post}$
	}
\end{algorithm}

{ \textbf{Disturbance set and model generalization:} In our experiment, we train and test a warm starter for a MadIoT scenario of  increasing the top K largest loads by the same amount (percentage) which is a severe scenario threatening the power grid. And the pre-disturbance load is randomly sampled within the range of 95\%-105\% base load, and topology is randomly sampled by disconnecting 1-2 random lines on the base case, to represent the different normal operating conditions of a power system. Such disturbance generation settings to a very specific setup of the dataset so that learning becomes more targeted: these disturbances are ”hard-to-solve” for an optimization solver, but a simple and interpretable learned
model might be able to easily extract the major relationships to provide good warm-start values. 
Obviously, this leads to limited generalization issues that the trained model can hardly apply to a different disturbance scenario where loads are manipulated on other locations and by another amount.
But in practice, this can be addressed with multiple models. The operators or decision makers can decide several other significant disturbance settings that are worth consideration and evaluation. And they can train a second model on a second dataset which describes another important disturbance setting. So that the different severe cyberthreat scenarios can be considered with learning performed in a targeted way.  
}

The experiment settings that are used for the data generation, model design, and model training are documented in Table \ref{tab:experiment settings}. 
{  Based on the idea of \textit{NN-node} and \textit{NN-edge} in Section \ref{sec: NN-node and NN-edge}, the neural networks used in this work aim to learn a low-dimensional mapping from local node / edge input features to local outputs to form $\Lambda, \eta$. This can be effectively done with simple and shallow neural network architectures. In our experiment, the model is designed with a shallow 3-layer NN architecture with 64 hidden layers in each, to save computation time and reduce overfitting. It also allows us to experiment on whether a simple model design can give good performance.
The training is then done with an Adam optimizer and step learning rate scheduler.}

\setlength{\tabcolsep}{6pt}
\begin{table}[htbp]
\small
\centering
	\caption{Experiment settings \label{tab:experiment settings}}
	\begin{tabular}{ @{}rl@{} }  
	\toprule
	\textbf{Settings} & \textbf{(see Table \ref{tab:notations} for definitions)} \\ \midrule
     $N_{data}$ & 
     case118: 1,000; ACTIVSg2000: 5,000. Data are split into train, val, test set by $8:1:1$\\
     \midrule
     \textit{NN-node} \& \textit{NN-edge}& shallow cylinder architecture, $(n\_layer,hidden\_ dim)=(3,64)$\\
     \midrule
     disturbance $c$ & $type:$ MadIoT\\
     & $location:$ randomly sampled 50\% loads\\
     & $parameter:$ case118 200\%,  ACTIVSg2000 120\%\\
     \midrule
     optimizer & $Adam,lr=0.001, scheduler=stepLR$\\
	\bottomrule
	\end{tabular} 
\end{table}

\subsubsection{Physical Interpretability of Gridwarm}

\begin{figure}[h]
	\centering
	\includegraphics[width=0.5\linewidth]{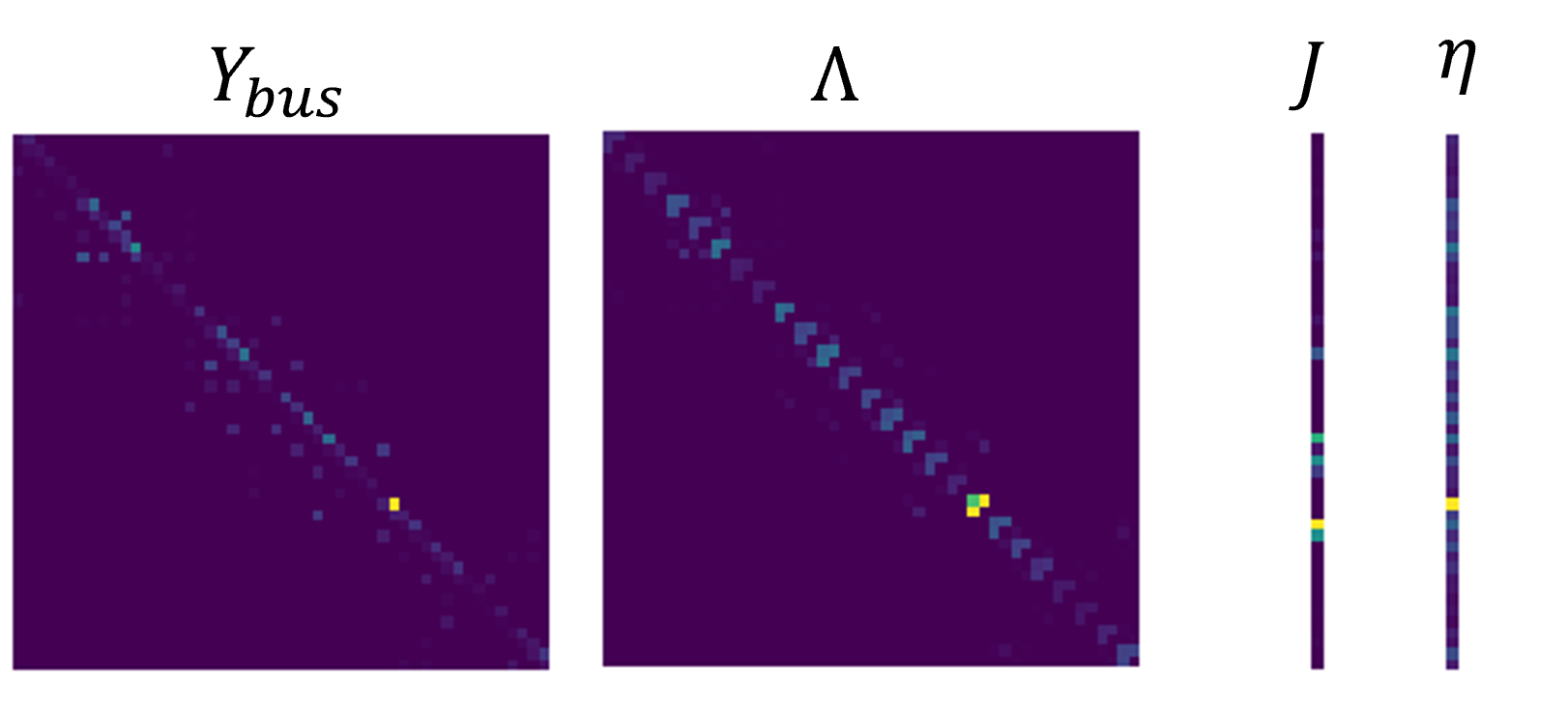}
	\caption{Physical interpretation:  This figure visualizes the values in matrices and vectors (the more yellow, the larger the value). Learned model parameters $\Lambda, \eta$ have some similarity with the true post-disturbance system admittance matrix $Y_{bus}$ and injection current vector $J$, in terms of the sparse structure and value distribution. This is because the learned parameters $\Lambda, \eta$ aims to form a linear model $\Lambda y= \eta$ which is a linear proxy of the true linearized system model $Y_{bus}y=J$.}
	\label{fig: interpretation}
\end{figure}
Figure \ref{fig: interpretation} validates our hypothesis in Section \ref{sec:cGRF} about \textit{the physical interpretation of our method as a linear system proxy}. In Figure \ref{fig: interpretation}, we visualize the result on a test sample to show the similarity between the linear proxy given by model parameters $\Lambda, \eta$ and the true post-disturbance system linearized admittance matrix $Y_{bus}$ and injection current vector $J$ at the solution; thus, validating that the model acts as a linear proxy for the post-disturbance operating condition.

\subsubsection{Gridwarm-aided simulation gives 3x faster speed}

To verify the effectiveness of the warm starter, we compare the convergence speed ($\#$ iterations) with three different initialization methods for the physical solver \cite{SUGAR-pf}: 
\begin{enumerate}
    \item flat start (flat)
    \item pre-disturbance solution ($V_{pre}$)
    \item physical solver warm-started by the three versions of the proposed method (cGRF) using different levels of domain knowledge as specified in Table \ref{tab: gridwarm variants}
\end{enumerate}

\begin{figure}[h]
	\centering
	\includegraphics[width=1\linewidth]{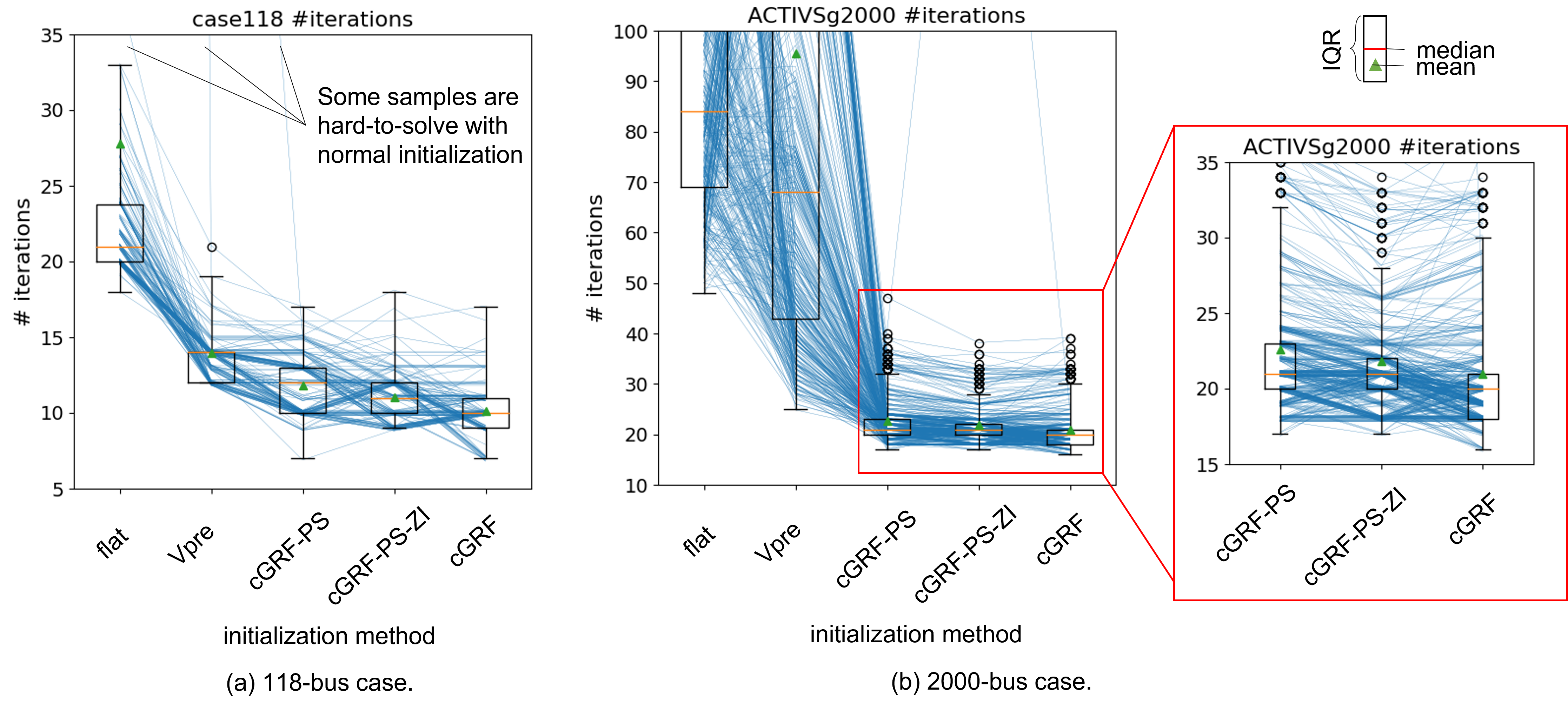}
	\caption{Result on test data: power flow simulation takes fewer iterations to converge with the proposed method, than traditional initialization methods:  i. \textbf{flat} start: starting with $(V_i, \delta_i) \leftarrow (1,0), \forall i \in \{1,...,n\}$; ii). \textbf{Vpre} start: warm-starting from pre-disturbance voltages.}
	\label{fig: gridwarm result}
\end{figure}

Figure \ref{fig: gridwarm result} shows the evaluation results on test data. These include the test samples  {   (we split the train, validate, and test set by 8:1:1 as illustrated in Table \ref{tab:experiment settings}) which include 100 unseen disturbance samples for case 118, and 500 unseen disturbance samples} for ACTIVSg2000.  For cGRF results, we feed the ML predictions into a power flow simulator: SUGAR \cite{SUGAR-pf}. The simulator is robust because it always converges. In case the general NR loop fails, the simulator uses homotopy \cite{SUGAR-pf} to ensure convergence, albeit at a computational cost.

The results show that simulation takes fewer iterations to converge with the proposed ML-based cGRF initialization when compared to traditional initialization methods (flat or $V_{pre}$). In particular, on ACTIVSg2000, many disturbance samples are hard-to-solve with traditional initialization (and may require the homotopy option in SUGAR). In contrast, our ML-based cGRF method significantly speeds up convergence (up to 5x improvement in speed) even with the shallow 3-layer NN architecture.

Moreover, due to the use of parameter sharing (PS) which enables NNs to share parameters, the \textit{lightweight} model cGRF-PS significantly reduces the total number of model parameters but achieves comparable results to the base model cGRF. In particular, while the average doesn't decrease significantly, the variation decreases due to the use of zero injection (ZI) knowledge. And cGRF-PS-ZI further shows that integrating more grid physics into the model through zero injection (ZI) knowledge can further improve convergence of the lightweight model.

\subsection{Discussion: potential upgrades to power grid contingency analysis}

By leveraging the synergy between Gridwarm and physics-based power flow simulators, we have developed an ML-aided simulation that significantly accelerates power flow simulations, achieving a 3x speed improvement (as shown in Figure \ref{fig: gridwarm result}). This advancement presents significant value for upgrading the contingency analysis units in modern grid operators, enhancing real-time situational awareness.

Contingency analysis \cite{contingency-analysis} is a simulation module in grid control rooms that evaluates numerous "what if" scenarios, evaluating the impact of selected disturbances and outages on key power grid health indicators (e.g., voltage and line flow). If the results from contingency analysis indicate that the grid is operating outside its operational limits, decision-makers can take remedial actions to maintain reliability, such as redispatching generation through security-constrained optimal power flow. Currently, real-time contingency analysis runs every 5-30 minutes (e.g., the Electric Reliability Council of Texas (ERCOT) runs it every 5 minutes \cite{ercot-rtca-5min}, and the North American Electric Reliability Corporation (NERC) requires it to run at least once every 30 minutes). However, due to computational constraints, operators typically simulate only a predefined set of N-1 contingencies (the loss of a single component), which are generally related to mechanical failures or natural disasters.

Now, with the faster ML-aided simulation, operators can proactively simulate a far greater number and variety of disturbances. This includes, but is not limited to, modern threats that could cause simultaneous outages and malicious disruptions at multiple locations, as well as a broad range of N-k contingencies with $k >> 1$. Consequently, future contingency analyses should also account for cyber events in their predefined set of contingencies. Simulating the power flow impact of these cyber events can help grid operators assess the system’s vulnerability to attacks and identify situations where preventive action is necessary. These improvements are made possible by the faster simulation tools we have developed.

To this end, it's worth noting that learning can be "targeted" in Gridwarm, making it directly applicable to address a large variety of threats, although this thesis specifically experimented on the MadIoT attack. In practice, different types of threats can be handled by training multiple Gridwarm models. Operators or decision-makers can define a set of significant disturbance types to evaluate. For instance, they can train one model to warm-start simulations for MadIoT attacks, where at most $50\%$ of loads are manipulated by less than $20\%$, and another model for more severe MadIoT incidents where up to $70\%$ of loads are manipulated. Additionally, they could also train several additional models for N-k line outage contingencies with $k=1,2,3...$. In this way, different severe threats can be considered with learning conducted in a targeted and focused way. And this trades generalization of each model for its high performance on specific disruption settings.

This opens opportunities to evaluate in real time the system’s vulnerability to numerous real-world and theoretical attacks, such as brute-force attacks on critical control devices (as seen in the Ukrainian power grid blackout \cite{ukrain2015}\cite{lee2017crashoverride}); hacking into a large set of grid-edge devices (e.g., Internet of Things-IoT devices), as described in MadIoT \cite{madiot}; and compromising the confidentiality and integrity of power grid data through false data injection attacks \cite{fdia-review}.

\section{Application 2: Physics-ML Synergy for estimation: robust to modern threat of falsified data}\label{sec: synergy for estimation}

\subsection{Identifying remaining gap of robustness}
Starting from this Section, we switch to another application of Physics-ML Synergy to fulfill a remaining robustness gap in estimation tools. 

Despite the advancements we have made on robustness of estimation methods, there still remains a limited robustness to modern threats. While the anomaly detectors can create alarms on modern threats, they do not interconnect with these estimation methods. Therefore we still face a limited situational awareness of the true system states (bus voltages and power flows) which is sometimes necessary to decide the best decisions to mitigate anomalies and recover the system to the normal state. 

For example, false data injection attack (FDIA)\cite{fdia-acse}\cite{fdia-review}\cite{FDIA_DC}\cite{grid-cybersecurity-Vyas} has been proved capable of leveraging spatial system constraints to inject falsified measurement data so that even the state-of-the-art (physics-based) estimation end up producing wrong estimates, resulting in erroneous financial or technical decisions made.
The attack has been theoretically proved destructive across various industries, including the electrical power grid, connected and automated vehicles, communication networks, etc.

To defend against the modern false data, this work aims at a synergy framework where time-series ML can augment physics-based estimation tools to make it robust. 

\subsection{A synergy of \method and state estimator}
How to mathematically bring them into a collaborative and interconnected design? The interconnection can be mathematically modeled with a probabilistic graphical model, more specifically a Bayesian Network, to describe the spatiotemporal interactions between power grid states, measurements, and anomalies at different time moments. Figure \ref{fig: pgm} shows the Bayesian Network which is a generative model depicting how a power grid evolves and how grid sensor data are generated over time.

\begin{figure}[htbp]
    \centering
\includegraphics[width=0.4\linewidth]{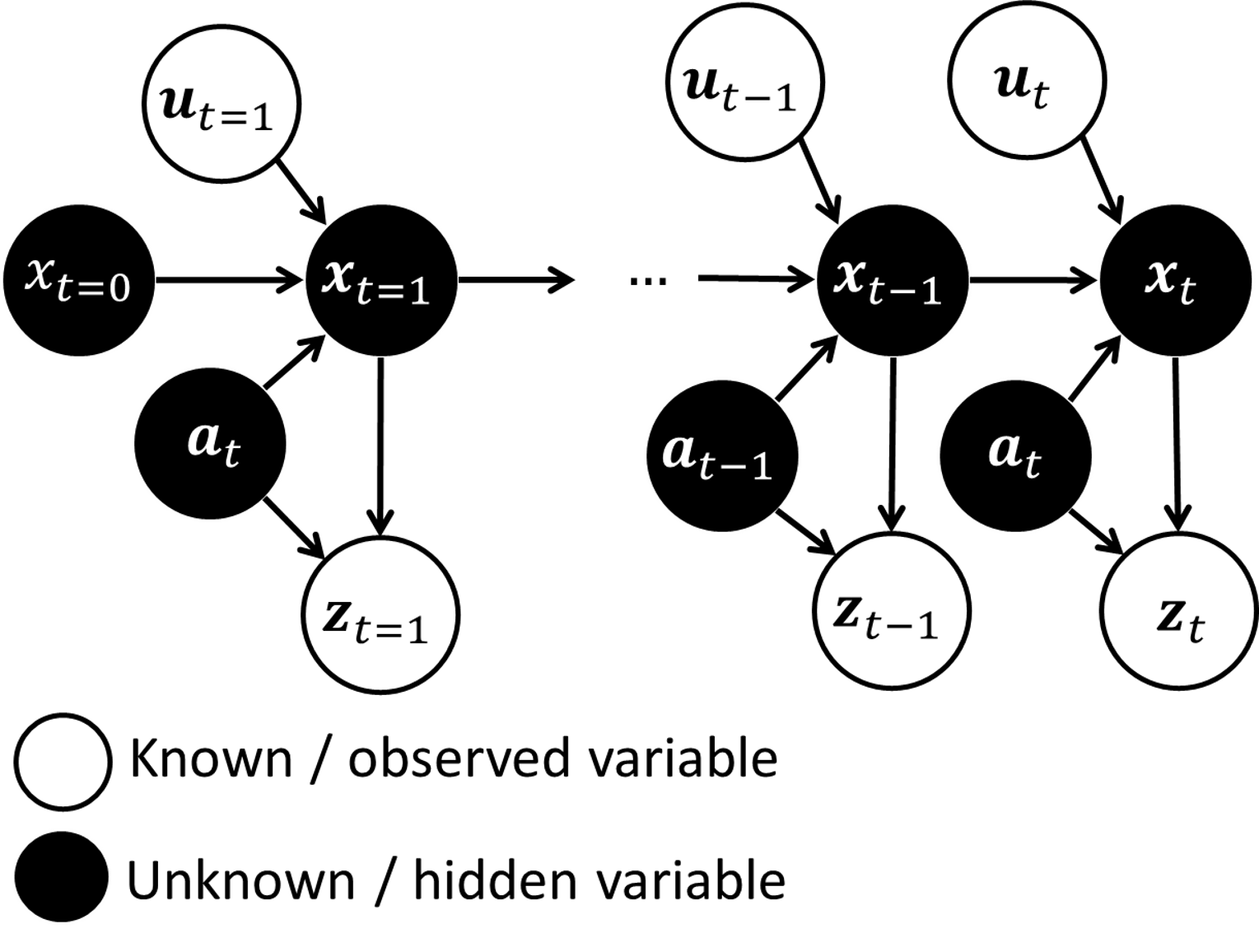}
    \caption{A Bayesian Network depicts the spatiotemporal interactions of variables in the electrical power grid. 
    }
    \label{fig: pgm}
\end{figure}

As illustrated in Figure \ref{fig: pgm}, the system starts from an initial state (vector) ${x}_{t=0}$ which includes the AC bus voltages $v_{t=0}$ and network topology $G_{t=0}$. At any time $t$, control action $u_t$ and stochastic process drive the system to change from $x_{t-1}$ to new state $x_t$; sensors collect real-time data (vector) $z_t$; and anomalies $a_t$ can affect the system's state and/or its data. Gaining situational awareness is to infer the unknown variables $x,a$ given the known or observed variables $z,u$.

The probabilistic graphical model enables a compact way of writing the conditional joint distribution and inferring the system variables thereafter, using observed data. Specifically, given a historical data of time-series measurements  $z_{t=1},...,z_{t-1},z_{t}$, a trace of (known) input control signals $u_{t=0},...,u_{t-1},u_{t}$ and a base system model $M$ which specifies the parameter information of all active and inactive nodes and edges, we can infer the AC states $v_{t=1},...,v_{t-1},v_{t}$ (which correspond to bus voltages), network topology $G_{t=1},...,G_{t-1},G_{t}$, and anomaly indicators $a_{t=1},...,a_{t-1},a_{t}$ from the conditional joint distribution $p(v_t,...,v_{t=1}, G_{t},...,G_{t=1}, a_{t},...,a_{t=1}| z_{t},...,z_{t=1},u_{t},...,u_{t=1})$. 

Here, this paper make the following assumptions on the variables:

\begin{assumption}[input control action]
We narrow down the action set to only include topology change actions, i.e., $u_t$ at $\forall t$ refers to control actions to change the circuit breaker statuses or other known topology changes.
\label{assumption: u}
\end{assumption}

\begin{assumption}[measurement data and observability]
    $z_t$ at $\forall t\in\{1,2.,,,,T\}$, includes both continuous data and discrete status data associated with time $t$. The continuous data come from a mixture of conventional SCADA remote terminal units (RTUs), and modern phasor measurement units (PMUs); the status data measures statuses of transmission lines, circuit breakers and other switching devices. I.e.,
\begin{equation}
    z_t=[z_{t,pmu}, z_{t,rtu}, z_{t,status}]
\end{equation}
where $z_{t,pmu}$ includes phasor data and $z_{t,rtu}$ includes $|V|, P, Q$ data (this work does not use frequency related data). 
And at time $t$, $z_t$ guarantees \textbf{observability} of the system, while the $z_{t,pmu}$ alone does not guarantee observability.
\label{assumption: z}
\end{assumption} 

For simplicity, let the 
 state $x_t$ at $\forall t$ include both AC states of bus voltages $v_t$ and network topology $G_t$, since they can be jointly estimated in a  (generalized) state estimation tool, i.e.,
 \begin{equation}
     x_t=[v_t,G_t]
     \label{eq: simplify with x}
 \end{equation}
 and let  $\widetilde{Z},\widetilde{U},\widetilde{X},\widetilde{A}$ denote the time-series of variables:
\begin{align}
    &\widetilde{Z}=[z_{t=1},...,z_{t-1},z_{t}],\notag\\ 
    & \widetilde{U}=[u_{t=1},...,u_{t-1},u_{t}],\notag\\
   &\widetilde{X}=[x_{t=1},...,x_{t-1},x_{t}],\notag\\
  & \widetilde{A}=[a_{t=1},...,a_{t-1},a_{t}]\notag
\end{align}
we estimate $\widetilde{V},\widetilde{G},\widetilde{A}$ via an Maximum A Posteriori (MAP) Estimation:
\begin{align}
\widetilde{X}^*, \widetilde{A}^*=&\arg\max p(\widetilde{X},\widetilde{A}|\widetilde{Z},\widetilde{U}, \underbrace{M}_\text{M is constant})\\
=&\arg\max p(\widetilde{X},\widetilde{A}|\widetilde{Z},\widetilde{U})
\end{align}
which is equivalent to maximizing the joint distribution $p(\widetilde{X},\widetilde{A},\widetilde{Z},\widetilde{U})$ which can be further factorized  into the product of priors and conditional likelihoods according to the conditional independence relationship defined in the Bayesian Network, i.e.,
\begin{align}
\widetilde{X}^*,\widetilde{A}^*
&=\arg\max p(\widetilde{X},\widetilde{A}|\widetilde{Z},\widetilde{U})\notag\\
&=\arg\max p(\widetilde{X},\widetilde{A},\widetilde{Z},\widetilde{U})\notag\\
&=\arg\max log p(\widetilde{X},\widetilde{A},\widetilde{Z},\widetilde{U})\notag\\
&=\arg\max log\Biggl(p(v_{t=0})\cdot\notag\\& \prod_{t=1}^T P(a_{t})P(u_{t})
p(x_t|x_{t-1},u_{t}, a_{t})p(z_t|x_t,a_t)\Biggr) \notag\\
&=\arg\max log p(v_{t=0}) \notag\\ &+\sum_{t=1}^T 
 log P(a_t) + \sum_{t=1}^T log P(u_t) \notag\\
 &+\sum_{t=1}^T log p(z_t|x_t,a_t) + \sum_{t=1}^T log p(x_t|x_{t-1},u_t,a_t) 
\end{align}

Then we make assumptions on the priors that 
\begin{assumption}[initial state prior distribution]
    for $\forall t$, prior $p(v_{t=0})$ is a continuous uniform distribution, because, intuitively, for any episode of state transition, the power grid can start at any random initial state.
    \label{assumption: prior P(v0)}
\end{assumption}
\begin{assumption}[anomaly and action prior distribution]
    for $\forall t$, prior $P(a_t)$ and $P(u_t)$ are discrete uniform distributions. For $P(a_t)$, the uniform distribution is assumed because this paper focuses on point outliers and treats anomaly $a_t$ as being independent from other time moments, and we assume no availability to other prior knowledge about the occurrence of anomalies. 
    Similarly, for $P(u_t)$, we assume no prior knowledge on the control scheme of topology, thus assuming a uniform distribution to allow for any random topology change to occur at any time. 
    \label{assumption: P(at),P(ut)}
\end{assumption}

These assumptions reduce the estimation to the follows:
\begin{align}
&x_{t=1}^*,...,x_{t}^*,a_{t=1}^*,...,a_{t}^*\\
=&\arg\max 
\underbrace{\sum_{t} log p(z_t|x_t,a_t) }_\text{spatial}
+ \underbrace{\sum_{t} log p(x_t|x_{t-1},u_t,a_t)}_\text{temporal}\notag
\end{align}
where the first term is a (generalized) state estimation problem which relies on the spatial relationship that depicts how the data $z_t$ at time $t$ can be generated, whereas the second is a temporal state transition describing how topology control actions and anomalies cause grid state to change.   

The first term $\sum_t p(z_t|x_t,a_t)$ can be easily written in (generalized) state estimation. Typically for $\forall t$, the measurement model for $z_t$ is defined as $z_t=h(x_t)+error/noise$ where the noise or error distribution determines $p(z_t|x_t,a_t)$. The details are given in Section \ref{sec: augmented (generalized) state estimation} where error is assumed to be sparsely distributed, resulting in a robust estimator.

On the other hand, the temporal term $\sum_t p(x_t|x_{t-1},u_t,a_t)$ is not directly available. Power flow simulation can give an accurate estimate, but only when $x_{t-1}$ is accurate, and it is a nonlinear programming problem that reduces time efficiency. Many existing works of state-space models have treated the temporal state transition as a Markov Chain $x_t=x_{t-1}+noise$, but these cannot account for dynamic graphs where the network topology changes with control actions $u_t$. 

This work, instead, adopts a fast data-driven time-series model which predicts a new distribution $q_{x_t}$ using a series of historical AC state and topology estimates, to approximate the true distribution $p(x_t|x_{t-1},u_t,a_t)$. I.e.,
\begin{align}
    q_{x_t}&=q(x_t|x_{t-1},...,x_{t=1},a_t,...,a_{t=1})\notag\\
    &=q(x_t|v_{t-1},...,v_{t=1},G_t,...,G_{t=1},a_t,...,a_{t=1})
\end{align} 

The probabilistic graphical model is thus reduced and split into two components:
\begin{enumerate}
    \item \textbf{time-series prediction} to output distribution $q_{x_t}$ as an approximation of $p(x_t|x_{t-1},u_t,a_t)$ for each $t$: the prediction is conditioned on the most updated $x_t,G_t$ for $\forall t$ from (generalized) state estimation. This paper uses \method as the base model due to its ability to consider topology changes in time-series data. The model is explained in Section \ref{sec: augmented dynwatch}.
    \item \textbf{augmented (generalized) state estimation}: given $q_{x_t}$, the estimation problem becomes:
    \begin{align}
&x_{t=1}^*,...,x_{t}^*,a_{t=1}^*,...,a_{t}^*\\
=&\arg\max 
\underbrace{\sum_{t} log p(z_t|x_t,a_t) }_\text{spatial ((generalized) state estimation)}
+ \underbrace{\sum_{t} log q_{x_{t}}}_\text{temporal (prior knowledge)}\notag
\end{align}
or equivalently, for $\forall t$,
\begin{equation}
    x_{t}^*,a_{t}^*
=\arg\max 
\underbrace{log p(z_t|x_t,a_t) }_\text{(generalized) state estimation}
+ \underbrace{log q_{x_{t}}}_\text{temporal prior}
\label{prob: augmented SI, probabilistic concept}
\end{equation}
This can be considered as a (generalized) state estimation augmented by prior knowledge from temporal patterns. This paper adopts (generalized) state estimation as the base model to be augmented, due to its advantages over other baseline estimators. The augmentation is explained in Section \ref{sec: augmented (generalized) state estimation}. 
\end{enumerate}

Further, Section \ref{sec: info transfer}  designs the criteria and alternating algorithm to manage the information transfer between two augmented components; illustrates how the final (generalized) state estimation, anomaly detection, root cause analysis decisions can be obtained from the interaction.

\subsection{ML-augmented ckt-GSE: robust circuit-based estimation augmented by prior knowledge}\label{sec: augmented (generalized) state estimation}

As illustrated in (\ref{prob: augmented SI, probabilistic concept}), the estimation problem:
\begin{equation}
    x_t^*,a_t^*
=\arg\max 
{log p(z_t|x_t,a_t) }
\end{equation}
can be seen as a snapshot-based (generalized) state estimation problem that exploits the spatial patterns in measurement models and power flow constraints at a certain time moment $t$.

According to assumptions made in this work, $x_t=[v_t, G_t]$ and $z_t=[z_{pmu},z_{rtu},z_{status}]$, so the above estimation is equivalent to:
\begin{equation}
    v_t^*, G_t^*,a_t^*
=\arg\max_{v_t,G_t, a_t} 
log p(z_{pmu},z_{rtu},z_{status}|v_t,G_t,a_t) 
\end{equation}

This requires a \textbf{robust estimator} for (generalized) state estimation which can:
\begin{itemize}
    \item  jointly estimate AC states of bus voltages $v_t$ and topology $G_t$ from a mixture of conventional RTU data, modern PMU data and discrete status data
    \item identify some anomalies $a_t$
\end{itemize}

As developed earlier in Section \ref{sec: ckt-GSE}, our robust circuit-based estimation (Problem \ref{def: ckt-GSE}) can fulfill this task.  However, without using any historical data beyond $z_t$, this method is vulnerable to targeted false data injected by cyber attack which exploits the spatial model of the power grid. 
Now we consider the augmented estimation in (\ref{prob: augmented SI, probabilistic concept}) which introduces temporal prior knowledge to advance its robustness against cyberattacks.  Specifically, consider a prior knowledge related to AC state vector $v_t$:

\begin{assumption}[state prior]
for a bus/node $i$ at time $t$, its AC state $v_{i,t}=[V^R_{i,t}, V^{I}_{i,t}]$ is expected to take the value around $\mu_{i,t}=[\mu^R_{i,t}, \mu^I_{i,t}]$ with uncertainty (standard deviation) $\delta_{i,t}$. Consider the prior distribution as Gaussian, we have 
\begin{equation}
    v_{i,t}=\begin{bmatrix}
V^R_{i,t}\\ V^{I}_{i,t}
\end{bmatrix}\sim N(\begin{bmatrix}
\mu^R_{i,t}\\ \mu^{I}_{i,t}
\end{bmatrix}, \begin{bmatrix}
\delta_{i,t}^2,0\\
0,\delta_{i,t}^2\\
\end{bmatrix})
\end{equation}
or, equivalently, the state vector $v_t$ at time $t$ has a Gaussian prior distribution $q_{v_t}$ such that
\begin{equation}
   v_t\sim N(\mu_t, \Delta_t) 
\end{equation}
with $\Delta_t$ being a diagonal matrix specifying the covariance matrix of the prior.
\end{assumption}

Taking this state prior, the augmented estimation becomes:
\begin{align}
     &v_{t}^*, G_t^*,a_{t}^*\notag\\
=&\arg\max 
\underbrace{log p(z_{t,pmu},z_{t,rtu},z_{t,status}|v_t,G_t,a_t) }_\text{estimation}
+ \underbrace{log q_{x_{t}}}_\text{temporal prior}\label{eq: ckt-GSE, p}\\
=&\arg\min \underbrace{||W_tn_t||_1}_\text{robust circuit-based estimation objective} + \underbrace{w_{prior} (v_t-\mu_t)^T\Delta_t^{-1}(v_t-\mu_t)}_\text{state prior}\notag\\
&\text{subject to } 
\underbrace{[Y_t, B_t]\begin{bmatrix}
    v_t\\n_t
\end{bmatrix} = b_t }_\text{robust circuit-based estimation constraints}\label{eq: ckt-GSE, constrained}
\end{align}
where $n_t$ includes slack variables ($n_{sw},n_{rtu},n_{pmu},n_{pmu}^v$) indicating the errors in corresponding measurements at time $t$.

Therefore, we have the following task definition:
\begin{problem}[augmented robust circuit-based estimation with state prior] 
\label{def: augmented ckt-GSE}
At time $t$, given  $q_{v_t}$ which is a prior knowledge of AC states $v_t$, the augmented estimator runs a convex constrained optimization problem:

\begin{align}
    &\min_{v_t,G_t,a_t} \underbrace{||W_tn_t||_1}_\text{robust circuit-based estimator objective} + \underbrace{w_{prior} (v_t-\mu_t)^T\Delta_t^{-1}(v_t-\mu_t)}_\text{state prior}\notag\\
    &\text{subject to:  } \underbrace{[Y_t, B_t]\begin{bmatrix}
    v_t\\n_t
\end{bmatrix} = b_t }_\text{KCL equations}
\end{align}
\end{problem}

Taking the model in Figure \ref{fig: case4 convert2EC} as an example, the KCL equations $[Y_t, B_t]\begin{bmatrix}
    v_t\\n_t
\end{bmatrix} = b_t$ at time $t$ are written as (\ref{eq: kcl example start})-(\ref{eq: kcl example end}), for any time $t$.

\subsection{Gaining state 
prior with uncertainty quantification }\label{sec: augmented dynwatch}

On the other hand, we need a time-series model to extract temporal patterns from the historical operational data. At any time $t$, let $q_t=N(\mu_t,\delta_t^2)$ denote the temporal behavior learned from historical data. This can be achieved with our time-series model in Section \ref{sec: dynwatch}, DynWatch, which adapts to dynamic graphs.

\begin{problem}[\method for distribution prediction]
At any time $t$, given time-series values $y_1,y_2,...,y_t$, graph (topology) series $G_{t=1}, G_{t=2},...G_{t}$ (these topologies come from estimation), and distance measure $d_k=d(G_k,G_t), k=1,...,t-1$, \method learns a distribution of normal behavior $\mathcal{N}_{t}=\mathcal{N}(\mu_t,\delta_t^2)$: 
\begin{align}
    &\mu_t=WeightedMedian(y_1,...,y_{t-1})\\
    &\delta_t=WeightedIQR(y_1,...,y_{t-1}) 
\end{align}
with weights $w_{t=1},w_{t=2},...w_{t-1}$ obtained from temporal weighting based on bias-variance trade-off optimization which minimizes the sum of bias and variance, to balance the goals of low bias (i.e. using data from similar graphs) and low variance (using sufficient data to form our estimates):
\begin{equation}
\min_{w}{\underbrace{\sum_{k}w_k d_k}_\text{bias} + c\cdot \underbrace{ \frac{1}{2}w^Tw}_\text{variance}}
\label{fobj: bias-var trade-off}
\end{equation}
\begin{equation} 
\text{subject to: }
\underbrace{w_k\geq 0,  \forall k; \sum_{k}w_k = 1, }_\text{non-negativeness and sum to 1}
\end{equation}
\end{problem}

For any data-driven model, a valid concern is \textit{when does it fail?} Quantifying uncertainty of the prediction is necessary to handle this concern. Generally, a model can be uncertain in two important ways 1) epistemic uncertainty which accounts for model uncertainty due to lack of  data (that fits in the target distribution), and 2) aleatoric uncertainty which corresponds to sensing uncertainty, or noises, and can be further divided into heteroscedastic and homoscedastic, depending on whether it stays constant for different inputs.
todo: Here we discuss two types of uncertainties:
\begin{itemize}
    \item \textbf{data uncertainty (aleatoric uncertainty or sensing uncertainty):} Unlike many other time-series models (e.g. auto-regression models, etc) which gives only point estimation of the predicted data values, \method provides a distribution estimation. The variance $\delta^2$ quantifies the uncertainty that comes of data variance, i.e., the dispersion of the weighted historical data.
    \begin{equation}
        \delta_t^2=\text{data uncertainty}
    \end{equation}
    Whenever the (positively weighted) historical data vary significantly (due to noise, power variance, etc), a 'flatter' distribution is obtained representing a larger uncertainty. 
    \item \textbf{bias (model uncertainty or epistemic uncertainty):} when does the predicted distribution fail? 
    In the \method model, a perturbation of the model parameters $w$ can affect the statistical error which consists of bias and variance, as in (\ref{fobj: bias-var trade-off}). In our dynamic graph setting, bias plays a dominant role in causing large statistical errors, and this happens when the power grid changes to a new (unseen) topology. 
    Therefore, we use the bias to quantify the uncertainty or trustworthiness of the predicted distribution. A low bias means the distribution is a reliable prediction; whereas a large bias indicates a high uncertainty such that the predicted distribution can be erroneous.  
    \begin{equation}
        bias(t)=\sum_{k=1}^{t} w_kd_k
    \end{equation}
    In a synergy design, the predicted distribution can be used as a reliable source of prior knowledge only when bias is low, which will be discussed later.  
\end{itemize}

\subsection{Management of Information Transfer}\label{sec: info transfer}

\begin{figure}[h]
     \centering
            \includegraphics[width=0.7\linewidth]{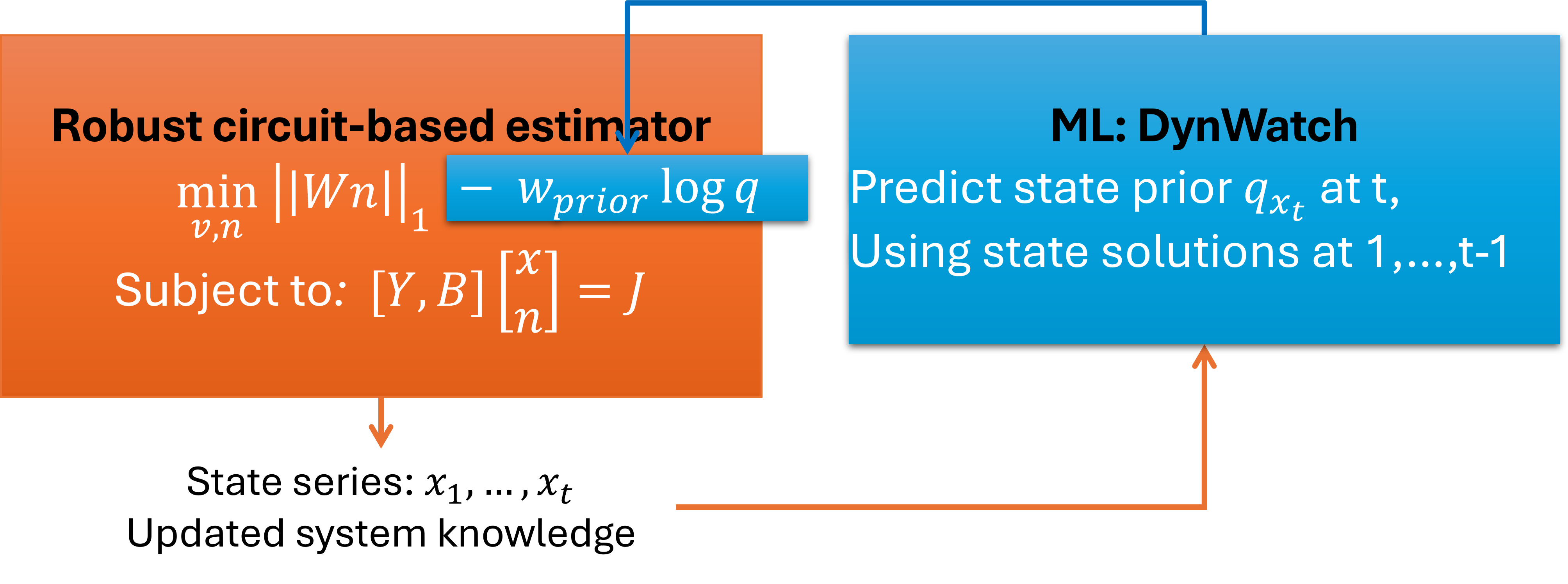}
    \caption{Physics-ML Synergy creates an interconnection: prior knowledge from ML comes in the form of regularization to augmented the estimation problem; whereas time-series state data and system knowledge are fed into ML to gain statistical prior capturing temporal patterns.  
    }
    \label{fig: SynSA interconnection}  
\end{figure}

The ML-augmented estimation, and the uncertainty-quantified calculation of a statistical prior, come together to form a loop of interconnection as illustrated in Figure \ref{fig: SynSA interconnection}. Central to this physics-ML synergy is to properly manage the information transfer between the two components. 
This section develops an iterative updating scheme as Algorithm \ref{alg: outerloop} shows. We design an adaptable interconnection scheme to activate and deactivate the interconnection link depending on trustworthiness of information. Within the outer loops, \method augments robust circuit-based estimation with temporal (state) prior, with the feeding of prior knowledge guided by tracking the local peak of bias;
whereas the robust circuit-based estimation benefits \method with the updated system model whose topology is corrected upon rigorous verification of topology errors so that \method proceeds with the best possible graph distances. 
This guarantees that the interconnection happens at the right time so that they augment each other with reliable sources of information.
\begin{algorithm}
	\caption{SynSA: Physics-ML Synergy for Estimation.} 
	\label{alg: outerloop}
	\KwIn{Time-series data $z_{t=1},...,z_{t-1},z_{t}$, system model $M$ (which contains node set, edge set and  parameter information)}  
	\KwOut{estimation $v_t^*, G_t^*, \forall t$, and root cause diagnosis (RCD) on anomalies $a_{t}^{type,location}, \forall t$,
 } 
 {\bf 1. Initialize graph structures $G_t$ }: if input system does not contain switches, add pseudo nodes and pseudo switches to build extended graph as in Figure \ref{fig:case4}, according to status data $z_{status,t}, \forall t$

{\bf 2. Initial anomaly detection: } run \method detector, get anomaly scores $[Y_t, B_t]^{ML}, \forall t$ 

{\bf 3. Initial estimation and root cause diagnosis: }

	\For{{$t \gets 1$ to $t$} }{
	 {\bf 1. Run robust circuit-based estimation }with $z_t$, get $\hat{v_t}$

        {\bf 2. Root cause diagnosis of random anomalies}:  
        
        1) identify random bad data and topology errors by sparse error indicators.
        \begin{align}
            &a_t^{BadData, PMUi}\gets n_{pmui}, 
            \text{ alarm if }>0.05  \\
            &a_t^{BadData, RTUi}\gets n_{rtui}, 
            \text{ alarm if }>0.05\\ 
            & a_t^{TopoErr,edge k}\gets n_{swk}, \text{ alarm if } >0.01, 
        \end{align}
  
 Also, check solved state and topology with operational bounds and limits to see if there is abnormal generation/load outage, and abnormal voltage conditions, etc.}
 

	{\bf 4. Synergy Interconnection:} 

\For{{loopCount $ \gets 1$ to $MaxLoop$} }{
	 {\bf 1. Run \method predictor} to get state prior $q_{v_t}, \text{for } \forall t$

  {\bf 2. Run augmented robust circuit-based estimation with state prior} if $bias(t)<30\% *$maximum bias on the same topology

  {\bf 3. Identify cyberattacks} using using infeasibility values in augmented robust circuit-based estimation solutions

    {\bf 4. Correctify graph structure $G_t, \forall t$}}
    
 {\bf 3. Output graph structure $G_t^*$}: remove pseudo nodes and pseudo switches, convert back to external node and edge indexes, for $\forall t$. 
\end{algorithm}

\subsection{Evaluating on power systems: accurate estimation and detection under a mixture of traditional and modern threats}

This Section conducts experiments to demonstrate the efficacy of Physics-ML Synergy for estimation. We compare a variety of methods whose abbreviations are as follows:
\begin{itemize}
    \item ckt-GSE: robust circuit-based generalized state estimation developed in Problem \ref{def: ckt-GSE} in Section \ref{sec: ckt-GSE}. This is a system physics-based model.
    \item ckt-SE: robust circuit-based state estimation, a reduced version of Problem \ref{def: ckt-GSE} in Section \ref{sec: ckt-GSE}. Compared with ckt-GSE, ckt-SE does not account for topology error and has an empty switch set; whereas ckt-GSE accounts for topology error and includes switching devices in the power system model. This is also a system physics-based model.
    \item ML-augmented ckt-GSE: ckt-GSE augmented by state prior as developed in Problem \ref{def: augmented ckt-GSE} in Section \ref{sec: augmented (generalized) state estimation}. This is a synergy of ckt-GSE and \method. 
\end{itemize}

To goal of our experiments is to answer the following questions:
\begin{enumerate}
    \item \textbf{Situational awareness capabilities:}  Does ML-augmented ckt-GSE jointly perform high-performance estimation, anomaly detection and root cause diagnosis as expected, outperforming the separate use of ckt-GSE or ckt-SE alone? 
    \item \textbf{Robustness:}
    Is ML-augmented ckt-GSE able to guarantee high-quality state estimates on different sized networks, under a mixture of traditional and modern threats? 
    \item \textbf{Speed and scalability:} What's the work time different sized networks?
\end{enumerate}
\noindent 

\textbf{Reproducibility:}  
And all experiments are run on a laptop computer with 11th Gen Intel(R) Core(TM) i7-1185G7 @ 3.00GHz   1.80 GHz processor and 32 GB RAM. 

\textbf{Anomaly  assumptions: } Table \ref{tab: anomaly simulation} lists the anomalies of interest and the experiment settings to simulation these anomalies. We consider both the traditional element outages, data errors, and a modern cyber attack.

\setlength{\tabcolsep}{6pt}
\begin{table}[htbp]
\small
\centering
	\caption{Anomaly Simulation \label{tab: anomaly simulation}}
	\begin{tabular}{ @{}ll@{} }  
	\toprule
	\textbf{Anomaly} & \textbf{How the anomaly is simulated} \\
 \midrule
     line outage & disconnect 1 transmission line at random location\\ 
     (contingency)& (system remains feasible after line outage)\\
     \midrule
     \begin{tabular}[c]{@{}l@{}}
          (random)  
          bad data
     \end{tabular} & 
     \begin{tabular}[c]{@{}l@{}}
          add large deviation (1 per unit) to continuous data values on 2 randomly selected locations
     \end{tabular}\\
     \midrule
     \begin{tabular}[c]{@{}l@{}}
          topology error
     \end{tabular} & 
     \begin{tabular}[c]{@{}l@{}}
          change status data at 2 random lines, making the operator have a wrong graph structure
     \end{tabular}
     \\
     \midrule
     \begin{tabular}[c]{@{}l@{}}
        FDIA
          (cyberattack)
     \end{tabular} &
    \begin{tabular}[c]{@{}l@{}}
        attackers simulate power flow of the power system with more or less demand at all loads, \\and then modify data according to the attackers' power flow result\\
     \end{tabular} \\
	\bottomrule
	\end{tabular} 
\end{table}

\textbf{Dynamic graph assumption:} The time-series data in this paper is created with known topology change actions, resulting in dynamic graphs. The topology changes occur every 60 time ticks. In each instance, a previously open transmission line becomes closed, and meanwhile a new (random) line becomes open.





\subsection{Situational awareness output: ML-augmented ckt-GSE is an all-in-one product}

{ Here, we evaluate how the interconnection of ckt-GSE and \method enabled by ML-augmented ckt-GSE can make a difference in the situational awareness output.  Here the ML-augmented ckt-GSE method is expected to jointly perform estimation, anomaly detection and root cause diagnosis. It's also expected to outperform the separate use of ckt-GSE, DynWatch, as well as other baselines, in these capabilities.

\textbf{Experiment settings:} Here we visualize the outputs on a small toy example of the IEEE 30-bus test case, which represents a simple approximation of the American Electric Power system. We simulate (steady-state) time-series data of 600 time ticks, with dynamic graph assumption described above. 
On this test case, we assume PMUs are installed on every generation bus measuring $V, I$ phasors; traditional SCADA RTUs are installed on every load bus measuring $|V|, P, Q$; RTU line meters are installed on  selected transmission lines measuring line power flow $P_{line}, Q_{line}$ and $|V|$ at one end of the line. Normal measurement data is generated by simulating power flow and adding  random Gaussian noise to the power flow results. Four types of traditional and modern anomalies are also added on randomly selected moments. Table \ref{tab: anomaly simulation} shows how these anomalies are created. Figure \ref{fig: toy measurements} visualizes the time-series measurement data, with anomalous time moments marked by red Vertical lines.

Results in \ref{fig: toy SI} demonstrates that  ML-augmented ckt-GSE is a  robust estimator under modern cyber attacks. We visualize and compare the accuracy of state solution using root mean squared error $RMSE=\sqrt{\sum_{\text{bus } i}(v_{pred,i}-v_{true,i})^2}$. Whenever a false data injection attack (FDIA) happens, the augmented ckt-GSE in ML-augmented ckt-GSE  provides an accurate solution with a small RMSE, while solutions of ckt-GSE, when used alone, are significantly perturbed by falsified data.

In terms of anomaly detection, results in Figure \ref{fig: toy AD}(a)-(c) shows that ML-augmented ckt-GSE outperforms the separate use of ckt-GSE, ckt-SE, \method and other anomaly detectors in the detectability of anomalies. 
The bad data detection (BDD) capabilities of ckt-GSE and ckt-SE are limited to detecting data errors. 

In terms of root cause diagnosis, results in Figure \ref{fig: toy AD}(d)-(g) shows that ML-augmented ckt-GSE can better identify the accurate type and location of anomalies. Physics-based estimation Ckt-GSE, when used alone, is limited to separating and localizing random bad data and topology errors. Ckt-SE method detects topology errors as random bad data. \method and other ML anomaly detectors are in general unable to recognize the specific type of anomaly.  

\begin{figure}[h]
     \centering
     \begin{subfigure}[b]{\linewidth}
         \centering \leftskip 0.5cm
         \includegraphics[width=0.92\textwidth]{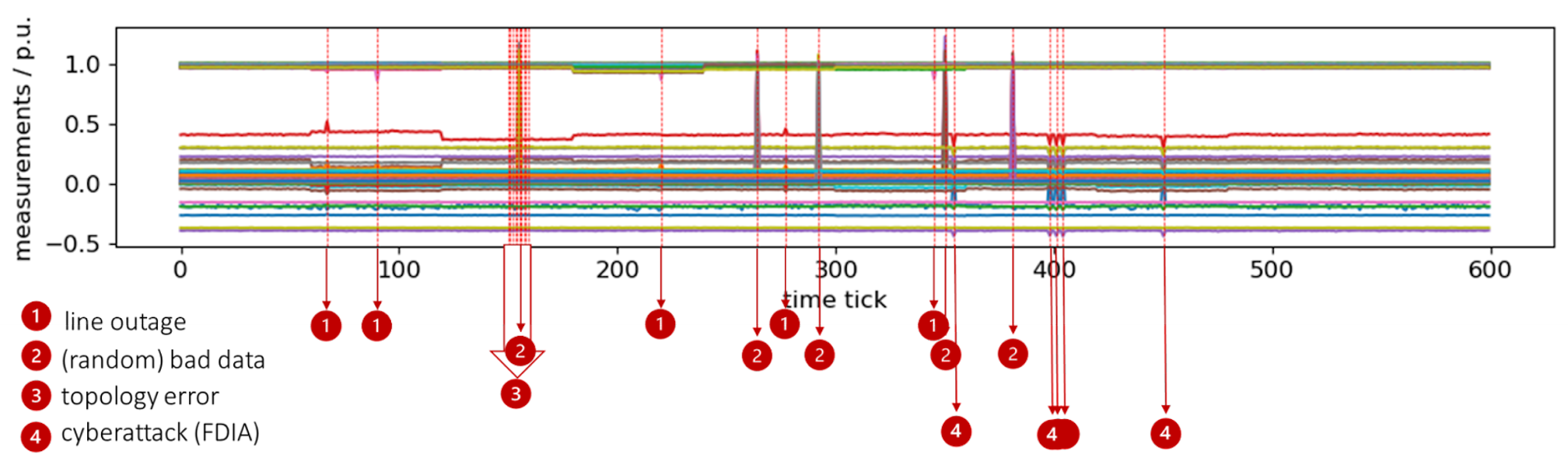}
         \caption{Time-series measurements on case30: 600 time ticks.}
         \label{fig: toy measurements}
     \end{subfigure}
     \begin{subfigure}[b]{\linewidth}
         \centering         \includegraphics[width=0.9\textwidth]{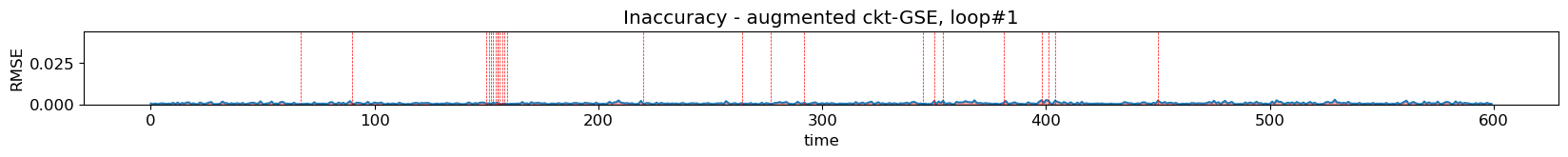}
         \caption{The augmented ckt-GSE in ML-augmented ckt-GSE provides an accurate solution under FDIA attacks.}
     \end{subfigure}
     \hfill
     \begin{subfigure}[b]{\linewidth}
         \centering         \includegraphics[width=0.9\textwidth]{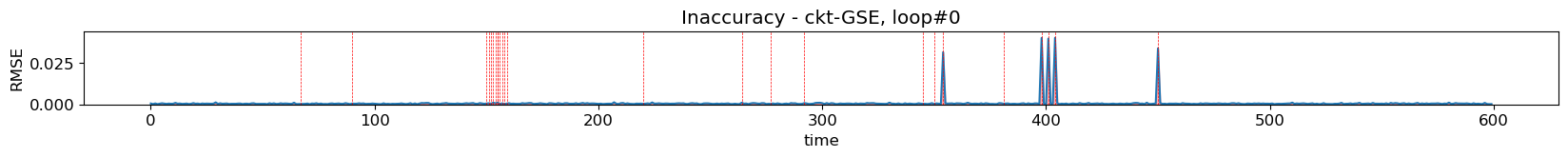}
         \caption{Ckt-GSE becomes inaccurate under FDIA attacks.}
     \end{subfigure}
     \hfill
     \begin{subfigure}[b]{\linewidth}
         \centering         \includegraphics[width=0.9\textwidth]{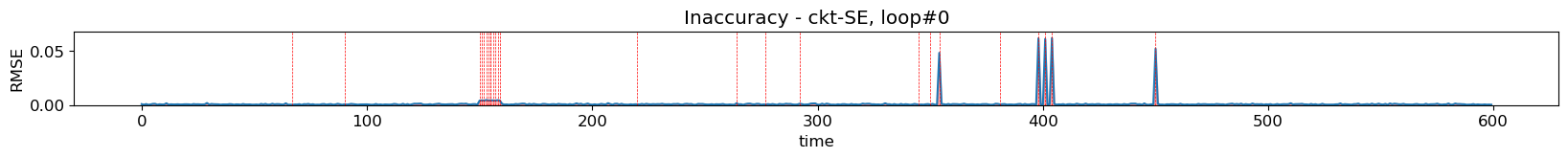}
         \caption{Both topology errors and FDIA attacks affect the accuracy of ckt-SE.}
     \end{subfigure}
     \caption{Case30: accuracy of estimation accuracy quantified by root mean squared error (RMSE). }
     \label{fig: toy SI}
     \end{figure}     
\begin{figure}[h]
     \centering
     \begin{subfigure}[b]{\linewidth}
         \centering         \includegraphics[width=0.9\textwidth]{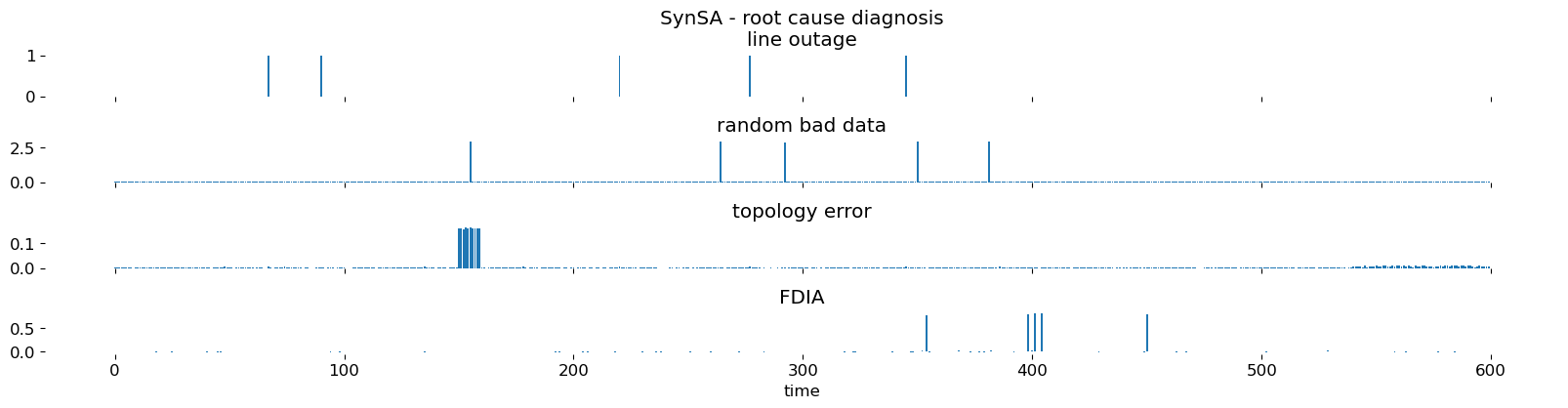}
        \caption{ML-augmented ckt-GSE not only detects all the four types of anomalies, but also identifies anomaly types.}
         \label{fig: wlav switch bdd}
     \end{subfigure}
     \hfill
     \begin{subfigure}[b]{\linewidth}
         \centering         \includegraphics[width=0.9\textwidth]{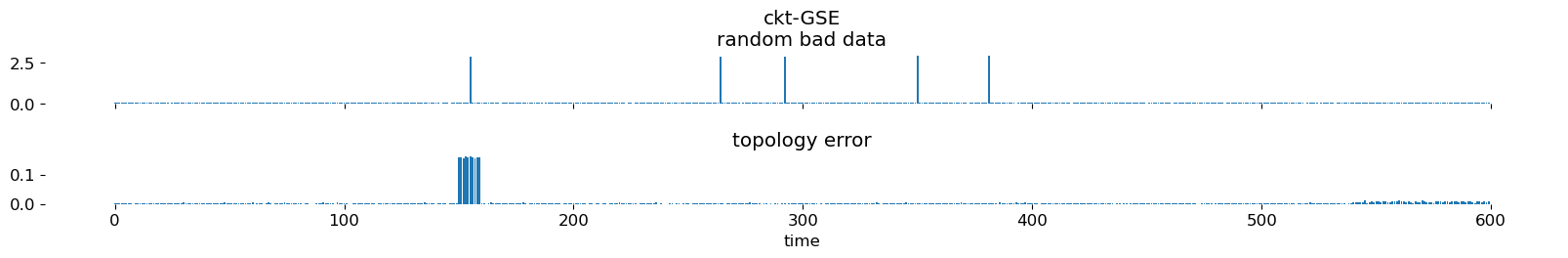}
         \caption{Ckt-GSE BDD can separate random bad data and topology error; but cannot detect FDIA directly from BDD. }
         \label{fig: wls switch bdd}
     \end{subfigure}
     \hfill
     \begin{subfigure}[b]{\linewidth}
         \centering         \includegraphics[width=0.9\textwidth]{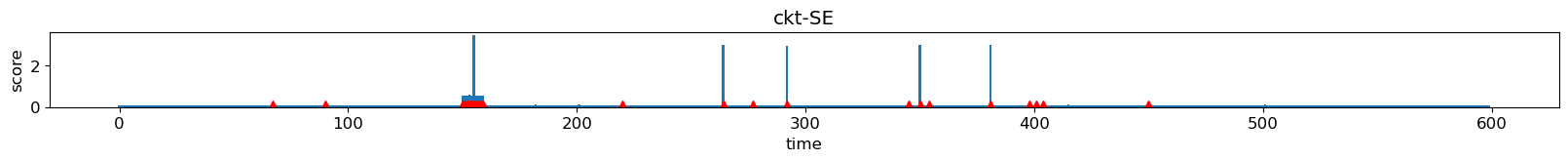}
         \caption{Ckt-SE BDD creates alarms on bad data and topology errors, but cannot separate them.}
         \label{fig: wls switch bdd}
     \end{subfigure}
     \ContinuedFloat
\rotatebox{90}{Synergy}
     \begin{subfigure}[h]{0.22\linewidth}
         \centering         
         \caption{t = 68,\\ line outage.}
\includegraphics[width=\textwidth]{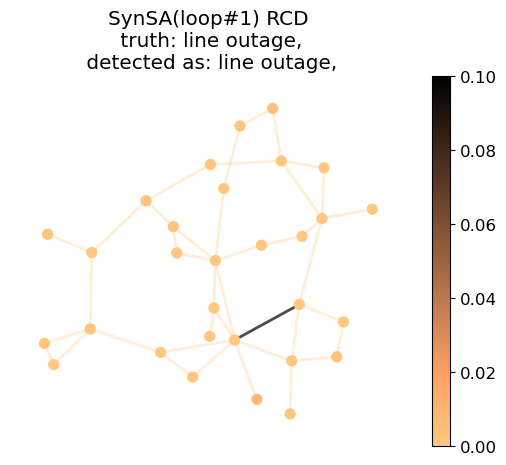}   
\end{subfigure}
     \hfill
     \begin{subfigure}[h]{0.22\linewidth}
         \centering        
         \caption{t = 293,\\random bad data}\includegraphics[width=\textwidth]{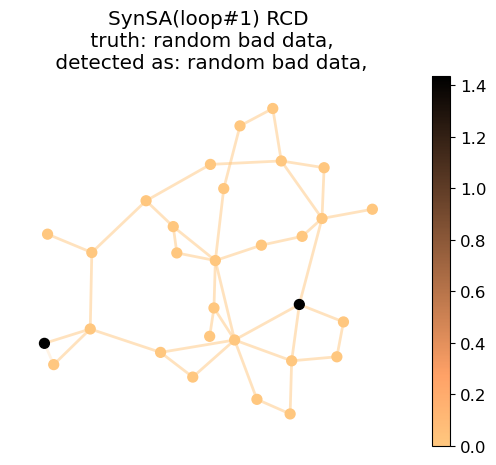}
     \end{subfigure}
     \hfill
     \begin{subfigure}[h]{0.22\linewidth}
         \centering       
         \caption{t = 156,\\ bad data+topo error }
\includegraphics[width=\textwidth]{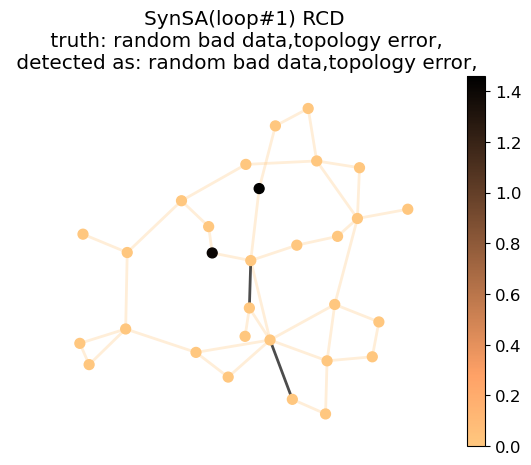}         
     \end{subfigure}
     \hfill
     \begin{subfigure}[h]{0.22\linewidth}
         \centering       
         \caption{t = 402,\\FDIA}
         \includegraphics[width=\textwidth]{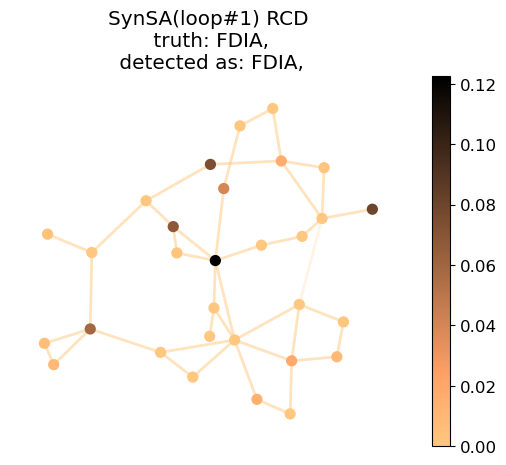}      
     \end{subfigure}\\
     \rotatebox{90}{ckt-GSE}
     \begin{subfigure}[h]{0.22\linewidth}
         \centering   
         \includegraphics[width=\textwidth]{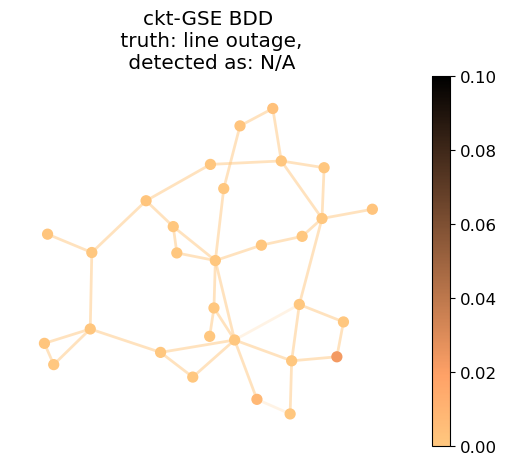}
         
     \end{subfigure}
     \hfill
     \begin{subfigure}[h]{0.22\linewidth}
         \centering         
         \includegraphics[width=\textwidth]{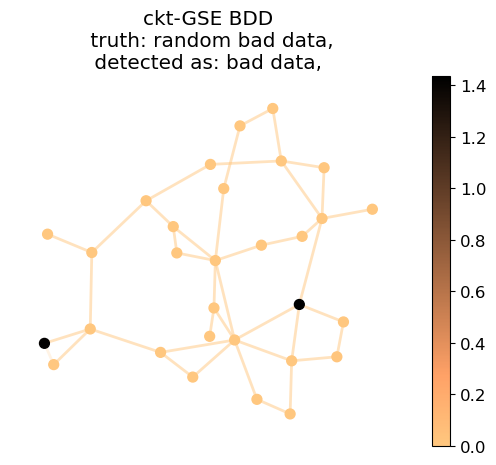}
         
     \end{subfigure}
     \hfill
     \begin{subfigure}[h]{0.22\linewidth}
         \centering         \includegraphics[width=\textwidth]{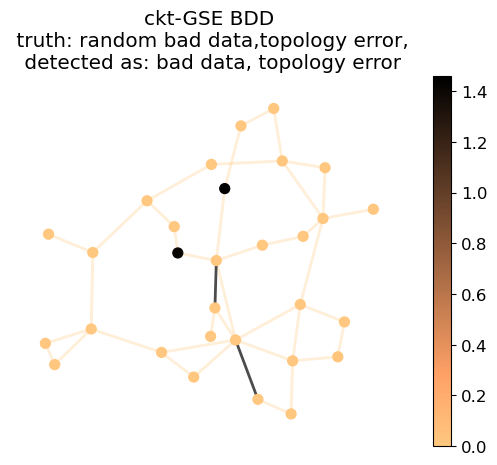}
     \end{subfigure}
     \hfill
     \begin{subfigure}[h]{0.22\linewidth}
         \centering         \includegraphics[width=\textwidth]{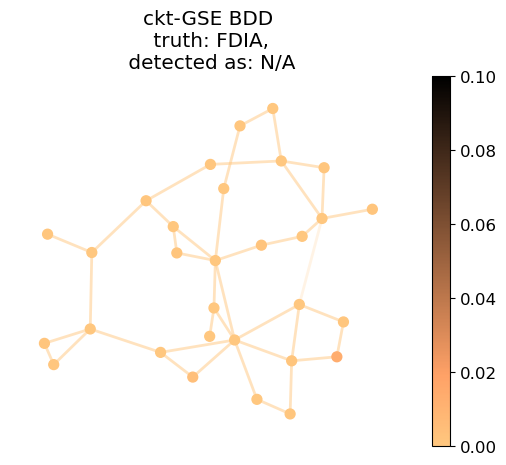}
     \end{subfigure}\\
 \rotatebox{90}{ckt-SE}
     \begin{subfigure}[h]{0.22\linewidth}
         \centering   
         \includegraphics[width=\textwidth]{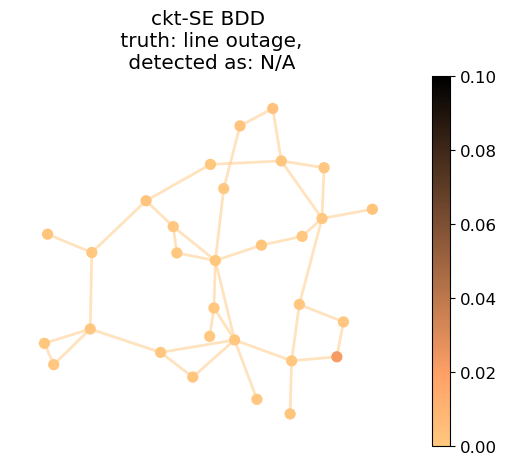}  
     \end{subfigure}
     \hfill
     \begin{subfigure}[h]{0.22\linewidth}
         \centering         
         \includegraphics[width=\textwidth]{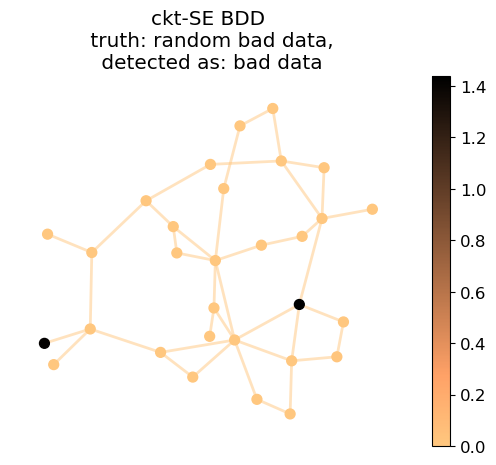}         
     \end{subfigure}
     \hfill
     \begin{subfigure}[h]{0.22\linewidth}
         \centering         \includegraphics[width=\textwidth]{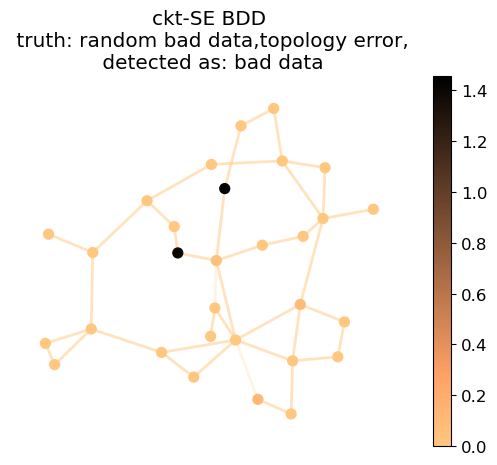}
     \end{subfigure}
     \hfill
     \begin{subfigure}[h]{0.22\linewidth}
         \centering         \includegraphics[width=\textwidth]{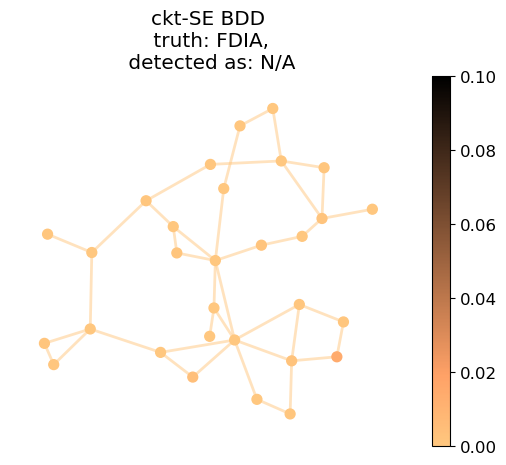}
     \end{subfigure}
     \caption{ML-augmented ckt-GSE identifies types and locations of mixed anomalies.}
     \label{fig: toy AD}
     \end{figure}

\subsection{Robustness of estimation solution}
To minimize the impact of modern threats, we need the estimation solution of power system states to be accurate when the system is subject to the threats. So here we evaluate the quality of the bus voltage estimates using the root mean squared error (RMSE) of the solution. To demonstrate efficacy, we evaluate the solution accuracy on networks of different sizes: 

\begin{enumerate}
\item \textbf{Case30}\cite{case30}: 30-bus test case which is a simple approximation of the American Electric Power system.
    \item \textbf{Case1354pegase}\cite{case1354pegase}: a 1354-bus network which accurately represents the size and complexity of part of the European high voltage transmission network.
   \item \textbf{Case2383wp}\cite{case2383wp}: part of the 7500+ bus 
   Europen UCTE system which represents the Polish 400, 220 and 110 kV networks during
   winter 1999-2000 peak conditions.
   \item \textbf{Case2869pegase} \cite{case2869pegase_1}\cite{case2869pegase_2}: a 2869-bus network representing the size and complexity of part of the
   European high voltage transmission network.
\end{enumerate}
Figure \ref{fig: synergy accuracy} shows the results.

\begin{figure}[h]
 \ContinuedFloat
     \begin{subfigure}[h]{0.49\linewidth}
         \centering         
\includegraphics[width=\textwidth]{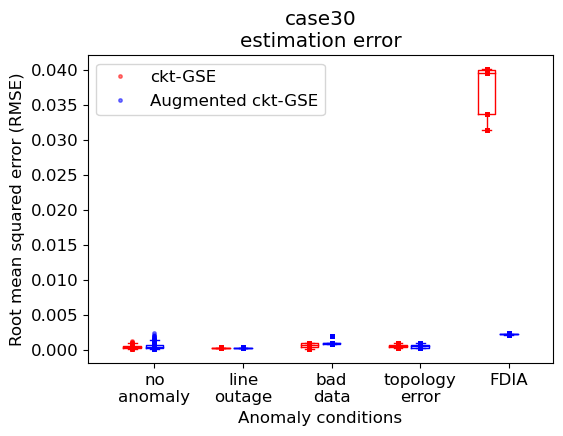}   \end{subfigure}
     \hfill
     \begin{subfigure}[h]{0.49\linewidth}
         \centering        
         \includegraphics[width=\textwidth]{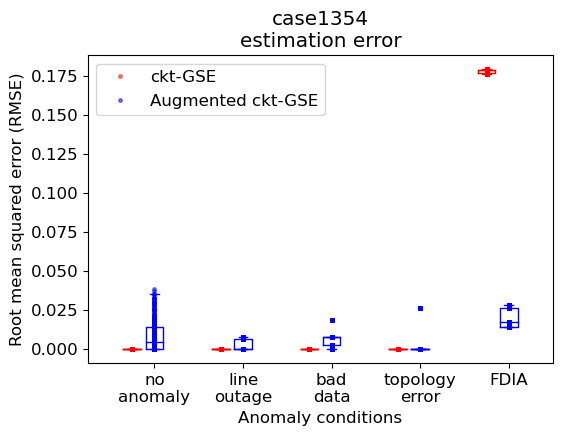}
     \end{subfigure}\\
     \begin{subfigure}[h]{0.49\linewidth}
         \centering       
\includegraphics[width=\textwidth]{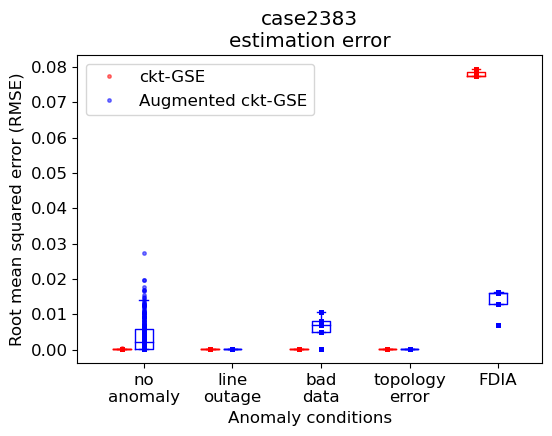}         
     \end{subfigure}
     \hfill
     \begin{subfigure}[h]{0.49\linewidth}
         \centering       
         \includegraphics[width=\textwidth]{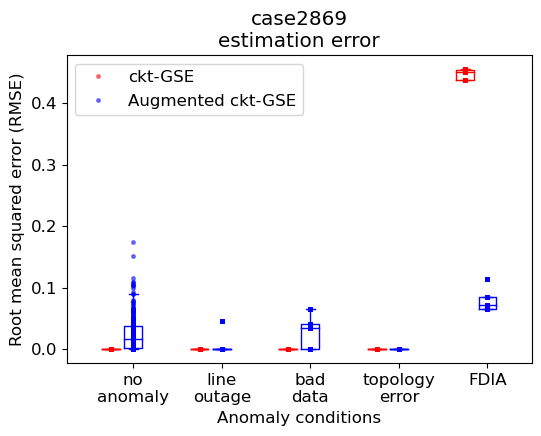}      
     \end{subfigure}\\
     \caption{Robustness: ML-augmented ckt-GSE advances ckt-GSE in robustness against modern false data injection attacks. It improves accuracy of state estimates, thereby minimizing the intended impact an adversary can have on the system. But this happens at the cost of degraded accuracy on normal and traditional (random) anomaly conditions. }
     \label{fig: synergy accuracy}
     \end{figure}

\subsection{Speed and scalability}
The ML-augmented ckt-GSE method needs to be time-efficient on large-scale networks to be applicable in real-world control rooms.
As designed in this paper, the ML-augmented ckt-GSE algorithm starts with ckt-GSE and \method in the initialization step; then alternates between augmented ckt-GSE and \method processing. Thus, here we evaluate the speed and scalability of ML-augmented ckt-GSE by evaluating its 3 essential steps: ckt-GSE, augmented ckt-GSE and \method processing. 

In our experiment, both ckt-GSE and augmented ckt-GSE are solved using the circuit-based linear programming (LP) solver developed in \cite{SUGAR_GSE_WLAV}. Prior work \cite{SUGAR_GSE_WLAV} has already shown that 
ckt-GSE, which uses our circuit-based LP solver, is significantly faster than standard interior-point (IP) solver in python CVXOPT toolbox and Simplex method in SciPy which solves min-max models, especially on large scale cases. Prior work in \cite{dynwatch} has also shown that the distance calculation and time-series processing in \method scales approximately linearly when used for anomaly detection. 

Figure \ref{fig:scalability} shows the work time of the 3 essential steps on different sized networks. Results show that these essential steps scale well. Meanwhile, the comparison shows that the speed of augmented ckt-GSE and ckt-GSE are approximately the same, meaning that the ML-augmented ckt-GSE interconnection advances estimation without introducing computation burdens. 

\begin{figure}[h]
	\centering
	\includegraphics[width=0.5\linewidth]{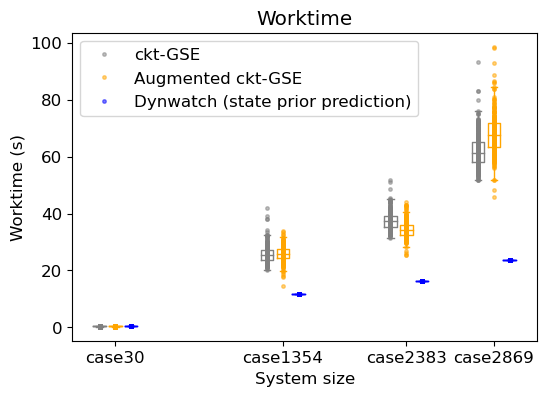}
	\caption{  Speed and scalability: worktime of the 3 essential procedures in the synergy design, as system size increases. System size is quantified by number of nodes $+$ number of edges in the network. The work time of \method is the average work time needed for per time tick to predicting state prior on all buses. Experiments are conducted using a time-series data of 300 ticks, topology changing every 60 ticks.
     } 
\label{fig:scalability}
\end{figure}

\chapter{Conclusion and future work}
This thesis addresses the grand challenge of improving situational awareness to make the future electrical power grid and related cyber-physical systems (CPS) resilient and reliable. To this end, this thesis made significant advancements in the state-of-the-art simulation and estimation tools which are key to situational awareness. The key contribution lies in overcoming efficiency and robustness gaps which have posed main obstacles to these tools. Within the scope of this thesis, we have focused on the steady-state horizon and achieved high-speed simulation and estimation on up to 80K-node network (the eastern interconnection); as well as a higher level of robustness that gives meaningful and actionable results under blackout failures, corrupted data, and a variety of cyberattack scenarios.

To realize these achievements, this thesis first addressed the inherent limitations within physics-based and data-driven approaches, which are commonly used for situational awareness. The key enablers in this part include circuit-based formulations and the sparsity-exploiting optimizations. Further, this work transcends conventional algorithmic design by introducing a new hybrid paradigm—Physics-ML Synergy—which merges the strengths of both worlds.

The chapters of this thesis were structured to form the different pieces of the complete "puzzle". 
First, Chapter \ref{ch: ECF} generalized the modeling of a steady-state power system into a circuit, for both simulation and real-time estimation purposes. We discussed from a circuit viewpoint that both simulation and estimation problems can be treated as solving the feasible solution to a circuit system on which the device models of transmission lines, generators, loads and shunts are equivalent circuit elements. We also illustrated that, for simulation, these device models come from system behaviors; whereas for estimation, device models can depend on real-time measurements. This Chapter further developed measurement-based device models with linear circuit modeling and demonstrated that circuit-based estimation for large-scale power systems can be formulated with inherent convexity, and solved with fast speed and scalability. This circuit approach addresses the inherent nonlinearity induced by traditional measurements and results in an efficient estimator applicable to combined transmission and distribution systems. 

Followed by the mapping of steady-state simulation and estimation to a circuit-solving problem, Chapter \ref{ch: physics} further introduced sparse optimization to bring intrinsic robustness into the methods. Intuitively, we look into system blackout and anomalous measurement data which lead to "infeasible circuit systems" (for simulation and estimation respectively). Independent current sources can be added to the circuit device models, restoring a feasible system, and serving as indicators of these threats. This Chapter developed sparse enforcing techniques towards formulating robust circuit-based simulation and robust circuit-based estimation problems. The result demonstrated that the resulting simulation produces sparse "infeasibility" indicators to pinpoint dominant sources of blackout and suggest actionable fix; similarly for estimation where our estimator produces sparse error indicators to identify and reject anomalous data. These methods solve efficiently on up to 80k-node networks (the Eastern Interconnection power grid); and advanced the overall robustness of situational awareness, and enhanced decision-making to mitigate random threats.

With some intrinsic efficiency and robustness to random threats built within the physics-based simulation and estimation tools, the second piece of the puzzle is to leverage data-driven predictors and detectors to handle more advanced threats. It was illustrated in Chapter \ref{ch: ML} that exploiting temporal and spatial sparse structures in the ML model design can contribute to lightweight ML methods which are simple-structured but scalable, generalizable and interpretable. In the first application, we exploit temporal sparse structure to develop \method, which is the first to adapt time-series anomaly detection to sensors placed on dynamic graphs in the context of large-scale electrical grids, and achieves real-time processing at the millisecond scale for 60K-node power systems (75\% the size of the Eastern Interconnection) per time tick per sensor. In the second application, we exploited spatial sparse graphical structures towards developing GridWarm, a highly lightweight ML predicting threat impact.

Last but not least, after addressing the inherent limitations in both the physics and ML world, Chapter \ref{ch: synergy} completed the final piece of the puzzle - Physics-ML Synergy where physics-based tools “chat with” ML to merge their benefits. We show that interconnecting these tools makes a transformative impact to accomplish tasks that are impossible for each separate tool. In this thesis, it was demonstrated that interconnecting GridWarm predictors with our circuit-based simulator, in the form of providing warm start points, leads to 3x faster simulation on hard-to-solve multiple-location load disturbances. And it was also shown that interconnecting \method prediction with our robust circuit-based estimator, in the form of feeding prior knowledge, results in diminishing the impact of false data injection attack (a cyberattack) and accurately identifying a mixture of random anomalies and targeted cyberattack events. These efforts provide strong evidence to demonstrate that Physics-ML Synergy is able to bring the efficiency and robustness of situational awareness further to defend against targeted cyberthreats.

For future works, we are seeing exciting opportunities of building on this prior foundation to extend Physics-ML Synergy designs for a broad range of estimation, simulation, optimization, and defense applications. This opens room to explore how collaborative efforts from the two worlds can enable secure, time-efficient, and robust operation and planning, initially in the context of electric power systems but with broad generalizability to other smart infrastructure systems. The proposed research is both fundamental and potentially transformative in practice, given the rapid electrification to support decarbonization in many industries and connected infrastructures.

Specifically, one possible direction for Physics-ML Synergy research is to address more complex anomalies, threats and uncertainties that are occurring today.
In the energy sector, voltage spikes and transients are dynamic anomaly conditions that can malfunction the electric devices; vulnerabilities in modern power electronics expand the attack surface; renewable energy introduces more uncertainties and instabilities; and new cyberattacks (e.g., MadIoT, CrashOverride) and adversarial attacks motivate continuous improvements on the information security and model robustness. Physics-ML Synergy provides a new solution to these challenges, by adaptively interconnecting distinct modeling and data-driven components (each focusing on specific types of threats) so as to comprehensively address the diverse threats.

Another potential direction is the Physics-ML Synergy can leverage ML to complement the missing physics in partially-known or partially-observable scenarios. One notable example of partial-observability is the distribution power grids where some local regions are merely measured "on the top", with limited or even no real-time information of what is happening inside. To handle partial observability, a lot of efforts in literature have been made on optimal sensor placement to make the system fully-observable \cite{sensor-placement}, learning the entire system through data-driven function approximators \cite{structure-learning}, as well as network tomography approaches to estimate hidden states from observations \cite{network-tomography}.
{Now Physics-ML Synergy provides a new possibility by investigating how ML can interconnect with physical tools to complement partial observability and derive something meaningful about the unobservable and unknown regions.}

Moreover, in the general field of engineering systems, future research also has the potential towards a strategic way and distributed implementation of Physics-ML Synergy. Based on user-defined physical models and ML, there potentially exists an optimal way of interconnecting these  tools to maximize the over performance for solving constrained optimization problems, system identification, simulation and problems for any given cyber-physical system.
Moreover, for systems that have a network nature, the implementation of Physics-ML Synergy can also be optimized to run in parallel via distributed computing. 
This includes developing a  federated distributed control scheme for components in the Physics-ML Synergy and their interconnections. 
As a result, a large-scale cyber-physical system can perform autonomous edge computing for local defense, simulation, and optimization within a small segment, while providing sufficient feedback information to the larger infrastructure for both spatial and time-based monitoring and analysis among these federated components.


\backmatter

\bibliographystyle{acm}
\bibliography{thesis.bib}

\end{document}